\documentclass{aa}  

\usepackage{graphicx}
\usepackage[space]{grffile}
\usepackage{txfonts}
\usepackage{float}
\usepackage{booktabs}
\usepackage[section]{placeins}
\usepackage{multicol}
\usepackage{subfig}
\usepackage{hyperref}
\usepackage{gensymb}
\usepackage{xcolor}

\begin{document}

\title{Abundance ties:\\Nephele and the globular cluster population accreted with $\omega$ Cen}
\subtitle{Based on APOGEE~DR17 and Gaia~EDR3}
  \titlerunning{The globular cluster family of $\omega$ Cen}

   \author{G.~Pagnini
          \inst{1}
          \and
          P.~Di Matteo\inst{1}
          \and
          M.~Haywood\inst{1}
          \and
          A.~Mastrobuono-Battisti\inst{1,2}
          \and
          F.~Renaud\inst{3,7}
          \and
          M.~Mondelin\inst{4}
          \and
          O.~Agertz\inst{5}
          \and
          P.~Bianchini\inst{3}
          \and 
          L.~Casamiquela\inst{1}
          \and
          S.~Khoperskov\inst{6}
          \and
          N.~Ryde\inst{5}
          }

   \institute{GEPI, Observatoire de Paris, PSL Research University, CNRS, Place Jules Janssen, 92195 Meudon, France\\
              \email{giulia.pagnini@obspm.fr}
        \and Dipartimento di Fisica e Astronomia “Galileo Galilei”, Università di Padova, Vicolo dell’Osservatorio 3, 35122 Padova, Italy
         \and Université de Strasbourg, CNRS, Observatoire Astronomique de Strasbourg, F-67000 Strasbourg, France 
          \and CEA, AIM, Université Paris-Saclay, 91191 Gif-sur-Yvette, France
          \and Division of Astrophysics, Department of Physics, Lund University, Box 118, SE-221 00 Lund, Sweden
          \and  Leibniz Institut für Astrophysik Potsdam (AIP), An der Sternwarte 16, D-14482, Potsdam, Germany
          \and University of Strasbourg Institute for Advanced Study, 5 all\'ee du G\'en\'eral Rouvillois, F-67083 Strasbourg, France}

   \date{Received xxx; accepted xxx}

  \abstract
  {The peculiar Galactic globular cluster $\omega$~Centauri (NGC 5139) has drawn attention for its unique features, such as an unusually high stellar mass compared to other Galactic globular clusters and a broad distribution of chemical elements. These features have led to the hypothesis that $\omega$~Centauri might be the nuclear remnant of an ancient dwarf galaxy accreted by the Milky Way, potentially bringing along its own globular cluster system.} {In this work, we adopt an innovative approach by examining the individual chemical abundances of Galactic globular clusters to identify shared patterns with $\omega$~Centauri.}{Applying Gaussian Mixture Models to globular cluster stars, whose membership is based on the analysis of the Gaia~EDR3 release, and whose chemical abundances have been obtained from  APOGEE DR17, we depart from traditional kinematic-based procedures and search for globular clusters that are chemically compatible with $\omega$~Centauri in a 8-dimensional space defined by [Fe/H], $\alpha$-elements as [Mg/Fe], [Si/Fe] and [Ca/Fe], light+odd-Z elements as [C/Fe], [Al/Fe] and [K/Fe], and an iron-peak element as [Mn/Fe]. With this approach, clusters that are chemically compatible with $\omega$~Centauri are clusters whose chemical patterns are contained in the abundance domain defined by $\omega$~Centauri stars.} {Our analysis leads to the identification of six globular clusters -- NGC 6752, NGC 6656, NGC 6809, NGC 6273, NGC 6205, and NGC 6254 -- that exhibit strong chemical similarities with $\omega$~Centauri, and which have metallicities that coincide with those of the two known peaks (primary and secondary) of $\omega$~Centauri's metallicity distribution. They all exhibit not null intrinsic [Fe/H] dispersions, ranging between 0.07 and 0.12 dex, unless the ASPCAP uncertainties had been severely underestimated, and three of them have statistically significant skewed [Fe/H] distributions.  Furthermore, the chemical patterns of these clusters lead to the exclusion that they were formed in progenitor galaxies with chemical enrichment histories similar to those of the Large and Small Magellanic Clouds, Sagittarius, and Fornax. Once placed in kinematic spaces such as the energy - angular momentum plane, these clusters result scattered across an extended region, which is predicted by \textit{N}-body simulations if their common progenitor was sufficiently massive compared to the Milky Way.} {Our novel approach suggests a common origin for NGC 6752, NGC 6656, NGC 6809, NGC 6273, NGC 6205, NGC 6254 and $\omega$~Centauri,  indicating that \textit{Nephele}, as we propose to call the progenitor in which all these clusters formed, played a substantial role in the Galaxy's history. The finding that a set of globular clusters can be associated to $\omega$~Centauri reinforces the hypothesis that this system is the remnant of a galaxy, and not, simply, an unusual globular cluster. This study also shows that the spectroscopic data at our disposal have reached the quality needed to compare chemical patterns of stellar systems, to reveal their common origins or exclude their association with specific progenitor galaxies.}
  

   \maketitle
%

\section{Introduction}

The Galactic globular cluster $\omega$~Centauri (NGC~5139) has been known to be peculiar in many aspects for several decades. Its total stellar mass, estimated to be $\rm{3.5 \pm 0.03} \times 10^6\,M_{\odot}$ \citep[][]{baumgardt18} is 10 times greater than the mean stellar mass of all known Galactic globular clusters (hereafter GCs). Since the early work of \citet{sistero70}, it has been known that this cluster shows a flattened shape, which is accompanied by a significant amount of rotation \citep{meylan86, merritt97, norris97, pancino07, bianchini13,kamann2018,sanna20} and also a counter-rotating core \citep{pechetti24}. It contains stars with an extended [Fe/H] distribution \citep{norris95, suntzeff96,   smith00, sollima05, villanova07, calamida09, johnson2010, pancino11, marino11, nitschai23} with values ranging from about -2.2 to -0.4 dex \citep{meszaros2021}. Not only [Fe/H], but all chemical abundance elements studied so far indicate that $\omega$~Cen had a complex formation \citep[among others, see][]{norris95, hilker04, johnson2010, dorazi11}, which possibly reflects also in an extended age range of its stars \citep{villanova07, villanova14}. Because of these peculiar properties, it has been suggested already more than 20 years ago \citep{lee99, majewski00, carraro00, bekki03, tsuchiya03, tsuchiya04} that  $\omega$~Cen could be the remnant of an ancient dwarf galaxy accreted by the Milky Way (MW) during the first billion years after its formation. In this scenario, the current cluster would be the stellar nucleus of the dwarf, whose stellar envelope would have been stripped during the accretion process, due to the tidal effects exerted by the Galactic potential. The stellar nucleus, much more compact and dense, would have survived the accretion, and continued orbiting the Milky Way since then.  \citet{bekki03} showed that this scenario may explain some of the properties of $\omega$~Cen (namely its current mass, and its orbital characteristics) if the host galaxy had an initial stellar mass of about $10^8 M_{\odot}$,  the nucleus an initial mass of about $6\times 10^6 M_{\odot}$ and a total-to-baryonic mass ratio of 10. Since then, the remnant galaxy and its nucleus would have continued to lose stars (and dark matter) and keep sinking in the inner Galaxy, due to the combined effect of the tidal forces and dynamical friction exerted by the Milky Way.  If such a merger occurred in the first few billion years of the formation of our Galaxy, the estimated mass of the $\omega$~Cen progenitor is sufficiently large \citep[at least a factor 1/100 with respect to the mass of the Milky Way, depending on the time of the accretion, see][]{snaith14} to consider it a significant accretion event in the early evolution of our Galaxy. 

Given the numbers above, such an accretion event not only would have contributed to increase the stellar mass budget of the Galaxy of an amount similar to that of the stellar halo \citep[in this respect, it is interesting to recall the recent discovery of the Fimbulthul stream, which is unequivocally showing that $\omega$~Cen is still losing part of its stars in the field, see][]{ibata19a}, as this dwarf galaxy should have also brought its own globular cluster system to the Milky Way. Galaxies in the local Universe with stellar masses of the order of $10^8 M_{\odot}$ can contain up to dozen clusters \citep{eadie22}; simulations-based estimates \citep{kruijssen20} of the number of globular clusters accreted by the Milky Way over time also reinforce that a dozen of clusters is a realistic number for a dwarf whose stellar mass was about a hundredth of that of the Milky Way at the time of the accretion. Where are then the globular clusters that the $\omega$~Cen progenitor brought with it, how are they redistributed in the Galaxy, how many are they and, first of all, how to distinguish them among all globular clusters populating our Galaxy today? 

One may be tempted to retrieve the population of globular clusters lost by the $\omega$~Cen progenitor galaxy by simply looking for the population of GCs which today share similar kinematic properties to those of this cluster. Following this approach, for example, \citet{myeong19} suggested that $\omega$~Cen may be associated to the Sequoia accretion event, while \citet{massari2019} were more in favour of an association to the Gaia Sausage Enceladus galaxy, of which $\omega$~Cen would have been the nuclear star cluster (NSC). In both cases, the similarity of the cluster's orbital properties with Sequoia (or Gaia Sausage Enceladus) was a decisive argument. An association of  $\omega$~Cen  to Sequoia would mean that this cluster shares a common origin with  NGC~5466, IC~4499, NGC~ 7006, Pal~13 and FSR~1758 \citep[these are indeed the clusters for which the association to Sequoia appears robust, according to ][]{massari2019}, while an association with Gaia Sausage Enceladus implies that $\omega$~Cen would be rather associated to the globular clusters NGC~288, NGC~362, NGC~1261, NGC~1851, NGC~1904, NGC~2298, NGC~2808, NGC~4147, NGC~4833, NGC~5286, NGC~5897, NGC~6205, NGC~6229, NGC~6235, NGC~6284, NGC~6341, IC~1257, Djorg~1, Terzan~10, ESO-SC06, NGC~6779, NGC~6864, NGC~7089, NGC~7099 and NGC~7492 \citep[still following][for the list of clusters robustly associated to Gaia Sausage Enceladus]{massari2019}. These two classifications are thus in conflict, unless Sequoia and Gaia Sausage Enceladus are not part of the same progenitor galaxy \citep{koppelman20, amarante22}. \\
In these recent years, however, a number of studies has shown that the association of a group of GCs, as well as of field stars, to a common progenitor on the basis of orbital criteria (namely, their kinematic coherence -- the approach followed in the above cited papers) is not straightforward at all, and even worst, it is not completely physical motivated, unless very specific assumptions are made on the accretion history of our Galaxy \citep[see][]{jeanbaptiste17, khoperskov23a, khoperskov23b}. 
Also the simulations by \citet{bekki03} show that the progenitor galaxy of $\omega$~Cen possibly left an extended trail of stars distributed over a wide range of orbital parameters (see Fig.~5 in their article). In \citet{pagnini2023}, we made use of $N$-body simulations to show that kinematic spaces are not a reliable tool for distinguishing globular clusters originating from a single accreted galaxy from the rest of the GC population. In fact, clusters that have different orbital features today may originate from the same progenitor galaxy, just as clusters with similar kinematic properties may actually have a different origin \citep[see][]{pagnini2023}.  

For all these reasons, in this work we have chosen to investigate the population of GCs associated with the progenitor of $\omega$~Cen by searching for all Galactic GCs that have chemical abundances compatible with those of $\omega$~Cen, in a sort of "chemical tagging" approach \citep{freeman02}, where we look for common chemical patterns between Galactic GCs, and $\omega$~Cen,  based on star individual abundances. Indeed, unlike the orbital characteristics of GCs which are generally not conserved during an accretion, their chemical properties are linked to those of the interstellar medium (hereafter ISM) in which these clusters were formed. 
We therefore take a new approach compared to those proposed so far in the literature, and take advantage of the wealth of new spectroscopic homogeneous data available for tens of Galactic GCs, to search for their common characteristics in abundance spaces. For this, we search for all GCs in the recently published catalogue of \citet{schiavon2023}, based on APOGEE~DR17 data  \citep{abdurro22}, that are chemically compatible with $\omega$~Cen, i.e. whose distribution in chemical abundance spaces is contained within that of $\omega$~Cen. Indeed, if $\omega$~Cen was the nucleus of a dwarf galaxy, its chemical abundance patterns should be representative (at least of part) of the chemical abundance patterns of its progenitor (as we will show to be the case for the M~54-Sagittarius dwarf system in Appendix \ref{NSC_dwarf}), which in turn reflect those of the ISM in which the progenitor stellar populations formed, at a given time. In other words, the chemical abundances of globular clusters, as well as of field stars,  formed in the  $\omega$~Cen progenitor should be representative of the chemical enrichment history of the progenitor galaxy itself, that is they should trace specific epochs of this evolution. We can use the nuclear star cluster of this progenitor galaxy, i.e. $\omega$~Cen itself, as representative of at least part of the progenitor chemical evolution, based on the chemical similarities that exist, as we will show, among the Sagittarius dwarf galaxy and M~54, its nuclear star cluster.

Our analysis allows us to identify six globular clusters -- namely NGC~6752, NGC~6656, NGC~6809, NGC~6273, NGC~6205, and NGC~6254 -- which have chemical patterns found also in $\omega$~Cen, and which also show some common characteristics in their metallicity distribution function -- all results that we interpret as evidence of their common origin. NGC~5024, NGC~6544, FSR~1758, and NGC~1904 may also be part of this same group, but -- as we shall see -- the analysis of these clusters is based on statistics that are too weak to allow us to derive any firm conclusions. According to the analysis presented in this paper,  \textit{Nephele}\footnote{In Greek mythology, Nephele is the mother of centaurs. We are aware that adding a new name to an already rich list of possible galaxies accreted by the MW over time may be confusing at first sight, but we believe it is important to assign a name to the set of clusters (and, in a future work, of field stars) associated with $\omega$~Cen on the basis of the individual abundances of stars,  thus making clear that this association is based on different criteria from the kinematically-based ones, usually adopted in the literature.}, as we propose to call the progenitor of $\omega$~Centauri, in which all these clusters formed, has potentially brought at least six clusters to the Milky Way, thus proving to be a significant accretion in the history of our Galaxy.

The paper is organised as follows. In Sect.~\ref{obsdata}, we describe the observational dataset used for this study; in Sect.~\ref{method} the method used to analyse it; in Sect.~\ref{results} we present the obtained results; finally, after a discussion in Sect. \ref{discussion},  in Sect.~\ref{concl}, we derive our conclusions.


\section{Observational data}\label{obsdata}

For this study, we make use of data from the APOGEE Value Added Catalogue (VAC) of Galactic globular cluster stars \citep[see][]{schiavon2023}. The catalogue comprises a total of 7737 entries for 6422 unique stars associated with 72 Galactic GCs and contains full APOGEE DR17 information \citep[][]{abdurro22} including radial velocities and abundances for up to 20 elements\footnote{The APOGEE project is based on H-band spectra with resolution of $R\sim 22,500$ and derives elemental abundances for up to 20 species at a precision of 0.1 dex.}.
As in \citet{horta2023}, among these stars, only those satisfying the following criteria have been used for this study:
\begin{enumerate}
\item a signal-to-noise ratio $\tt{SNREV} > 70$;
\item temperatures in the range $\rm 3500\,K < T_{eff} < 5500\,K$ and surface gravities  $\rm logg < 3.6$; 
\item \tt{APOGEE STARFLAG} and \tt{APOGEE STARBAD} $= 0$;
\rm \item  stars which have, according to \citet{vasiliev21}, a high probability of being members of the cluster (\tt{VB\_PROB} $\geq 0.9$)
\end{enumerate}

With these selections, the number of GC stars reduces from 6422 (before selection) to 3223 (after selection) corresponding to 57 Galactic GCs.

\section{Methods}\label{method}

To assess whether a given GC is compatible, in the chemical abundance space, with $\omega$~Cen, we make use of a Gaussian Mixture Model (GMM) approach. We consider an 8-dimensional high-precisions abundance space defined by [Fe/H], $\alpha-$elements as  [Mg/Fe], [Si/Fe] and [Ca/Fe], light+odd-Z elements as  [C/Fe], [Al/Fe] and [K/Fe], and an iron-peak element as [Mn/Fe]\footnote{Note that the abundances of elements like K or Mn are based on a few weak lines.  Towards the low metallicity end ([Fe/H] <~ -1.5), ASPCAP may be delivering upper limits only, depending on the star's T$\rm_{eff}$ and log~g.}, 
and we keep only stars in the \citet{schiavon2023} catalogue which have all the abundance flags of these elements equal to zero.  In this way, we are covering several nucleosynthetic channels. The different $\alpha$-elements have slightly different origins and Mg is the best determined APOGEE-element. Al and Mg are anti-correlated in GCs at low metallicities, and Carbon, although its abundance changes due to internal processes within red giant stars, could be further evidence in favour of a common origin.
With this choice, the number of stars used for the GMM analysis described below reduces to 2077, associated to 54 different GCs. For all the elements listed above, the median uncertainty is smaller than 0.07. 
Note that the choice to use 8 elements for the GMM analysis is a compromise between two different needs. On the one hand, the higher the number of abundances used, the higher the probability of identifying clusters that are chemically similar to $\omega$~Cen (two clusters may have similar abundances of some elements, but not in others, so using enough abundances allows us to better discriminate clusters from each other, for a discussion on this point see Sect~\ref{no_ocen-like_gcs}). On the other hand, the higher the number of elements used, the lower the number of stars per cluster (the flags imposed on the quality of each chemical abundance imply that the number of stars meeting the quality criteria on all abundances decreases as the number of elements used increases). We therefore chose to use 8 elements as a compromise, but still tested the algorithm with up to 12 elements (namely [Fe/H], [C/Fe], [N/Fe], [O/Fe], [Al/Fe], [Mg/Fe], [Si/Fe], [Ca/Fe], [K/Fe], [Mn/Fe], [Ni/Fe]) and found that the results presented in the following of this paper were consistent. In particular, as we will show in Appendix~\ref{ocen-like_other_elems} the chemical compatibility between $\omega$~Cen and the clusters found using this approach extends also to chemical elements not used for the GMM analysis, as [Na/Fe] and [Ce/Fe] for which the corresponding uncertainties are large ($\simeq0.17$ and $\simeq0.07$),
but which are available in the  \citet{schiavon2023} catalogue.

\begin{figure}
\includegraphics[width=\columnwidth]{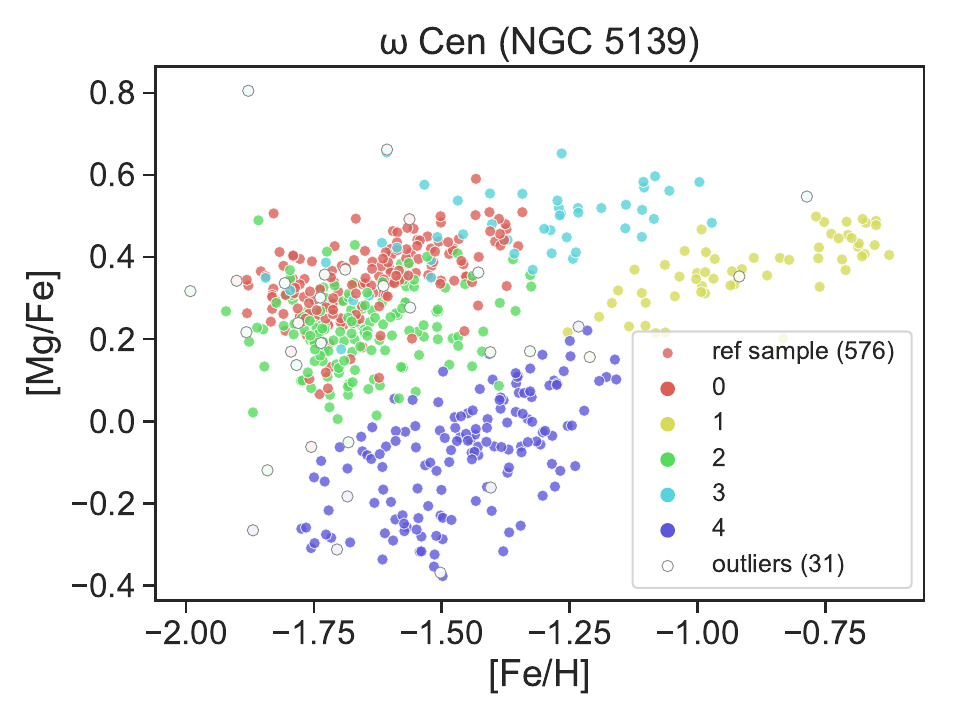}
    \caption{[Mg/Fe] vs [Fe/H] distribution of stars belonging to $\omega$~Cen colour-coded according to the different components retrieved when minimising the BIC criterion in the GMM. Filled symbols represent the reference sample of stars of $\omega$~Cen (i.e. stars with a probability density above a threshold defined in Sect.\ref{results}), while empty symbols are the outliers (i.e. stars below this threshold).}
    \label{ocen_components}
\end{figure}

The distribution of $\omega$~Cen in the 8-dimensional abundance space defined by [Fe/H], [Mg/Fe], [Si/Fe], [Ca/Fe], [C/Fe], [Al/Fe], [K/Fe], and [Mn/Fe] is then fitted by using an increasing number of gaussian components and the optimal number of components for our dataset is then determined by minimising the Bayesian information criterion (BIC). With this procedure, we find that the number of components that best reproduces the 8-dimensional distribution of  $\omega$~Cen  is five (see  Figure \ref{ocen_components} for their distribution in the [Mg/Fe]-[Fe/H] plane, where also outliers of this distribution are reported, these latter being stars with a probability density below a threshold defined in Sect.\ref{results}). Note that, as we checked, the number of gaussian components tends to decrease with the number of chemical elements used in the GMM and we advise the reader against giving a physical meaning to this specific number. Depending on the chemical abundances under study, as well as on the cluster algorithm used, this number may also vary. For example, \citet{johnson2010} found that the metallicity distribution function (hereafter MDF) of $\omega$~Cen could be fitted by 4 components.
\citet{meszaros2021}, on the other hand, considered a 3-dimensional chemical space (including the [Fe/H], [Al/Fe], and [Mg/Fe] abundances), and found 7 groups. The difference in the retrieved number of components may also be due to the different algorithm used to perform the clustering analysis. In particular, the \textit{k-means} algorithm used in \citet{meszaros2021} relies on centroid-based clustering, which performs particularly well in identifying clusters in a set of data points with spherical-like structure, while it may perform less well in representing more complex patterns, such as those describing chemical abundances.

For each of the 54 GCs included in this study, we then estimate the fraction of stars that have a high probability of belonging to the GMM model obtained for $\omega$~Cen, i.e. the fraction of stars whose distribution in the 8-dimensional abundance space falls within that of $\omega$~Cen. To estimate the uncertainties on the derived fractions, we have repeated this procedure a hundred times, each time bootstrapping the data of both  $\omega$~Cen and each GC: for each bootstrap realisation, we have also taken into account the individual uncertainties of the chemical abundances in the \citet{schiavon2023} catalogue using a Monte-Carlo sampling.  As an estimator of the compatibility of each GC in the sample with $\omega$~Cen, we used the fraction of compatible stars - i.e. the fraction of stars that fall within the distribution of  $\omega$~Cen in the 8-dimensional chemical spaces according to the GMM - out of the total number of stars in the GC. Compatible stars are defined as those whose log-likelihood exceeds or is equal to a threshold, which we established as the 5th percentile of the log-likelihood distribution of the training dataset, namely $\omega$~Cen. In contrast, outliers are stars with a probability density below the established threshold. The sample of $\omega$~Cen stars with a log-likelihood above the 5th percentile threshold is referred to as the ``reference sample''.

\section{Results}\label{results}

\begin{table}
\centering
\caption{Fraction of stars chemically compatible with $\omega$~Cen. }\label{OCenGCs_table_VAC}
\resizebox{0.95\columnwidth}{!}{
\begin{tabular}{lcccc}
\toprule
GC name &  Fraction (\%)  &  \# stars & [Fe/H]$_{median}$ & [Fe/H]$_{mode}$ \\
\midrule
  Ter10 &   98 $\pm$   14 &   1 & -1.62 $\pm$ 0.02 & -1.60 $\pm$ 0.05\\
NGC2298 &   94 $\pm$   22 &   2 & -1.83 $\pm$ 0.02 & -1.83 $\pm$ 0.05\\
\bf NGC5139 &   \bf89 $\pm$    3 & \bf607 & \bf-1.57 $\pm$ 0.01 & \bf-1.66 $\pm$ 0.04\\
\bf NGC6752 &   \bf 84 $\pm$     8 &  \bf 83 & \bf-1.47 $\pm$ 0.01 & \bf-1.47 $\pm$ 0.02\\
\bf NGC6656 &   \bf 81 $\pm$    8 &  \bf 68 & \bf-1.66 $\pm$ 0.02 & \bf-1.64 $\pm$ 0.05\\
\bf NGC6809 &   \bf 79 $\pm$    13 &  \bf 18 & \bf-1.73 $\pm$ 0.02 & \bf-1.73 $\pm$ 0.05\\
\bf NGC6273 &   \bf 76 $\pm$    9 &  \bf 40 & \bf-1.67 $\pm$ 0.02 & \bf-1.69 $\pm$ 0.07\\
\bf NGC6205 &   \bf 69 $\pm$   14 &  \bf 26 & \bf-1.48 $\pm$ 0.02& \bf-1.48 $\pm$ 0.04\\
NGC5024 &   64 $\pm$   23 &   5 & -1.80 $\pm$ 0.02 & -1.80 $\pm$ 0.03\\
\bf NGC6254 &   \bf 64 $\pm$    15 &  \bf 50 & \bf-1.49 $\pm$ 0.01 & \bf-1.49 $\pm$ 0.03\\
NGC6544 &   56 $\pm$   20 &  15 & -1.47 $\pm$ 0.02 & -1.48 $\pm$ 0.05\\
FSR1758 &   56 $\pm$   26 &   7 & -1.40 $\pm$ 0.05 & -1.41 $\pm$ 0.07\\
NGC1904 &   52 $\pm$    18 &  26 & -1.51 $\pm$ 0.03 & -1.53 $\pm$ 0.07 \\
NGC6093 &   47 $\pm$    50 &   1& -1.61 $\pm$ 0.01 & -1.59 $\pm$ 0.04\\
NGC7089 &   44 $\pm$   23 &  15 & -1.48 $\pm$ 0.02 & -1.48 $\pm$ 0.05\\
NGC6218 &   33 $\pm$    19 &  40& -1.26 $\pm$ 0.01 & -1.26 $\pm$ 0.02 \\
NGC0288 &   31 $\pm$    22 &  37 & -1.27 $\pm$ 0.01 & -1.26 $\pm$ 0.03\\
NGC6380 &   31 $\pm$    28 &   9 & -0.75 $\pm$ 0.03 & -0.76 $\pm$ 0.05\\
   Ter4 &   30 $\pm$    46 &   1 & -1.45 $\pm$ 0.02 & -1.40 $\pm$ 0.02\\
NGC6121 &   29 $\pm$    25 & 169 & -1.05 $\pm$ 0.00 & -1.04 $\pm$ 0.01\\
Djorg\_2 &   24 $\pm$    36 &   4 & -1.07 $\pm$ 0.02 & -1.07 $\pm$ 0.03\\
NGC6171 &   24 $\pm$    21 &  23 & -0.97 $\pm$ 0.03 & -0.97 $\pm$ 0.05\\
NGC0104 &   20 $\pm$    19 & 224 & -0.75 $\pm$ 0.00 & -0.74 $\pm$ 0.01\\
NGC6522 &   20 $\pm$    34 &   2 & -1.15 $\pm$ 0.07 & -1.16 $\pm$ 0.10\\
NGC6715 &   16 $\pm$    10 &  26 & -1.41 $\pm$ 0.05 & -1.52 $\pm$ 0.06\\
    HP1 &   15 $\pm$    20 &  10 & -1.16 $\pm$ 0.05 & -1.17 $\pm$ 0.08\\
NGC6838 &   14 $\pm$    15 &  45 & -0.76 $\pm$ 0.01 & -0.77 $\pm$ 0.04\\
NGC6397 &   13 $\pm$    15 &  10 & -2.01 $\pm$ 0.02 & -2.02 $\pm$ 0.03\\
NGC5272 &   11 $\pm$    7 &  71 & -1.38 $\pm$ 0.01 & -1.37 $\pm$ 0.04\\
NGC6723 &   11 $\pm$    18 &   7 & -1.04 $\pm$ 0.02 & -1.06 $\pm$ 0.05\\
NGC6558 &   11 $\pm$    25 &   3 & -1.15 $\pm$ 0.05 & -1.15 $\pm$ 0.05\\
NGC6569 &   11 $\pm$    19 &   6 & -0.99 $\pm$ 0.02 & -1.00 $\pm$ 0.03\\
   Ter9 &   10 $\pm$    15 &   9 & -1.39 $\pm$ 0.04 & -1.40 $\pm$ 0.06\\
NGC3201 &    8 $\pm$     8 &  98 & -1.35 $\pm$ 0.01 & -1.34 $\pm$ 0.02\\
NGC6642 &    8 $\pm$   21 &   6 & -1.03 $\pm$ 0.29 & -1.04 $\pm$ 0.23 \\
   Ter2 &    5 $\pm$    17 &   2 & -0.84 $\pm$ 0.04 & -0.86 $\pm$ 0.06\\
NGC5904 &    3 $\pm$     5 &  79 & -1.20 $\pm$ 0.01 & -1.20 $\pm$ 0.04\\
NGC6316 &    2 $\pm$    10 &   6 & -0.76 $\pm$ 0.02 & -0.77 $\pm$ 0.03\\
NGC6717 &    2 $\pm$    14 &   2 & -1.17 $\pm$ 0.04 & -1.16 $\pm$ 0.06\\
NGC6760 &    1 $\pm$    10 &   3 & -0.73 $\pm$ 0.02 & -0.73 $\pm$ 0.03\\
NGC6229 &    1 $\pm$     7 &   3 & -1.27 $\pm$ 0.02 & -1.27 $\pm$ 0.03\\
NGC1851 &    1 $\pm$     3 &  31 & -1.12 $\pm$ 0.02 & -1.14 $\pm$ 0.02\\
NGC7078 &    0 $\pm$     0 &   1 & -2.24 $\pm$ 0.01 & -2.19 $\pm$ 0.01\\
   Ter5 &    0 $\pm$     0 &   2 & -0.61 $\pm$ 0.01 & -0.59 $\pm$ 0.03\\
   Pal6 &    0 $\pm$     0 &   1 & -0.92 $\pm$ 0.01 & -0.89 $\pm$ 0.05\\
NGC6341 &    0 $\pm$     0 &   3 & -2.20 $\pm$ 0.04 & -2.19 $\pm$ 0.04\\
NGC6553 &    0 $\pm$     0 &   1 & -0.22 $\pm$ 0.01 & -0.20 $\pm$ 0.05 \\
NGC6388 &    0 $\pm$     0 &  24 & -0.51 $\pm$ 0.02 & -0.53 $\pm$ 0.05\\
NGC6304 &    0 $\pm$     0 &   5 & -0.49 $\pm$ 0.05 & -0.51 $\pm$ 0.06\\
NGC6293 &    0 $\pm$     0 &   1 & -2.08 $\pm$ 0.02 & -2.03 $\pm$ 0.02 \\
NGC4590 &    0 $\pm$     0 &   1 & -2.12 $\pm$ 0.01 & -2.07 $\pm$ 0.01\\
NGC2808 &    0 $\pm$     0 &  98 & -1.09 $\pm$ 0.01 & -1.10 $\pm$ 0.02\\
NGC0362 &    0 $\pm$    1 &  48 & -1.11 $\pm$ 0.01 & -1.11 $\pm$ 0.01\\
   Ton2 &    0 $\pm$     5 &   2 & -0.61 $\pm$ 0.01 & -0.61 $\pm$ 0.04\\
\bottomrule
\end{tabular}}\tablefoot{Clusters for which this fraction is higher than 60\%, and which -- after the selections described in Sects.~\ref{obsdata} and \ref{method} -- contain at least 15 stars, are marked in bold. For each cluster, the number of stars used for the analysis, the median, and the mode of [Fe/H] distribution are reported.}
\end{table}

 \begin{figure*}[h!]
 \hspace{-22pt}\includegraphics[clip=true, trim = 3mm 0mm 0mm 3mm, width=0.75\columnwidth]{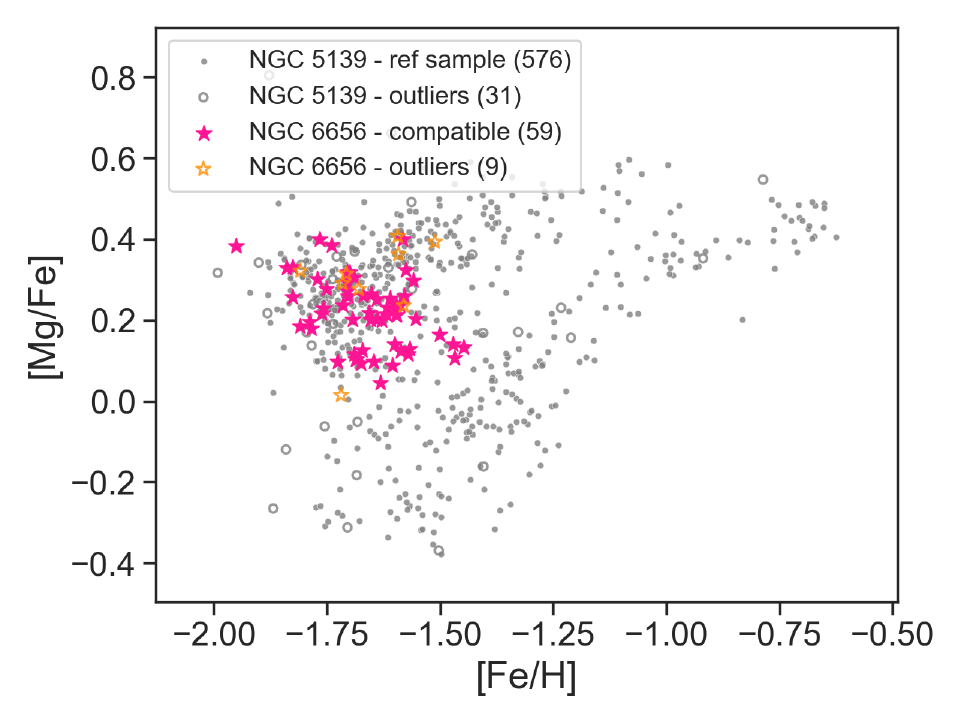}\hspace{-8.5pt}
\includegraphics[clip=true, trim = 3mm 0mm 0mm 3mm, width=0.75\columnwidth]{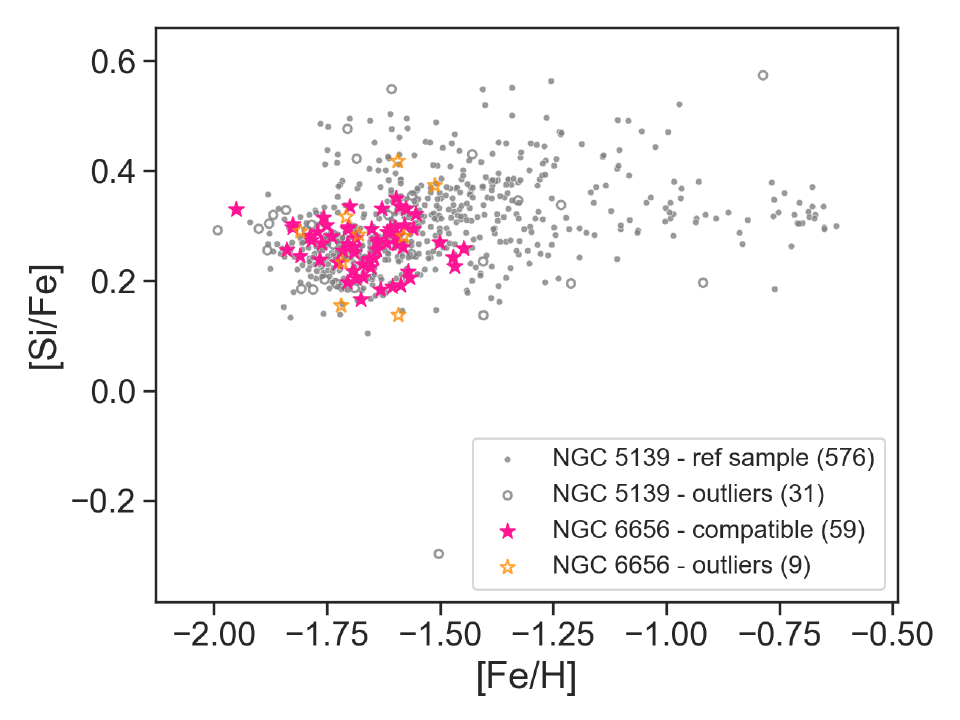}\hspace{-6pt}
\includegraphics[clip=true, trim = 3mm 0mm 0mm 3mm, width=0.75\columnwidth]{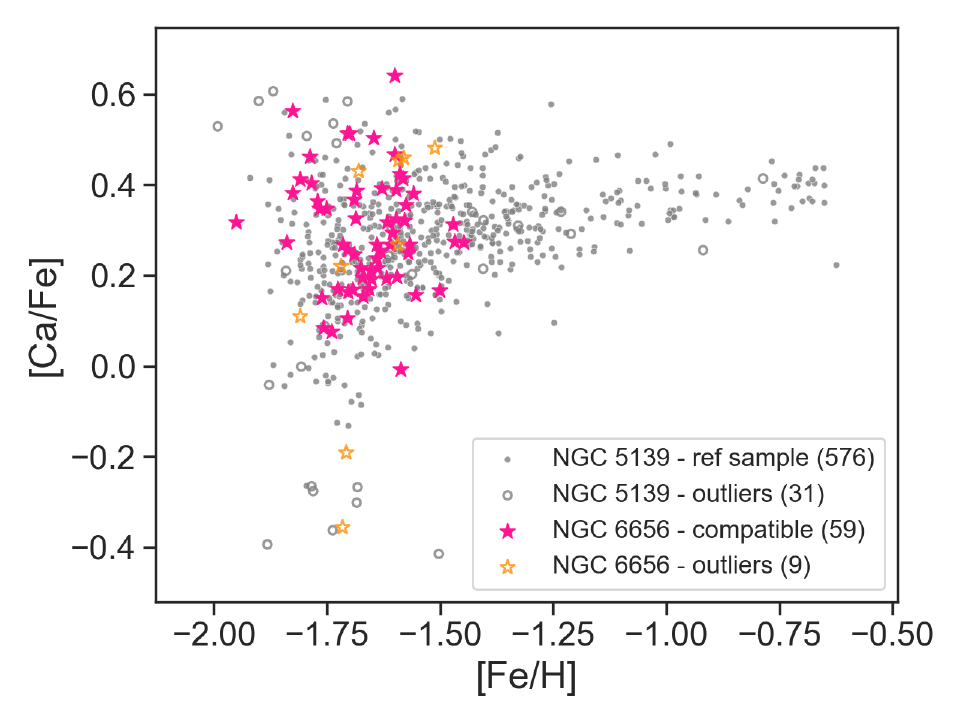}\par
\hspace{-25pt}\includegraphics[clip=true, trim = 3mm 0mm 0mm 2mm, width=0.76\columnwidth]{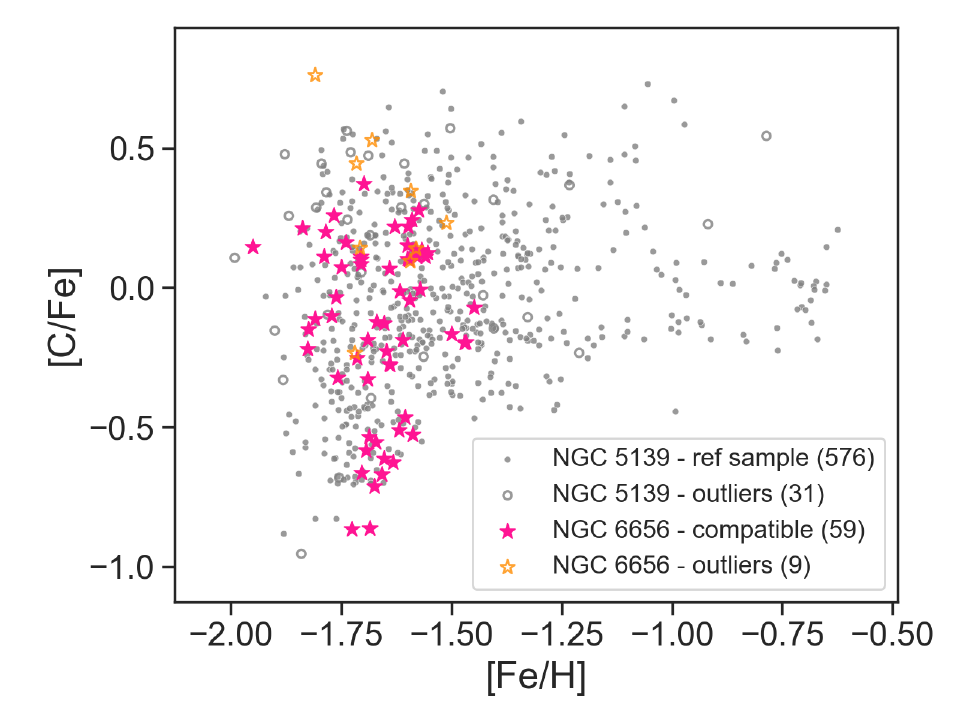}\hspace{-8pt}
\includegraphics[clip=true, trim = 2mm 0mm 0mm 1mm, width=0.76\columnwidth]{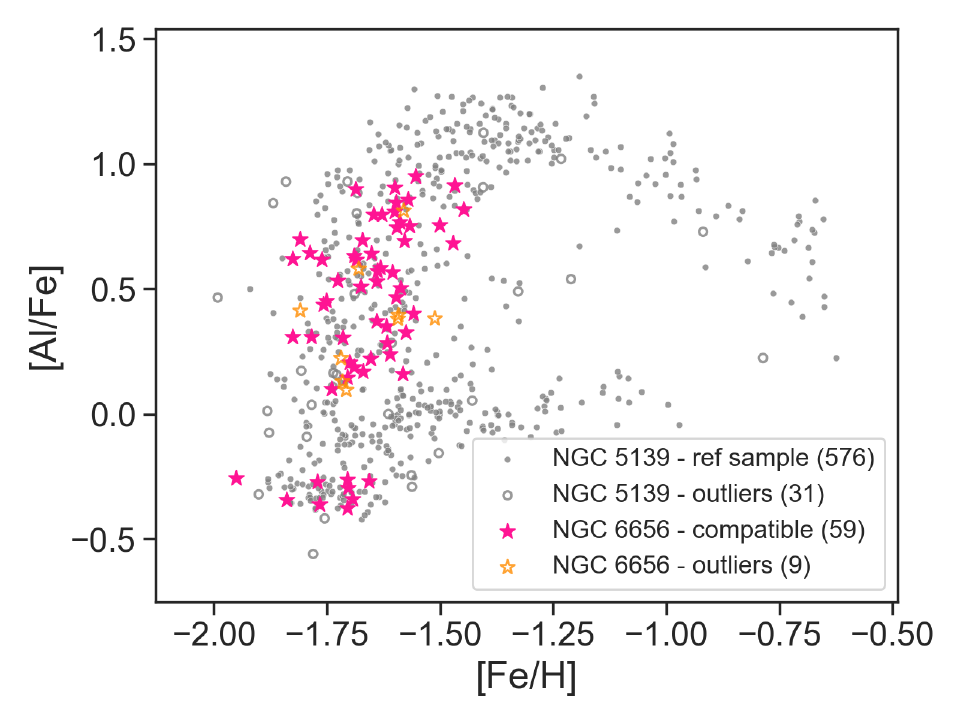}\hspace{-8pt}
\includegraphics[clip=true, trim = 2mm 0mm 0mm 1mm, width=0.76\columnwidth]{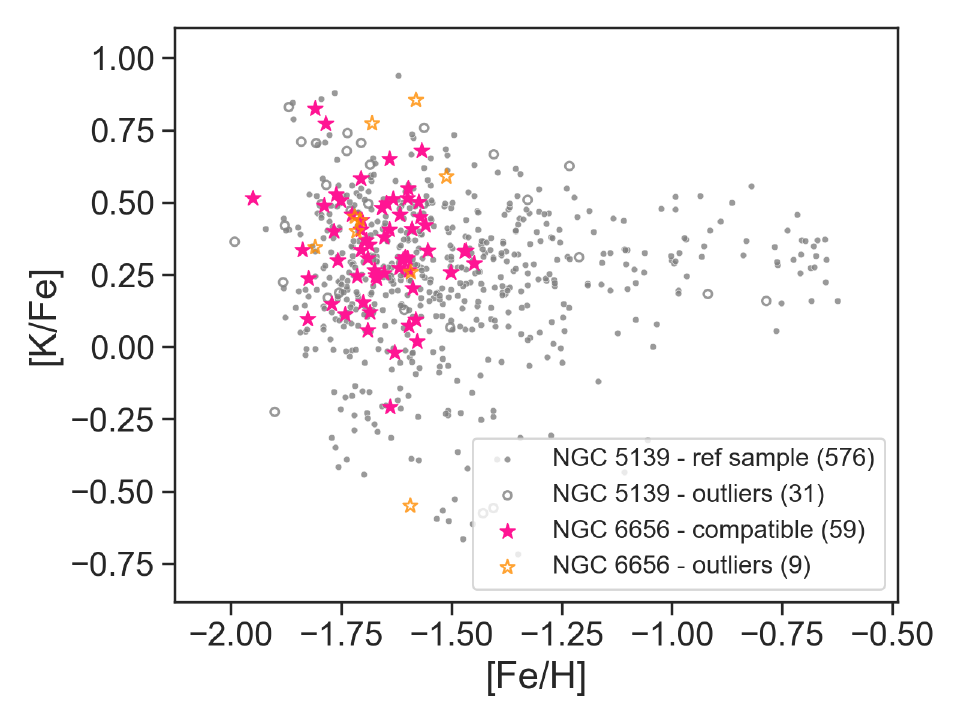}\par
\centering
\includegraphics[clip=true, trim = 1mm 0mm 0mm 1mm, width=0.75\columnwidth]{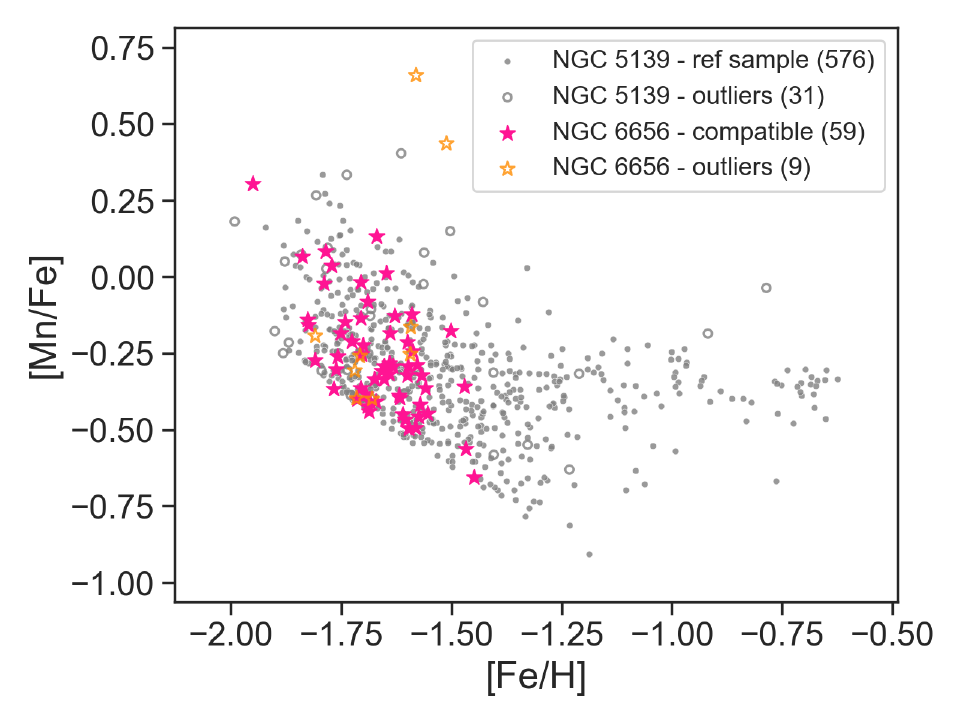}

  \caption{Chemical abundance relations for members of NGC~6656 (colour) and $\omega$~Cen (grey). The filled symbols show the reference sample of $\omega$~Cen (grey) and the stars of NGC~6656 (magenta) chemically compatible with it according to the GMM (see Sect. \ref{results}), while the empty ones (grey and orange colours) correspond to their outliers. The number of stars in each category is reported in parentheses.}
              \label{NGC6656}%
    \end{figure*}
    
\begin{figure*}[h!]
 \hspace{-20pt}\includegraphics[clip=true, trim = 3mm 0mm 0mm 3mm, width=0.75\columnwidth]{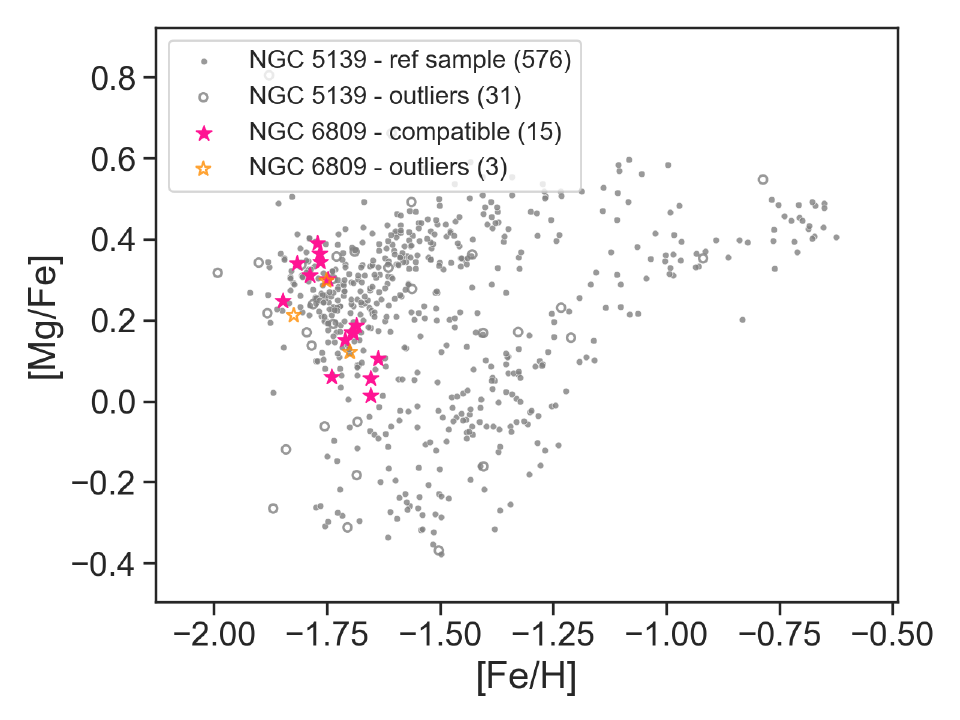}\hspace{-8pt}
\includegraphics[clip=true, trim = 3mm 0mm 0mm 3mm, width=0.75\columnwidth]{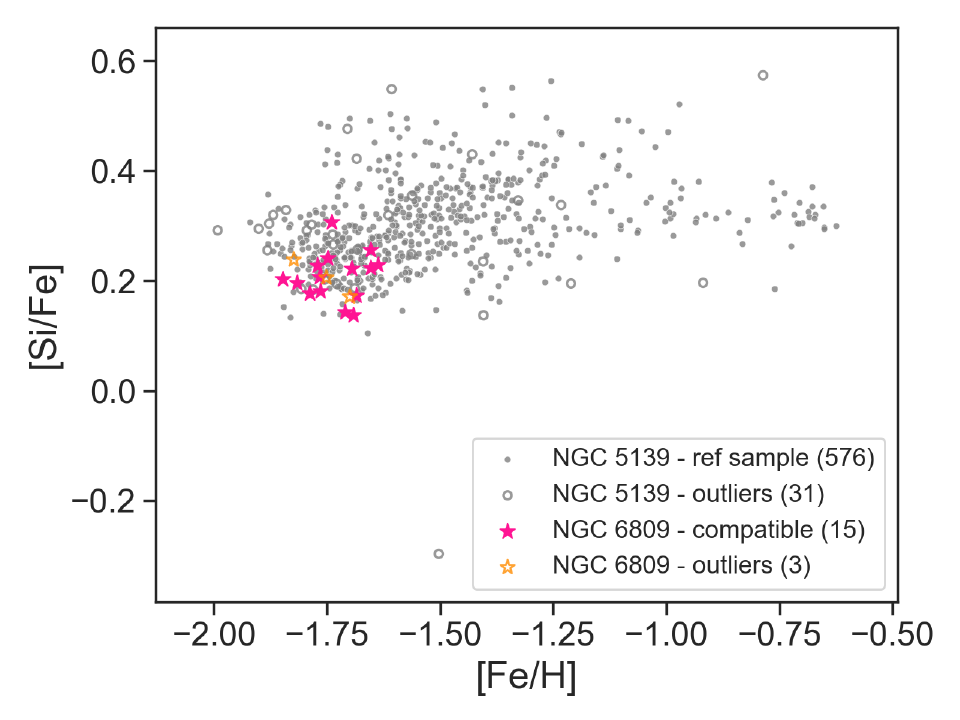}\hspace{-6pt}
\includegraphics[clip=true, trim = 3mm 0mm 0mm 3mm, width=0.75\columnwidth]{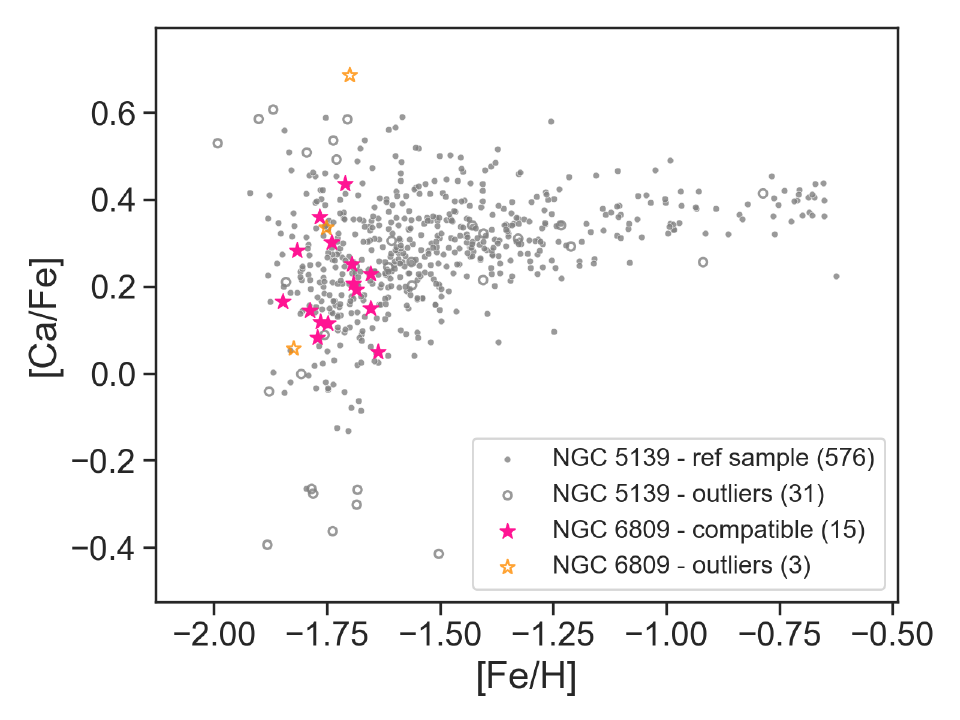}\par
\hspace{-25pt}\includegraphics[clip=true, trim = 1mm 0mm 0mm 2mm, width=0.76\columnwidth]{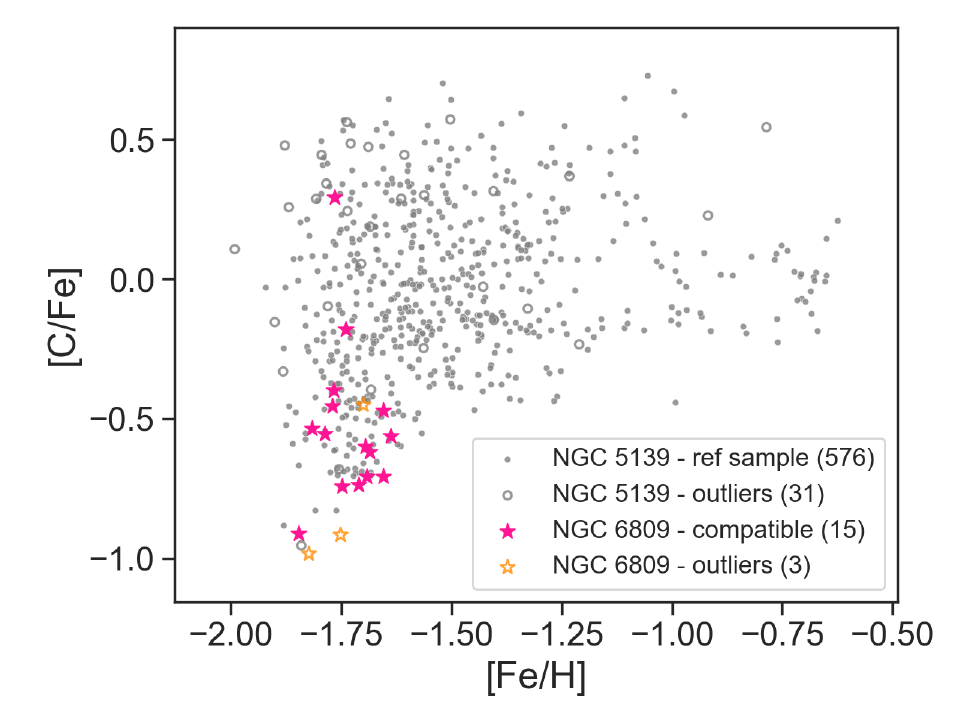}\hspace{-8pt}
\includegraphics[clip=true, trim = 1mm 0mm 0mm 1mm, width=0.76\columnwidth]{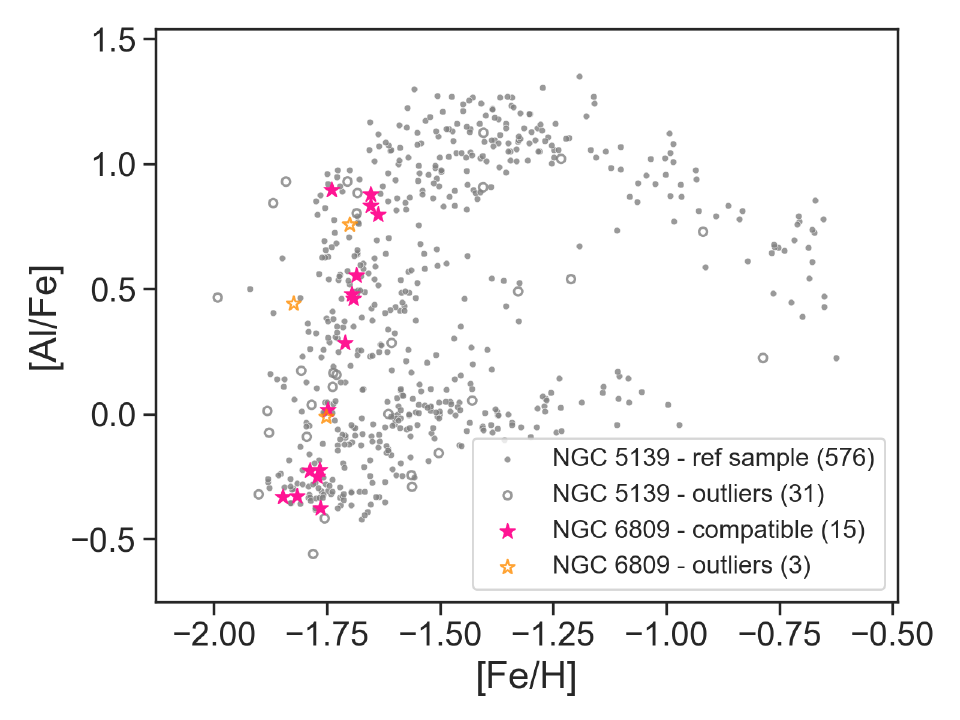}\hspace{-8pt}
\includegraphics[clip=true, trim = 1mm 0mm 0mm 1mm, width=0.76\columnwidth]{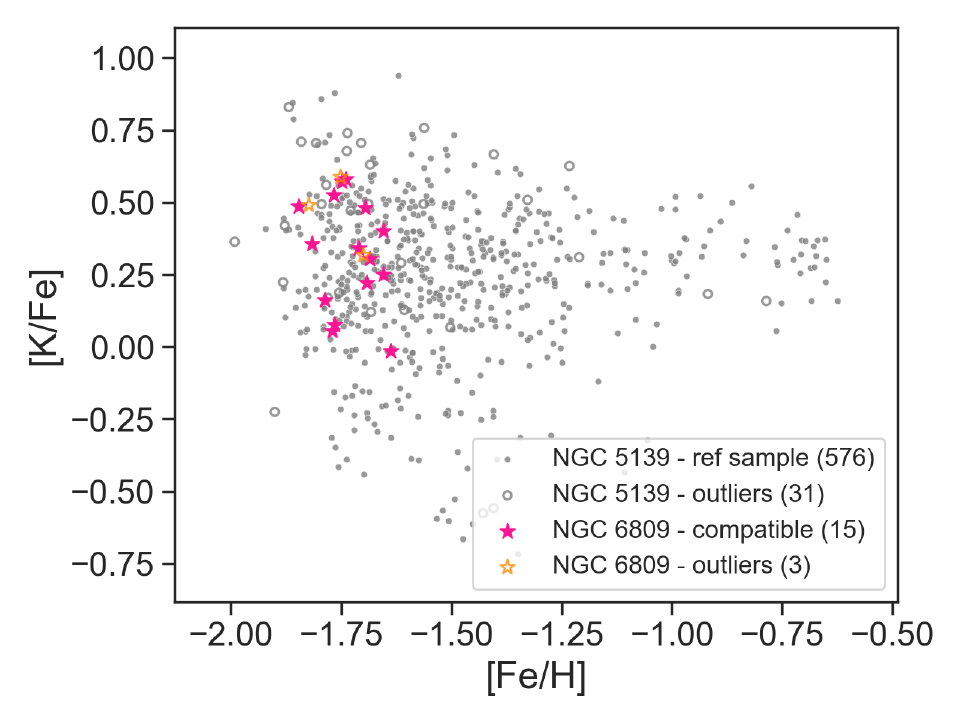}\par
\centering
\includegraphics[clip=true, trim = 1mm 0mm 0mm 1mm, width=0.75\columnwidth]{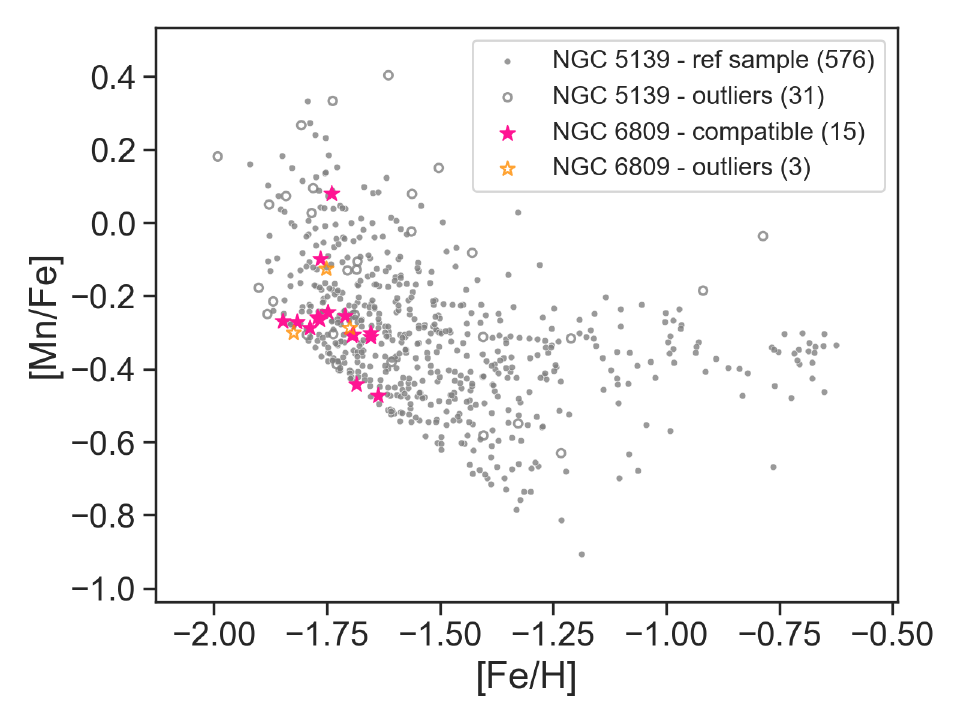}
  \caption{Same as Fig.~\ref{NGC6656} for NGC~6809.}
              \label{NGC6809}%
    \end{figure*}

 \begin{figure*}
\hspace{-20pt}\includegraphics[clip=true, trim = 3mm 0mm 0mm 3mm, width=0.75\columnwidth]{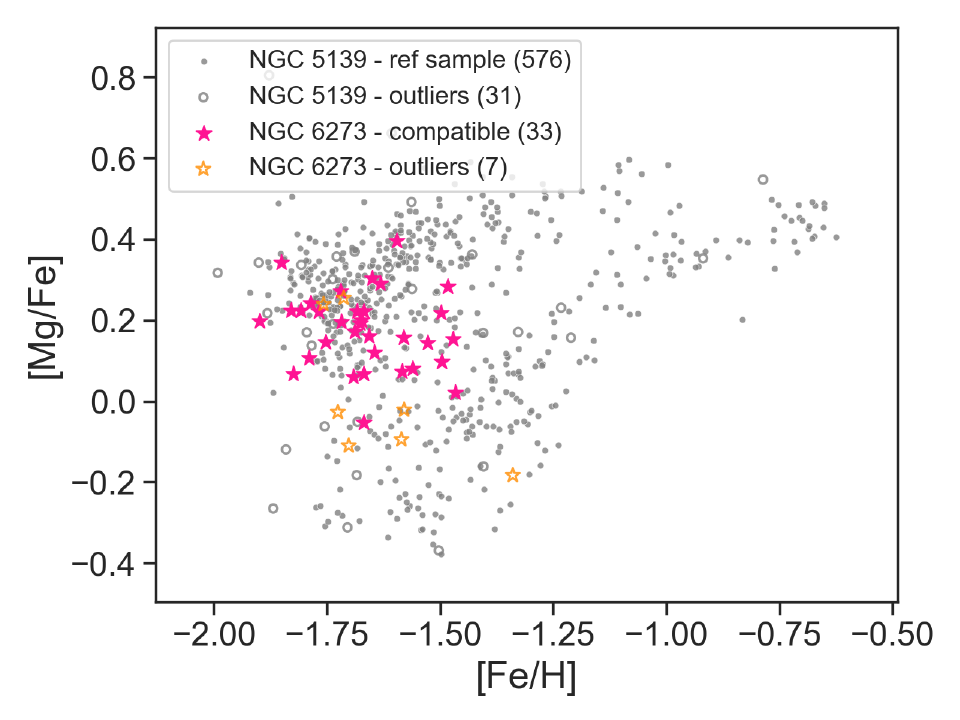}\hspace{-8pt}
\includegraphics[clip=true, trim = 3mm 0mm 0mm 3mm, width=0.75\columnwidth]{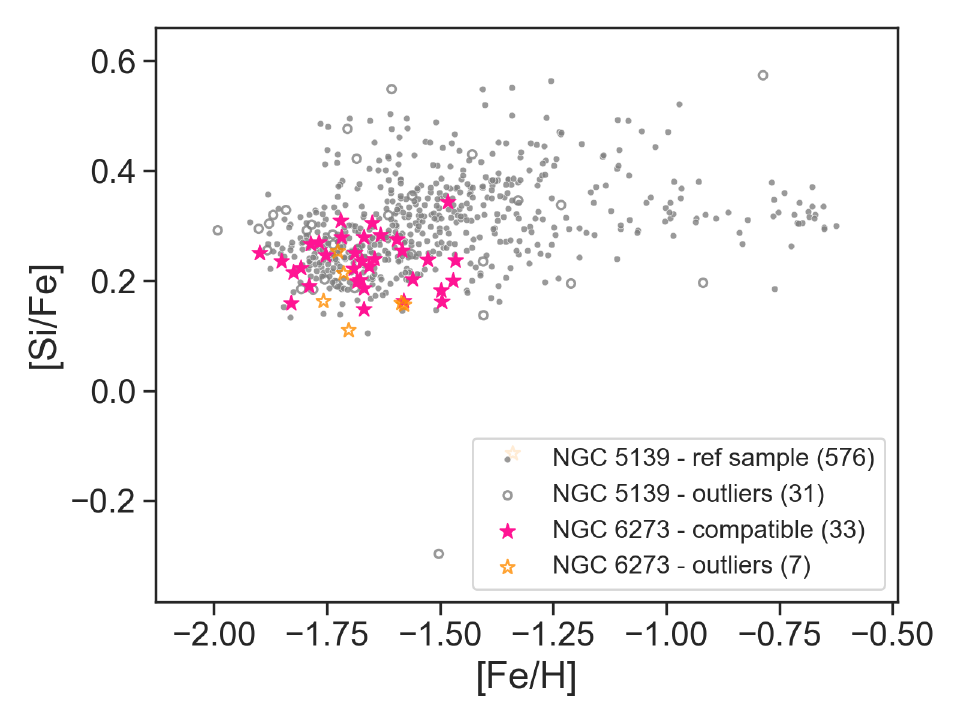}\hspace{-6pt}
\includegraphics[clip=true, trim = 3mm 0mm 0mm 3mm, width=0.75\columnwidth]{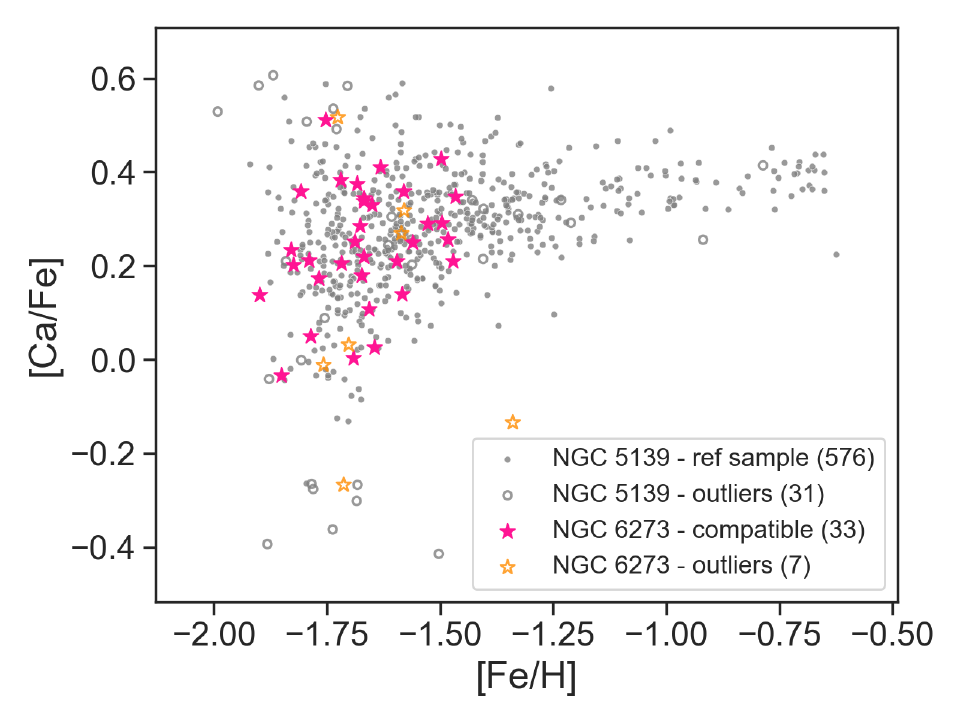}\par
\hspace{-25pt}\includegraphics[clip=true, trim = 1mm 0mm 0mm 2mm, width=0.76\columnwidth]{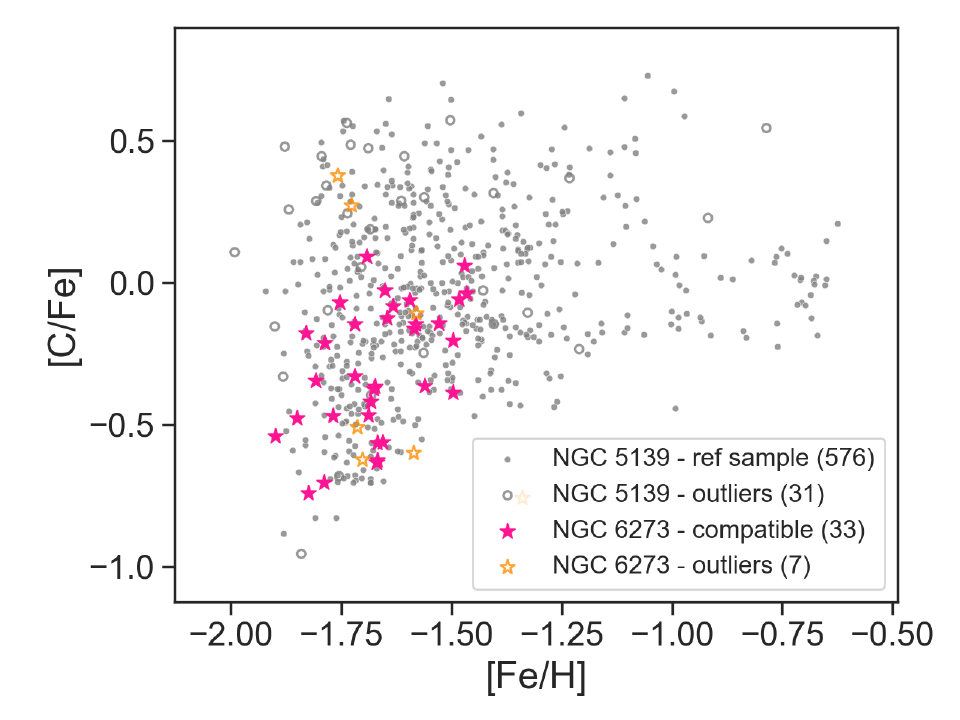}\hspace{-8pt}
\includegraphics[clip=true, trim = 1mm 0mm 0mm 1mm, width=0.76\columnwidth]{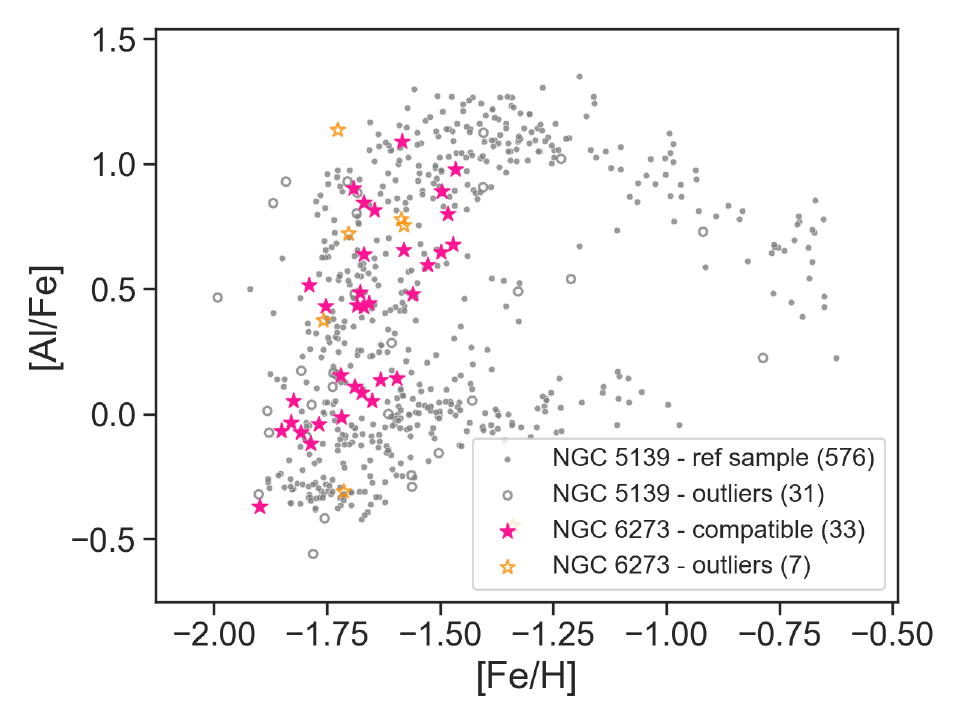}\hspace{-8pt}
\includegraphics[clip=true, trim = 1mm 0mm 0mm 1mm, width=0.76\columnwidth]{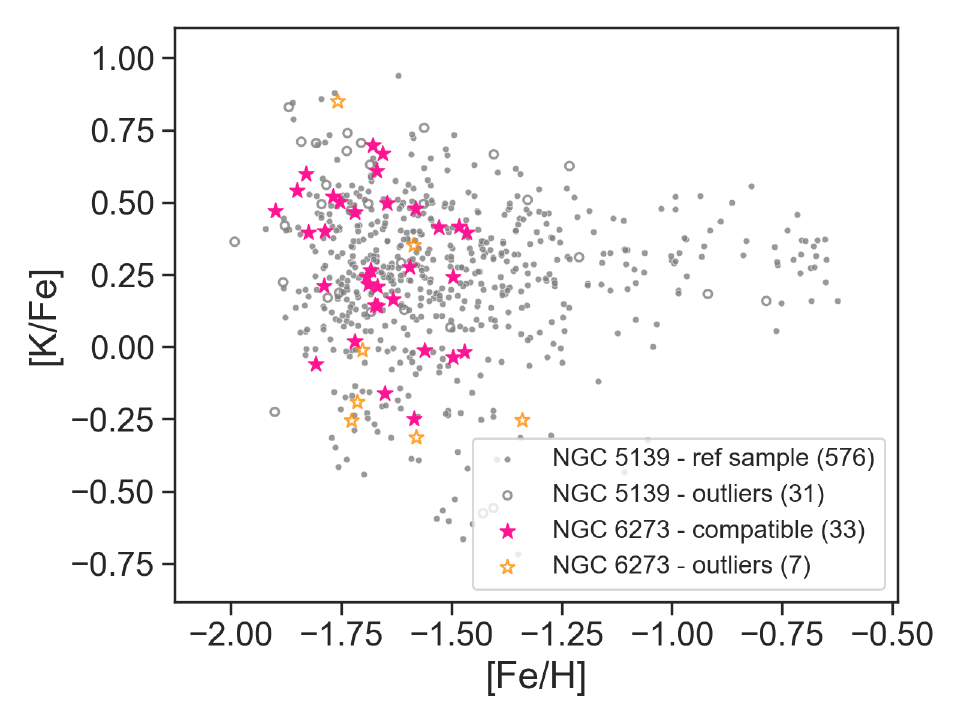}\par
\centering
\includegraphics[clip=true, trim = 1mm 0mm 0mm 1mm, width=0.75\columnwidth]{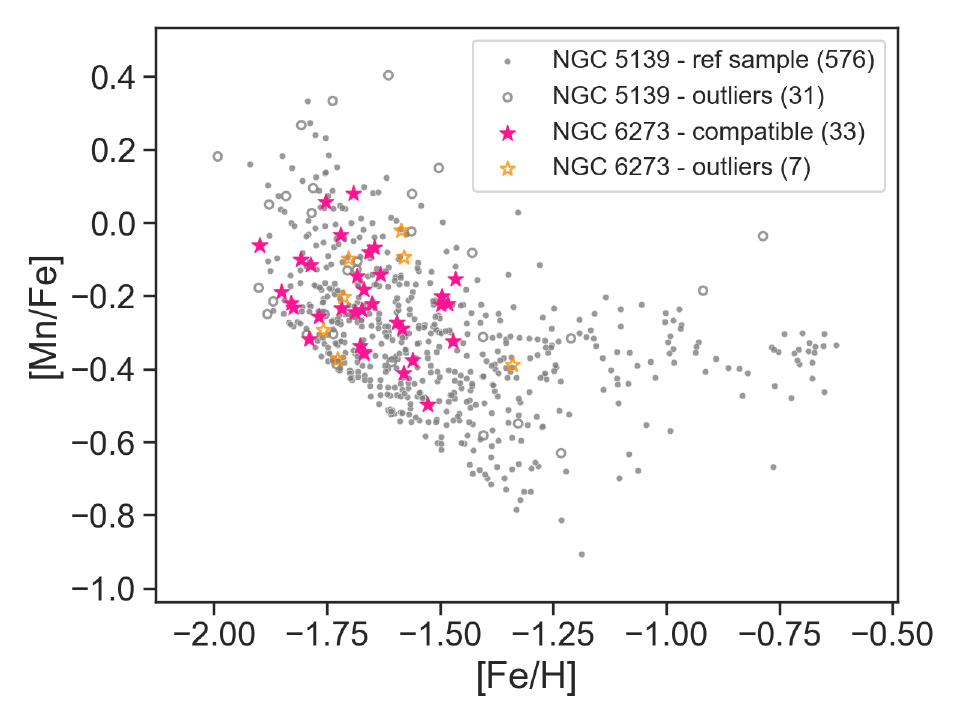}

  \caption{Same as Fig.~\ref{NGC6656} for NGC~6273.}
              \label{NGC6273}%
    \end{figure*}

 \begin{figure*}[h!]
 \hspace{-20pt}\includegraphics[clip=true, trim = 3mm 0mm 0mm 3mm, width=0.75\columnwidth]{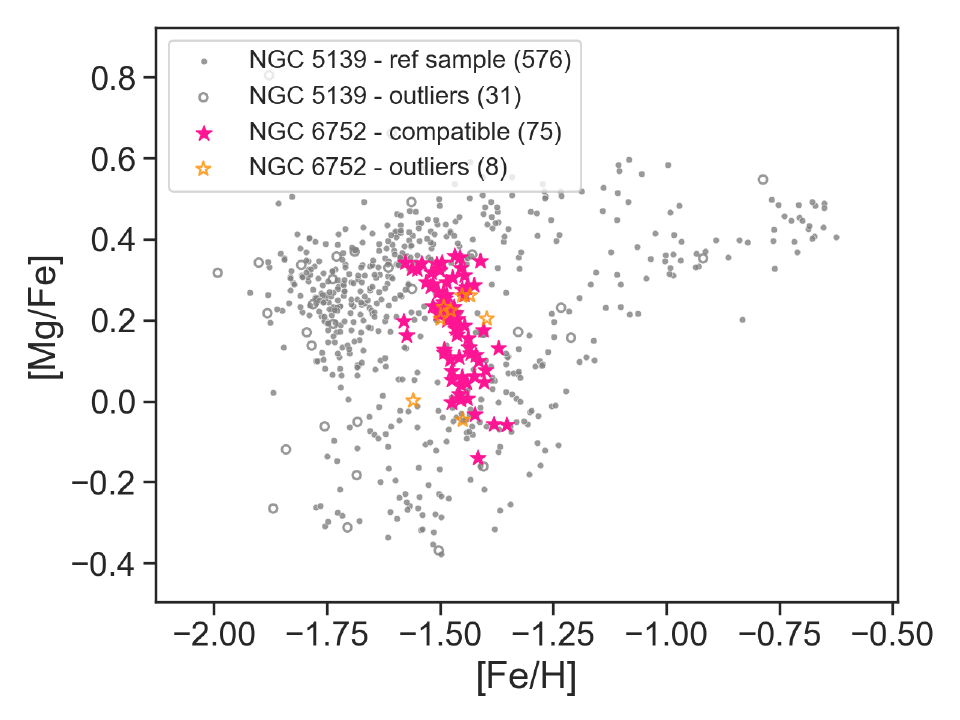}\hspace{-8pt}
\includegraphics[clip=true, trim = 3mm 0mm 0mm 3mm, width=0.75\columnwidth]{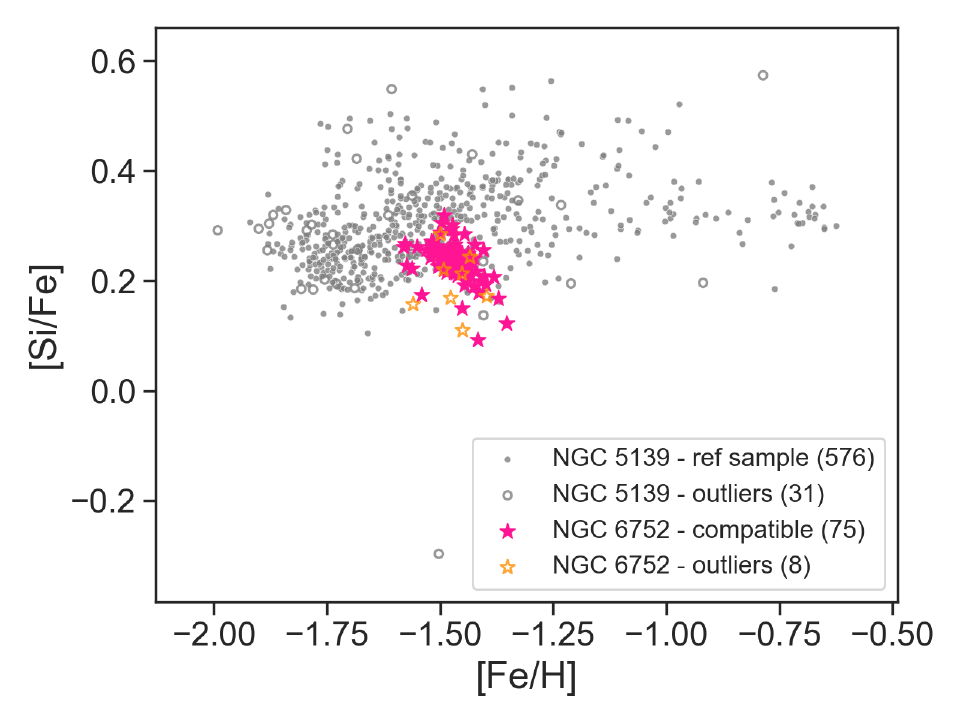}\hspace{-6pt}
\includegraphics[clip=true, trim = 3mm 0mm 0mm 3mm, width=0.75\columnwidth]{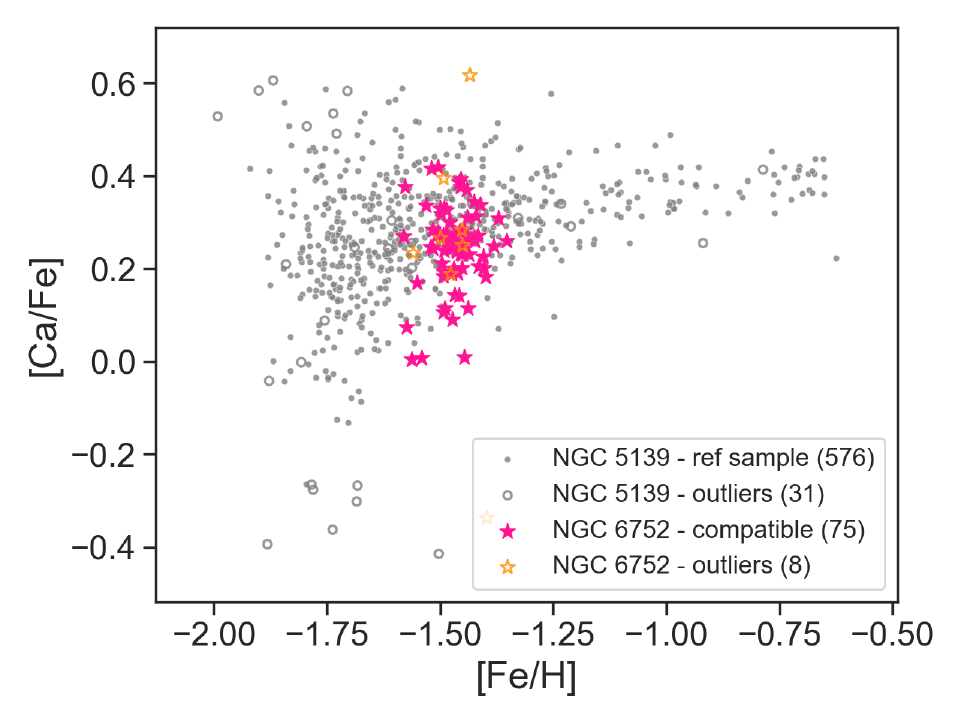}\par
\hspace{-25pt}\includegraphics[clip=true, trim = 1mm 0mm 0mm 2mm, width=0.76\columnwidth]{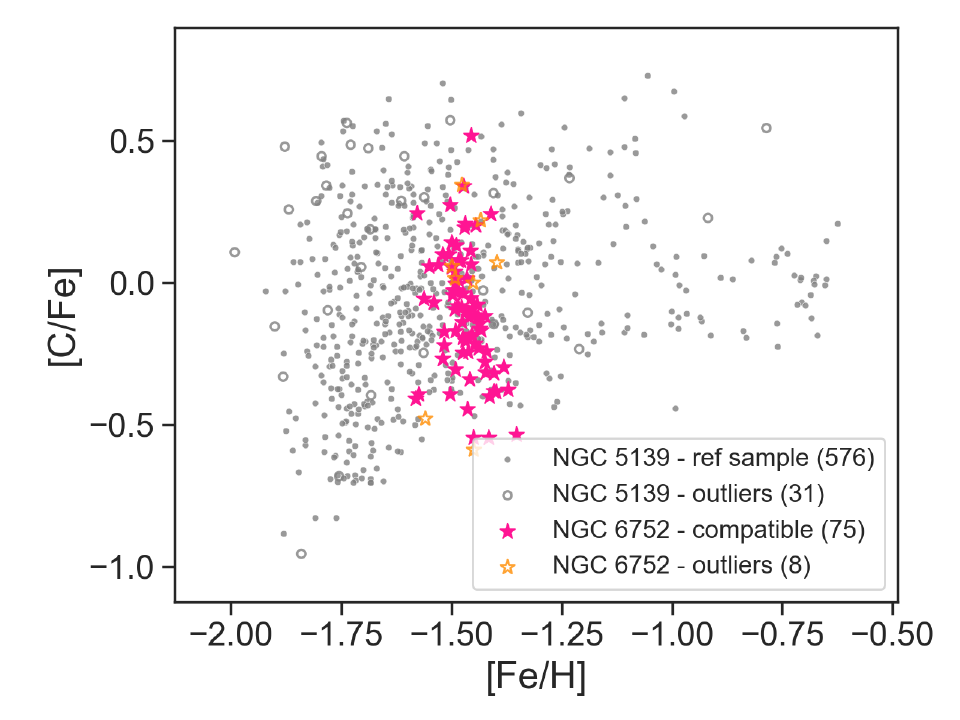}\hspace{-8pt}
\includegraphics[clip=true, trim = 1mm 0mm 0mm 1mm, width=0.76\columnwidth]{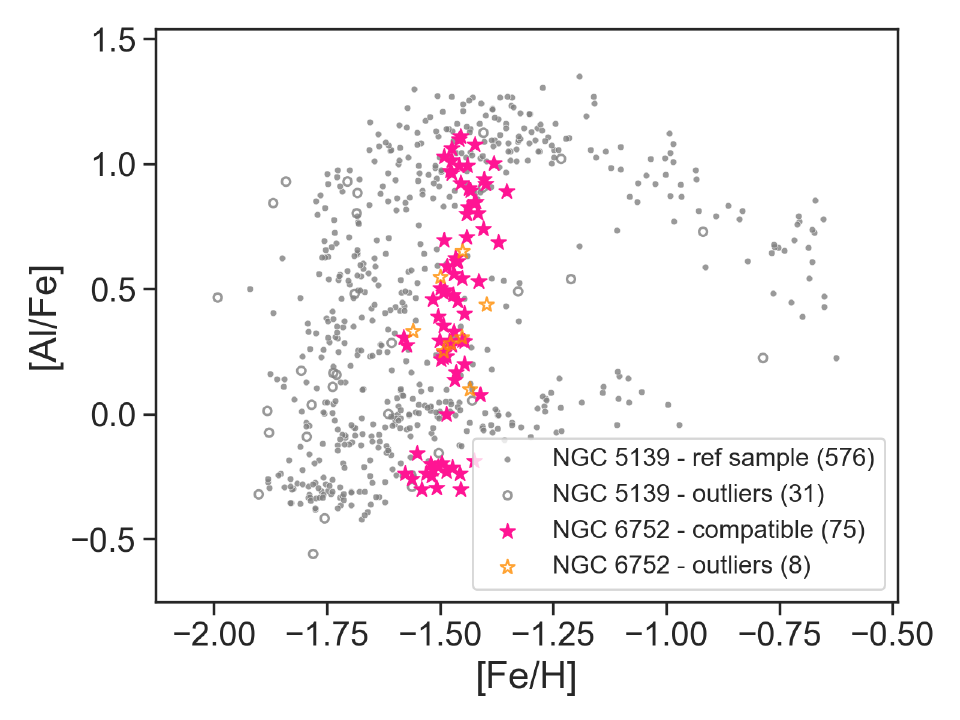}\hspace{-8pt}
\includegraphics[clip=true, trim = 1mm 0mm 0mm 1mm, width=0.76\columnwidth]{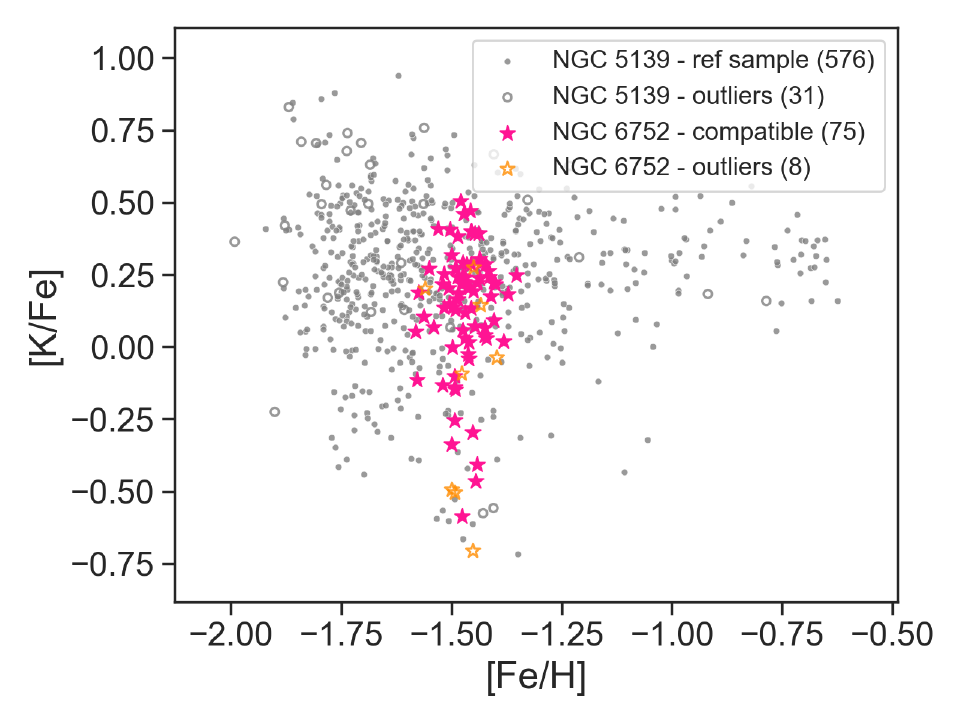}\par
\centering
\includegraphics[clip=true, trim = 1mm 0mm 0mm 1mm, width=0.75\columnwidth]{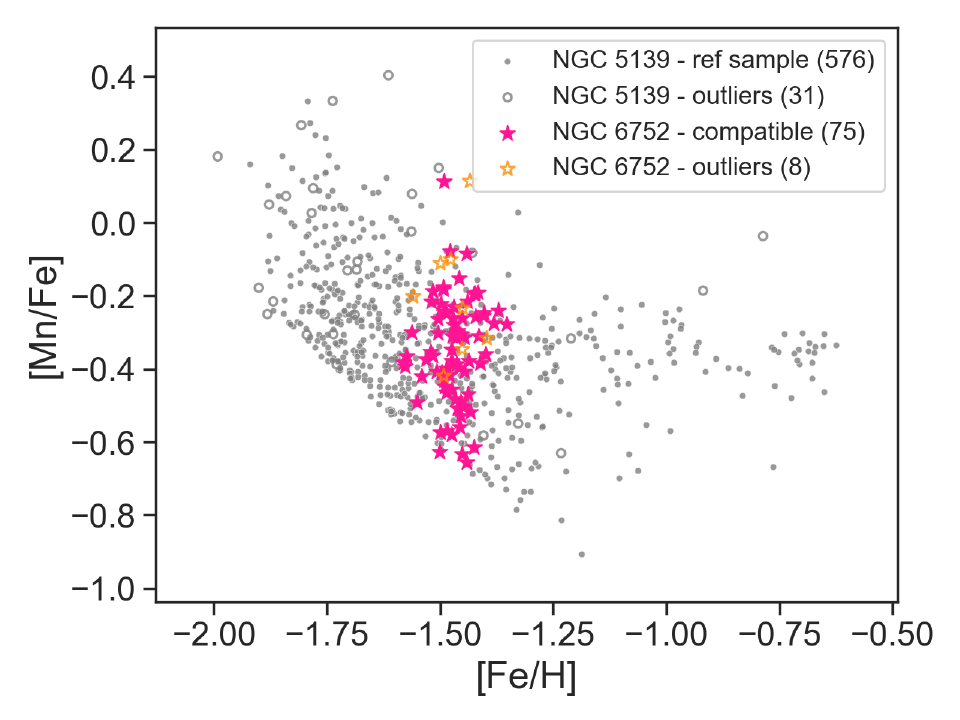}
  \caption{Same as Fig.~\ref{NGC6656} for NGC~6752.}
              \label{NGC6752}%
    \end{figure*}

\begin{figure*}
 \hspace{-20pt}\includegraphics[clip=true, trim = 3mm 0mm 0mm 3mm, width=0.75\columnwidth]{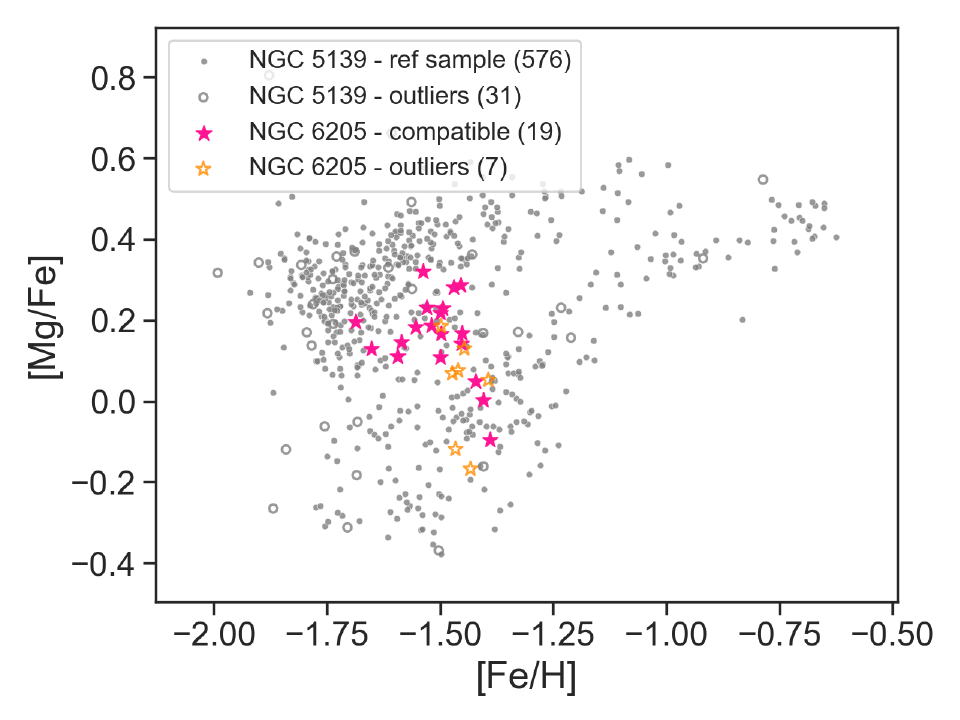}\hspace{-8pt}
\includegraphics[clip=true, trim = 3mm 0mm 0mm 3mm, width=0.75\columnwidth]{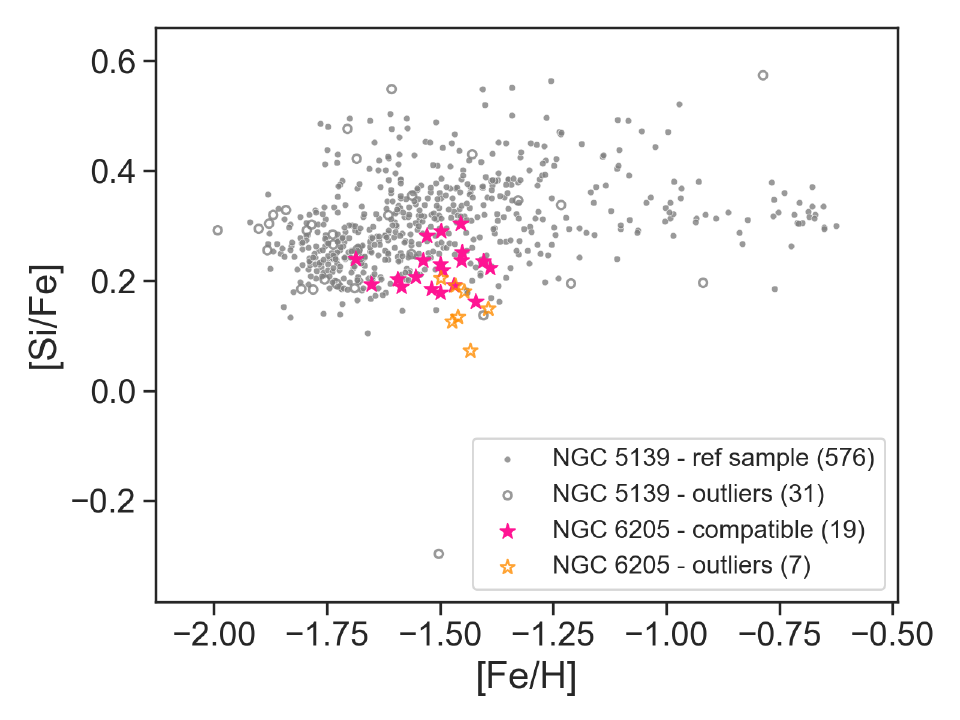}\hspace{-6pt}
\includegraphics[clip=true, trim = 3mm 0mm 0mm 3mm, width=0.75\columnwidth]{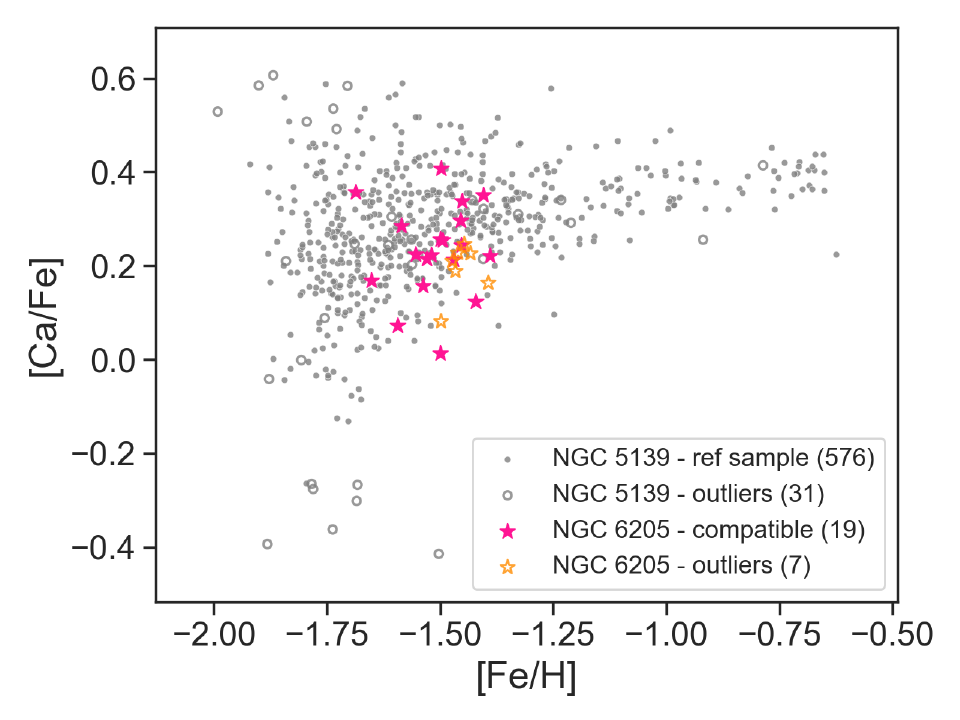}\par
\hspace{-25pt}\includegraphics[clip=true, trim = 1mm 0mm 0mm 2mm, width=0.76\columnwidth]{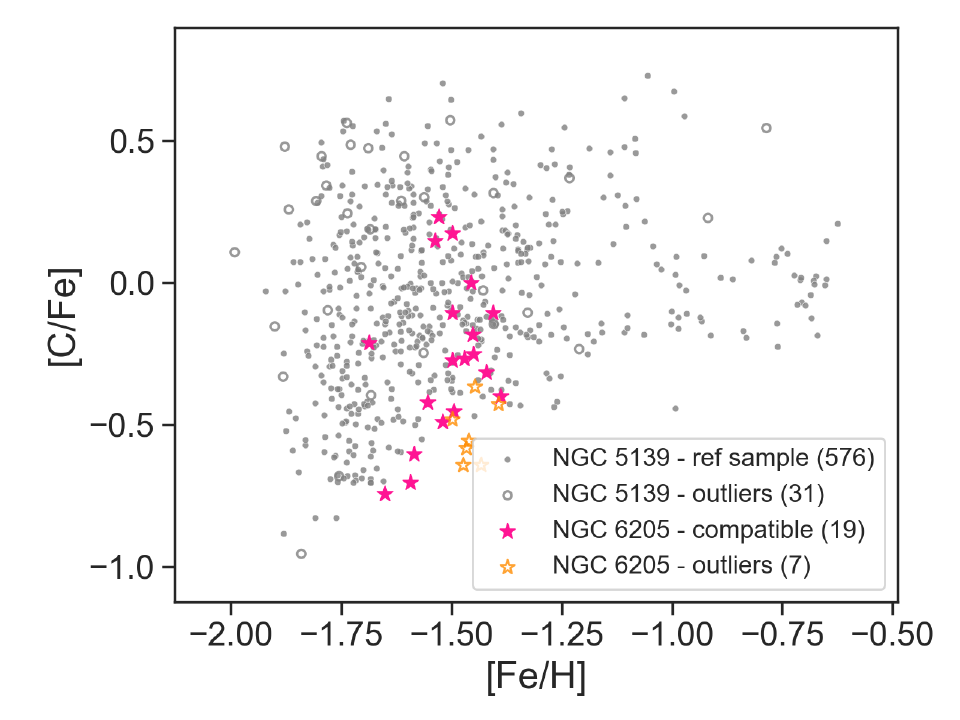}\hspace{-8pt}
\includegraphics[clip=true, trim = 1mm 0mm 0mm 1mm, width=0.76\columnwidth]{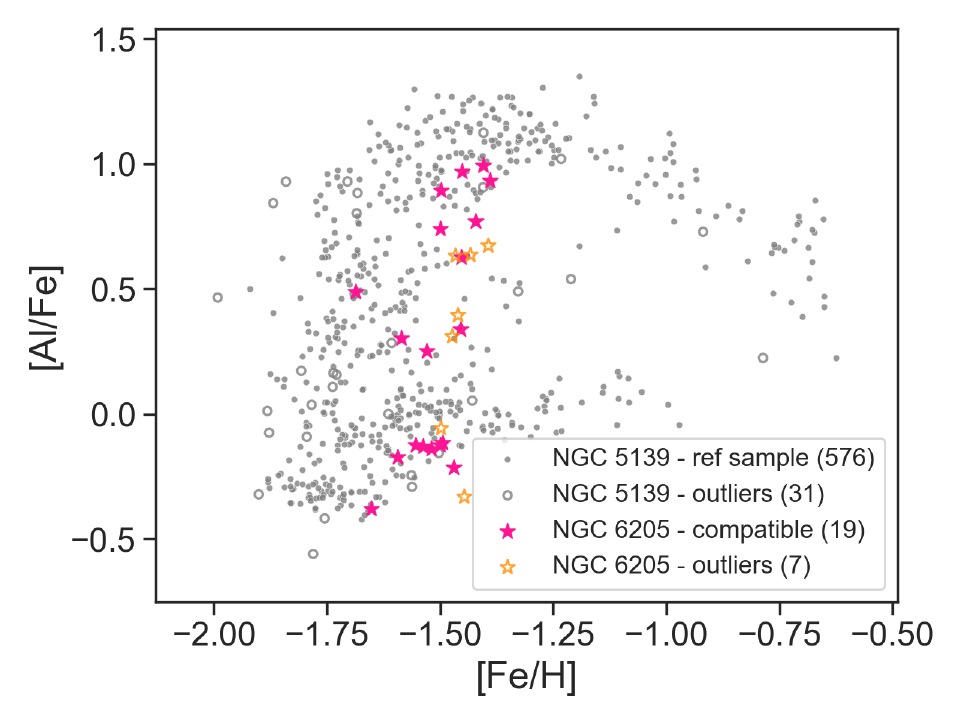}\hspace{-8pt}
\includegraphics[clip=true, trim = 1mm 0mm 0mm 1mm, width=0.76\columnwidth]{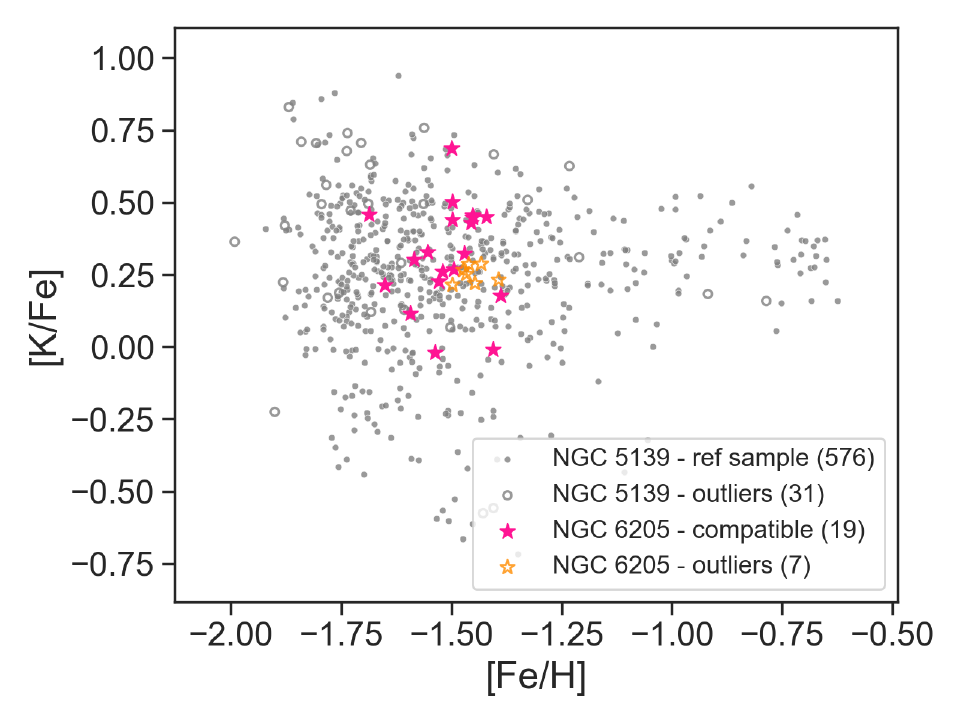}\par
\centering
\includegraphics[clip=true, trim = 1mm 0mm 0mm 1mm, width=0.75\columnwidth]{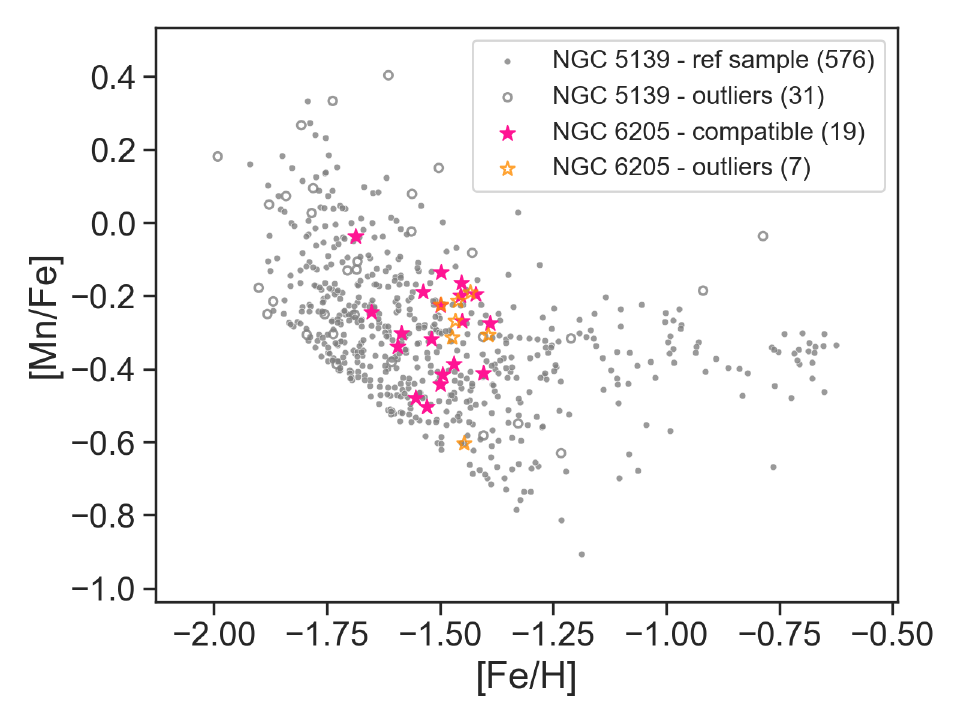}

  \caption{Same as Fig.~\ref{NGC6656} for NGC~6205.}
              \label{NGC6205}%
    \end{figure*}


 \begin{figure*}
 \hspace{-20pt}\includegraphics[clip=true, trim = 3mm 0mm 0mm 3mm, width=0.75\columnwidth]{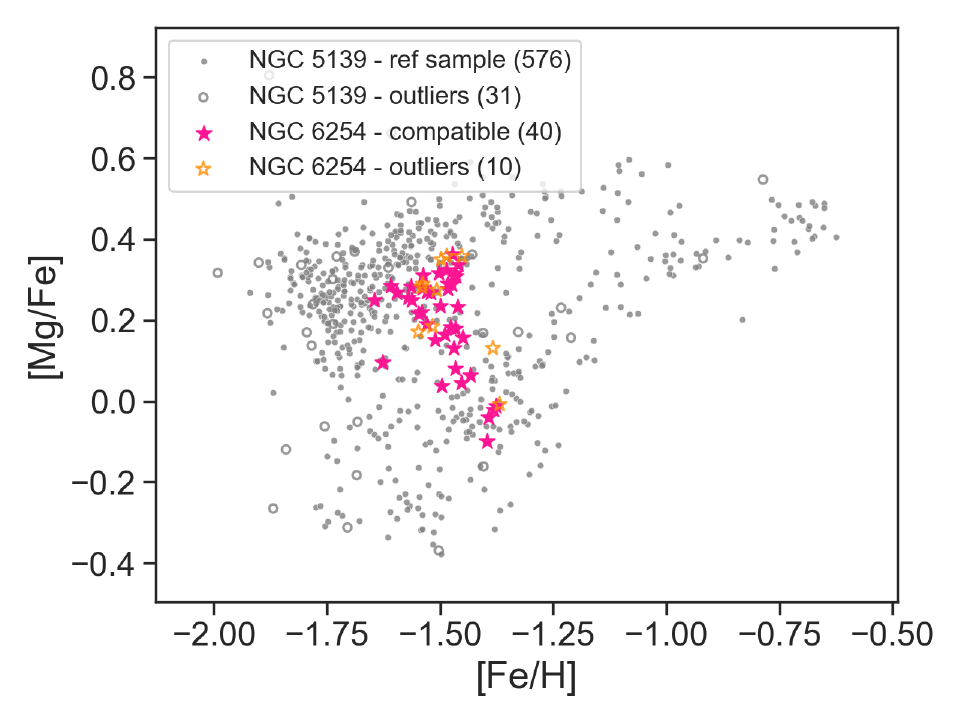}\hspace{-8pt}
\includegraphics[clip=true, trim = 3mm 0mm 0mm 3mm, width=0.75\columnwidth]{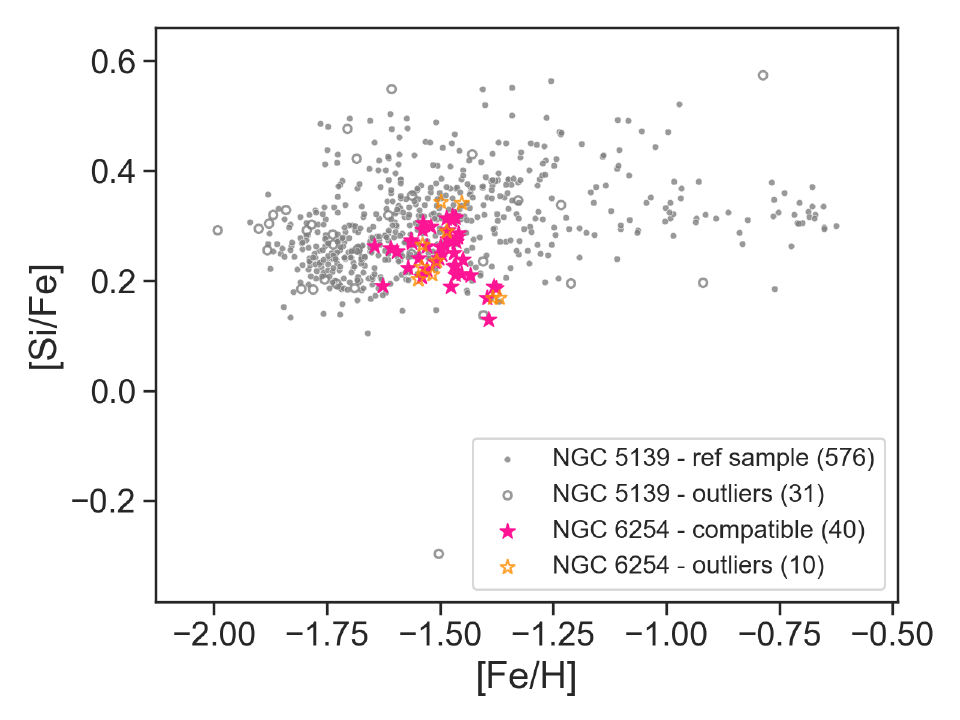}\hspace{-6pt}
\includegraphics[clip=true, trim = 3mm 0mm 0mm 3mm, width=0.75\columnwidth]{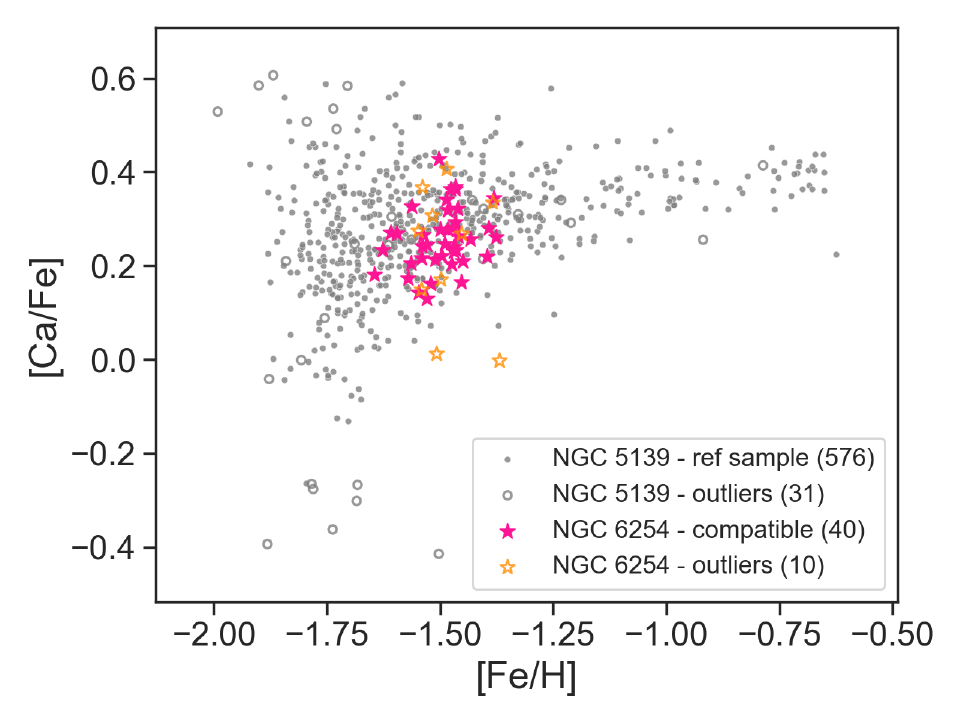}\par
\hspace{-25pt}\includegraphics[clip=true, trim = 1mm 0mm 0mm 2mm, width=0.76\columnwidth]{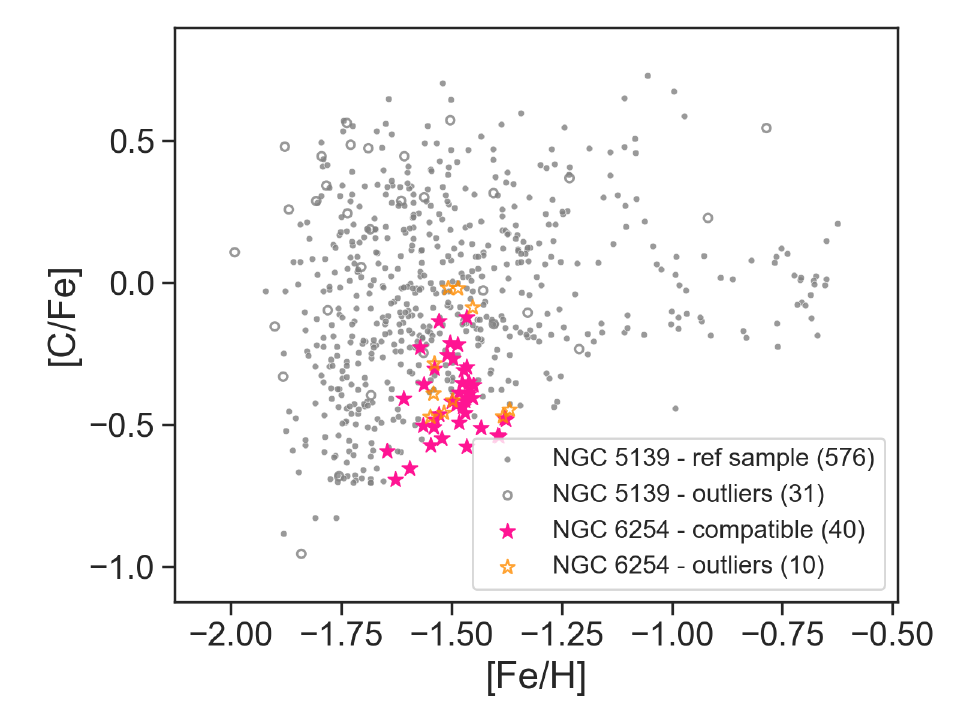}\hspace{-8pt}
\includegraphics[clip=true, trim = 1mm 0mm 0mm 1mm, width=0.76\columnwidth]{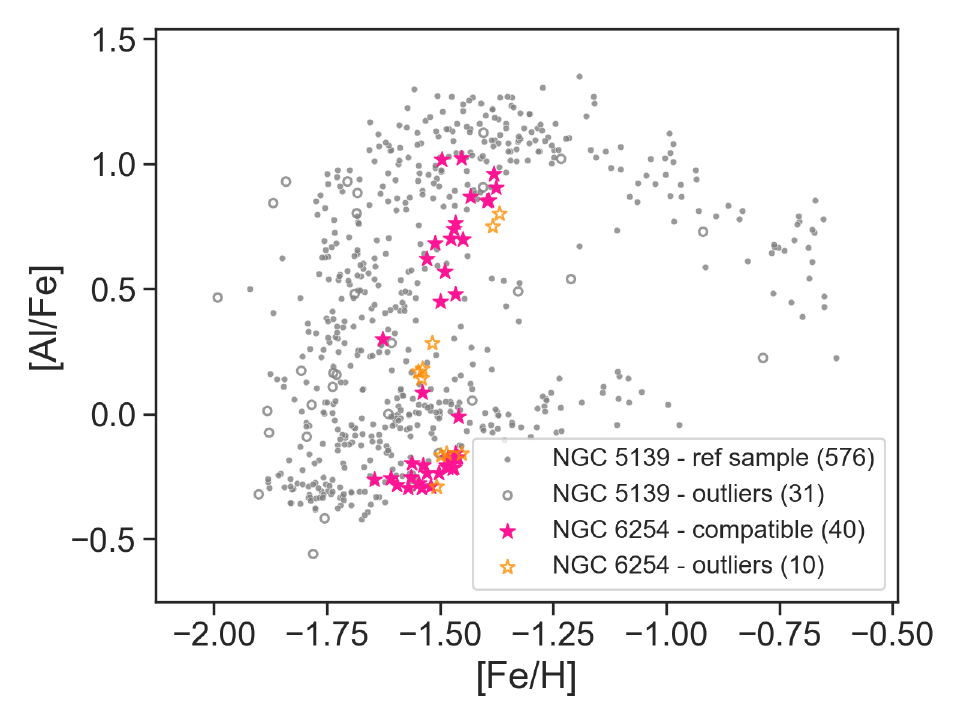}\hspace{-8pt}
\includegraphics[clip=true, trim = 1mm 0mm 0mm 1mm, width=0.76\columnwidth]{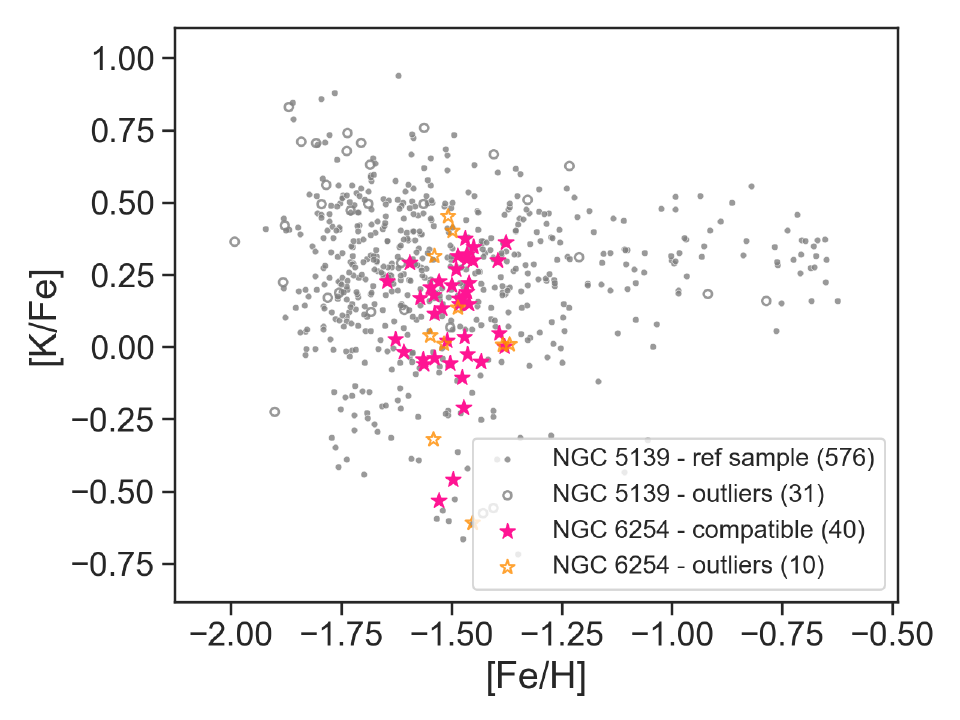}\par
\centering
\includegraphics[clip=true, trim = 1mm 0mm 0mm 1mm, width=0.75\columnwidth]{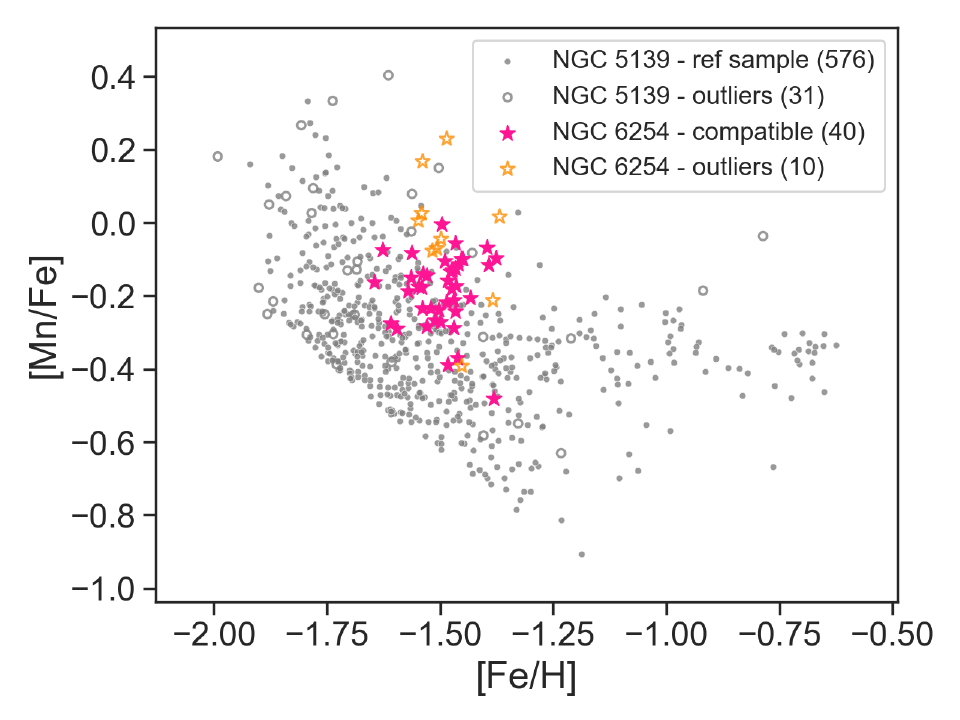}
  \caption{Same as Fig.~\ref{NGC6656} for NGC~6254.}
              \label{NGC6254}%
    \end{figure*}

For each cluster, the fraction of stars that are chemically compatible with $\omega$~Cen (in the 8-dimensional space defined in the previous section) out of the total is shown in Table~\ref{OCenGCs_table_VAC} with the corresponding uncertainty, respectively obtained by averaging over the fractions derived from the 100 GMM repetitions and computing the standard deviation. The total number of stars in each GC after the selections made (see Section \ref{obsdata}) is also listed, together with the median and mode of its [Fe/H] distribution. In this table, clusters are ranked according to the fraction of compatible stars, that is from clusters with the highest level of compatibility to clusters with the lowest. Note that we have proven that this ranking is robust even when changing the threshold or the number of elements considered in the GMM. Apart from the two first GCs reported in Table \ref{OCenGCs_table_VAC} (Ter~10 and NGC~2298), whose large fraction of stars which overlaps with $\omega$~Cen is based on the analysis of only 1 and 2 stars, respectively, result which we do not consider meaningful given the low statistics on which it is based, the third most compatible GC in the analysis is $\omega$~Cen itself. This result appears trivial, and indeed it is, but we checked this compatibility firstly to test that the whole procedure is correct and secondly to establish the compatibility fraction of two distributions (the training $\omega$~Cen 8-dimensional dataset and the ``GC=$\omega$~Cen"  8-dimensional dataset) drawn from the same initial set of stars through the bootstrap process described in the previous section. With a threshold for outliers detection fixed at the 5th percentile of the log-likelihood distribution of the $\omega$~Cen GMM model, we see that the fraction of the reference sample of stars in $\omega$~Cen, averaged over 100 bootstraps, is 89\%. This compatibility fraction sets essentially the upper limit that we can expect to find for all other clusters, compared to $\omega$~Cen.

\begin{figure*}[h!]
\includegraphics[width=0.7\columnwidth]{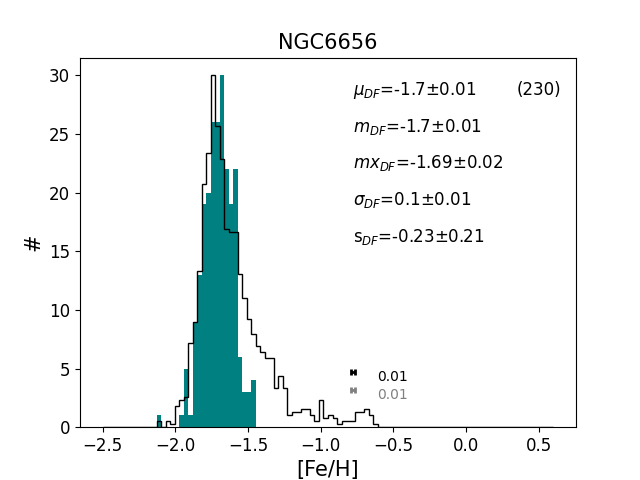}
\includegraphics[width=0.7\columnwidth]{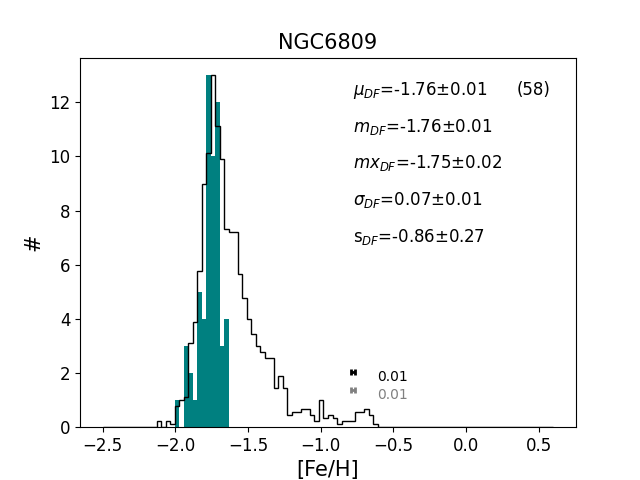}
\includegraphics[width=0.7\columnwidth]{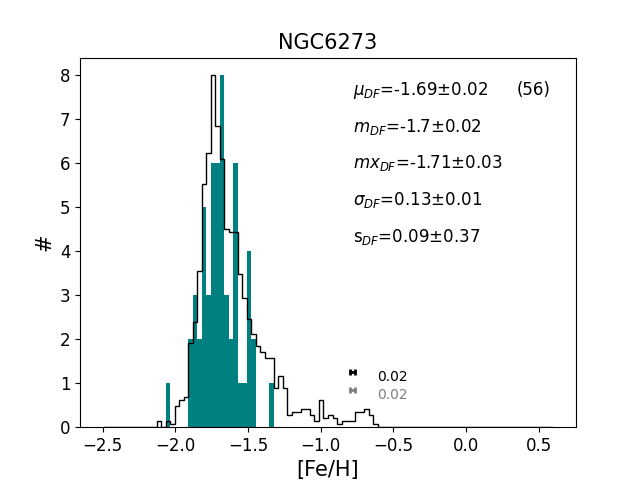}
\includegraphics[width=0.7\columnwidth]{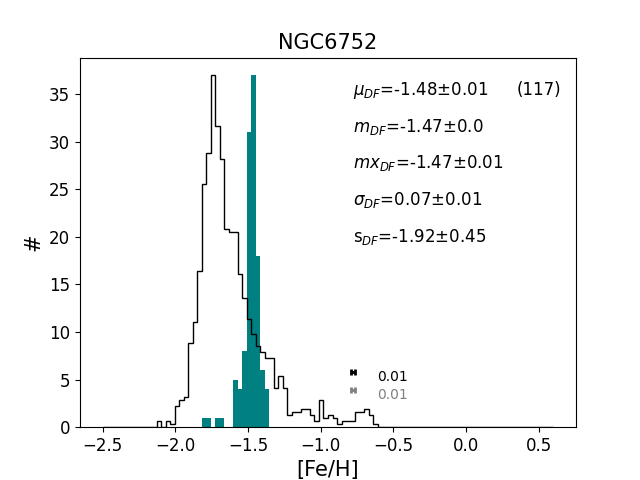}
\includegraphics[width=0.7\columnwidth]{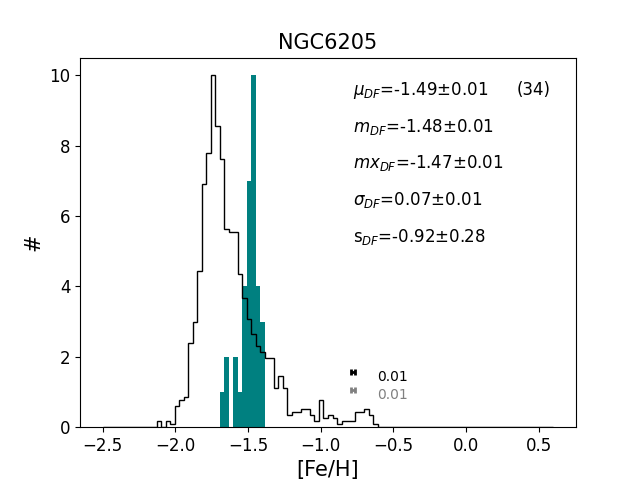}
 \includegraphics[width=0.7\columnwidth]{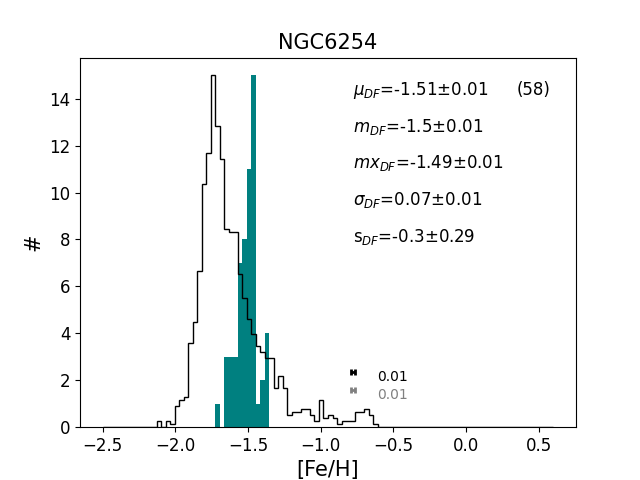}
  \includegraphics[width=0.7\columnwidth]{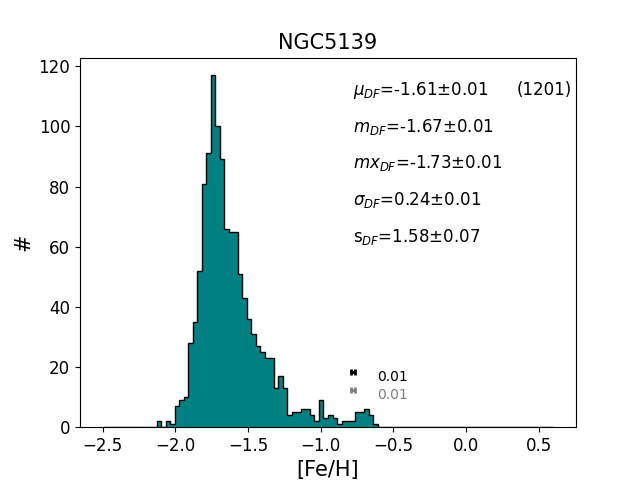}
\caption{[Fe/H] distribution of the Galactic globular clusters NGC~6656, NGC~6809, NGC~6273 (the metal-poor clusters, first row),  NGC~6752, NGC~6205 and NGC~6254 (the metal-rich clusters, second row), which -- according to our analysis -- are chemically compatible with $\omega$~Cen. For each cluster, the values of the mean, median, mode, dispersion, and skewness of the distribution, and their relative uncertainties, are also reported. The black and grey horizontal error bars in the bottom of each plot show the median and mean uncertainties in [Fe/H], as derived from the \citet{schiavon2023} catalogue. In each plot, the value reported in parenthesis gives the number of stars used to trace the histogram. For comparison, in each plot, the [Fe/H] distribution of $\omega$~Cen (NGC~5139) is also reported (black step-like histogram). The latter has been normalised in such a way that the maximum value coincides with the maximum of the [Fe/H] distribution of the cluster to which it is compared (hence the normalisation varies from plot to plot). The absolute [Fe/H] distribution of $\omega$~Cen (i.e. without the normalisation of the peak value) is reported in the last row, with the corresponding mean, median, mode, dispersion, and skewness of its distribution. }\label{70f_MDF}
\end{figure*}

\begin{figure*}[h!]\hspace{-20pt}
\includegraphics[width=0.7\columnwidth]{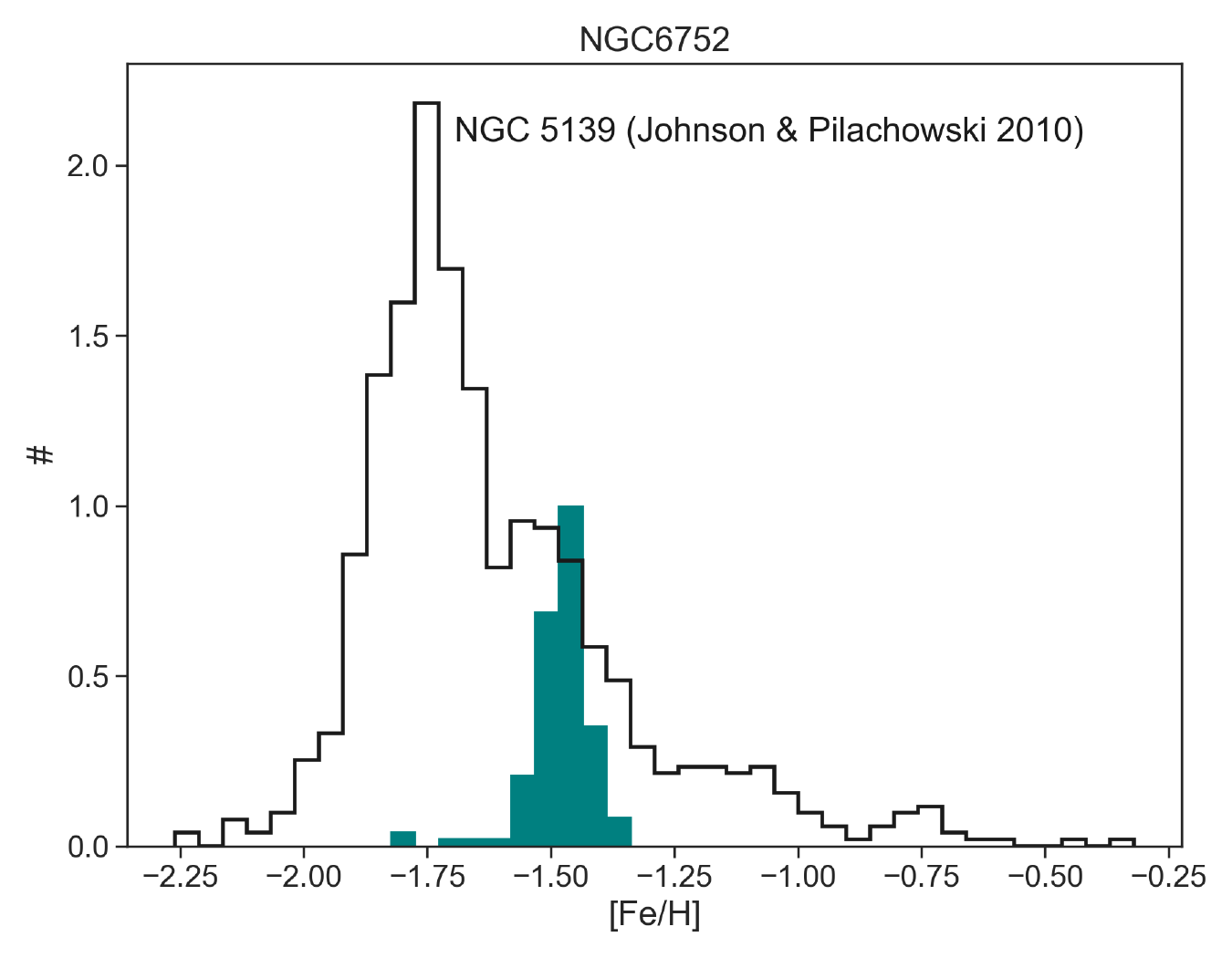}
\includegraphics[width=0.7\columnwidth]{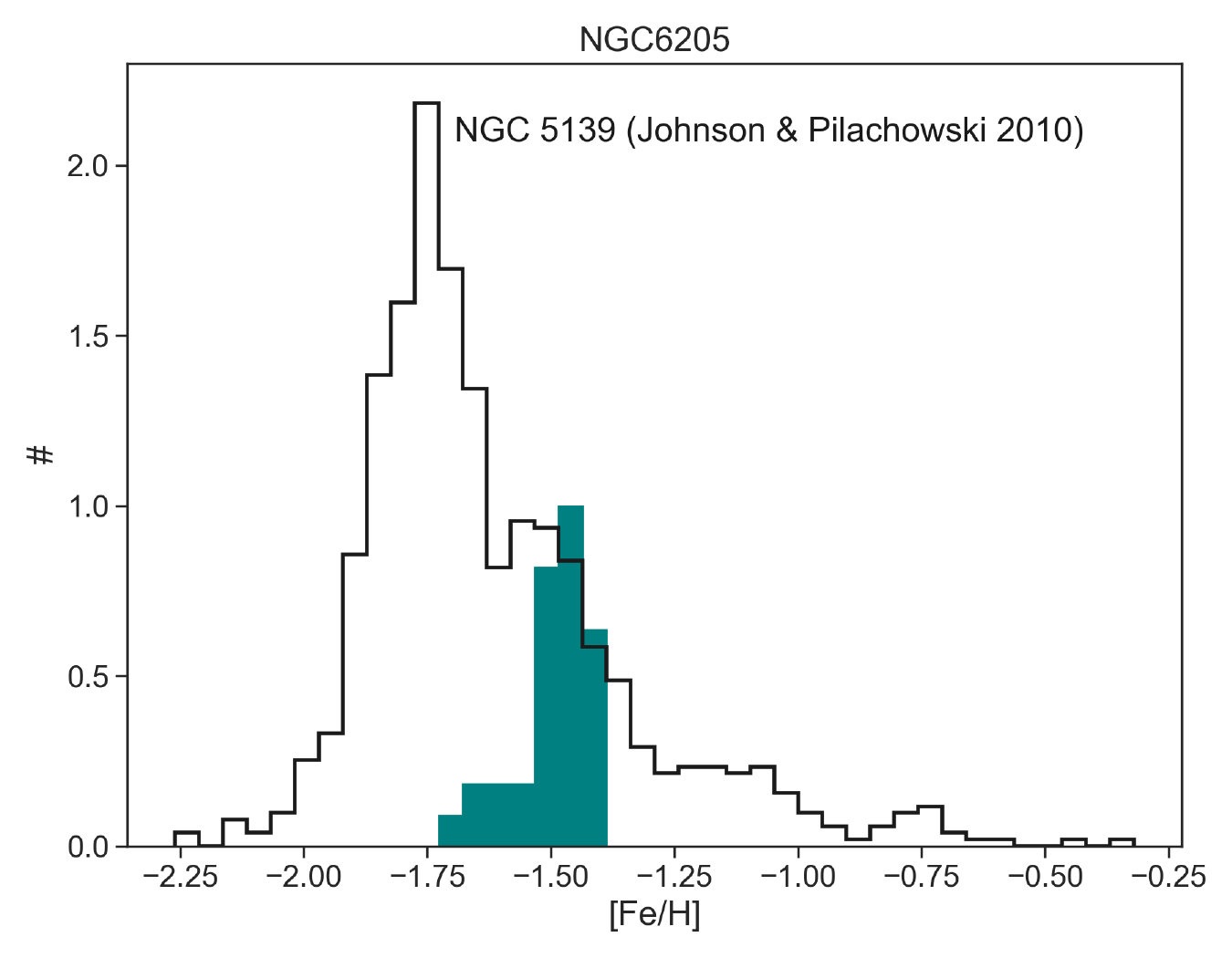}
\includegraphics[width=0.7\columnwidth]{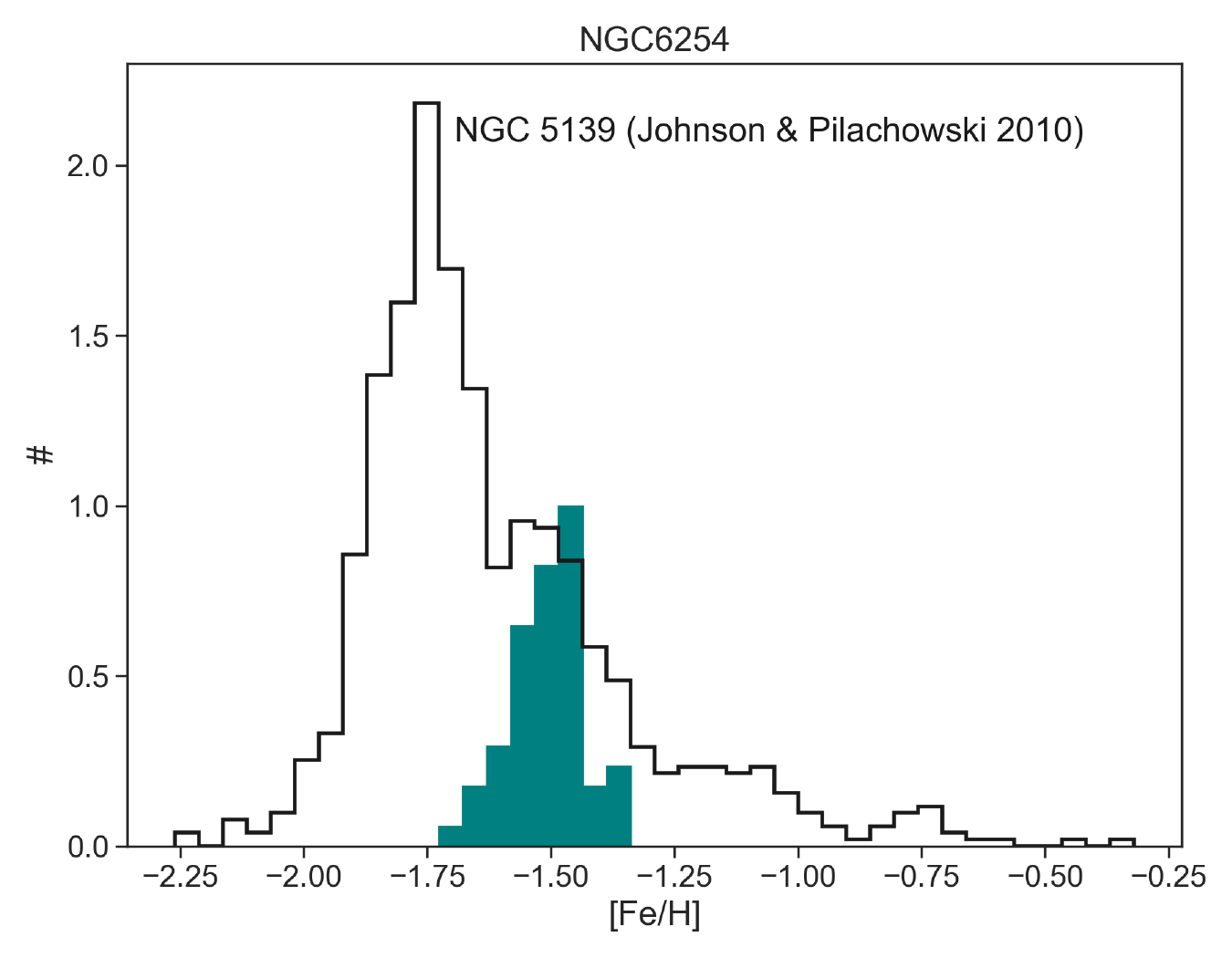}
\caption{[Fe/H] distribution of the metal-rich globular clusters that are chemically compatible with $\omega$~Cen, namely NGC 6752, NGC 6205 and NGC 6254.  For comparison, in each plot, the [Fe/H] distribution of $\omega$~Cen (NGC 5139) as reported in \citet{johnson2010} is also shown (black step-like histogram). The latter has been normalised in such a way that the secondary peak value coincides with the maximum of the [Fe/H] distribution of the cluster to which it is compared.}\label{mdfs_johnson}
\end{figure*}

\subsection{[X/Fe] versus [Fe/H] planes for GCs chemically compatible with $\omega$~Cen}

Apart from $\omega$~Cen itself, the GMM finds four GCs (with a total number of stars greater than 15) that have a fraction of stars compatible with $\omega$~Cen greater than 70\%, namely: NGC~6752, NGC~6809 (M 55), NGC~6273 (M 19), and NGC~6656 (M 22). Three additional clusters -- namely NGC~6254 (M 10), NGC~5024, and NGC~6205 (M 13) -- have a fraction of stars chemically compatible with $\omega$~Cen between 60\% and 70\%, however, the limited number of stars (5) available for NGC~5024 prevents us from drawing strong conclusions for this cluster.
In Figs.~\ref{NGC6656}, \ref{NGC6752}, \ref{NGC6809}, \ref{NGC6273},  \ref{NGC6205},  and  \ref{NGC6254}, we report the distributions in [X/Fe] versus [Fe/H] spaces for all these clusters (except for NGC~5024 for the reason given above), and for all [X/Fe] abundances used to build the $\omega$~Cen GMM model. We emphasise that while these figures report 2D projections of abundance spaces, the GMM model has been built by making use of the whole 8-dimensional abundance space, and not by estimating the chemical compatibility for each of these projections separately. In all panels of these figures, the corresponding distribution of $\omega$~Cen stars is also reported for comparison. For all clusters in these figures, compatible stars and outliers are shown with different colours or symbols, to visually show the result of the GMM classification.\\
Examination of Figs.~\ref{NGC6656} to \ref{NGC6254} reveals some interesting characteristics that lead us to classify the clusters chemically compatible with $\omega$~Cen into two groups:

\begin{enumerate}
\item the first group consisting of NGC~6656, NGC~6809, and NGC~6273:  stars of these GCs are distributed in the abundance spaces in the same regions where the density of $\omega$~Cen is highest, i.e. these clusters share the chemical properties of most of the stars today in $\omega$~Cen. Because, as we will discuss in Sect.~\ref{mdf},  the MDF of all these GCs peaks at [Fe/H]$\simeq -1.7$ (see upper row of Fig.~\ref{70f_MDF}), in the following of this paper we will refer to these GCs as  ``metal-poor clusters".
\item the second group consisting of NGC~6752, NGC~6205, and NGC~6254: stars of these GCs are distributed in the abundance spaces in regions where $\omega$~Cen stars are still present (otherwise these GCs would not appear as chemically compatible with $\omega$~Cen), but in smaller proportions. Their MDF peaks at [Fe/H]$\simeq -1.5$ (see middle row of Fig.~\ref{70f_MDF}) and for this reason in the following we will refer to these GCs as ``metal-rich clusters''. Note that, as we will discuss in Sect.~\ref{mdf}, the peak of the MDF of these clusters coincides with one of the secondary peaks of $\omega$~Cen MDF, which is often referred to in the literature as one of the ``metal-intermediate'' components of $\omega$~Cen \cite[RGB-Int1 or MI1 or M-int1, see for example][]{sollima05, calamida2020, alvarez2024}. 
This group of GCs is in no way associated to the metal-rich population of $\omega$~Cen, which is mostly at [Fe/H]$\gtrsim -1$ \citep[see, for example,][]{alvarez2024}. 
\end{enumerate}

Before proceeding with a more detailed description of the characteristics of these clusters in the abundance spaces used, we would like to point out that not all GCs in the metallicity range of $\omega$~Cen are chemically compatible with it. For example, NGC~5272, Ter~9, and NGC~3201 have a very low degree of compatibility with $\omega$~Centauri (respectively 11$\pm$7\%, 10$\pm$15\% and
8$\pm$8\%, see Table~\ref{OCenGCs_table_VAC}), despite the modes and medians of their [Fe/H] distributions are in the  range of $\omega$~Cen
stars — more specifically in between the 1st (-1.83~dex) and 9th (-1.31~dex) deciles of
the $\omega$~Cen MDF. A comparison of the distributions in the [X/Fe] - [Fe/H] spaces of these clusters and $\omega$~Cen is given in Sect~\ref{no_ocen-like_gcs}.\\

\subsubsection{Metal-poor clusters}
Stars of the metal-poor GCs NGC~6656 and NGC~6273 are distributed along a very extended and ``thick"\footnote{By ``thick" in this context we mean that the distribution is extended in [Fe/H] at all values of [Al/Fe]} branch in the [Al/Fe]-[Fe/H] plane, and their distributions in the [Mg/Fe], [Si/Fe], [Ca/Fe], [C/Fe], and [K/Fe] versus [Fe/H] planes are visually quite broad.  NGC~6809 shows very similar characteristics with the two above cited clusters, except for the distribution in the [Al/Fe]-[Fe/H] plane, which still results very extended, however not as ``thick" as those of NGC~6656 and NGC~6273.
All these metal-poor GCs appear to have a very extended distribution in the [C/Fe] versus [Fe/H] plane, with a significant fraction of their stars at sub-solar [C/Fe] values, as expected for red giants during their evolution \citep[see][]{iben67}.  In NGC~6809, in particular, only a star has a super-solar [C/Fe]. Both the ``thinness" of the distribution in the [Al/Fe]-[Fe/H] plane and the absence of [C/Fe]-rich stars in this cluster may be due to the lower number of stars available for the analysis in this GC (18 -- this number rises respectively to 40 and 68, for NGC~6273 and  NGC~6656, see Table~\ref{OCenGCs_table_VAC}). \\
$\omega$~Cen shows a handful of stars (1.3\% of the total) at [Ca/Fe]$\lesssim -0.2$~dex for [Fe/H] values similar to those of the metal-poor GCs. Interestingly, a couple of stars with such low values of [Ca/Fe] are found also in NGC~6656 and NGC~6273, in a proportion similar to that of the [Ca/Fe] deficient population of $\omega$~Cen. Note that these stars are classified as outliers by the GMM analysis (orange empty asterisks in Figs.~\ref{NGC6656} and \ref{NGC6273}), since, given the threshold chosen to define the outliers population, also stars in $\omega$~Cen with similar low values of [Ca/Fe] result to be outliers (empty grey circles in Figs.~\ref{NGC6656} and \ref{NGC6273}). 
Among the other $\alpha$-elements used for the GMM analysis, it is worth commenting on the [Mg/Fe] trends with [Fe/H] for these metal-poor GCs. None of these clusters shows a correlation between the two abundances, except for NGC~6809 which seems to show a decrease of the [Mg/Fe] ratio with increasing [Fe/H] (see top left panel in Fig.~\ref{NGC6809}). It is worth noting that $\omega$~Cen has a fraction of 15\% of stars with sub-solar [Mg/Fe] ratios. Except for NGC~6273, for which about 10\% of stars has [Mg/Fe]$<0$, neither NGC~6656 nor NGC~6809 contains stars with such low values of [Mg/Fe].  Finally, all the metal-poor GCs show the same, nearly flat trend in the [Si/Fe] versus [Fe/H] plane, as observed in $\omega$~Cen. 

\subsubsection{Metal-rich clusters}
NGC~6752, NGC~6205, and NGC~6254, the metal-rich clusters, show a number of abundance versus [Fe/H] trends similar to those found for the metal-poor GCs (see for example the extended distribution in the [Al/Fe]-[Fe/H] plane, which ranges from sub-solar [Al/Fe] ratios to [Al/Fe]$\sim$ 1 in all three clusters), despite all these GCs being more metal-rich than the previous ones. However, they also show some peculiar features that it is worth mentioning:
\begin{itemize}
\item a decreasing trend of [Mg/Fe] with increasing [Fe/H], which is particularly prominent in NGC~6752, for which we estimate\footnote{By making use of the scipy spearmanr function, see \url{https://docs.scipy.org/doc/scipy/reference/generated/scipy.stats.spearmanr.html}} a Spearman correlation coefficient of $-0.42$, and a p-value of $2\times 10^{-6}$. In this cluster, an anti-correlation is found also for the [Si/Fe] versus [Fe/H] relation (Spearman coefficient equal $-0.34$ and p-value equal to $4\times 10^{-4}$). For NGC~6205 and NGC~6254, milder [Mg/Fe]-[Fe/H] and [Si/Fe]-[Fe/H] anti-correlations are present\footnote{For NGC~6205: Spearman correlation coefficients respectively equal to $-0.40$ and $-0.17$, with p-values of $0.018$ and $0.33$; for NGC~6254:  Spearman correlation coefficients respectively equal to $-0.26$ and $-0.21$, and corresponding p-values of $0.05$ and $0.12$}. For these two GCs, in fact, two trends seem to co-exist, a rising [Mg/Fe] (and [Si/Fe])-[Fe/H] relation (at [Fe/H]$\lesssim-1.55$) and a decreasing one (at [Fe/H]$\gtrsim-1.55$). That is, the trends in [Si/Fe] versus [Fe/H] (but also of [Mg/Fe] versus [Fe/H]) for these clusters are more complex than a simple monotonic description.   Besides the strength of the anti-correlations, however, in these clusters the most metal-poor stars\footnote{Defined as the 1st decile of the distribution} have median [Mg/Fe] and [Si/Fe] abundance ratios usually between 0.1 and 0.2~dex greater than those of the most metal-rich stars\footnote{Defined as the 9th decile of the distribution} (see first and second panels, top rows, of Figs.~\ref{NGC6752}, \ref{NGC6205} and \ref{NGC6254}). 
\item both NGC~6752 and  NGC~6254 show an extremely [K/Fe]-poor population ([K/Fe]$\lesssim -0.3$~dex) which constitutes less than 10\% of the total, and which is present, at similar [Fe/H] values, also in $\omega$~Cen.
\end{itemize}

It is worth mentioning that while we have based the GMM analysis on 8 elemental abundances, as discussed in Sect.~\ref{method}, the \citet{schiavon2023} catalogue includes several additional abundance ratios whose analysis allows to test and indeed strengthens the results presented here. This supplementary discussion is presented in Appendix~\ref{ocen-like_other_elems}, both for metal-poor and metal-rich GCs.\\

Finally, it is also important at this point to clarify an element of nomenclature. In this section, we have sometimes referred to distributions that appear ``broad'' in certain [X/Fe] ratios. Obviously, these distributions are not necessarily inherently broad. This depends on the associated ASPCAP uncertainties. We refer the interested reader to Appendix~\ref{spreads}, for an in-depth analysis on [X/Fe] distributions and their intrinsic dispersions, while we discuss the [Fe/H] distributions of the metal-poor and metal-rich clusters below, because - as we shall see - they show interesting similarities between each other and with $\omega$~Centauri itself.

\subsection{[Fe/H] distributions of GCs chemically compatible with $\omega$~Cen: their rise and fall}\label{mdf}

In this section, we analyse the distribution in [Fe/H] of stars in metal-poor and metal-rich clusters, for which we found a strong chemical compatibility with $\omega$~Centauri. Before proceeding with the presentation of this analysis, and the associated discussion, it is necessary to clarify one point. The approach used to establish (or not) a chemical compatibility between the clusters of the \citet{schiavon2023} catalogue and $\omega$~Cen is based on an analysis of the multi-dimensional chemical domain of $\omega$~Cen. Once this domain is defined (via the GMM), we define as compatible those clusters for which a high fraction of their stars falls within the  $\omega$~Cen chemical domain. At no point in this procedure does a density criterion in chemical spaces come into account. That is, it is sufficient for the 8-dimensional pattern of a cluster to be contained in the domain of $\omega$~Cen for this cluster to be compatible with it, regardless of the density of this pattern with respect to that of $\omega$~Cen itself. At no stage in this procedure, therefore, is it required, for example, that two distributions in chemical abundances - that of a chemically compatible cluster, and that of $\omega$~Cen - be similar. Yet, as we shall see below, interesting similarities do exist.\\

Fig.~\ref{70f_MDF} shows the [Fe/H] distributions of the six GCs which are chemically compatible with $\omega$~Cen, according to our analysis. For each of these clusters, we report also the mean, median, mode, dispersion, and skewness of the distributions, with their corresponding uncertainties, estimated by bootstrapping all stars of each cluster 100 times, and taking into account in the bootstrap also a Monte-Carlo sampling of the [Fe/H] uncertainties.
To draw these distributions, we have included only stars in these six clusters which satisfy the conditions described in Sect.~\ref{obsdata} and which have the APOGEE \texttt{FE\_H\_FLAG=0}. In each panel of Fig.~\ref{70f_MDF}, we also report the corresponding  [Fe/H] distribution of $\omega$~Cen stars, rescaled in such a way that its normalised maximum value coincides with the maximum value of the [Fe/H] distribution of the corresponding GC. \\
There are several characteristics common to many, sometimes all, these clusters:
\begin{itemize}
    \item all of them show total [Fe/H] dispersions, $\sigma_{\rm{[Fe/H], tot}}$, comprised between 0.07 and 0.13~dex. Given the median of [Fe/H] uncertainties for these clusters,  $\epsilon_{\rm{[Fe/H]}}$, this implies intrinsic [Fe/H] dispersions\footnote{For each cluster, the intrinsic dispersions have been estimated by subtracting the median of the measurement uncertainties, $\epsilon_{\rm{[Fe/H]}}$, in quadrature.}, $\sigma_{\rm{[Fe/H], int}}$, ranging between 0.07 and 0.12~dex. For the intrinsic [Fe/H] dispersions to be statistically non-significant, this would require median ASPCAP uncertainties in [Fe/H] of the order of $\sigma_{\rm{[Fe/H], tot}}/\sqrt{2}$ (for median uncertainties, $\epsilon_{\rm{[Fe/H]}}$, of this order, one would in fact get - by quadrature - comparable values of $\sigma_{\rm{[Fe/H], int}}$).  That is, ASPCAP uncertainties should have been underestimated by a factor $A$ between 4 and 5 for the intrinsic dispersions in [Fe/H] to be not significant (see Table~\ref{tab:FeH_spreads} for all these values). 
    \item Three out of six GCs, namely NGC~6809, NGC~6752 and NGC~6205, present negatively skewed [Fe/H] distributions, at least at $> 1\sigma$ level. The cases of NGC~6752 and NGC~6809 are particularly remarkable since the skewness of their [Fe/H] distribution is respectively equal to $-1.92\pm0.45$ and $-0.86\pm0.27$, that is the absolute value of the skewness is large and respectively nearly 5 and 4 times the corresponding uncertainty. A skewness test, performed by making use of the scipy skewtest function\footnote{\url{https://docs.scipy.org/doc/scipy/reference/generated/scipy.stats.skewtest.html\#scipy.stats.skewtest}}, confirms that for these three clusters there is a low probability that their [Fe/H] distributions have been drawn from normal distributions (the z-scores and p-values for these GCs are, respectively: -2.86 and 0.004, -6.66 and less than 0.0001,  -2.44 and 0.014)  \footnote{Note that NGC~6752 has four stars with [Fe/H] < -1.65 which, when removed, obviously imply a decrease in [Fe/H] dispersion (which decreases from 0.07 to 0.05) and skewness (which increases from -1.92 to -0.17). However, after checking their proper motions and radial velocities, these stars appear to be typical of the cluster and we could not find any particular reason to disregard them}. 
    Two clusters, NGC~6656 and NGC~6254, have marginally skewed MDFs, at $\sim 1\sigma$ level\footnote{For these two GCs the skewtest is also not conclusive, since it results in z-scores and p-values respectively equal to -1.43 and 0.15 (for NGC~6656) and -1.25 and 0.21 (for NGC~6254).}, while NGC~6273 is the only GC with a skewness compatible with 0, given the corresponding uncertainty.
    \item the significant negative skewness of the [Fe/H] distribution of the above cited GCs reflects a characteristic shape, with an initial rise in the number of stars at a given [Fe/H] for increasing values of [Fe/H], followed by a sharp truncation at the highest [Fe/H] values. This behaviour is found for all metal-rich GCs and for one of the metal-poor GCs, namely NGC~6809. 
\end{itemize}
About metal-poor GCs, it is very interesting to note that:
\begin{enumerate}
\item all of them have the peak of the MDF coinciding with the peak of the $\omega$~Cen MDF; 
\item the rising part of their MDF  is comparable to the rising part of the  (normalised) MDF of $\omega$~Cen (compare dark green and black step-like histogram in the top row panels of Fig.~\ref{70f_MDF}).
\end{enumerate}
These two findings indicate that the rate of iron formation in these GCs and in $\omega$~Cen was the same until these GCs and $\omega$~Cen itself reached a maximum  (corresponding to the peak of their MDF)  after which a more or less rapid decrease in iron production follows.\\

Among metal-rich GCs, it is interesting to note that:
\begin{enumerate}
    \item their MDFs peak at the same [Fe/H] value, given the uncertainties;
    \item their most metal-poor stars have metallicities similar to those at the peak of the $\omega$~Cen MDF  (at about [Fe/H]$\simeq -1.73$). 
\end{enumerate}
By assuming the metallicity as a measure of time, these two findings could suggest that these clusters started forming when the star formation and metal production in $\omega$~Cen was at its maximum and that in all of them, the star formation ceased at the same time. Additional investigation will be needed to confirm or reject this scenario (see also Sect.~\ref{discussion}, for a discussion on the ages of these clusters).\\
With regard to the population of metal-rich GCs, it is interesting to further investigate the comparison with the MDF of $\omega$~Cen stars. Several works indeed report that, in addition to the primary peak at [Fe/H]  $\sim -1.7$~dex, the MDF of $\omega$~Cen has a series of secondary peaks, one of which at [Fe/H]$\sim -1.5$~dex \citep[see, for example][]{johnson2010, alvarez2024}. This secondary peak is not clearly visible in the APOGEE data, possibly due to the spatial coverage of the $\omega$~Cen stars observed by APOGEE, with a deficiency of stars in the central regions of the cluster \citep[see Figs.~10 and 11 in][]{alvarez2024}. 
In Figure~\ref{mdfs_johnson}, we compare the MDF of $\omega$~Cen, derived from the data of \citet{johnson2010}, with that of the group of metal-rich clusters that we have identified as chemically compatible with it. The comparison is striking: these three GCs all have MDFs that peak where the secondary maximum of the $\omega$~Cen MDF is located. \\ 
The two groups of clusters found therefore have peaks that coincide with either the primary peak of metallicity of $\omega$~Cen or with its secondary peak. This result is not a consequence of the method we used to quantify the chemical compatibility between the clusters in question and $\omega$~Cen (see introduction to this section). It is rather a proof that the similarity between these clusters and $\omega$~Cen is not coincidental. Indeed, for a cluster to be chemically compatible with $\omega$~Cen it is sufficient that it falls within its chemical domain, but its mean (or median or mode) metallicity could be any value contained between $\omega$~Cen's minimum and maximum metallicity. That is, there is no reason for the metallicities of compatible clusters to clump into two distinct groups, which in addition overlap with the two main metallicity peaks of $\omega$~Cen itself. These similarities in the MDFs of these clusters is thus \textit{a posteriori} additional confirmation of the relevance of our findings, and allow us to strongly anchor NGC~6752, NGC~6205, and NGC~6254 to $\omega$~Cen and its progenitor galaxy.

\begin{table}
\centering
\caption{Total and intrinsic [Fe/H] dispersions for the metal-poor and the metal-rich globular clusters chemically compatible with $\omega$~Centauri.}
\label{tab:FeH_spreads}
\begin{tabular}{lrrrrc}
\toprule
 GC name &  \# stars &  $\sigma_{\rm{[Fe/H], tot}}$ & $\epsilon_{\rm{[Fe/H]}}$  &  $\sigma_{\rm{[Fe/H], int}}$ & $A$ \\
\midrule

NGC~6656 &     230 &      0.10 & 0.01 &      0.10 &   5.36 \\
NGC~6809 &      58 &      0.07 & 0.01 &      0.07 &   4.29 \\
NGC~6273 &      56 &      0.13 & 0.02 &      0.12 &   5.18 \\
\\
NGC~6752 &     117 &      0.07 & 0.01 &      0.07 &   4.13 \\
NGC~6205 &      34 &      0.07 & 0.01 &      0.07 &   4.05 \\
NGC~6254 &      58 &      0.07 & 0.01 &      0.07 &   4.18 \\
\\
NGC5139 &    1201 &      0.24 & 0.01 &      0.24 &  12.87 \\
\bottomrule
\end{tabular}\tablefoot{For each cluster, we also report the median of the [Fe/H] uncertainties of its stars, $\epsilon_{\rm{[Fe/H]}}$, and the factor $A$ by which these uncertainties would have to be underestimated for the intrinsic dispersions not to be statistically significant. The same quantities for $\omega$~Cen (NGC~5139) are also reported. }
\end{table}

\begin{figure*}[h!]
 \hspace{-22pt}\includegraphics[clip=true, trim = 3mm 0mm 0mm 3mm, width=0.75\columnwidth]{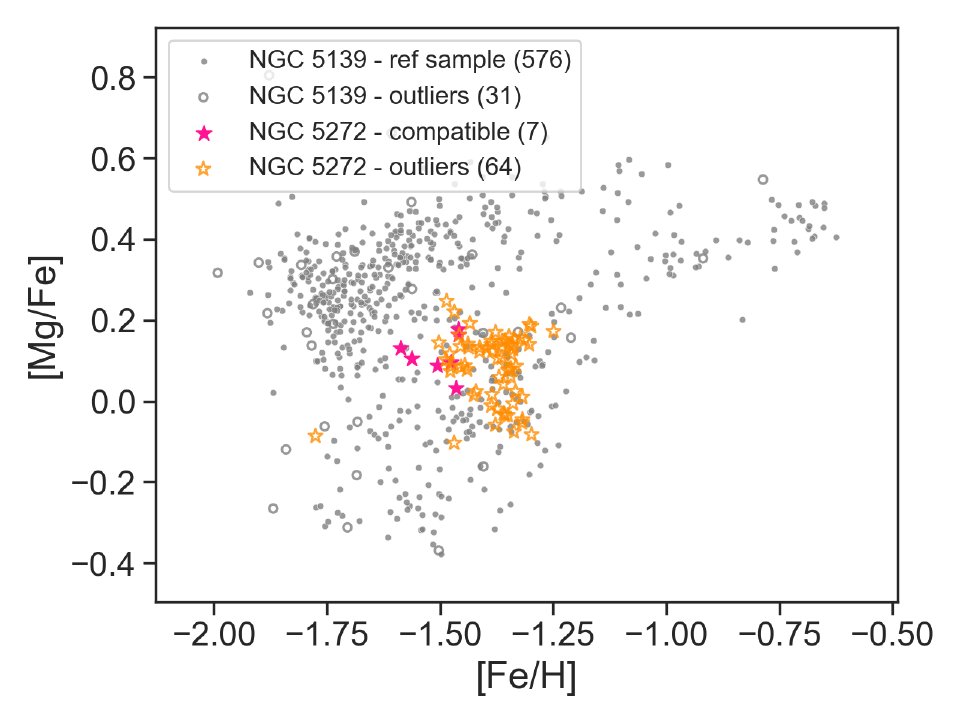}\hspace{-8.5pt}
\includegraphics[clip=true, trim = 3mm 0mm 0mm 3mm, width=0.75\columnwidth]{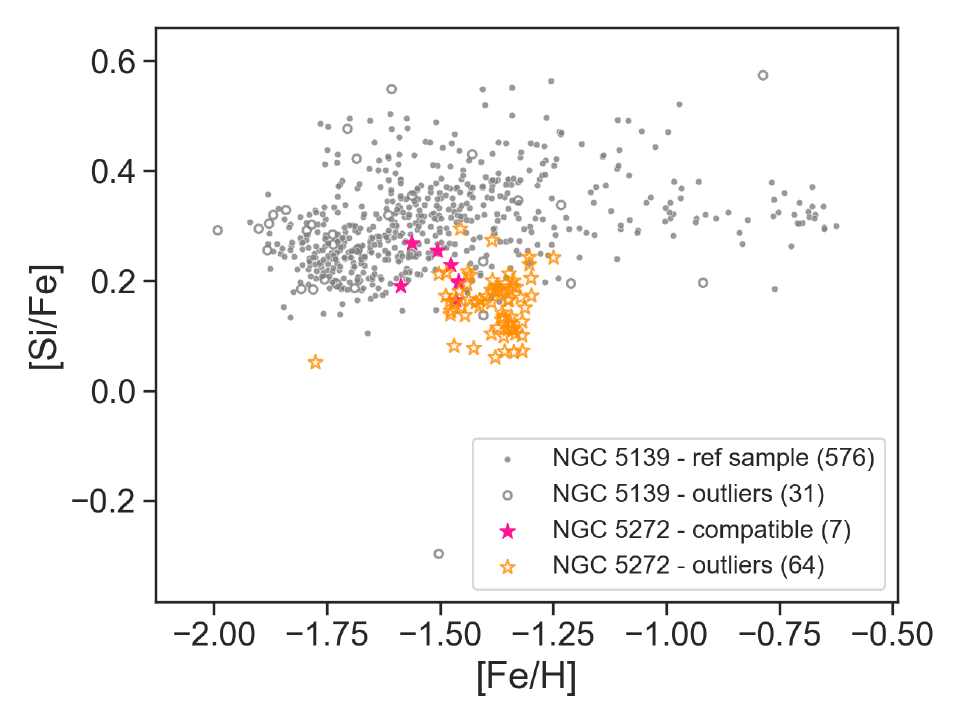}\hspace{-6pt}
\includegraphics[clip=true, trim = 3mm 0mm 0mm 3mm, width=0.75\columnwidth]{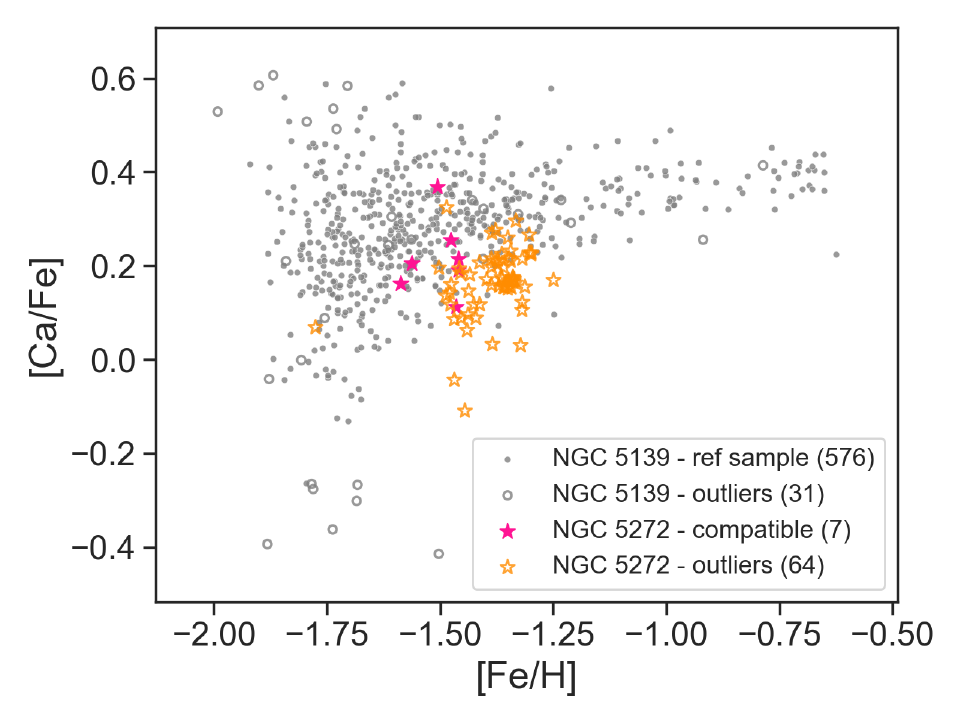}\par
\hspace{-25pt}\includegraphics[clip=true, trim = 3mm 0mm 0mm 2mm, width=0.76\columnwidth]{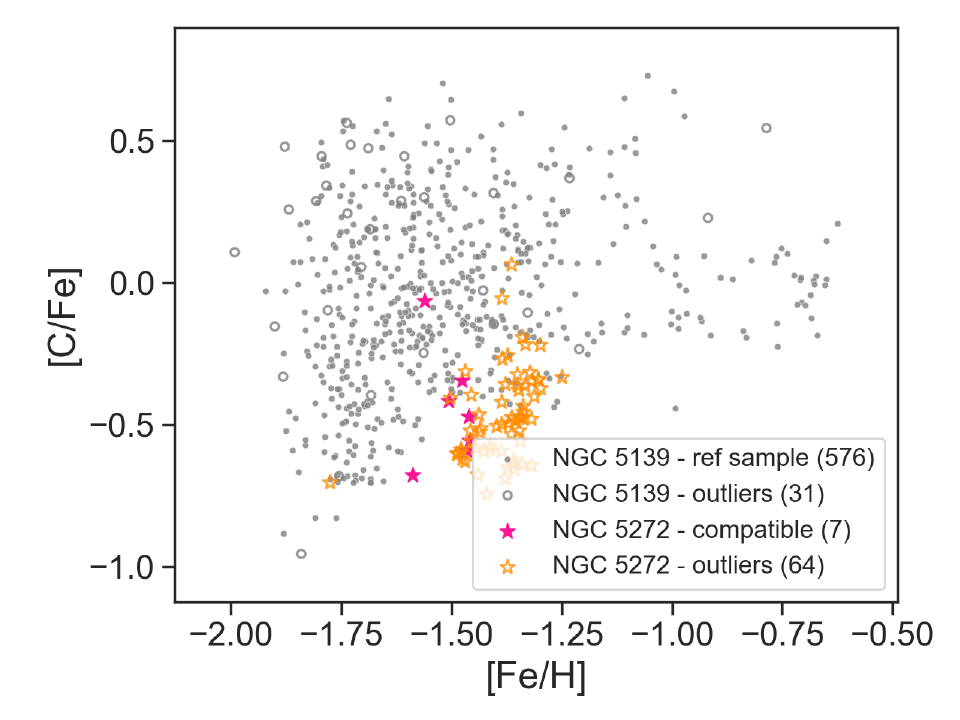}\hspace{-8pt}
\includegraphics[clip=true, trim = 2mm 0mm 0mm 1mm, width=0.76\columnwidth]{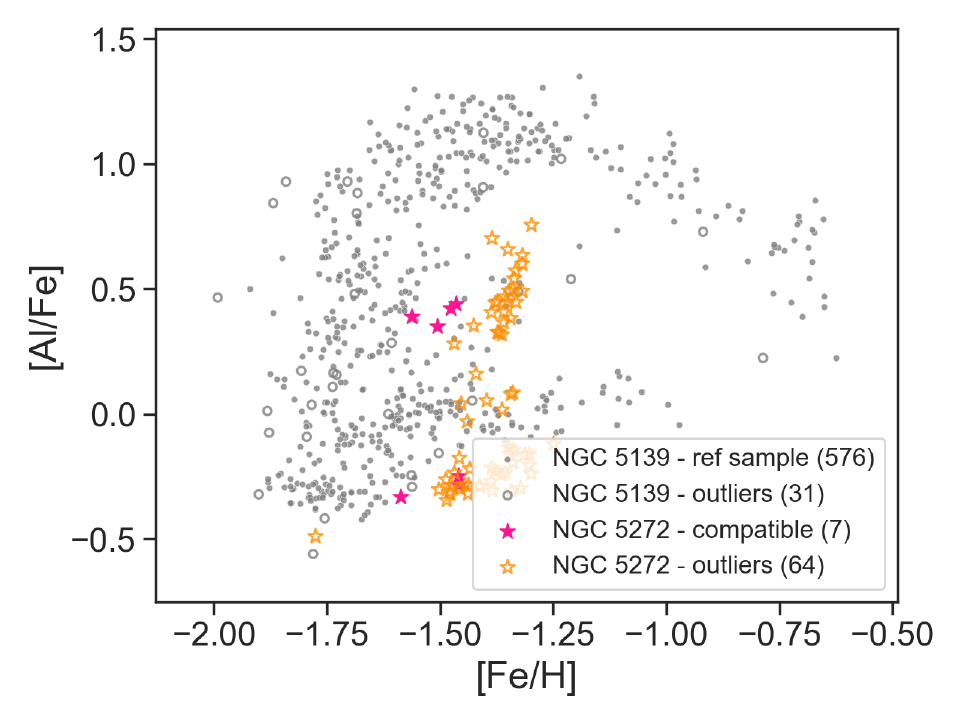}\hspace{-8pt}
\includegraphics[clip=true, trim = 2mm 0mm 0mm 1mm, width=0.76\columnwidth]{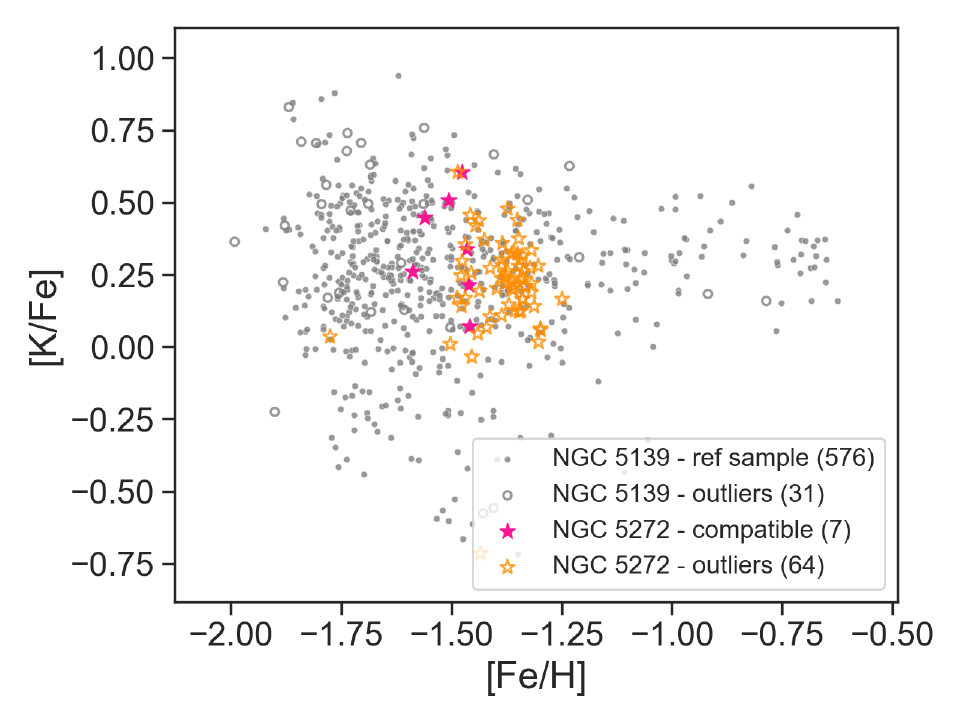}\par
\centering
\includegraphics[clip=true, trim = 1mm 0mm 0mm 1mm, width=0.75\columnwidth]{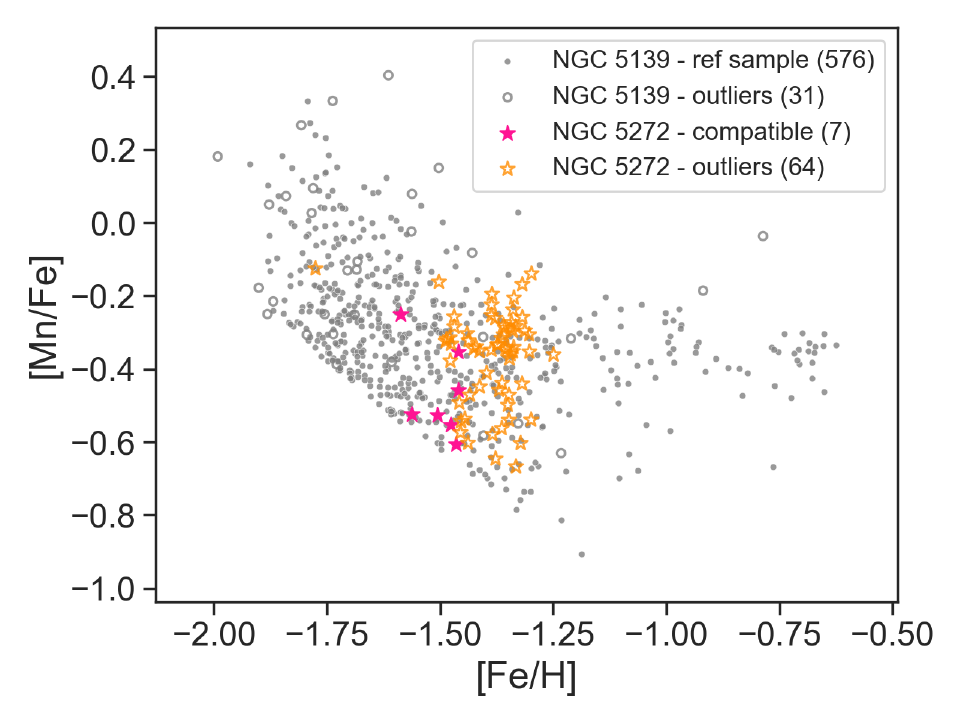}

  \caption{Chemical abundance relations for members of NGC~5272 (colour) and $\omega$~Cen (grey). The filled symbols show the reference sample of $\omega$~Cen (grey) and the stars of NGC~5272 (magenta) chemically compatible with it according to the GMM (see Sect. \ref{results}), while the empty ones (grey and orange colours) correspond to their outliers. The number of stars in each category is reported in parentheses. NGC~5272 is an example of a globular cluster in the metallicity interval of $\omega$~Cen, which has a very low chemical compatibility with it (see also Figs.~\ref{Ter9} and \ref{NGC3201}). }             \label{NGC5272}%
    \end{figure*}

\subsection{In the range of $\omega$~Cen metallicities, do all stellar systems have similar chemical patterns?\label{no_ocen-like_gcs}}

Previous results have allowed us to show the existence of a group of globular clusters with the same chemical patterns found within $\omega$~Centauri cluster. Is such a result obvious, i.e. is this similarity simply due to the fact at the metallicities of $\omega$~Cen, all clusters have similar chemical compositions?\\

\begin{figure*}[h!]
\hspace{-22pt}\includegraphics[clip=true, trim = 3mm 0mm 0mm 3mm, width=0.75\columnwidth]{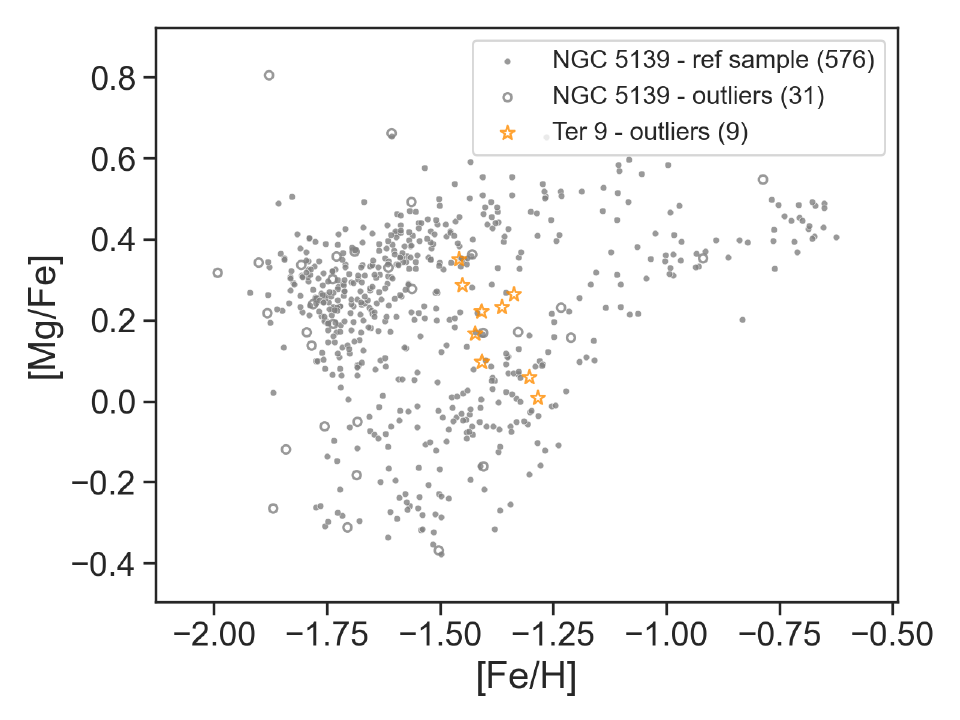}\hspace{-8.5pt}
\includegraphics[clip=true, trim = 3mm 0mm 0mm 3mm, width=0.75\columnwidth]{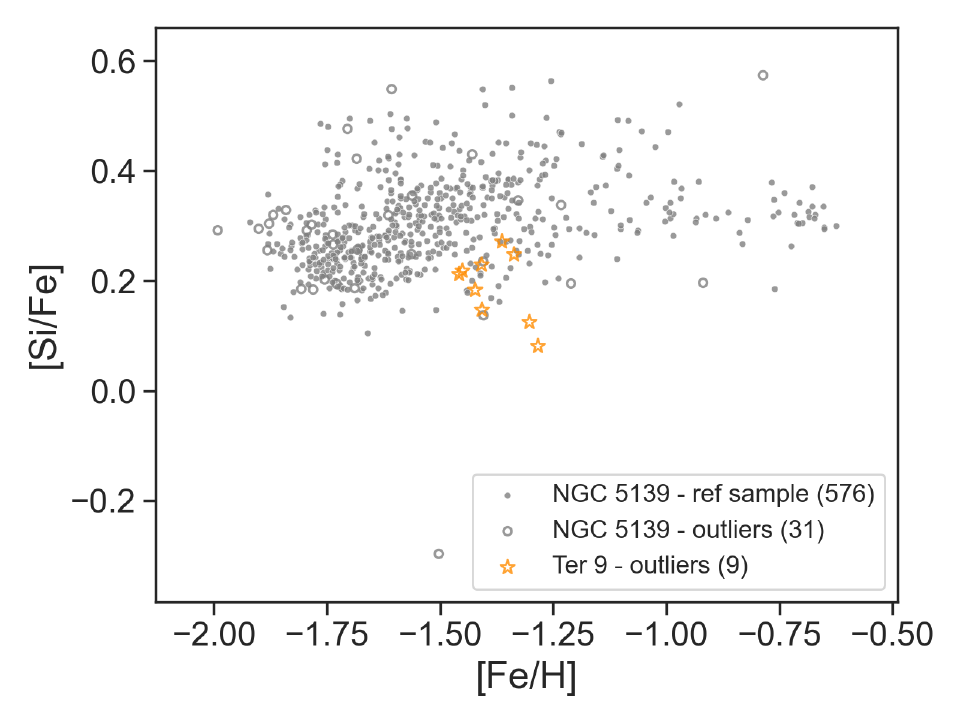}\hspace{-6pt}
\includegraphics[clip=true, trim = 3mm 0mm 0mm 3mm, width=0.75\columnwidth]{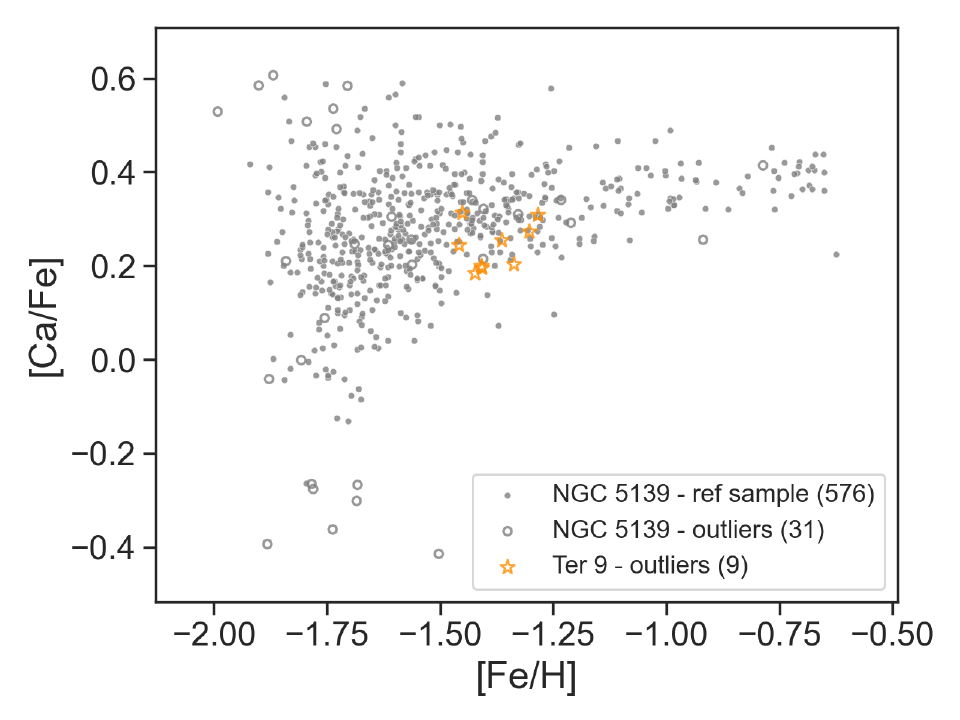}\par
\hspace{-25pt}\includegraphics[clip=true, trim = 3mm 0mm 0mm 2mm, width=0.76\columnwidth]{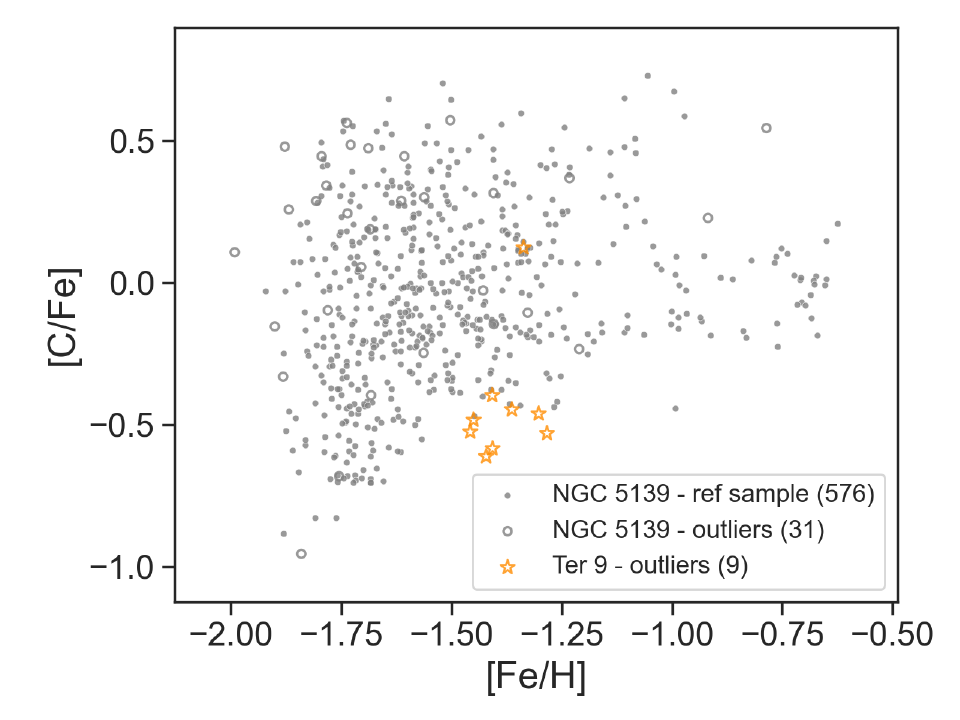}\hspace{-8pt}
\includegraphics[clip=true, trim = 2mm 0mm 0mm 1mm, width=0.76\columnwidth]{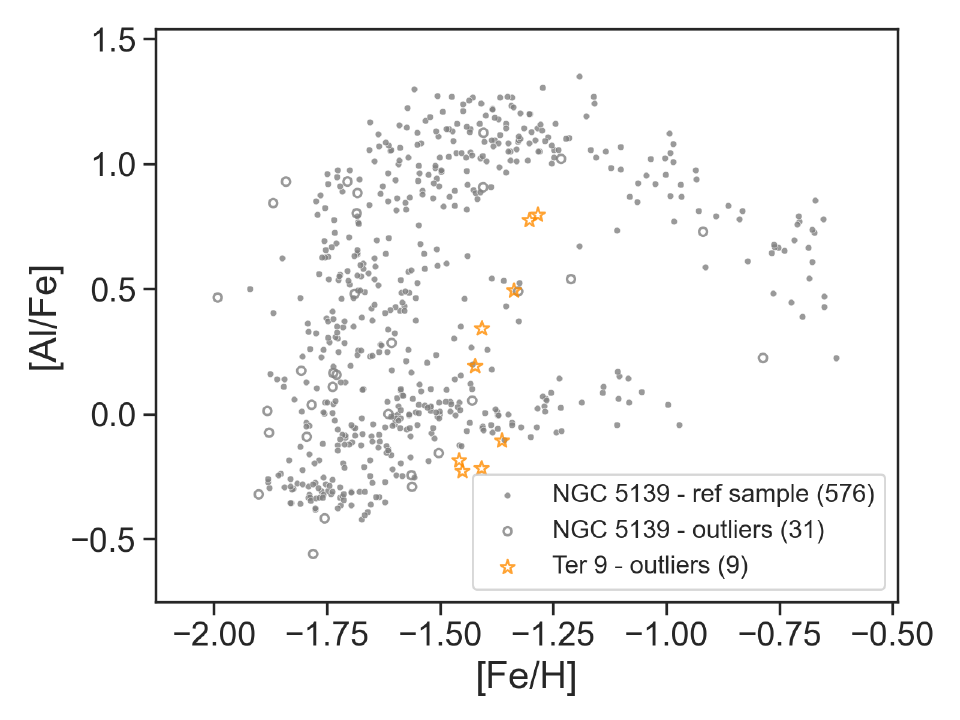}\hspace{-8pt}
\includegraphics[clip=true, trim = 2mm 0mm 0mm 1mm, width=0.76\columnwidth]{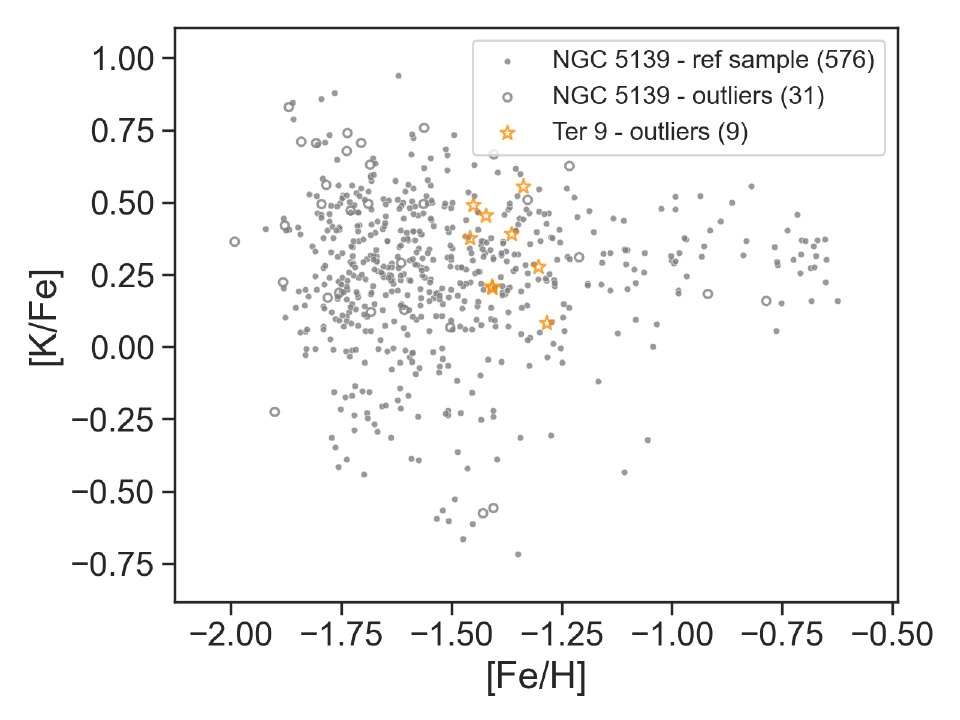}\par
\centering
\includegraphics[clip=true, trim = 1mm 0mm 0mm 1mm, width=0.75\columnwidth]{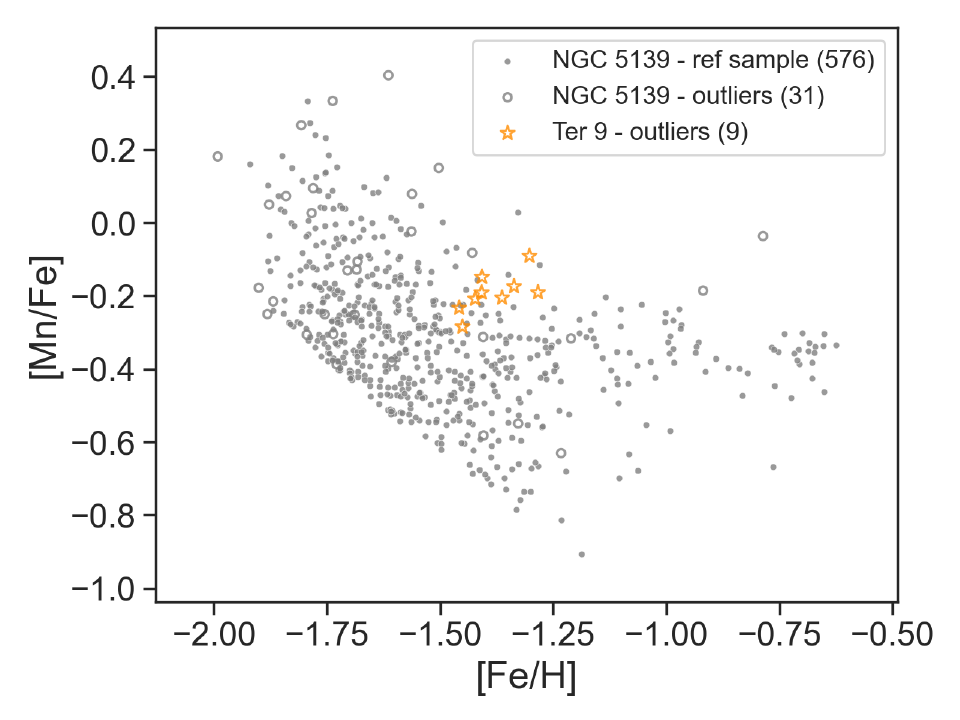}
  \captionof{figure}{Same as  Fig.~\ref{NGC5272}, for Ter~9. }             \label{Ter9}%
    \end{figure*}

\begin{figure*}[h!]
 \hspace{-22pt}\includegraphics[clip=true, trim = 3mm 0mm 0mm 3mm, width=0.75\columnwidth]{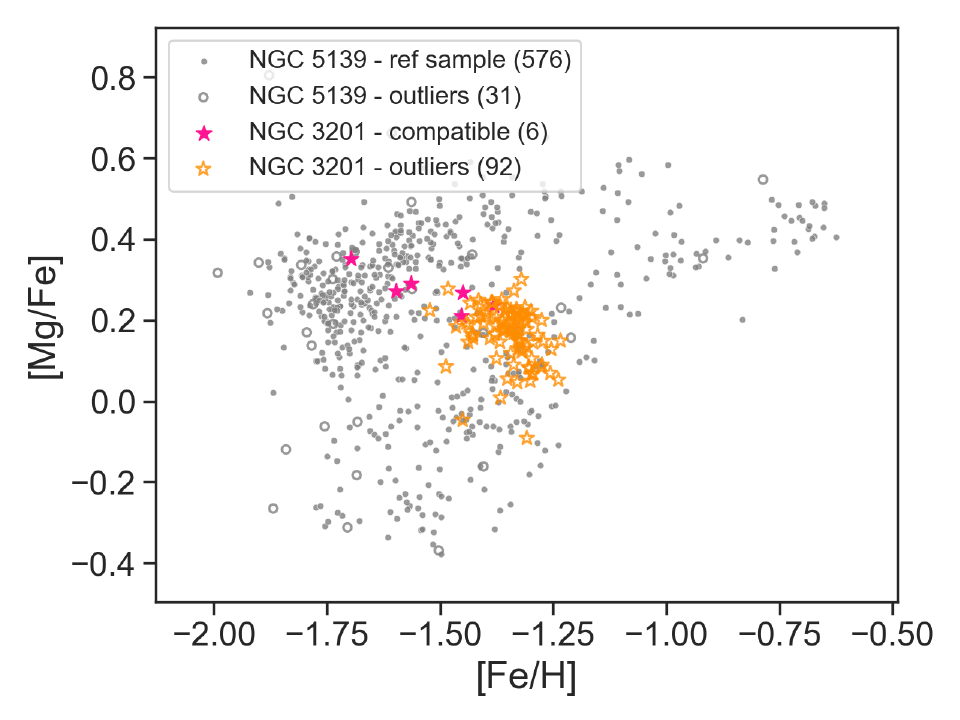}\hspace{-8.5pt}
\includegraphics[clip=true, trim = 3mm 0mm 0mm 3mm, width=0.75\columnwidth]{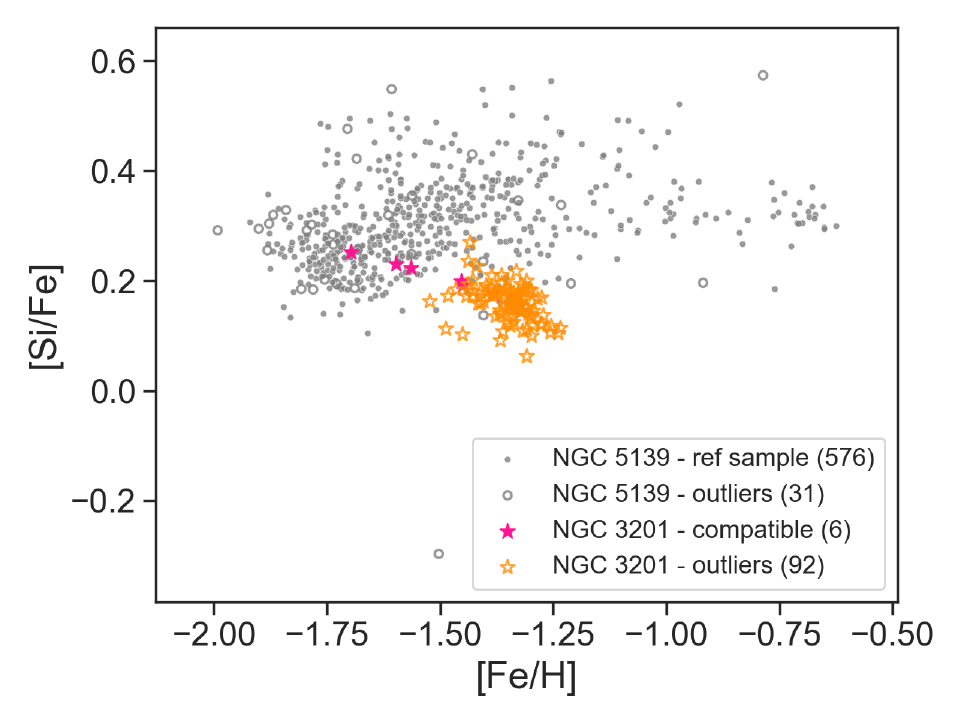}\hspace{-6pt}
\includegraphics[clip=true, trim = 3mm 0mm 0mm 3mm, width=0.75\columnwidth]{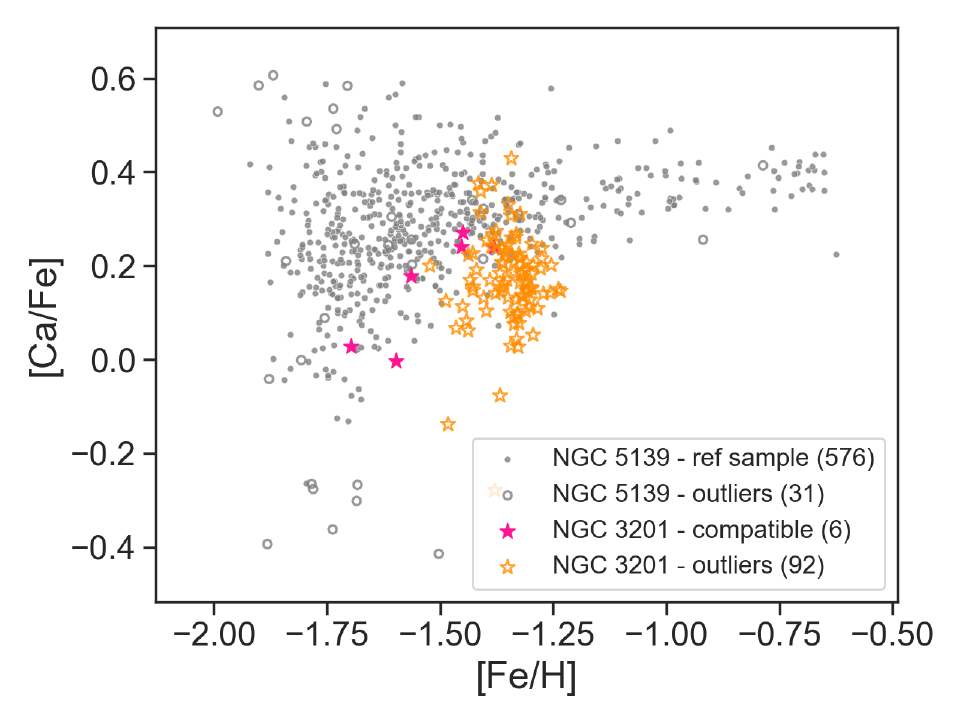}\par
\hspace{-25pt}\includegraphics[clip=true, trim = 3mm 0mm 0mm 2mm, width=0.76\columnwidth]{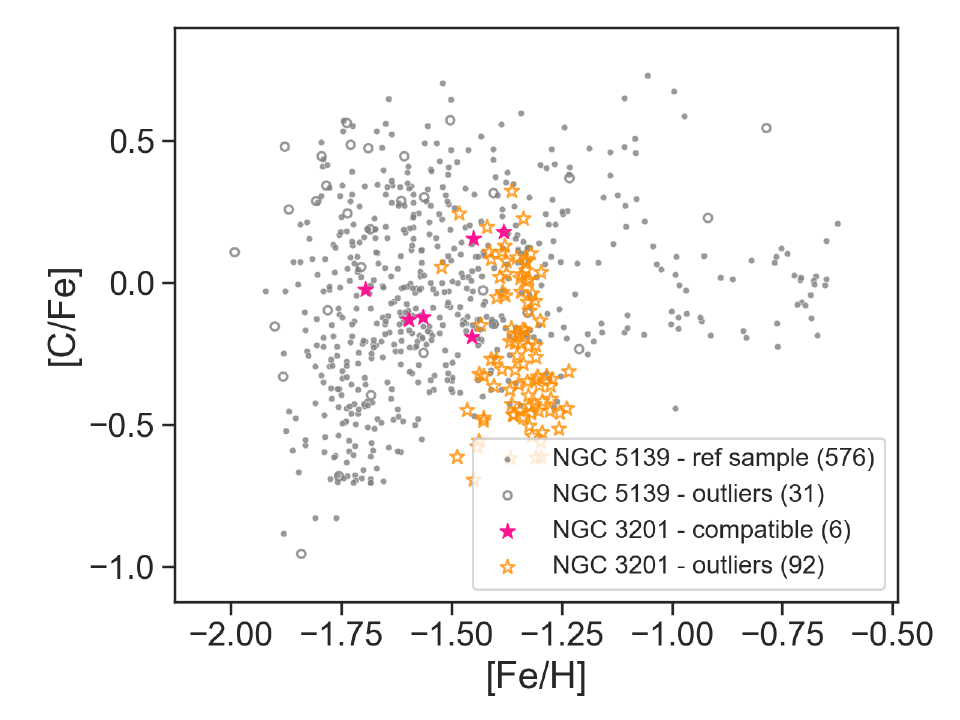}\hspace{-8pt}
\includegraphics[clip=true, trim = 2mm 0mm 0mm 1mm, width=0.76\columnwidth]{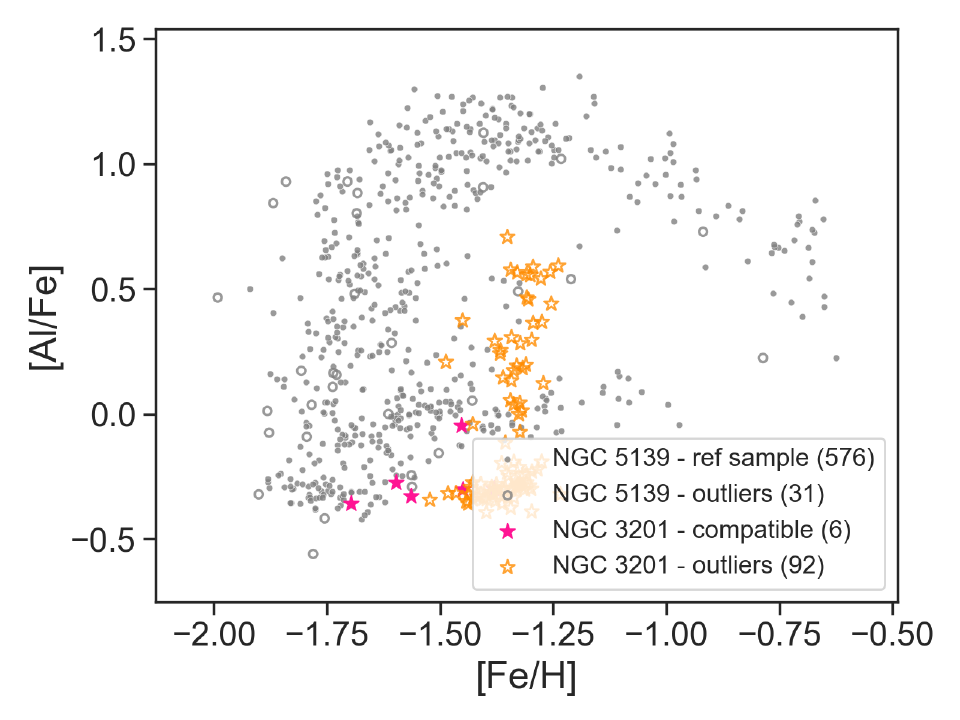}\hspace{-8pt}
\includegraphics[clip=true, trim = 2mm 0mm 0mm 1mm, width=0.76\columnwidth]{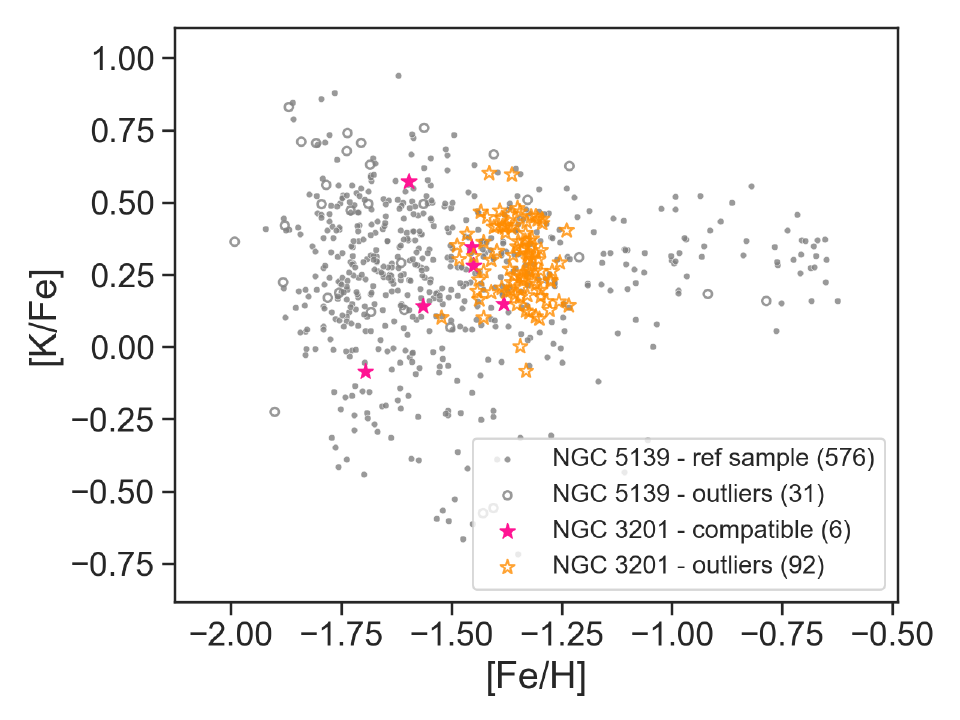}\par
\centering
\includegraphics[clip=true, trim = 1mm 0mm 0mm 1mm, width=0.75\columnwidth]{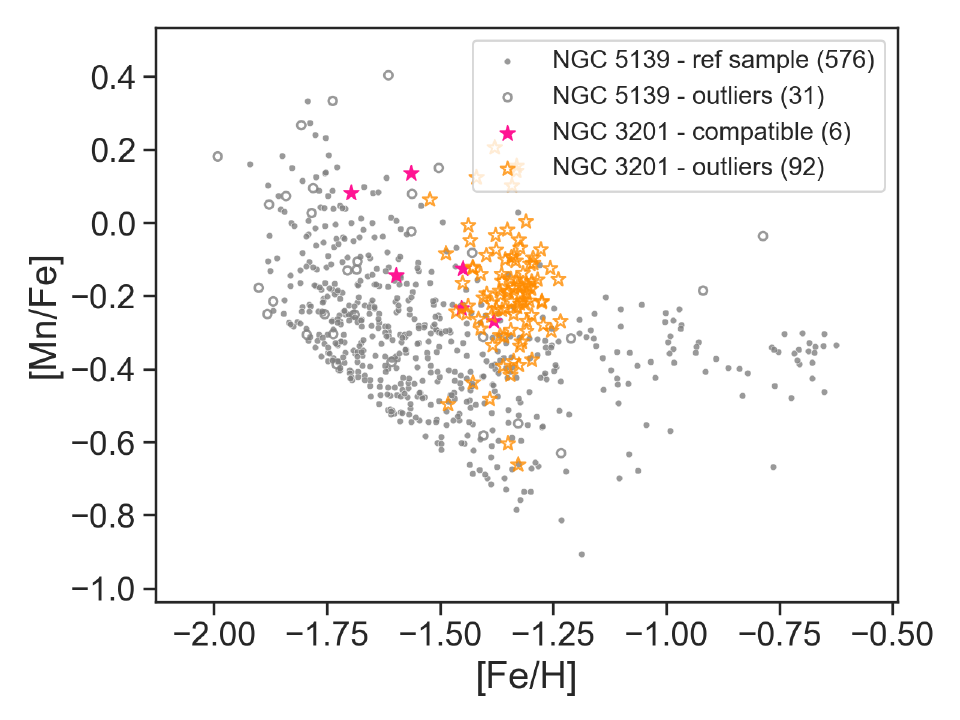}

  \caption{Same as  Fig.~\ref{NGC5272}, for NGC~3201.}    \label{NGC3201}%
    \end{figure*}

Among the clusters listed in Table~\ref{OCenGCs_table_VAC}, clusters like NGC~5272, Ter~9, and NGC~3201 have a very low fraction of stars compatible with $\omega$~Cen, despite being in its metallicity range. Abundance ratios for these clusters, as a function of [Fe/H], are shown in Figs.~\ref{NGC5272}, \ref{Ter9} and \ref{NGC3201}. In these figures, similar to what is shown in Figs.~\ref{NGC6656} to \ref{NGC6254}, for each cluster both stars that the GMM classifies as compatible with $\omega$~Centauri (magenta symbols) and the outliers (orange symbols) are shown. As can be seen, for all these clusters, some chemical patterns (e.g. [Mg/Fe] versus [Fe/H], [K/Fe] versus [Fe/H]) overlap with those of $\omega$~Cen, while others (e.g [Si/Fe] versus [Fe/H], or [Ca/Fe] versus [Fe/H] or [C/Fe] versus [Fe/H]) do not. Thus, when a sufficient number of chemical abundances are used, differences between clusters emerge, even for clusters that have similar metallicities.
This shows that the chemical compatibility found between NGC~6656, NGC~6273, NGC~6809, NGC~6254, NGC~6205, and NGC~6752 and $\omega$~Cen is not merely the result of the fact that at these metallicities all globular clusters should resemble each other. NGC~5272, Ter~9, and NGC~3201 are indeed proof that there differences in the chemistry of stellar systems can be found even at these metallicities.
The chemical compatibility of NGC~6656, NGC~6273, NGC~6809, NGC~6254, NGC~6205, and NGC~6752 and $\omega$~Cen, combined with the common characteristics of their MDFs discussed above, is rather proof that these clusters share the same formation environment, which is different from that of other Galactic globular clusters with similar metallicities.\\

Not only are there globular clusters in the metallicity range of $\omega$~Cen that are not chemically compatible with it. By comparing the chemical patterns of the most massive Milky Way satellites -- namely Large Magellanic Cloud (LMC), Small Magellanic Cloud (SMC),  Sagittarius (Sgr), and Fornax (Fnx) -- for which data are available in APOGEE~DR17, we can quantify their degree of compatibility with $\omega$~Cen, applying the same procedure used for the globular clusters.  For these galaxies, we used the APOGEE data presented in \citet{hasselquist21}. In particular, their Table~2 provides the identifiers of stars belonging to these galaxies which we cross-identified with the APOGEE~DR17 catalogue to obtain their chemical abundances. From the retrieved sample, for the LMC, SMC and Sag, we then selected only stars with:
\begin{enumerate}
\item a signal-to-noise ratio $\tt{SNREV} > 70$;
\item temperatures in the range $\rm 3500\,K < T_{eff} < 5500\,K$ and surface gravities  $\rm logg < 3.6$; 
\item \tt{APOGEE STARFLAG} and \tt{APOGEE STARBAD} $= 0$
\end{enumerate}
For Fornax, we applied less strict selection criteria, than those used above, and made use of all Fornax stars in \citet{hasselquist21}, in order to keep stars at [Fe/H]$<-1$. 

As for NGC~5139 ($\omega$~Cen), we used the \citet{schiavon2023} catalogue applying the same selections discussed in Sect~\ref{obsdata}.

By making use of the same 8-dimensional space of chemical abundances used for the analysis of globular clusters, in Table~\ref{SatellitesGCs_table_VAC} we report the fraction of stars in these galaxies that are chemically compatible with $\omega$~Cen, for different cuts in metallicity. Even for the most restrictive cuts, where we limit the analysis only to stars with [Fe/H] lower than $-1.5$, we see that the fraction of stars of these galaxies compatible with $\omega$~Cen chemistry is 32\%, at the most (see Appendix~\ref{SatvsoCen} for the corresponding plots). That is, none of the aforementioned dwarf galaxies is chemically compatible with $\omega$~Cen, in its range of metallicities. This result has two implications: 
\begin{enumerate}
\item the progenitor galaxy of $\omega$~Cen must have had a different chemical evolution from that which LMC, SMC, Sagittarius, and Fornax had, at the same metallicities; 
\item the chemical non-compatibility of $\omega$~Cen with LMC, SMC, Sagittarius and Fornax also implies the chemical non-compatibility of   NGC~6656, NGC~6273, NGC~6809, NGC~6254, NGC~6205 and NGC~6752  with them. We can therefore exclude that these globular clusters formed in galaxies that had similar chemical evolution to the most massive satellites of the Milky Way, at the same metallicities.\\
\end{enumerate}

\begin{table}[h!]
\centering
\caption{Fraction of stars in the most massive satellites of the Milky Way (LMC, SMC, Sagittarius and Fornax) chemically compatible with $\omega$~Cen.}
\label{SatellitesGCs_table_VAC}
\resizebox{1.\columnwidth}{!}{
\begin{tabular}{lccc}
\toprule
Satellite&  [Fe/H]$< -1$  &  [Fe/H]$< -1.3$ & [Fe/H]$< -1.5$ \\
\midrule
LMC & 4\% (8 out of 211) & 13\% (8 out of 62) & 32\% (8 out of 25)\\
SMC & 0\% (0 out of 322) & 0\% (0 out of 40) & 0\% (0 out of 14)\\
Sgr & 4\% (2 out of 81) & 6\% (2 out of 34) & 25\% (2 out of 8)\\
Fnx & 0\% (0 out of 43) & 0\% (0 out of 6) & 0\% (0 out of 2)
\end{tabular}}\tablefoot{This fraction has been estimated for three different [Fe/H] intervals, as reported in the different columns.}
\end{table}

\begin{figure*}[h!]
\includegraphics[width=\columnwidth]{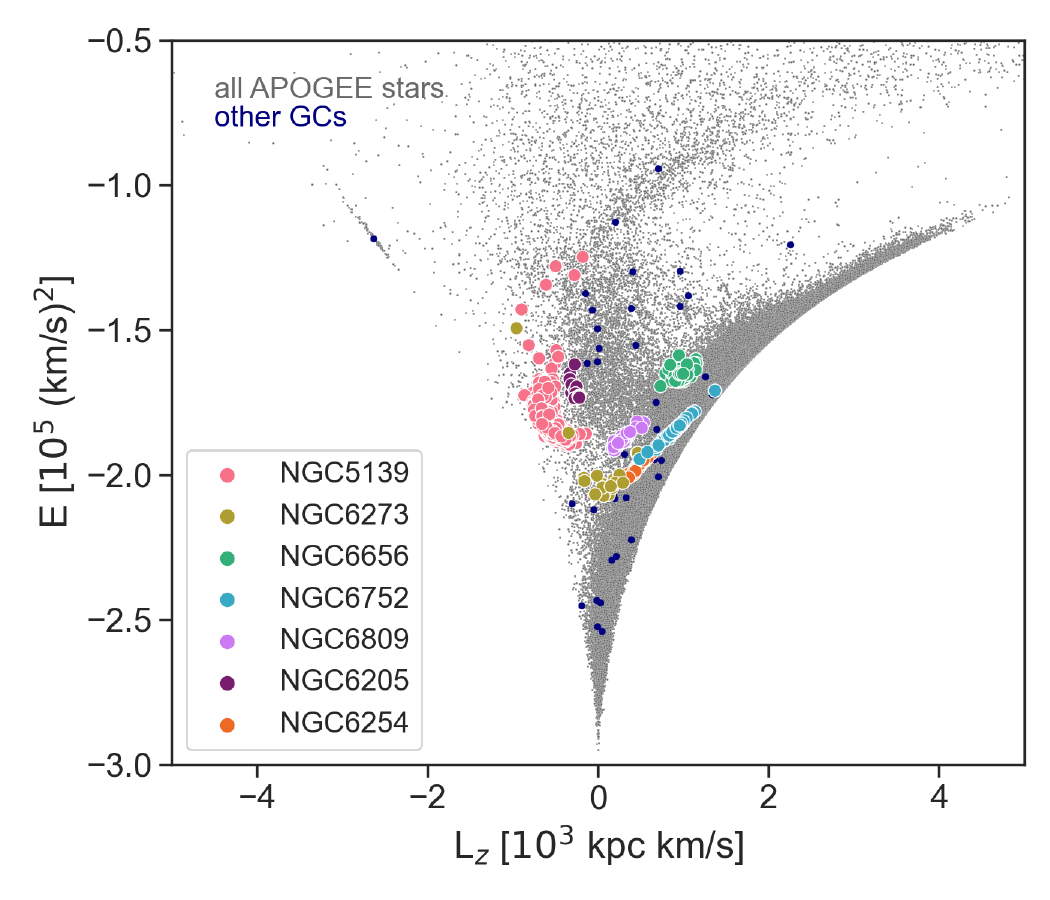}
\includegraphics[width=\columnwidth]{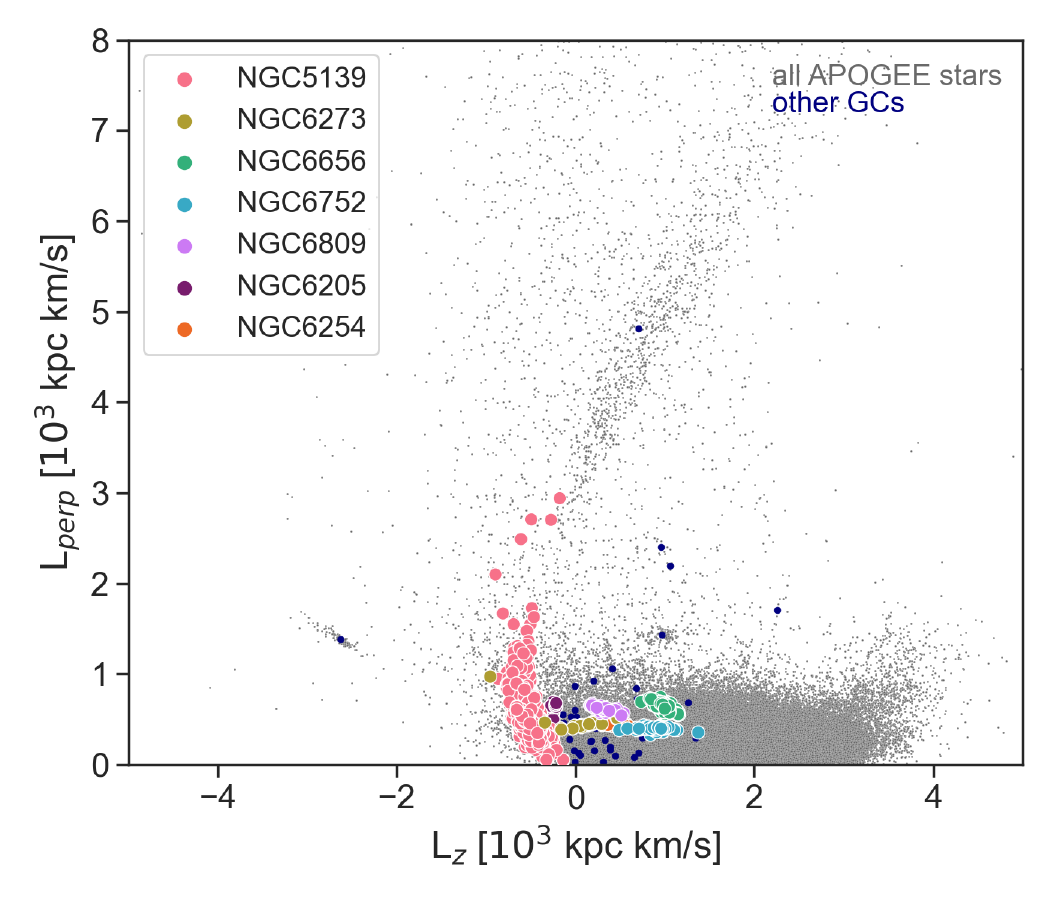}\par
\includegraphics[width=\columnwidth]{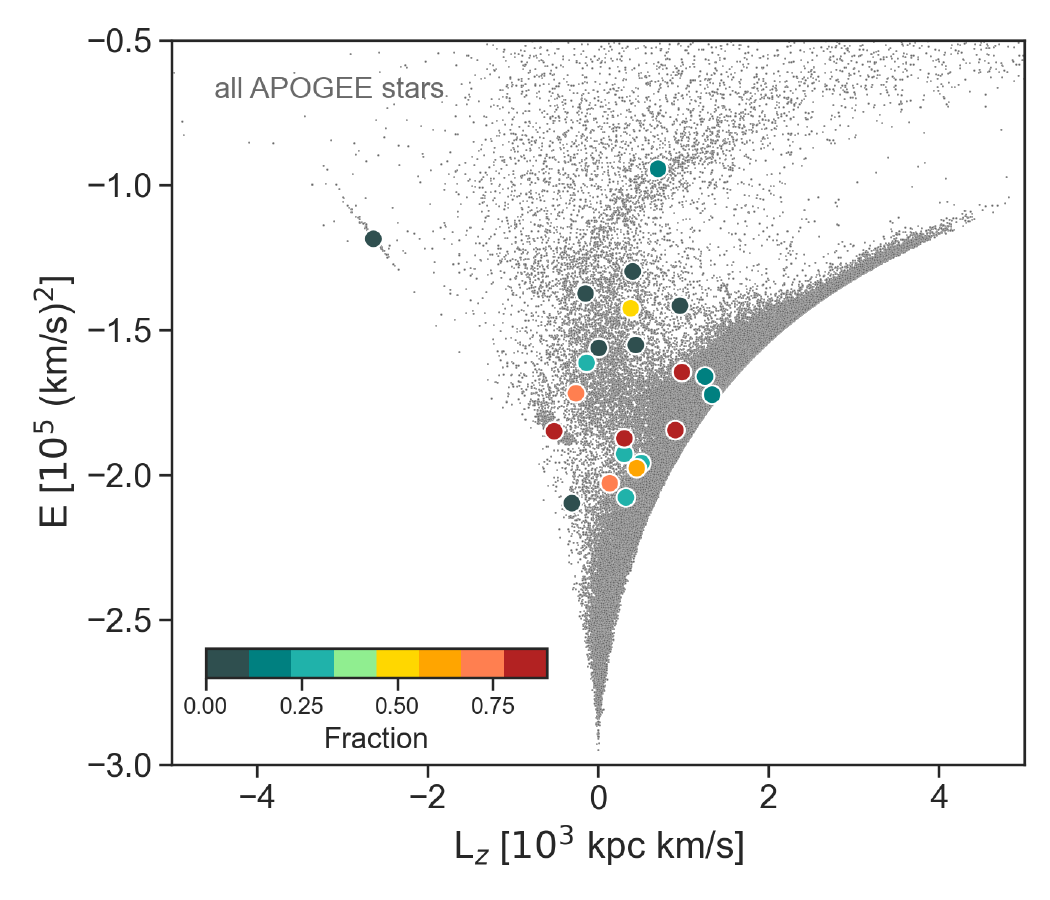}
\includegraphics[width=\columnwidth]{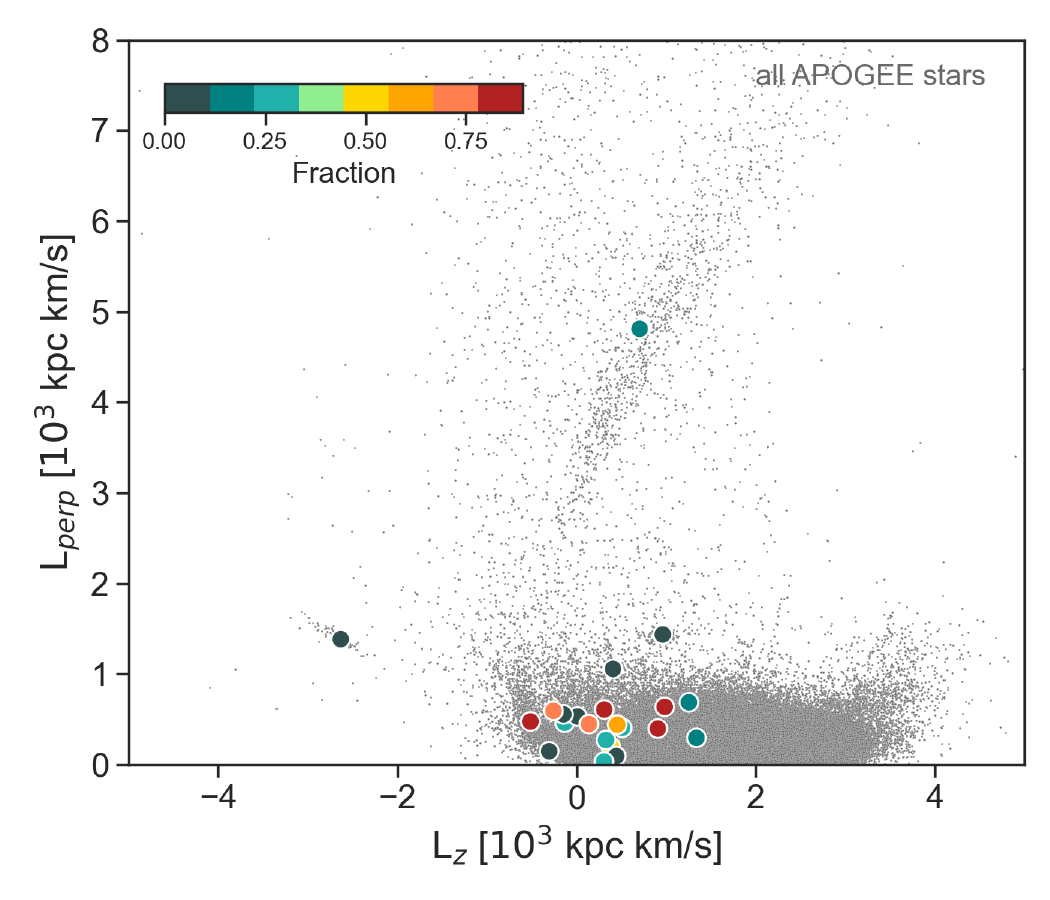}
    \caption{Distribution of GCs and field stars in kinematic spaces: orbital energy (E) and projection of the total angular momentum onto the Galactic plane (L$_{\rm perp}$) versus the $z$-component of the angular momentum (L$_z$). Top panel: Star by star distribution in the E - L$_z$ and L$_{\rm perp}$ - L$_z$ spaces of GCs chemically compatible - with at least 60\% of their stars - with $\omega$~Cen. Other GCs in our sample are shown as blue points (mean values). Bottom panel: Distribution in the E - L$_z$ and L$_{\rm perp}$ - L$_z$ spaces of GCs which are colour-coded according to their fraction of stars chemically compatible with $\omega$ Cen (see Tab.~\ref{OCenGCs_table_VAC}). Only GCs with a total number of stars greater than 15 are shown. For comparison, in both panels, the distribution of all APOGEE stars is shown as grey points. Note that the sign of L$_z$ has been inverted from the usual sign of L$_z$ in a right-handed system to enable a better comparison with other works in the literature.}
    \label{e_lz_ocenlike}
\end{figure*}

\subsection{Orbital properties of GCs chemically compatible with $\omega$~Cen}
Once clusters chemically compatible with $\omega$~Cen have been identified on the basis of their abundances only, it is interesting to go back to kinematic spaces and investigate their orbital properties. For estimating orbital parameters, we made use of the publicly available code \texttt{galpy} \citep[][]{bovy2015}, adopting the \citet{mcmillan2017} Galactic potential. The positions and proper motions of all Galactic GCs and field stars studied in this section are taken from Gaia EDR3 data, whilst the radial velocities are taken from APOGEE DR17. Distances are the ones generated by \citet{leung2019a} for the APOGEE DR17 catalogue, using the \texttt{astroNN} python package \citep[see][]{leung2019b}.
These distances are determined using a re-trained \texttt{astroNN} neural-network software, which predicts stellar luminosity from spectra using a training set comprised of stars with both APOGEE DR17 spectra and Gaia EDR3 parallax measurements. We use a right-handed Galactocentric frame that leads to a 3D velocity of the Sun equal to $\rm [U_{\odot}, V_{\odot}, W_{\odot}] = [11.1, 248.0, 8.5] \,km\, s^{-1}$ \citep[][]{horta2023}. We assume the distance between the Sun and the Galactic Centre to be $\rm R_{\odot} = 8.178 \,kpc$ \citep[][]{gravity2019}, and the vertical height of the Sun above the midplane $\rm z_{\odot} = 0.02 \,kpc$ \citep[][]{bennett2019}. \\

\begin{figure}
\includegraphics[width=\columnwidth]{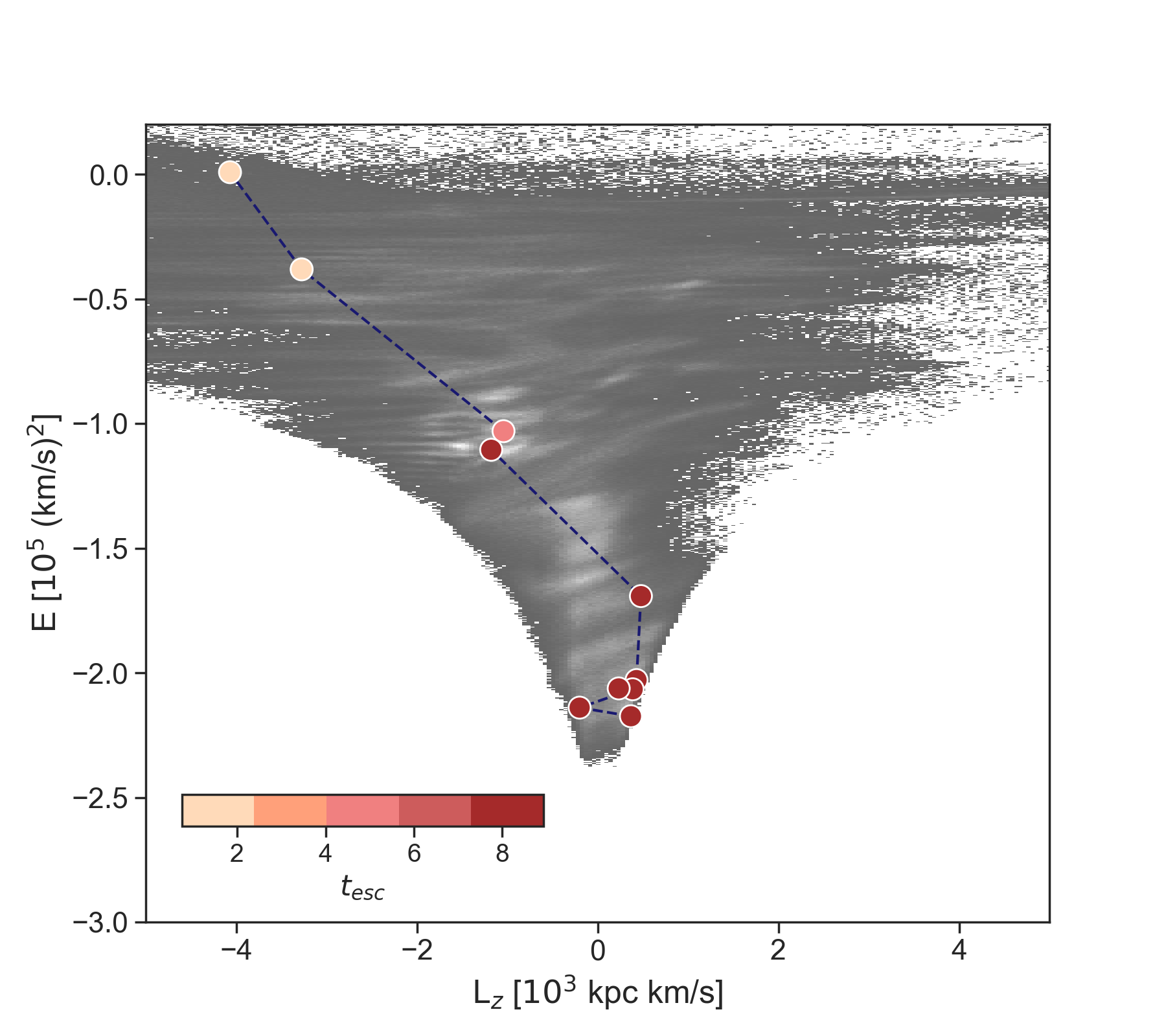}
    \caption{Distribution in E - L$_z$ of a set of 10 clusters accreted together with their progenitor galaxy on a MW-type galaxy through one of the $N$-body simulations presented in \citet{pagnini2023} (identified in that paper with the ID=$\rm MWsat\_n1\_\Phi180$). Clusters are colour-coded according to their escape time from the progenitor satellite which is defined as the time when the distance between the GC and the satellite's centre of mass is larger than 15 kpc (see Sect.~3 in \citet{pagnini2023} for the details). In the background, the distribution of field stars of the accreted satellite is also shown as a density map. Energies and angular momenta have been rescaled to obtain a distribution comparable with that of the Galactic GCs and field stars presented in Fig.~\ref{e_lz_ocenlike}.}
    \label{sim_rescaled}
\end{figure}

Figure \ref{e_lz_ocenlike} shows the distribution in the E - L$_z$ and L$_{\rm perp}$ - L$_z$ spaces of our sample of GCs in comparison to field stars in APOGEE DR17.
These are the two kinematic spaces most commonly used in the literature to derive the origin of Galactic GCs where E is the total orbital energy, L$_z$ is the $z$ component of the angular momentum space in a reference frame with the Galactic disc in the $xy$ plane, and L$_{\rm perp}$ is the projection of the total angular momentum onto the Galactic plane.
The bottom panel of Figure \ref{e_lz_ocenlike} shows the distribution of all Galactic GCs  (with at least 15 stars) in the \citet{schiavon2023} catalogue in the $E-L_z$ plane (left panel) and in the L$_{\rm perp}$- L$_z$  plane,  in comparison to all field stars in the APOGEE~DR17 catalogue. For the GCs, the values of energies and angular momenta reported in these panels are the mean values, averaged over all stars which have a high probability of being members of each cluster (see Sect. \ref{obsdata}). In the same panel of Fig.~\ref{e_lz_ocenlike}, clusters are colour-coded according to the fraction of their stars compatible with $\omega$~Cen, as listed in Tab.~\ref{OCenGCs_table_VAC}.
We can see that GCs with a high fraction of stars chemically compatible with $\omega$~Cen are distributed over an extended region of these kinematic planes and are mixed with GCs with a lower fraction. The upper panel of Figure \ref{e_lz_ocenlike} focuses on the E - L$_z$ and L$_{\rm perp}$ - L$_z$ of GCs chemically compatible - with at least 60\% of their stars - with $\omega$~Cen in comparison to the other GCs in the sample and the field stars in APOGEE DR17. As we can see, these GCs occupy a large region in the kinematic diagrams due to the uncertainties on these quantities and the cluster's intrinsic velocity dispersion. In fact, the more massive the cluster, the greater its dispersion, as we can see from the $\omega$~Cen extension in the kinematic spaces. Despite this, we can clearly see that the GCs which are chemically compatible with $\omega$~Cen have different kinematic properties, spanning a wide range of L$_z$ ($\rm -10^3 kpc\,km/s \lesssim L_z \lesssim  10^3 kpc\,km/s$) and E ($\rm -2.1 \times 10^5 (km/s)^2 \lesssim E \lesssim -1.6 \times 10^5 (km/s)^2$ ). In particular, we emphasize the fact that $\omega$~Cen and NGC~6205 appear to have a retrograde orbit while the other compatible clusters have both prograde orbits (NGC~6656 and NGC~6752) and orbits with $\rm L_z \sim 0$ (NGC~6273, NGC~6809, and NGC~6254). In particular, it is interesting to note that the clusters most compatible with $\omega$~Cen, namely NGC~6752 and NGC~6656, are actually the most prograde, meaning that the kinematic criteria would suggest they are not associated to $\omega$~Cen.\\
Given the wide spread found in energy and angular momenta for GCs chemically compatible with $\omega$~Cen, it is questionable whether it is realistic to imagine a common origin for them. 
In Fig.~\ref{sim_rescaled}, we show the simulated distribution in energy and angular momenta of a set of 10 clusters, accreted, together with their progenitor galaxy, on a MW-type galaxy. This simulation is one of those presented in \citet{pagnini2023} (identified in that paper with the ID=$\rm MWsat\_n1\_\Phi180$). In this simulation, the orbital plane of the satellite is initially inclined by 180 degrees with respect to the disc of the MW-type galaxy -- meaning that the orbit is initially retrograde in the plane of the MW-type galaxy --  and the mass of the satellite is one-tenth of that of the MW. In Fig.~\ref{sim_rescaled}, energies and angular momenta have been renormalized in order to obtain a distribution comparable with that of the Galactic GCs and field stars presented in the previous figure. For this, we have multiplied positions, velocities, and masses respectively by factors of 1.89, 1.06, and 2.12, which implies also rescaling the time scale by a factor of 1.79, to keep the virial ratio = 1. We emphasize that this simulation does not intend to provide the most probable scenario for the accretion of $\omega$~Cen and its related GCs in the Milky Way, and therefore neither the mass ratio nor the orbital parameters used in this simulation should be taken literally as representative of this accretion history. Incidentally, no nuclear star cluster is included in the simulated satellite. This simulation simply serves us to show that, even in the case of an initially retrograde orbit, a sufficiently massive accretion on a MW-type galaxy can lead to accreted clusters on prograde orbits similarly to what is shown in Fig.~\ref{e_lz_ocenlike}.  An interesting point emerging from this simulation is that a number of accreted GCs are found on retrograde orbits at high energy levels (at  $\rm E \gtrsim -1 \times 10^5 (km/s)^2 $). These high-energy GCs are those, in the simulations, lost by the satellite in the early phases of its accretion. This is a region not probed by the GCs data used for this work, so it is not possible for us to test whether GCs chemically compatible with $\omega$~Cen are present also at these energy values, but it would be extremely interesting to conduct such kind of analysis in the future, once sufficient spectroscopic data will be available, because -- if found -- the location of these high-energy GCs in the $E-L_z$ plane may help to constrain the early history of this accretion.


\section{Discussion}
\label{discussion}
The chemical similarity between $\omega$~Cen and some of the clusters discussed in this paper has already been studied and emphasised in the literature. \\
The first comparison, to our knowledge, of chemical abundances in $\omega$~Cen, NGC~6656 (M~22), and NGC~6752 has been presented by 
 \citet{norris83}, who, after comparing CN, Ca, and Al in these three clusters, suggested ``common abundance enhancement mechanisms which increase in degree in going from NGC~6752, through M~22, to $\omega$~Cen.'' The high levels of carbon enhancement found in $\omega$~Cen and NGC~6656, but not in NGC~6752, led the authors to point out the exceptionality of the former two GCs with respect to NGC~6752 and other, more normal, clusters. \\
Extensive work has been presented in more recent years on the chemical peculiarities and abundance variations in NGC~6656 \citep[see, for example,][]{marino09, mckenzie2022} and the striking similarities with $\omega$~Cen  \citep{dacosta2011, marino15}. In particular,  \citet{dacosta2011} provide a number of elements of comparison between the two clusters, and it is worth discussing here the similarities and differences between our conclusions and interpretations and those presented in their paper. \\One of the first striking similarities between $\omega$~Cen and NGC~6656 is, of course, the discovery of a significant iron spread in both of them. While the existence of an iron spread in $\omega$~Cen has been known for quite a long time now (see Introduction), the discovery of a spread in Fe-content in NGC~6656 is more recent \citep[][but see also \citet{pilachowski82, lehnert91} for earlier evidence of Fe-variations in this cluster]{dacosta09, marino09}.  Since then this discovery has been subject to some debate \citep[see, for example][and, in particular, Table~1 in this latter work for a summary of the publications on this matter]{mucciarelli15, bailin19, meszaros2021, mckenzie2022}.  Fig.~1 by \citet{dacosta2011} shows the comparison between the [Fe/H] distribution of $\omega$~Cen and that of NGC~6656, based on which the authors emphasise a considerable degree of similarity between these two clusters. They indeed observe that both distributions rise rapidly on the metal-poor side to a well-defined peak, which,  however, they find to differ of about 0.09~dex (the peak in the NGC~6656 MDF results to be more metal-poor than that in $\omega$~Cen).
Our analysis confirms \citet{dacosta2011}'s findings, and allows us to push these similarities between the two MDFs even further: the two peaks result indeed mostly superposed, with the MDF of $\omega$~Cen peaking at [Fe/H]$=-1.73\pm 0.01$ and that of NGC~6656 peaking at [Fe/H]$=-1.69\pm 0.02$. Moreover, the rate of increase on the metal-poor side of the MDF is also impressively similar, as discussed in Sect.~\ref{mdf}. The higher difference found by \citet{dacosta2011} in the location of the peaks compared to ours may be due to the way the [Fe/H] distribution for $\omega$~Cen was derived in their work (from the [Ca/H] distribution presented in \citet{norris96b}, by assuming a constant [Ca/Fe] ratio of 0.4~dex, see details in their paper). \\ Concerning $\alpha-$elements, we confirm what already discussed by \citet{dacosta2011} about the lack of Mg-depleted stars in NGC~6656, which are instead present, at similar metallicity values, in $\omega$~Cen (see Fig.~\ref{NGC6656}). Some Mg-depleted stars are however present in another of the metal-poor clusters chemically compatible with $\omega$~Cen, that is NGC~6273, which is also a cluster strikingly similar to $\omega$~Cen in many aspects, as shown in the previous sections, and which shares, with NGC~6656, the same similarities to $\omega$~Cen as for their MDFs. While NGC~6656 does not show, in the current data, any presence of Mg-depleted stars, it does show the presence of a few Ca-depleted stars, another characteristic found also in NGC~6273 as well as in $\omega$~Cen itself (see Fig.~\ref{NGC6656} and Fig.~\ref{NGC6273}). The similar abundance patterns shared by $\omega$~Cen, NGC~6656, and NGC~6273 have been also investigated by \citet{johnson15}, who pointed out, among other similarities, that the shape and slope of the [La/Fe] and [La/Eu] distributions are nearly identical in all these clusters \citep[see also][for further analysis]{johnson2017}. \\
Of the three metal-poor GCs chemically compatible with $\omega$~Cen, NGC~6809 is the only cluster for which we could not find in the literature any direct comparison of its abundances with those of $\omega$~Cen. NGC~6809 is actually considered a ``normal" cluster, with the usual light-element variations found in other clusters, but with no iron spread \citep[][but see also \citet{bailin19} who reported a small, but nonzero spread for this cluster]{carretta09, rain19}. We do find an iron spread in the APOGEE data,  which is significant, given the uncertainties on the [Fe/H] estimates reported in the catalogue (for further details, see discussion in Sect.~\ref{mdf}). In this regard, it is interesting to note that in the work by \citet{rain19}, based on the abundance analysis of a sample of 11 red giant branch stars in NGC~6809, a star with a difference of -0.2~dex compared to the average iron content is reported\footnote{Note that this star was removed by \citet{bailin19} in estimating the spread of NGC~6809.}. To explain the presence of such a star, the authors evoke the possibility that it could be a pulsating variable (but note that, as the authors themselves acknowledge, they lack the photometry necessary to probe the statement). While this possibility cannot be ruled out, it is also possible that the iron content of this star is intrinsically different from the rest of the stars analysed in their work. We do also find a star in NGC~6809 with a [Fe/H] of -0.2~dex lower than the average value for the cluster, which represents 2\% of our NGC~6809 sample. As we checked, this ratio rises to 14\% for stars with a difference of -0.1~dex compared to the average iron content of the cluster.\\
As for the metal-rich clusters chemically compatible with $\omega$~Cen, we have already reported the early work by \citet{norris83}, who emphasised a common abundance enhancement mechanism increasing in strength from NGC~6752 to $\omega$~Cen. The lack of stars in NGC~6752 as Ca-enhanced as in $\omega$~Cen led the authors to consider the similarity between the two clusters with some caution. Our analysis confirms that $\omega$~Cen contains stars with [Ca/Fe] > 0.5~dex  which are rare in NGC~6752 -- but still one of the 83 stars in our sample shows such a high Ca-content (see Fig.~\ref{NGC6752}). However, it is important to emphasize that $\omega$~Cen stars with extreme Ca-abundances are found for [Fe/H]$\lesssim -1.6$~dex, so essentially at metallicities lower than those spanned by NGC~6752 stars. Coming back to Fig.~\ref{NGC6752}, it is indeed striking to see that NGC~6752 stars do cover the entire [Ca/Fe] range spanned by $\omega$~Cen at similar metallicities (note in particular the presence of a Ca-depleted star with a [Ca/Fe] ratio similar to those found for the (rare) Ca-depleted population present in $\omega$~Cen, as well as the presence of the stars with [Ca/Fe] > 0.5~dex, already mentioned above). As for the iron spread,  we do find one, even if smaller than the spread found for clusters such as NGC~6656 or NGC~6273. Note that a weak iron spread for NGC~6752 has been reported also by \citet{yong13}.\\

The fact that different clusters have chemical patterns also observed in $\omega$~Cen should not be surprising in itself. One hypothesis for the formation of NSCs is that they are the product of the orbital decay of globular clusters in the central regions of their galaxy \citep[e.g.][]{Tremaine1975, Capuzzo1993, antonini2012, mastrobuono-battisti14, gnedin2014, perets2014, antonini2015, arca-sedda2015, tsatsi2017, abbate2018}. Thus, finding the imprint of globular clusters in a nuclear star cluster is in itself a result in line with one of the theoretical hypotheses for the formation of these systems. This would be, to our knowledge, the first time that such a correspondence is highlighted in observational data and this in itself constitutes an important result of our study \citep[see][for a similar conclusion on M~54]{alfaro-cuello2019, alfaro-cuello2020, kacharov2022}. In this scenario, clusters with chemical patterns similar to those of NGC~6656, NGC~6273, NGC~6809, NGC~6205, NGC~6254, and NGC~6752  would have decayed into the centre of the progenitor galaxy of $\omega$~Cen, forming a part (at least the metal-poor part) of it. However, this hypothesis does not fully represent what we find, or rather, what we find seems to indicate a more global scenario, which this hypothesis alone cannot explain. For example, why did clusters decay in the inner regions of the progenitor galaxy should share such chemical similarities? In this scenario, it would have been sufficient to find a few clusters chemically compatible with $\omega$~Cen, but what our data show is that not only is each of these GCs chemically compatible with $\omega$~Cen, but that they all share some important chemical similarities. \\

Indeed, beyond the individual comparison of each cluster with $\omega$~Cen, what our analysis reveals is that all these clusters have properties in common. The similarities of the MDF of the metal-poor clusters, on the one hand, and of the metal-rich clusters, on the other, reveal that these groups of clusters have similar global properties, which point to a formation and an early evolution tightly linked to that of their environment (i.e. the progenitor galaxy itself or part of it). For example: 1) the similarities in the rising part of the MDFs of NGC~6656, NGC~6273, NGC~6809 and $\omega$~Cen itself point to a simultaneous origin of these clusters, which must have experienced a similar star formation in the early phases of their evolution; 2) the similarities in the MDF of NGC~6205, NGC~6254, and NGC~6752 suggest a formation that started, reached its maximum, and ceased at the same time in these clusters and which is moreover clearly linked to one of the populations found in $\omega$~Cen \citep[i.e. the intermediate~1 population, see for example][]{sollima05, calamida2020, alvarez2024}; 3) the correlation found in the [Mg/Fe]-[Fe/H] and in the [Si/Fe]-[Fe/H] plane for these three clusters may suggest that the ISM in which these GCs formed was being enriched by SNeIa, and that possibly the star formation process went on for some time, in order to produce the declining [Mg/Fe] and [Si/Fe] ratios with [Fe/H] found in these clusters\footnote{Note that we have checked that this trend is robust, that is it does not depend either on the surface gravity nor on the effective temperature of the analysed stars}. The possibility that SNeIa may play a role in the star formation and chemical enrichment of Galactic globular clusters has been recently addressed, by means of 3D hydrodynamical simulations by \citet{lacchin21}. Interestingly, some of their models (those in which SNeIa ejecta are retained by the cluster, and mix with AGB ejecta and with high-density gas accreted by the cluster during its motion in the galaxy) produce negatively skewed metallicity distributions (see Figs.~10 and 11 in their paper), as we do find for NGC~6205, NGC~6254, and NGC~6752.\\ It is also interesting to comment on the ages of these two groups of clusters. The three metal-rich GCs and two out of three metal-poor GCs have their ages measured in \citet{marin09} and \citet{vandenberg13}. In \citet{marin09} the three metal-rich GCs are found to belong to the "young" GC group while the two metal-poor GCs belong to the old group classification introduced by these authors, with an overall age difference of about 1 Gyr. Similarly \citet{vandenberg13} find an age of 12.00, 11.75 and 12.50 Gyr for NGC~6205, NGC~6254 and NGC~6752, respectively, and 12.50 and 13.00 for NGC~6656 and NGC~6809. So there is a consistent indication of an age difference between the metal-poor and metal-rich groups, suggestive of a possible chemical enrichment with time from one group to the other. Whether this is compatible with a corresponding evolution in $\omega$~Cen remains to be clarified given the complexity of the relation between age and metallicity of this system.

The common chemical abundance properties of these clusters (with each other and with $\omega$~Cen) also lead us to rule out the possibility that these clusters are the nuclear star clusters of galaxies that have been accreted in the past by MW. This is one of the most accepted hypotheses to explain the origin of clusters with high metallicity spreads, including NGC~6656 and NGC~6273 \citep[see, for example][]{dacosta2016, Pfeffer2014, pfeffer2021}. While we think that among the clusters proposed by \citet{dacosta2016} some are indeed NSCs of dwarf galaxies - besides $\omega$~Cen and NGC~6715, NGC~2808 is another interesting case that deserves additional investigation \citep[see][]{lardo2023} - we believe that NGC~6656 and NGC~6273 are not ancient NSCs of galaxies, for the following reasons:
\begin{itemize}
    \item 
the chemical similarities of these clusters with each other and with $\omega$~Cen, and the similar properties of their MDFs, would imply that they were formed in galaxies with a strikingly similar chemical evolution. Elements that would be difficult to explain if they were truly independent systems.
\item Among the clusters for which we have quantified their chemical compatibility with $\omega$~Cen there is NGC~6715 (M54), which is believed to be the NSC of the Sagittarius dwarf galaxy. The comparison between $\omega$~Cen and the latter shows very low chemical similarity (at the 16\% level, see Table~\ref{OCenGCs_table_VAC}), which, for us, reinforces the argument that NSCs formed in different galaxies should have different chemical properties. As a further check, we applied the GMM restricting only to the metallicity range shared by $\omega$~Cen and NGC~6715, again obtaining a low level of compatibility (15\%) with [C/Fe] and [Si/Fe] contributing more to this difference between the two. 
\item Finally, if these clusters were all ancient NSCs from some galaxy, the basic question would remain: where are the clusters associated with $\omega$~Cen and its progenitor? According to simulations \citep{pfeffer2018, pagnini2023}, massive enough progenitors should deposit part of their GCs in the inner Galaxy, that is in the interval of orbital energies explored in this study, and it is hard to envisage that the clusters arrived in the Galaxy with $\omega$~Cen are among the least chemically compatible with it.
\end{itemize}

With this work, we have shown that a possible way to proceed in finding clusters sharing the same origin (i.e. formed in the same progenitor) is to look for similarities in individual chemical abundances of stars. Star individual abundances keep traces of the different sources (SNeII, AGB stars, SNeIa, etc) which may have concurred in producing a stellar system with given chemical characteristics, which are only partially reflected in means, medians, or modes of the distributions in abundance planes (see Appendix~\ref{median_abundances}). The underlying motivation for looking for similarities in the chemical abundances of different clusters formed in the same progenitor galaxy is that their abundances should, to some extent, reflect those of the ISM in which these clusters were formed, that is they should share the same genetic heritage. While we do not have direct evidence of how the chemical evolution proceeded in the progenitor galaxy of $\omega$~Cen, we do know the chemical abundance patterns of $\omega$~Cen itself, which in the hypothesis that  $\omega$~Cen is the NSC of its host galaxy, provide a good representation of the chemical evolution of its host. In Appendix~\ref{NSC_dwarf} we show indeed that M~54 (NGC~6715), the nuclear star cluster of the Sagittarius dwarf galaxy, shows chemical patterns strikingly similar to those of the Sagittarius galaxy itself. 
A similar result seems to hold also for the innermost regions of the Milky Way (i.e. the inner degree), whose chemical abundances compare remarkably well with those of the inner disc \citep[][]{nandakumar2024}{}{}. 
\\ Once a common chemical link is established between a NSC and its host, it is justified to use the chemical patterns of the NSC - if known - as representative of those of their host galaxy and hence establish a link with the globular cluster population.  This is exactly what we have done in this work, reaching the conclusion that NGC~6656, NGC~6273, NGC~6809, NGC~6205, NGC~6254, and NGC~6752 were all formed in the same progenitor galaxy of which $\omega$~Cen is the remnant nucleus. We propose to name this galaxy \textit{Nephele} -- in Greek mythology the mother of Centaurs --  to emphasise a common origin for clusters that otherwise would be difficult to associate with each other. Indeed, the most popular way currently used in the literature to associate globular clusters with their progenitor galaxies and thus with episodes of accretion onto the Milky Way is to use their actual kinematic properties.
Following such an approach,  \citet{massari2019} classified the above mentioned GCs as follows: NGC~6205 as one of the clusters associated to the Gaia Sausage Enceladus accretion, of which $\omega$~Cen  would have been the NSC, according to their analysis;   NGC~6273, NGC~6809, and NGC~6254  as part of the low-energy group (later identified with Kraken by \citet{kruijssen20} or Koala by \citet{forbes2020});  NGC~6752 and NGC~6656 as disc (i.e. in-situ) clusters.  
\citet{callingham2022} using a multi-component chemo-dynamical model to split the GC populations, also favoured the Kraken origin for NGC~6809, NGC~6273, and NGC~6254 with the addition of NGC~6752 while assigned to NGC~6656 a Gaia Sausage Enceladus origin. Making also use of ``kinematic-based''  motivations \citet{myeong19} associated  $\omega$~Cen with the Sequoia galaxy, which should have brought to our Galaxy other GCs, such as FSR~1758, NGC~3201, NGC~6101, NGC~5635, and NGC~6388. Of the above listed GCs, we could quantify the chemical similarity with $\omega$~Cen for three of them: FSR~1758  shows a fraction of stars chemically compatible with $\omega$~Cen which is significant, even if affected by large uncertainties ($56\%\pm26\%$), while for the other two clusters, namely NGC~3201 and NGC~6388, our analysis shows no chemical similarity with $\omega$~Cen.

In the introduction we have recalled the limitations and lack of physical motivations behind the assumption of ``kinematic coherence'' that clusters (and field stars) accreted onto the MW over time should have. Here we make a step forward by showing that the kinematic-based classification leads to establish connections between GCs that do not share common chemical characteristics and to miss connections among GCs that show strong chemical similarities. A possible further reason, other than those already given by \citet{jeanbaptiste17, pagnini2023}, to go beyond this approach.\\

\section{Conclusions}
\label{concl}
In this paper, we made use of data from the APOGEE Value Added Catalogue (VAC) of Galactic globular cluster stars \citep[see][]{schiavon2023} that contains full APOGEE DR17 information \citep[][]{abdurro22} for a total of 6422 unique stars associated with 72 Galactic GCs. Following the prediction that $\omega$~Cen (NGC~5139) should be the nuclear star cluster of a galaxy accreted by the Milky Way in the past, our analysis aimed to search for the globular clusters brought by the $\omega$~Cen progenitor by looking for common chemical patterns between galactic GCs, and $\omega$~Cen, based on star individual abundances.
For this purpose, we made use of a Gaussian Mixture Model (GMM) approach considering an 8-dimensional abundance space defined by [Fe/H], [Mg/Fe], [Si/Fe], [Ca/Fe], [C/Fe], [Al/Fe], [K/Fe] and [Mn/Fe]. After having fitted the distribution of $\omega$~Cen in this 8-dimensional abundance space, we considered the other GCs in the sample and for each of them we estimated the fraction of stars with a high probability of belonging to the GMM model obtained for $\omega$~Cen. 
Apart from $\omega$~Cen itself, our analysis allowed us to identify six globular clusters -- namely NGC~6752, NGC~6656, NGC~6809, NGC~6273, NGC~6205, and NGC~6254 -- which have a fraction of stars compatible with $\omega$~Cen greater than 60\%. We suggest that these clusters have been potentially brought into the Milky Way by the progenitor of $\omega$~Cen, in this paper referred to as \textit{Nephele}, the mother of Centaurs. Also NGC~5024, NGC~6544, FSR~1758, and NGC~1904 - GCs with a fraction greater than 50\%, but affected by large uncertainties because of the limited number of available stars -  may also be associated to \textit{Nephele} although additional work will be needed to understand whether or not they are truly chemically similar with $\omega$~Cen.
We have divided the clusters that are chemically compatible with $\omega$~Cen into two classes, namely the metal-poor (NGC~6656, NGC~6809, and NGC~6273) and the metal-rich one (NGC 6752, NGC 6205, and NGC 6254), as these two groups share other features in addition to the [Fe/H] range. \\
For metal-poor clusters, these are:
\begin{itemize}
    \item the peak of the MDF coinciding with the peak of the $\omega$~Cen MDF;
    \item the rising part of their MDF comparable to the rising part of the (normalised) MDF of $\omega$~Cen;
    \item no trend of [Mg/Fe] with [Fe/H] (except for NGC 6809 which seems to show a slight decrease of the [Mg/Fe] with increasing [Fe/H]);
    \item a nearly flat trend in the [Si/Fe] versus [Fe/H] plane;
    \item a broad distribution in [Mg/Fe], [Si/Fe], [Ca/Fe], [C/Fe], and [K/Fe] with intrinsic dispersions that appear statistically significant in [Mg/Fe], [Ca/Fe], [C/Fe] and [K/Fe], given the reported ASPCAP uncertainties. Note, however, that this result crucially depends on their underestimation (if any).
\end{itemize}
Metal-rich GCs instead show these peculiarities:
\begin{itemize}
    \item a decreasing trend of [Mg/Fe] and [Si/Fe] with increasing [Fe/H];
    \item an extremely [K/Fe]-poor population (except for NGC~6205) present also in $\omega$~Cen at similar [Fe/H] values;
    \item their MDFs peak at the same [Fe/H] value, given the uncertainties, and this peak coincides with one of the secondary peaks found in $\omega$~Cen MDF \citep[see, for example][]{johnson2010};
    \item two out of three GCs -- namely NGC~6752 and NGC~6205 -- have negatively skewed MDFs;
    \item their most metal-poor stars have metallicities similar to those at the peak of the $\omega$~Cen MDF (at about [Fe/H]$\simeq$ -1.73).
\end{itemize}
Finally, both metal-poor and metal-rich GCs show significant, or nonzero, [Fe/H] intrinsic dispersions, comprised between 0.07 and 0.12~dex. For these dispersions not to be significant, the ASPCAP uncertainties in [Fe/H] should have been underestimated of a factor between 4 and 5, at the metallicities of these clusters.
These global characteristics, common to the aforementioned clusters and $\omega$~Cen, suggest that these clusters did not evolve chemically independently of each other. Their chemical evolution, on the contrary, must have been closely linked to the environment in which they originated and thus to the chemical evolution of their galactic progenitor or part of it.
In this respect, we would like to comment on the possibility that the evolutionary history of $\omega$~Cen is more complex than the one assumed in this paper. For example, \citet{calamida2020} suggest that $\omega$~Cen may be the result of a merger of at least two stellar systems, characterised by different mean metallicities. Our results are not in contradiction with these suggestions, in the sense that, given the current chemical space occupied by  $\omega$~Cen stars, we derive the sample of clusters chemically compatible with it. It is still possible that the two groups of GCs (metal-poor and metal-rich) that we have identified have been formed in different systems, later merged to form the current $\omega$~Cen cluster. \\
 Based on the above cited similarities, we also tend to rule out the possibility that these GCs are all NSCs of former accreted dwarfs, since this would require that all these NSCs (and their dwarfs) had a very similar chemical evolution, which should be unlikely. As additional support to this conclusion, our analysis shows that  M~54 (NGC~6715), which is known to be the NSC of the Sagittarius galaxy, is not chemically compatible with $\omega$~Cen.\\
Finally, on the basis of their chemical characteristics, we also exclude that $\omega$~Cen and its associated clusters formed in galaxies with a chemical enrichment history similar to that experienced, at early times (i.e. similar metallicities), by the Large and Small Magellanic clouds, Sagittarius and Fornax.\\

Once placed in kinematic spaces such as the E - L$_z$ space, these GCs turn out to be spread out over an extended region, with some having retrograde, some other direct orbits. For this reason, they have been linked to different galactic progenitors by kinematic-based classifications \citep[see for instance][]{massari2019, myeong19, kruijssen20, forbes2020, callingham2022, belokurov2023}. 
This spread is actually expected since, if \textit{Nephele} was massive enough compared to the Milky Way (with a mass ratio of about 1/10, as order of magnitude), simulations show that the energy and angular momentum of its stars and globular clusters should have not been conserved during its fall into the Milky Way \citep[][]{jeanbaptiste17, amarante22, pagnini2023, khoperskov23b, khoperskov23a}, and indeed in one of our \textit{N}-body simulations the accreted GCs have a distribution similar to that observed in the E - L$_z$ space for $\omega$~Cen and its related GCs. The accretion of other satellite galaxies as Gaia Sausage Enceladus, onto the Milky Way has also probably contributed to change the kinematics of these GCs, and hence their position in the E - L$_z$ space.\\
Concerning Gaia Sausage Enceladus, it is also interesting to mention that while the set of clusters commonly associated (through kinematic methods) with it does not coincide with those that we associate with \textit{Nephele}, one or two of the \textit{Nephele} clusters (NGC~6205 and NGC~6656) have been associated in the literature with Gaia Sausage Enceladus \citep[][as already mentioned]{massari2019, callingham2022}. For $\omega~$Cen itself a possible affiliation with Gaia Sausage Enceladus has been suggested \citep{massari2019}. The distribution in E-L$_z$ space of \textit{Nephele} clusters shows an overlap in kinematic spaces between these two systems. Such an overlap is not surprising if \textit{Nephele} and Gaia Sausage Enceladus were both massive galaxies relative to the MW, at the time of their accretion. Simulations indeed show that even independent systems, once accreted onto the MW, can be redistributed in similar regions of the kinematic spaces \citep[see e.g.][]{pfeffer2018, pagnini2023}. It is also possible, however, that this overlap hides closer links between \textit{Nephele} and Gaia Sausage Enceladus, which need further analysis. 

Overall, our work opens the possibility to connect different Galactic GCs to the same origin in terms of progenitor galaxy, suggesting for the first time a different procedure to achieve this goal, namely that of exploiting the chemical similarities shown by individual chemical abundances.

Our results also open new possibilities (and stimulate new questions) to understand the formation of globular clusters, whose chemical evolution and related star formation histories appear to be tightly linked to that of the environment in which they formed. A similar conclusion stands also for the Sagittarius dwarf, and its nuclear star cluster NGC~6715 (M~54). Consequently, the finding that the formation of star clusters is closely related to the environment in which they formed and that star clusters keep traces of the chemical evolution of their host galaxy seem both results of general significance, and not exclusively valid for \textit{Nephele} and its globular clusters.

\begin{acknowledgements}
The authors are grateful to the referee for their report, which much improved the presentation of the results.
The authors wish to thank R. Schiavon for his valuable comments on this work.
P.D.M. and M.H. acknowledge the support of the French Agence Nationale de la Recherche (ANR), under grant ANR-13-BS01-0005 (project ANR-20- CE31-0004-01 MWDisc). A.M.B. acknowledges funding from the European Union's Horizon 2020 research and innovation programme under the Marie Sk\l{}odowska-Curie grant agreement No 895174. F.R. acknowledges support provided by the University of Strasbourg Institute for Advanced Study (USIAS), within the French national programme Investment for the Future (Excellence Initiative) IdEx-Unistra. O.A. acknowledges support from the Knut and Alice Wallenberg Foundation, the Swedish Research Council (grant 2019-04659), and the Swedish National Space Agency (SNSA Dnr 2023-00164). N.R. acknowledges support from the Swedish Research Council (grant 2023-04744) and the Royal Physiographic Society in Lund through the Stiftelsen Walter Gyllenbergs and M{\"a}rta och Erik Holmbergs donations. Funding for the Sloan Digital Sky Survey IV has been provided by the Alfred P. Sloan Foundation, the U.S. Department of Energy Office of Science, and the Participating Institutions. SDSS-IV acknowledges support and resources from the Center for High Performance Computing at the University of Utah. The SDSS
website is \href{}{www.sdss.org}. SDSS-IV is managed by the
Astrophysical Research Consortium for the Participating Institutions of the SDSS Collaboration including the Brazilian Participation Group, the Carnegie Institution for Science, Carnegie Mellon University, Center for Astrophysics | Harvard \&
Smithsonian, the Chilean Participation Group, the French Participation Group, Instituto de Astrofisica de Canarias, The Johns Hopkins University, Kavli Institute for the Physics and Mathematics of the Universe (IPMU) / University of Tokyo, the Korean Participation Group, Lawrence Berkeley National Laboratory, Leibniz Institut f{\"u}r Astrophysik Potsdam (AIP),  Max-Planck-Institut f{\"u}r Astronomie (MPIA Heidelberg), Max-Planck-Institut f{\"u}r Astrophysik (MPA Garching), Max-Planck-Institut f{\"u}r Extraterrestrische Physik (MPE), National Astronomical Observatories of China, New Mexico State University, New York University, University of Notre Dame, Observatário Nacional / MCTI, The Ohio State University, Pennsylvania State University, Shanghai Astronomical Observatory, United Kingdom Participation Group, Universidad Nacional Autónoma de México, University of Arizona, University of Colorado Boulder, University of Oxford, University of Portsmouth, University of Utah, University of Virginia, University of Washington, University of Wisconsin, Vanderbilt University, and Yale University. This work has made use of data from the European Space Agency (ESA) mission Gaia (\href{}{https://www.cosmos.esa.int/gaia}), processed by the Gaia Data Processing and Analysis Consortium (DPAC, \href{}{https://www.cosmos.esa.int/web/gaia/dpac/consortium}). Funding for the DPAC has been provided by national institutions, in particular the institutions participating in the Gaia Multilateral Agreement. This work has made use of the computational resources obtained through the DARI grant A0120410154. 
\end{acknowledgements}

\bibliographystyle{aa}
\bibliography{biblio}

\begin{thebibliography}{105}
\expandafter\ifx\csname natexlab\endcsname\relax\def\natexlab#1{#1}\fi

\bibitem[{{Abbate} {et~al.}(2018){Abbate}, {Mastrobuono-Battisti}, {Colpi},
  {Possenti}, {Sippel}, \& {Dotti}}]{abbate2018}
{Abbate}, F., {Mastrobuono-Battisti}, A., {Colpi}, M., {et~al.} 2018, \mnras,
  473, 927

\bibitem[{{Abdurro'uf} {et~al.}(2022){Abdurro'uf}, {Accetta}, {Aerts}, {Silva
  Aguirre}, {Ahumada}, {Ajgaonkar}, {Filiz Ak}, {Alam}, {Allende Prieto},
  {Almeida}, {Anders}, {Anderson}, {Andrews}, {Anguiano}, {Aquino-Ort{\'\i}z},
  {Arag{\'o}n-Salamanca}, {Argudo-Fern{\'a}ndez}, {Ata}, {Aubert},
  {Avila-Reese}, {Badenes}, {Barb{\'a}}, {Barger}, {Barrera-Ballesteros},
  {Beaton}, {Beers}, {Belfiore}, {Bender}, {Bernardi}, {Bershady}, {Beutler},
  {Bidin}, {Bird}, {Bizyaev}, {Blanc}, {Blanton}, {Boardman}, {Bolton},
  {Boquien}, {Borissova}, {Bovy}, {Brandt}, {Brown}, {Brownstein}, {Brusa},
  {Buchner}, {Bundy}, {Burchett}, {Bureau}, {Burgasser}, {Cabang}, {Campbell},
  {Cappellari}, {Carlberg}, {Wanderley}, {Carrera}, {Cash}, {Chen}, {Chen},
  {Cherinka}, {Chiappini}, {Choi}, {Chojnowski}, {Chung}, {Clerc}, {Cohen},
  {Comerford}, {Comparat}, {da Costa}, {Covey}, {Crane}, {Cruz-Gonzalez},
  {Culhane}, {Cunha}, {Dai}, {Damke}, {Darling}, {Davidson}, {Davies},
  {Dawson}, {De Lee}, {Diamond-Stanic}, {Cano-D{\'\i}az}, {S{\'a}nchez},
  {Donor}, {Duckworth}, {Dwelly}, {Eisenstein}, {Elsworth}, {Emsellem},
  {Eracleous}, {Escoffier}, {Fan}, {Farr}, {Feng}, {Fern{\'a}ndez-Trincado},
  {Feuillet}, {Filipp}, {Fillingham}, {Frinchaboy}, {Fromenteau}, {Galbany},
  {Garc{\'\i}a}, {Garc{\'\i}a-Hern{\'a}ndez}, {Ge}, {Geisler}, {Gelfand},
  {G{\'e}ron}, {Gibson}, {Goddy}, {Godoy-Rivera}, {Grabowski}, {Green},
  {Greener}, {Grier}, {Griffith}, {Guo}, {Guy}, {Hadjara}, {Harding},
  {Hasselquist}, {Hayes}, {Hearty}, {Hern{\'a}ndez}, {Hill}, {Hogg},
  {Holtzman}, {Horta}, {Hsieh}, {Hsu}, {Hsu}, {Huber}, {Huertas-Company},
  {Hutchinson}, {Hwang}, {Ibarra-Medel}, {Chitham}, {Ilha}, {Imig}, {Jaekle},
  {Jayasinghe}, {Ji}, {Johnson}, {Jones}, {J{\"o}nsson}, {Katkov}, {Khalatyan},
  {Kinemuchi}, {Kisku}, {Knapen}, {Kneib}, {Kollmeier}, {Kong}, {Kounkel},
  {Kreckel}, {Krishnarao}, {Lacerna}, {Lane}, {Langgin}, {Lavender}, {Law},
  {Lazarz}, {Leung}, {Leung}, {Lewis}, {Li}, {Li}, {Lian}, {Liang}, {Lin},
  {Lin}, {Lin}, {Lintott}, {Long}, {Longa-Pe{\~n}a}, {L{\'o}pez-Cob{\'a}},
  {Lu}, {Lundgren}, {Luo}, {Mackereth}, {de la Macorra}, {Mahadevan},
  {Majewski}, {Manchado}, {Mandeville}, {Maraston}, {Margalef-Bentabol},
  {Masseron}, {Masters}, {Mathur}, {McDermid}, {Mckay}, {Merloni},
  {Merrifield}, {Meszaros}, {Miglio}, {Di Mille}, {Minniti}, {Minsley},
  {Monachesi}, {Moon}, {Mosser}, {Mulchaey}, {Muna}, {Mu{\~n}oz}, {Myers},
  {Myers}, {Nadathur}, {Nair}, {Nandra}, {Neumann}, {Newman}, {Nidever},
  {Nikakhtar}, {Nitschelm}, {O'Connell}, {Garma-Oehmichen}, {Luan Souza de
  Oliveira}, {Olney}, {Oravetz}, {Ortigoza-Urdaneta}, {Osorio}, {Otter},
  {Pace}, {Padilla}, {Pan}, {Pan}, {Parikh}, {Parker}, {Peirani}, {Pe{\~n}a
  Ram{\'\i}rez}, {Penny}, {Percival}, {Perez-Fournon}, {Pinsonneault},
  {Poidevin}, {Poovelil}, {Price-Whelan}, {B{\'a}rbara de Andrade Queiroz},
  {Raddick}, {Ray}, {Rembold}, {Riddle}, {Riffel}, {Riffel}, {Rix}, {Robin},
  {Rodr{\'\i}guez-Puebla}, {Roman-Lopes}, {Rom{\'a}n-Z{\'u}{\~n}iga}, {Rose},
  {Ross}, {Rossi}, {Rubin}, {Salvato}, {S{\'a}nchez}, {S{\'a}nchez-Gallego},
  {Sanderson}, {Santana Rojas}, {Sarceno}, {Sarmiento}, {Sayres}, {Sazonova},
  {Schaefer}, {Schiavon}, {Schlegel}, {Schneider}, {Schultheis}, {Schwope},
  {Serenelli}, {Serna}, {Shao}, {Shapiro}, {Sharma}, {Shen}, {Shetrone}, {Shu},
  {Simon}, {Skrutskie}, {Smethurst}, {Smith}, {Sobeck}, {Spoo}, {Sprague},
  {Stark}, {Stassun}, {Steinmetz}, {Stello}, {Stone-Martinez},
  {Storchi-Bergmann}, {Stringfellow}, {Stutz}, {Su}, {Taghizadeh-Popp},
  {Talbot}, {Tayar}, {Telles}, {Teske}, {Thakar}, {Theissen}, {Tkachenko},
  {Thomas}, {Tojeiro}, {Hernandez Toledo}, {Troup}, {Trump}, {Trussler},
  {Turner}, {Tuttle}, {Unda-Sanzana}, {V{\'a}zquez-Mata}, {Valentini},
  {Valenzuela}, {Vargas-Gonz{\'a}lez}, {Vargas-Maga{\~n}a}, {Alfaro},
  {Villanova}, {Vincenzo}, {Wake}, {Warfield}, {Washington}, {Weaver},
  {Weijmans}, {Weinberg}, {Weiss}, {Westfall}, {Wild}, {Wilde}, {Wilson},
  {Wilson}, {Wilson}, {Wolf}, {Wood-Vasey}, {Yan}, {Zamora}, {Zasowski},
  {Zhang}, {Zhao}, {Zheng}, {Zheng}, \& {Zhu}}]{abdurro22}
{Abdurro'uf}, {Accetta}, K., {Aerts}, C., {et~al.} 2022, \apjs, 259, 35

\bibitem[{{Alfaro-Cuello} {et~al.}(2020){Alfaro-Cuello}, {Kacharov},
  {Neumayer}, {Bianchini}, {Mastrobuono-Battisti}, {L{\"u}tzgendorf}, {Seth},
  {B{\"o}ker}, {Kamann}, {Leaman}, {Watkins}, \& {van de
  Ven}}]{alfaro-cuello2020}
{Alfaro-Cuello}, M., {Kacharov}, N., {Neumayer}, N., {et~al.} 2020, \apj, 892,
  20

\bibitem[{{Alfaro-Cuello} {et~al.}(2019){Alfaro-Cuello}, {Kacharov},
  {Neumayer}, {L{\"u}tzgendorf}, {Seth}, {B{\"o}ker}, {Kamann}, {Leaman}, {van
  de Ven}, {Bianchini}, {Watkins}, \& {Lyubenova}}]{alfaro-cuello2019}
{Alfaro-Cuello}, M., {Kacharov}, N., {Neumayer}, N., {et~al.} 2019, \apj, 886,
  57

\bibitem[{{Alvarez Garay} {et~al.}(2024){Alvarez Garay}, {Mucciarelli},
  {Bellazzini}, {Lardo}, \& {Ventura}}]{alvarez2024}
{Alvarez Garay}, D.~A., {Mucciarelli}, A., {Bellazzini}, M., {Lardo}, C., \&
  {Ventura}, P. 2024, \aap, 681, A54

\bibitem[{{Amarante} {et~al.}(2022){Amarante}, {Debattista}, {Beraldo e Silva},
  {Laporte}, \& {Deg}}]{amarante22}
{Amarante}, J. A.~S., {Debattista}, V.~P., {Beraldo e Silva}, L., {Laporte}, C.
  F.~P., \& {Deg}, N. 2022, \apj, 937, 12

\bibitem[{{Antonini} {et~al.}(2015){Antonini}, {Barausse}, \&
  {Silk}}]{antonini2015}
{Antonini}, F., {Barausse}, E., \& {Silk}, J. 2015, \apj, 812, 72

\bibitem[{{Antonini} {et~al.}(2012){Antonini}, {Capuzzo-Dolcetta},
  {Mastrobuono-Battisti}, \& {Merritt}}]{antonini2012}
{Antonini}, F., {Capuzzo-Dolcetta}, R., {Mastrobuono-Battisti}, A., \&
  {Merritt}, D. 2012, \apj, 750, 111

\bibitem[{{Arca-Sedda} {et~al.}(2015){Arca-Sedda}, {Capuzzo-Dolcetta},
  {Antonini}, \& {Seth}}]{arca-sedda2015}
{Arca-Sedda}, M., {Capuzzo-Dolcetta}, R., {Antonini}, F., \& {Seth}, A. 2015,
  \apj, 806, 220

\bibitem[{{Bailin}(2019)}]{bailin19}
{Bailin}, J. 2019, \apjs, 245, 5

\bibitem[{{Baumgardt} \& {Hilker}(2018)}]{baumgardt18}
{Baumgardt}, H. \& {Hilker}, M. 2018, \mnras, 478, 1520

\bibitem[{{Bekki} \& {Freeman}(2003)}]{bekki03}
{Bekki}, K. \& {Freeman}, K.~C. 2003, \mnras, 346, L11

\bibitem[{{Bellazzini} {et~al.}(2003){Bellazzini}, {Ferraro}, \&
  {Ibata}}]{bellazzini2003}
{Bellazzini}, M., {Ferraro}, F.~R., \& {Ibata}, R. 2003, \aj, 125, 188

\bibitem[{{Bellazzini} {et~al.}(2020){Bellazzini}, {Ibata}, {Malhan}, {Martin},
  {Famaey}, \& {Thomas}}]{bellazzini2020}
{Bellazzini}, M., {Ibata}, R., {Malhan}, K., {et~al.} 2020, \aap, 636, A107

\bibitem[{{Bellazzini} {et~al.}(2008){Bellazzini}, {Ibata}, {Chapman},
  {Mackey}, {Monaco}, {Irwin}, {Martin}, {Lewis}, \&
  {Dalessandro}}]{bellazzini2008}
{Bellazzini}, M., {Ibata}, R.~A., {Chapman}, S.~C., {et~al.} 2008, \aj, 136,
  1147

\bibitem[{{Belokurov} \& {Kravtsov}(2024)}]{belokurov2023}
{Belokurov}, V. \& {Kravtsov}, A. 2024, \mnras, 528, 3198

\bibitem[{Bennett \& Bovy(2019)}]{bennett2019}
Bennett, M. \& Bovy, J. 2019, Monthly Notices of the Royal Astronomical
  Society, 482, 1417

\bibitem[{{Bensby} {et~al.}(2010){Bensby}, {Feltzing}, {Johnson}, {Gould},
  {Ad{\'e}n}, {Asplund}, {Mel{\'e}ndez}, {Gal-Yam}, {Lucatello}, {Sana},
  {Sumi}, {Miyake}, {Suzuki}, {Han}, {Bond}, \& {Udalski}}]{bensby10}
{Bensby}, T., {Feltzing}, S., {Johnson}, J.~A., {et~al.} 2010, \aap, 512, A41

\bibitem[{{Bensby} {et~al.}(2021){Bensby}, {Gould}, {Asplund}, {Feltzing},
  {Mel{\'e}ndez}, {Johnson}, {Lucatello}, {Udalski}, \& {Yee}}]{bensby21}
{Bensby}, T., {Gould}, A., {Asplund}, M., {et~al.} 2021, \aap, 655, A117

\bibitem[{{Bianchini} {et~al.}(2013){Bianchini}, {Varri}, {Bertin}, \&
  {Zocchi}}]{bianchini13}
{Bianchini}, P., {Varri}, A.~L., {Bertin}, G., \& {Zocchi}, A. 2013, \apj, 772,
  67

\bibitem[{{Bovy}(2015)}]{bovy2015}
{Bovy}, J. 2015, \apjs, 216, 29

\bibitem[{{Calamida} {et~al.}(2009){Calamida}, {Bono}, {Stetson}, {Freyhammer},
  {Piersimoni}, {Buonanno}, {Caputo}, {Cassisi}, {Castellani}, {Corsi},
  {Dall'Ora}, {Degl'Innocenti}, {Ferraro}, {Grundahl}, {Hilker}, {Iannicola},
  {Monelli}, {Nonino}, {Patat}, {Pietrinferni}, {Prada Moroni}, {Primas},
  {Pulone}, {Richtler}, {Romaniello}, {Storm}, \& {Walker}}]{calamida09}
{Calamida}, A., {Bono}, G., {Stetson}, P.~B., {et~al.} 2009, \apj, 706, 1277

\bibitem[{{Calamida} {et~al.}(2020){Calamida}, {Zocchi}, {Bono}, {Ferraro},
  {Mastrobuono-Battisti}, {Saha}, {Iannicola}, {Rest}, {Strampelli}, \&
  {Zenteno}}]{calamida2020}
{Calamida}, A., {Zocchi}, A., {Bono}, G., {et~al.} 2020, \apj, 891, 167

\bibitem[{{Callingham} {et~al.}(2022){Callingham}, {Cautun}, {Deason}, {Frenk},
  {Grand}, \& {Marinacci}}]{callingham2022}
{Callingham}, T.~M., {Cautun}, M., {Deason}, A.~J., {et~al.} 2022, \mnras, 513,
  4107

\bibitem[{{Capuzzo-Dolcetta}(1993)}]{Capuzzo1993}
{Capuzzo-Dolcetta}, R. 1993, \apj, 415, 616

\bibitem[{{Carraro} \& {Lia}(2000)}]{carraro00}
{Carraro}, G. \& {Lia}, C. 2000, \aap, 357, 977

\bibitem[{{Carretta} {et~al.}(2009){Carretta}, {Bragaglia}, {Gratton},
  {D'Orazi}, \& {Lucatello}}]{carretta09}
{Carretta}, E., {Bragaglia}, A., {Gratton}, R., {D'Orazi}, V., \& {Lucatello},
  S. 2009, \aap, 508, 695

\bibitem[{{Da Costa}(2016)}]{dacosta2016}
{Da Costa}, G.~S. 2016, in The General Assembly of Galaxy Halos: Structure,
  Origin and Evolution, ed. A.~{Bragaglia}, M.~{Arnaboldi}, M.~{Rejkuba}, \&
  D.~{Romano}, Vol. 317, 110--115

\bibitem[{{Da Costa} {et~al.}(2009){Da Costa}, {Held}, {Saviane}, \&
  {Gullieuszik}}]{dacosta09}
{Da Costa}, G.~S., {Held}, E.~V., {Saviane}, I., \& {Gullieuszik}, M. 2009,
  \apj, 705, 1481

\bibitem[{{Da Costa} \& {Marino}(2011)}]{dacosta2011}
{Da Costa}, G.~S. \& {Marino}, A.~F. 2011, \pasa, 28, 28

\bibitem[{{D'Orazi} {et~al.}(2011){D'Orazi}, {Gratton}, {Pancino}, {Bragaglia},
  {Carretta}, {Lucatello}, \& {Sneden}}]{dorazi11}
{D'Orazi}, V., {Gratton}, R.~G., {Pancino}, E., {et~al.} 2011, \aap, 534, A29

\bibitem[{{Eadie} {et~al.}(2022){Eadie}, {Harris}, \& {Springford}}]{eadie22}
{Eadie}, G.~M., {Harris}, W.~E., \& {Springford}, A. 2022, \apj, 926, 162

\bibitem[{{Forbes}(2020)}]{forbes2020}
{Forbes}, D.~A. 2020, \mnras, 493, 847

\bibitem[{{Freeman} \& {Bland-Hawthorn}(2002)}]{freeman02}
{Freeman}, K. \& {Bland-Hawthorn}, J. 2002, \araa, 40, 487

\bibitem[{{Gnedin} {et~al.}(2014){Gnedin}, {Ostriker}, \&
  {Tremaine}}]{gnedin2014}
{Gnedin}, O.~Y., {Ostriker}, J.~P., \& {Tremaine}, S. 2014, \apj, 785, 71

\bibitem[{{GRAVITY Collaboration} {et~al.}(2019){GRAVITY Collaboration},
  {Abuter}, {Amorim}, {Baub{\"o}ck}, {Berger}, {Bonnet}, {Brandner},
  {Cl{\'e}net}, {Coud{\'e} Du Foresto}, {de Zeeuw}, {Dexter}, {Duvert},
  {Eckart}, {Eisenhauer}, {F{\"o}rster Schreiber}, {Garcia}, {Gao}, {Gendron},
  {Genzel}, {Gerhard}, {Gillessen}, {Habibi}, {Haubois}, {Henning}, {Hippler},
  {Horrobin}, {Jim{\'e}nez-Rosales}, {Jocou}, {Kervella}, {Lacour},
  {Lapeyr{\`e}re}, {Le Bouquin}, {L{\'e}na}, {Ott}, {Paumard}, {Perraut},
  {Perrin}, {Pfuhl}, {Rabien}, {Rodriguez Coira}, {Rousset}, {Scheithauer},
  {Sternberg}, {Straub}, {Straubmeier}, {Sturm}, {Tacconi}, {Vincent}, {von
  Fellenberg}, {Waisberg}, {Widmann}, {Wieprecht}, {Wiezorrek}, {Woillez}, \&
  {Yazici}}]{gravity2019}
{GRAVITY Collaboration}, {Abuter}, R., {Amorim}, A., {et~al.} 2019, \aap, 625,
  L10

\bibitem[{{Hasselquist} {et~al.}(2021){Hasselquist}, {Hayes}, {Lian},
  {Weinberg}, {Zasowski}, {Horta}, {Beaton}, {Feuillet}, {Garro}, {Gallart},
  {Smith}, {Holtzman}, {Minniti}, {Lacerna}, {Shetrone}, {J{\"o}nsson},
  {Cioni}, {Fillingham}, {Cunha}, {O'Connell}, {Fern{\'a}ndez-Trincado},
  {Mu{\~n}oz}, {Schiavon}, {Almeida}, {Anguiano}, {Beers}, {Bizyaev},
  {Brownstein}, {Cohen}, {Frinchaboy}, {Garc{\'\i}a-Hern{\'a}ndez}, {Geisler},
  {Lane}, {Majewski}, {Nidever}, {Nitschelm}, {Povick}, {Price-Whelan},
  {Roman-Lopes}, {Rosado}, {Sobeck}, {Stringfellow}, {Valenzuela}, {Villanova},
  \& {Vincenzo}}]{hasselquist21}
{Hasselquist}, S., {Hayes}, C.~R., {Lian}, J., {et~al.} 2021, \apj, 923, 172

\bibitem[{{Haywood} {et~al.}(2018){Haywood}, {Di Matteo}, {Lehnert}, {Snaith},
  {Fragkoudi}, \& {Khoperskov}}]{haywood18}
{Haywood}, M., {Di Matteo}, P., {Lehnert}, M., {et~al.} 2018, \aap, 618, A78

\bibitem[{{Hilker} {et~al.}(2004){Hilker}, {Kayser}, {Richtler}, \&
  {Willemsen}}]{hilker04}
{Hilker}, M., {Kayser}, A., {Richtler}, T., \& {Willemsen}, P. 2004, \aap, 422,
  L9

\bibitem[{Horta {et~al.}(2023)Horta, Schiavon, Mackereth, Weinberg,
  Hasselquist, Feuillet, O’Connell, Anguiano, Allende-Prieto, Beaton,
  {et~al.}}]{horta2023}
Horta, D., Schiavon, R.~P., Mackereth, J.~T., {et~al.} 2023, Monthly Notices of
  the Royal Astronomical Society, 520, 5671

\bibitem[{{Ibata} {et~al.}(2019){Ibata}, {Bellazzini}, {Malhan}, {Martin}, \&
  {Bianchini}}]{ibata19a}
{Ibata}, R.~A., {Bellazzini}, M., {Malhan}, K., {Martin}, N., \& {Bianchini},
  P. 2019, Nature Astronomy, 3, 667

\bibitem[{{Iben}(1967)}]{iben67}
{Iben}, Icko, J. 1967, \araa, 5, 571

\bibitem[{{Jean-Baptiste} {et~al.}(2017){Jean-Baptiste}, {Di Matteo},
  {Haywood}, {G{\'o}mez}, {Montuori}, {Combes}, \& {Semelin}}]{jeanbaptiste17}
{Jean-Baptiste}, I., {Di Matteo}, P., {Haywood}, M., {et~al.} 2017, \aap, 604,
  A106

\bibitem[{{Johnson} {et~al.}(2017){Johnson}, {Caldwell}, {Rich}, {Mateo},
  {Bailey}, {Clarkson}, {Olszewski}, \& {Walker}}]{johnson2017}
{Johnson}, C.~I., {Caldwell}, N., {Rich}, R.~M., {et~al.} 2017, \apj, 836, 168

\bibitem[{{Johnson} \& {Pilachowski}(2010)}]{johnson2010}
{Johnson}, C.~I. \& {Pilachowski}, C.~A. 2010, \apj, 722, 1373

\bibitem[{{Johnson} {et~al.}(2015){Johnson}, {Rich}, {Pilachowski}, {Caldwell},
  {Mateo}, {Bailey}, \& {Crane}}]{johnson15}
{Johnson}, C.~I., {Rich}, R.~M., {Pilachowski}, C.~A., {et~al.} 2015, \aj, 150,
  63

\bibitem[{{Kacharov} {et~al.}(2022){Kacharov}, {Alfaro-Cuello}, {Neumayer},
  {L{\"u}tzgendorf}, {Watkins}, {Mastrobuono-Battisti}, {Kamann}, {van de Ven},
  {Seth}, {Voggel}, {Georgiev}, {Leaman}, {Bianchini}, {B{\"o}ker}, \&
  {Mieske}}]{kacharov2022}
{Kacharov}, N., {Alfaro-Cuello}, M., {Neumayer}, N., {et~al.} 2022, \apj, 939,
  118

\bibitem[{Kamann {et~al.}(2018)Kamann, Husser, Dreizler, Emsellem, Weilbacher,
  Martens, Bacon, den Brok, Giesers, Krajnovi{\'c}, {et~al.}}]{kamann2018}
Kamann, S., Husser, T.-O., Dreizler, S., {et~al.} 2018, Monthly Notices of the
  Royal Astronomical Society, 473, 5591

\bibitem[{{Khoperskov} {et~al.}(2023{\natexlab{a}}){Khoperskov}, {Minchev},
  {Libeskind}, {Haywood}, {Di Matteo}, {Belokurov}, {Steinmetz}, {Gomez},
  {Grand}, {Hoffman}, {Knebe}, {Sorce}, {Spaare}, {Tempel}, \&
  {Vogelsberger}}]{khoperskov23a}
{Khoperskov}, S., {Minchev}, I., {Libeskind}, N., {et~al.} 2023{\natexlab{a}},
  \aap, 677, A90

\bibitem[{{Khoperskov} {et~al.}(2023{\natexlab{b}}){Khoperskov}, {Minchev},
  {Libeskind}, {Haywood}, {Di Matteo}, {Belokurov}, {Steinmetz}, {Gomez},
  {Grand}, {Hoffman}, {Knebe}, {Sorce}, {Spaare}, {Tempel}, \&
  {Vogelsberger}}]{khoperskov23b}
{Khoperskov}, S., {Minchev}, I., {Libeskind}, N., {et~al.} 2023{\natexlab{b}},
  \aap, 677, A89

\bibitem[{{Koppelman} {et~al.}(2020){Koppelman}, {Bos}, \&
  {Helmi}}]{koppelman20}
{Koppelman}, H.~H., {Bos}, R. O.~Y., \& {Helmi}, A. 2020, \aap, 642, L18

\bibitem[{{Kruijssen} {et~al.}(2020){Kruijssen}, {Pfeffer}, {Chevance},
  {Bonaca}, {Trujillo-Gomez}, {Bastian}, {Reina-Campos}, {Crain}, \&
  {Hughes}}]{kruijssen20}
{Kruijssen}, J.~M.~D., {Pfeffer}, J.~L., {Chevance}, M., {et~al.} 2020, \mnras,
  498, 2472

\bibitem[{{Lacchin} {et~al.}(2021){Lacchin}, {Calura}, \&
  {Vesperini}}]{lacchin21}
{Lacchin}, E., {Calura}, F., \& {Vesperini}, E. 2021, \mnras, 506, 5951

\bibitem[{{Lardo} {et~al.}(2023){Lardo}, {Salaris}, {Cassisi}, {Bastian},
  {Mucciarelli}, {Cabrera-Ziri}, \& {Dalessandro}}]{lardo2023}
{Lardo}, C., {Salaris}, M., {Cassisi}, S., {et~al.} 2023, \aap, 669, A19

\bibitem[{{Lee} {et~al.}(1999){Lee}, {Joo}, {Sohn}, {Rey}, {Lee}, \&
  {Walker}}]{lee99}
{Lee}, Y.~W., {Joo}, J.~M., {Sohn}, Y.~J., {et~al.} 1999, \nat, 402, 55

\bibitem[{{Lehnert} {et~al.}(1991){Lehnert}, {Bell}, \& {Cohen}}]{lehnert91}
{Lehnert}, M.~D., {Bell}, R.~A., \& {Cohen}, J.~G. 1991, \apj, 367, 514

\bibitem[{{Leung} \& {Bovy}(2019{\natexlab{a}})}]{leung2019b}
{Leung}, H.~W. \& {Bovy}, J. 2019{\natexlab{a}}, \mnras, 483, 3255

\bibitem[{{Leung} \& {Bovy}(2019{\natexlab{b}})}]{leung2019a}
{Leung}, H.~W. \& {Bovy}, J. 2019{\natexlab{b}}, \mnras, 489, 2079

\bibitem[{{Majewski} {et~al.}(2000){Majewski}, {Patterson}, {Dinescu},
  {Johnson}, {Ostheimer}, {Kunkel}, \& {Palma}}]{majewski00}
{Majewski}, S.~R., {Patterson}, R.~J., {Dinescu}, D.~I., {et~al.} 2000, in
  Liege International Astrophysical Colloquia, Vol.~35, Liege International
  Astrophysical Colloquia, ed. A.~{Noels}, P.~{Magain}, D.~{Caro}, E.~{Jehin},
  G.~{Parmentier}, \& A.~A. {Thoul}, 619

\bibitem[{{Mar{\'\i}n-Franch} {et~al.}(2009){Mar{\'\i}n-Franch}, {Aparicio},
  {Piotto}, {Rosenberg}, {Chaboyer}, {Sarajedini}, {Siegel}, {Anderson},
  {Bedin}, {Dotter}, {Hempel}, {King}, {Majewski}, {Milone}, {Paust}, \&
  {Reid}}]{marin09}
{Mar{\'\i}n-Franch}, A., {Aparicio}, A., {Piotto}, G., {et~al.} 2009, \apj,
  694, 1498

\bibitem[{{Marino}(2015)}]{marino15}
{Marino}, A.~F. 2015, Highlights of Astronomy, 16, 234

\bibitem[{{Marino} {et~al.}(2009){Marino}, {Milone}, {Piotto}, {Villanova},
  {Bedin}, {Bellini}, \& {Renzini}}]{marino09}
{Marino}, A.~F., {Milone}, A.~P., {Piotto}, G., {et~al.} 2009, \aap, 505, 1099

\bibitem[{{Marino} {et~al.}(2011){Marino}, {Milone}, {Piotto}, {Villanova},
  {Gratton}, {D'Antona}, {Anderson}, {Bedin}, {Bellini}, {Cassisi}, {Geisler},
  {Renzini}, \& {Zoccali}}]{marino11}
{Marino}, A.~F., {Milone}, A.~P., {Piotto}, G., {et~al.} 2011, \apj, 731, 64

\bibitem[{{Massari} {et~al.}(2019){Massari}, {Koppelman}, \&
  {Helmi}}]{massari2019}
{Massari}, D., {Koppelman}, H.~H., \& {Helmi}, A. 2019, \aap, 630, L4

\bibitem[{{Mastrobuono-Battisti} {et~al.}(2014){Mastrobuono-Battisti},
  {Perets}, \& {Loeb}}]{mastrobuono-battisti14}
{Mastrobuono-Battisti}, A., {Perets}, H.~B., \& {Loeb}, A. 2014, \apj, 796, 40

\bibitem[{{McKenzie} {et~al.}(2022){McKenzie}, {Yong}, {Marino}, {Monty},
  {Wang}, {Karakas}, {Milone}, {Legnardi}, {Roederer}, {Martell}, \&
  {Horta}}]{mckenzie2022}
{McKenzie}, M., {Yong}, D., {Marino}, A.~F., {et~al.} 2022, \mnras, 516, 3515

\bibitem[{{McMillan}(2017)}]{mcmillan2017}
{McMillan}, P.~J. 2017, \mnras, 465, 76

\bibitem[{{McWilliam} {et~al.}(2013){McWilliam}, {Wallerstein}, \&
  {Mottini}}]{mcwilliam13}
{McWilliam}, A., {Wallerstein}, G., \& {Mottini}, M. 2013, \apj, 778, 149

\bibitem[{{Merritt} {et~al.}(1997){Merritt}, {Meylan}, \& {Mayor}}]{merritt97}
{Merritt}, D., {Meylan}, G., \& {Mayor}, M. 1997, \aj, 114, 1074

\bibitem[{{M{\'e}sz{\'a}ros} {et~al.}(2021){M{\'e}sz{\'a}ros}, {Masseron},
  {Fern{\'a}ndez-Trincado}, {Garc{\'\i}a-Hern{\'a}ndez}, {Szigeti}, {Cunha},
  {Shetrone}, {Smith}, {Beaton}, {Beers}, {Brownstein}, {Geisler}, {Hayes},
  {J{\"o}nsson}, {Lane}, {Majewski}, {Minniti}, {Munoz}, {Nitschelm},
  {Roman-Lopes}, \& {Zamora}}]{meszaros2021}
{M{\'e}sz{\'a}ros}, S., {Masseron}, T., {Fern{\'a}ndez-Trincado}, J.~G.,
  {et~al.} 2021, \mnras, 505, 1645

\bibitem[{{Meylan} \& {Mayor}(1986)}]{meylan86}
{Meylan}, G. \& {Mayor}, M. 1986, \aap, 166, 122

\bibitem[{{Mucciarelli} {et~al.}(2015){Mucciarelli}, {Lapenna}, {Massari},
  {Pancino}, {Stetson}, {Ferraro}, {Lanzoni}, \& {Lardo}}]{mucciarelli15}
{Mucciarelli}, A., {Lapenna}, E., {Massari}, D., {et~al.} 2015, \apj, 809, 128

\bibitem[{{Myeong} {et~al.}(2019){Myeong}, {Vasiliev}, {Iorio}, {Evans}, \&
  {Belokurov}}]{myeong19}
{Myeong}, G.~C., {Vasiliev}, E., {Iorio}, G., {Evans}, N.~W., \& {Belokurov},
  V. 2019, \mnras, 488, 1235

\bibitem[{{Nandakumar} {et~al.}(2024){Nandakumar}, {Ryde}, {Mace}, {Kaplan},
  {Nieuwmunster}, {Jaffe}, {Rich}, {Schultheis}, {Agertz}, {Andersson},
  {Sneden}, {Strickland}, \& {Thorsbro}}]{nandakumar2024}
{Nandakumar}, G., {Ryde}, N., {Mace}, G., {et~al.} 2024, arXiv e-prints,
  arXiv:2401.13991

\bibitem[{{Nitschai} {et~al.}(2023){Nitschai}, {Neumayer}, {Clontz},
  {H{\"a}berle}, {Seth}, {Husser}, {Kamann}, {Alfaro-Cuello}, {Kacharov},
  {Bellini}, {Dotter}, {Dreizler}, {Feldmeier-Krause}, {Latour}, {Libralato},
  {Milone}, {Pechetti}, {van de Ven}, {Voggel}, \& {Weisz}}]{nitschai23}
{Nitschai}, M.~S., {Neumayer}, N., {Clontz}, C., {et~al.} 2023, \apj, 958, 8

\bibitem[{{Norris} \& {Freeman}(1983)}]{norris83}
{Norris}, J. \& {Freeman}, K.~C. 1983, \apj, 266, 130

\bibitem[{{Norris} \& {Da Costa}(1995)}]{norris95}
{Norris}, J.~E. \& {Da Costa}, G.~S. 1995, \apj, 447, 680

\bibitem[{{Norris} {et~al.}(1997){Norris}, {Freeman}, {Mayor}, \&
  {Seitzer}}]{norris97}
{Norris}, J.~E., {Freeman}, K.~C., {Mayor}, M., \& {Seitzer}, P. 1997, \apjl,
  487, L187

\bibitem[{{Norris} {et~al.}(1996){Norris}, {Freeman}, \& {Mighell}}]{norris96b}
{Norris}, J.~E., {Freeman}, K.~C., \& {Mighell}, K.~J. 1996, \apj, 462, 241

\bibitem[{{Pagnini} {et~al.}(2023){Pagnini}, {Di Matteo}, {Khoperskov},
  {Mastrobuono-Battisti}, {Haywood}, {Renaud}, \& {Combes}}]{pagnini2023}
{Pagnini}, G., {Di Matteo}, P., {Khoperskov}, S., {et~al.} 2023, \aap, 673, A86

\bibitem[{{Pancino} {et~al.}(2007){Pancino}, {Galfo}, {Ferraro}, \&
  {Bellazzini}}]{pancino07}
{Pancino}, E., {Galfo}, A., {Ferraro}, F.~R., \& {Bellazzini}, M. 2007, \apjl,
  661, L155

\bibitem[{{Pancino} {et~al.}(2011){Pancino}, {Mucciarelli}, {Sbordone},
  {Bellazzini}, {Pasquini}, {Monaco}, \& {Ferraro}}]{pancino11}
{Pancino}, E., {Mucciarelli}, A., {Sbordone}, L., {et~al.} 2011, \aap, 527, A18

\bibitem[{{Pechetti} {et~al.}(2024){Pechetti}, {Kamann}, {Krajnovi{\'c}},
  {Seth}, {van de Ven}, {Neumayer}, {Dreizler}, {Weilbacher}, {Martens}, \&
  {Wragg}}]{pechetti24}
{Pechetti}, R., {Kamann}, S., {Krajnovi{\'c}}, D., {et~al.} 2024, \mnras, 528,
  4941

\bibitem[{{Perets} \& {Mastrobuono-Battisti}(2014)}]{perets2014}
{Perets}, H.~B. \& {Mastrobuono-Battisti}, A. 2014, \apjl, 784, L44

\bibitem[{{Pfeffer} {et~al.}(2014){Pfeffer}, {Griffen}, {Baumgardt}, \&
  {Hilker}}]{Pfeffer2014}
{Pfeffer}, J., {Griffen}, B.~F., {Baumgardt}, H., \& {Hilker}, M. 2014, \mnras,
  444, 3670

\bibitem[{{Pfeffer} {et~al.}(2018){Pfeffer}, {Kruijssen}, {Crain}, \&
  {Bastian}}]{pfeffer2018}
{Pfeffer}, J., {Kruijssen}, J.~M.~D., {Crain}, R.~A., \& {Bastian}, N. 2018,
  \mnras, 475, 4309

\bibitem[{{Pfeffer} {et~al.}(2021){Pfeffer}, {Lardo}, {Bastian}, {Saracino}, \&
  {Kamann}}]{pfeffer2021}
{Pfeffer}, J., {Lardo}, C., {Bastian}, N., {Saracino}, S., \& {Kamann}, S.
  2021, \mnras, 500, 2514

\bibitem[{{Pilachowski} {et~al.}(1982){Pilachowski}, {Leep}, {Wallerstein}, \&
  {Peterson}}]{pilachowski82}
{Pilachowski}, C., {Leep}, E.~M., {Wallerstein}, G., \& {Peterson}, R.~C. 1982,
  \apj, 263, 187

\bibitem[{Rain {et~al.}(2018)Rain, Villanova, Munõz, \&
  Valenzuela-Calderon}]{rain19}
Rain, M.~J., Villanova, S., Munõz, C., \& Valenzuela-Calderon, C. 2018,
  Monthly Notices of the Royal Astronomical Society, 483, 1674

\bibitem[{{Sanna} {et~al.}(2020){Sanna}, {Pancino}, {Zocchi}, {Ferraro}, \&
  {Stetson}}]{sanna20}
{Sanna}, N., {Pancino}, E., {Zocchi}, A., {Ferraro}, F.~R., \& {Stetson}, P.~B.
  2020, \aap, 637, A46

\bibitem[{{Schiavon} {et~al.}(2024){Schiavon}, {Phillips}, {Myers}, {Horta},
  {Minniti}, {Allende Prieto}, {Anguiano}, {Beaton}, {Beers}, {Brownstein},
  {Cohen}, {Fern{\'a}ndez-Trincado}, {Frinchaboy}, {J{\"o}nsson}, {Kisku},
  {Lane}, {Majewski}, {Mason}, {M{\'e}sz{\'a}ros}, \&
  {Stringfellow}}]{schiavon2023}
{Schiavon}, R.~P., {Phillips}, S.~G., {Myers}, N., {et~al.} 2024, \mnras, 528,
  1393

\bibitem[{{Sistero} \& {Fourcade}(1970)}]{sistero70}
{Sistero}, R.~F. \& {Fourcade}, C.~R. 1970, \aj, 75, 34

\bibitem[{{Smith} {et~al.}(2000){Smith}, {Suntzeff}, {Cunha}, {Gallino},
  {Busso}, {Lambert}, \& {Straniero}}]{smith00}
{Smith}, V.~V., {Suntzeff}, N.~B., {Cunha}, K., {et~al.} 2000, \aj, 119, 1239

\bibitem[{{Snaith} {et~al.}(2014){Snaith}, {Haywood}, {Di Matteo}, {Lehnert},
  {Combes}, {Katz}, \& {G{\'o}mez}}]{snaith14}
{Snaith}, O.~N., {Haywood}, M., {Di Matteo}, P., {et~al.} 2014, \apjl, 781, L31

\bibitem[{{Sollima} {et~al.}(2005){Sollima}, {Pancino}, {Ferraro},
  {Bellazzini}, {Straniero}, \& {Pasquini}}]{sollima05}
{Sollima}, A., {Pancino}, E., {Ferraro}, F.~R., {et~al.} 2005, \apj, 634, 332

\bibitem[{{Suntzeff} \& {Kraft}(1996)}]{suntzeff96}
{Suntzeff}, N.~B. \& {Kraft}, R.~P. 1996, \aj, 111, 1913

\bibitem[{{Tremaine} {et~al.}(1975){Tremaine}, {Ostriker}, \&
  {Spitzer}}]{Tremaine1975}
{Tremaine}, S.~D., {Ostriker}, J.~P., \& {Spitzer}, L., J. 1975, \apj, 196, 407

\bibitem[{{Tsatsi} {et~al.}(2017){Tsatsi}, {Mastrobuono-Battisti}, {van de
  Ven}, {Perets}, {Bianchini}, \& {Neumayer}}]{tsatsi2017}
{Tsatsi}, A., {Mastrobuono-Battisti}, A., {van de Ven}, G., {et~al.} 2017,
  \mnras, 464, 3720

\bibitem[{{Tsuchiya} {et~al.}(2003){Tsuchiya}, {Dinescu}, \&
  {Korchagin}}]{tsuchiya03}
{Tsuchiya}, T., {Dinescu}, D.~I., \& {Korchagin}, V.~I. 2003, \apjl, 589, L29

\bibitem[{{Tsuchiya} {et~al.}(2004){Tsuchiya}, {Korchagin}, \&
  {Dinescu}}]{tsuchiya04}
{Tsuchiya}, T., {Korchagin}, V.~I., \& {Dinescu}, D.~I. 2004, \mnras, 350, 1141

\bibitem[{{VandenBerg} {et~al.}(2013){VandenBerg}, {Brogaard}, {Leaman}, \&
  {Casagrande}}]{vandenberg13}
{VandenBerg}, D.~A., {Brogaard}, K., {Leaman}, R., \& {Casagrande}, L. 2013,
  \apj, 775, 134

\bibitem[{{Vasiliev} \& {Baumgardt}(2021)}]{vasiliev21}
{Vasiliev}, E. \& {Baumgardt}, H. 2021, \mnras, 505, 5978

\bibitem[{{Villanova} {et~al.}(2014){Villanova}, {Geisler}, {Gratton}, \&
  {Cassisi}}]{villanova14}
{Villanova}, S., {Geisler}, D., {Gratton}, R.~G., \& {Cassisi}, S. 2014, \apj,
  791, 107

\bibitem[{{Villanova} {et~al.}(2007){Villanova}, {Piotto}, {King}, {Anderson},
  {Bedin}, {Gratton}, {Cassisi}, {Momany}, {Bellini}, {Cool}, {Recio-Blanco},
  \& {Renzini}}]{villanova07}
{Villanova}, S., {Piotto}, G., {King}, I.~R., {et~al.} 2007, \apj, 663, 296

\bibitem[{{Yong} {et~al.}(2013){Yong}, {Mel{\'e}ndez}, {Grundahl}, {Roederer},
  {Norris}, {Milone}, {Marino}, {Coelho}, {McArthur}, {Lind}, {Collet}, \&
  {Asplund}}]{yong13}
{Yong}, D., {Mel{\'e}ndez}, J., {Grundahl}, F., {et~al.} 2013, \mnras, 434,
  3542

\end{thebibliography}

\begin{appendix}
\counterwithin{figure}{section}

\section{Are the chemical abundances of a nuclear cluster representative of that of the host galaxy? The case of M~54 and the Sagittarius dwarf}\label{NSC_dwarf}

 \begin{figure*}[h!]
   \centering
\includegraphics[clip=true, trim = 3mm 0mm 0mm 3mm, width=0.69\columnwidth]{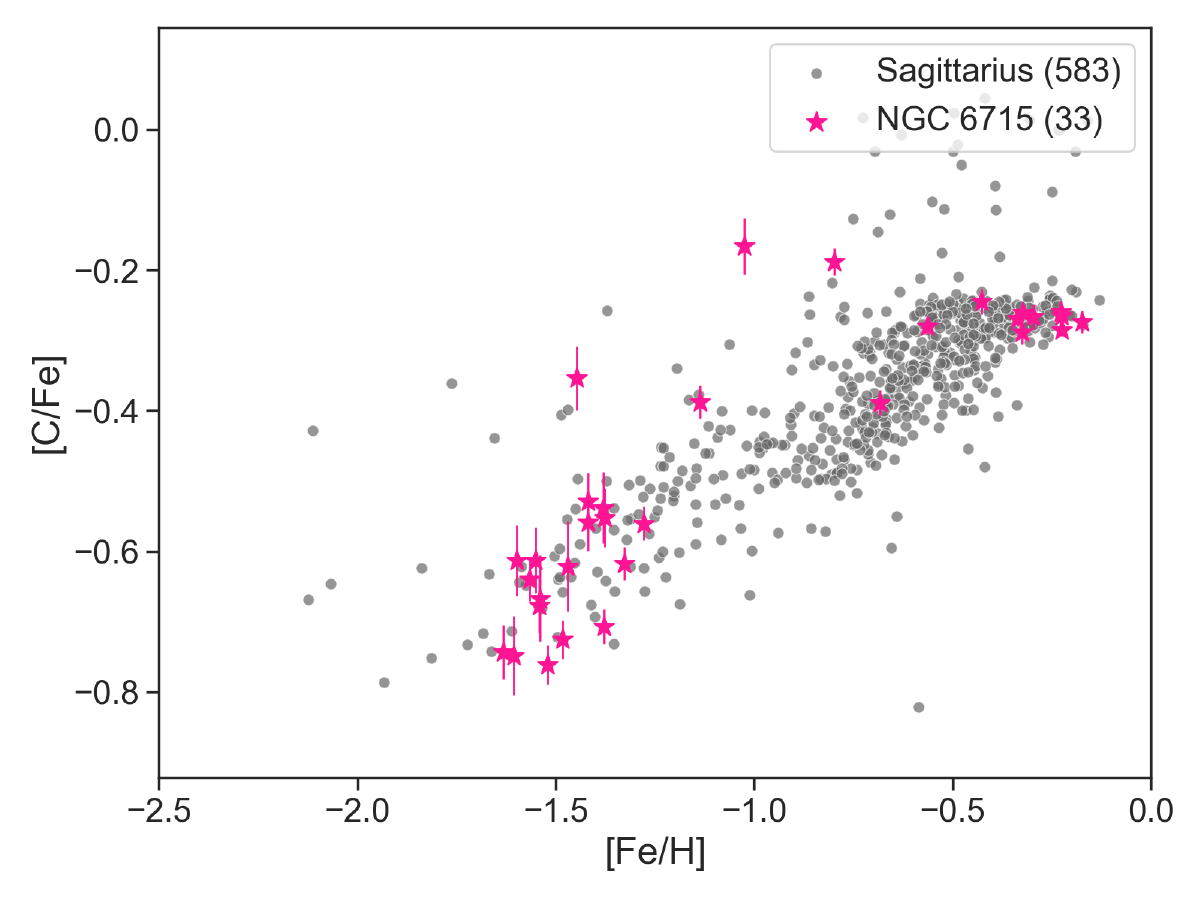}\hspace{-6pt}
\includegraphics[clip=true, trim = 3mm 0mm 0mm 3mm, width=0.69\columnwidth]{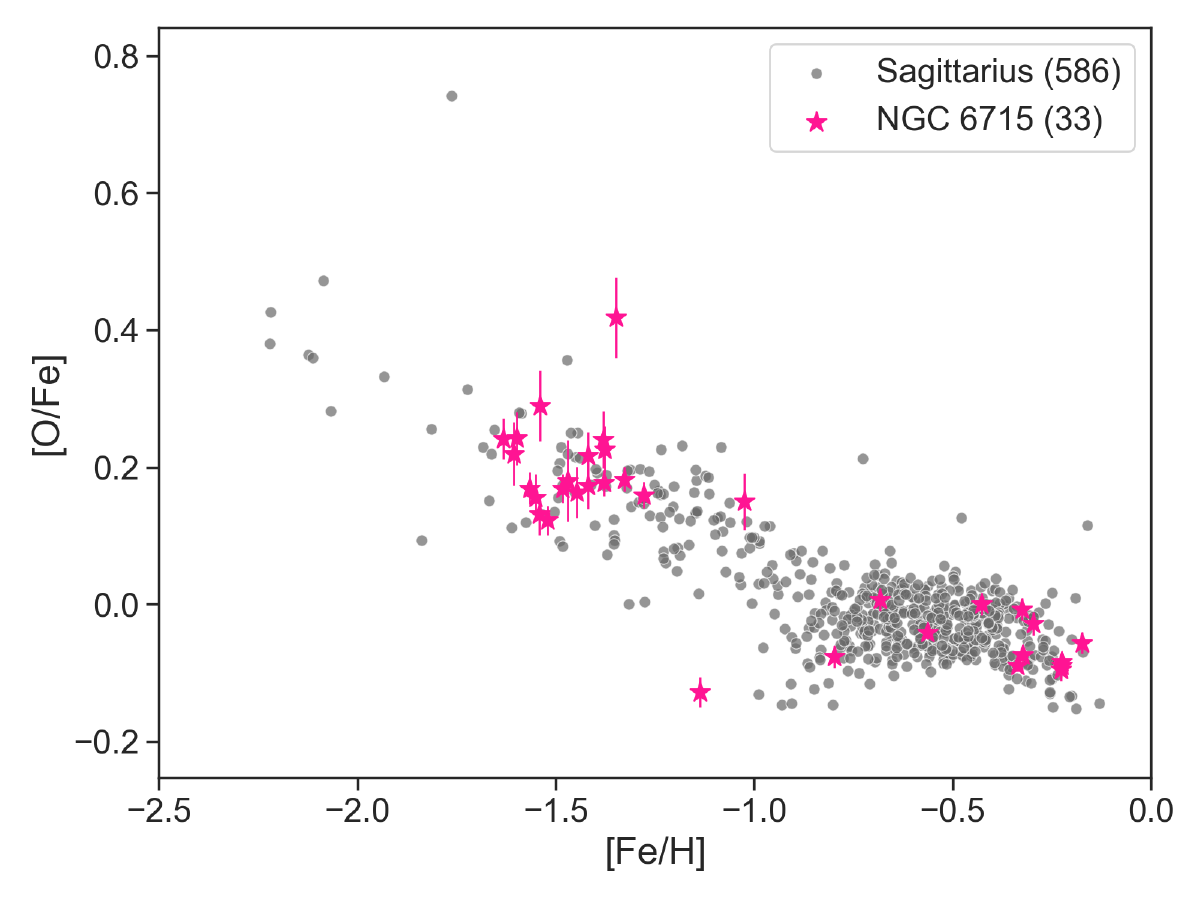}\hspace{-6pt}
\includegraphics[clip=true, trim = 3mm 0mm 0mm 3mm, width=0.69\columnwidth]{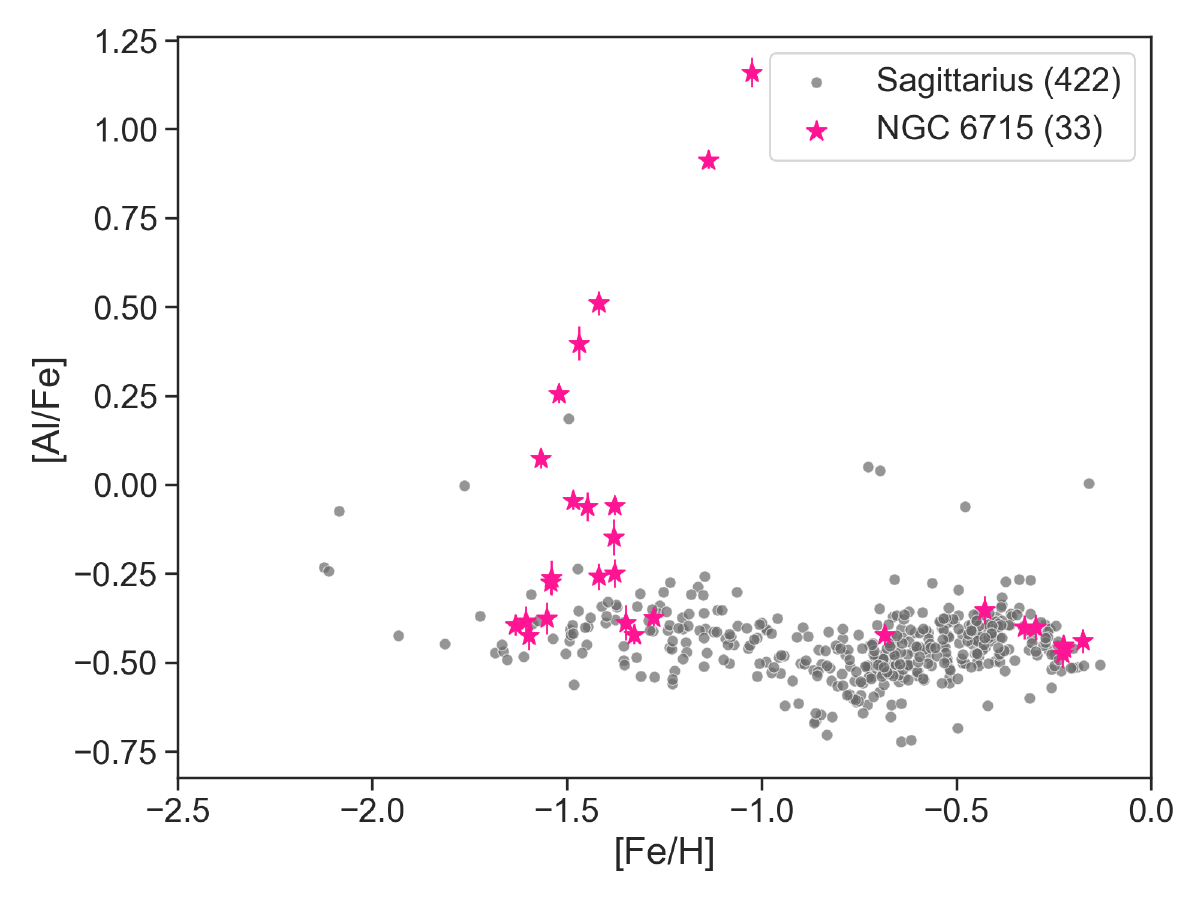}\hspace{-6pt}
\includegraphics[clip=true, trim = 3mm 0mm 0mm 3mm, width=0.69\columnwidth]{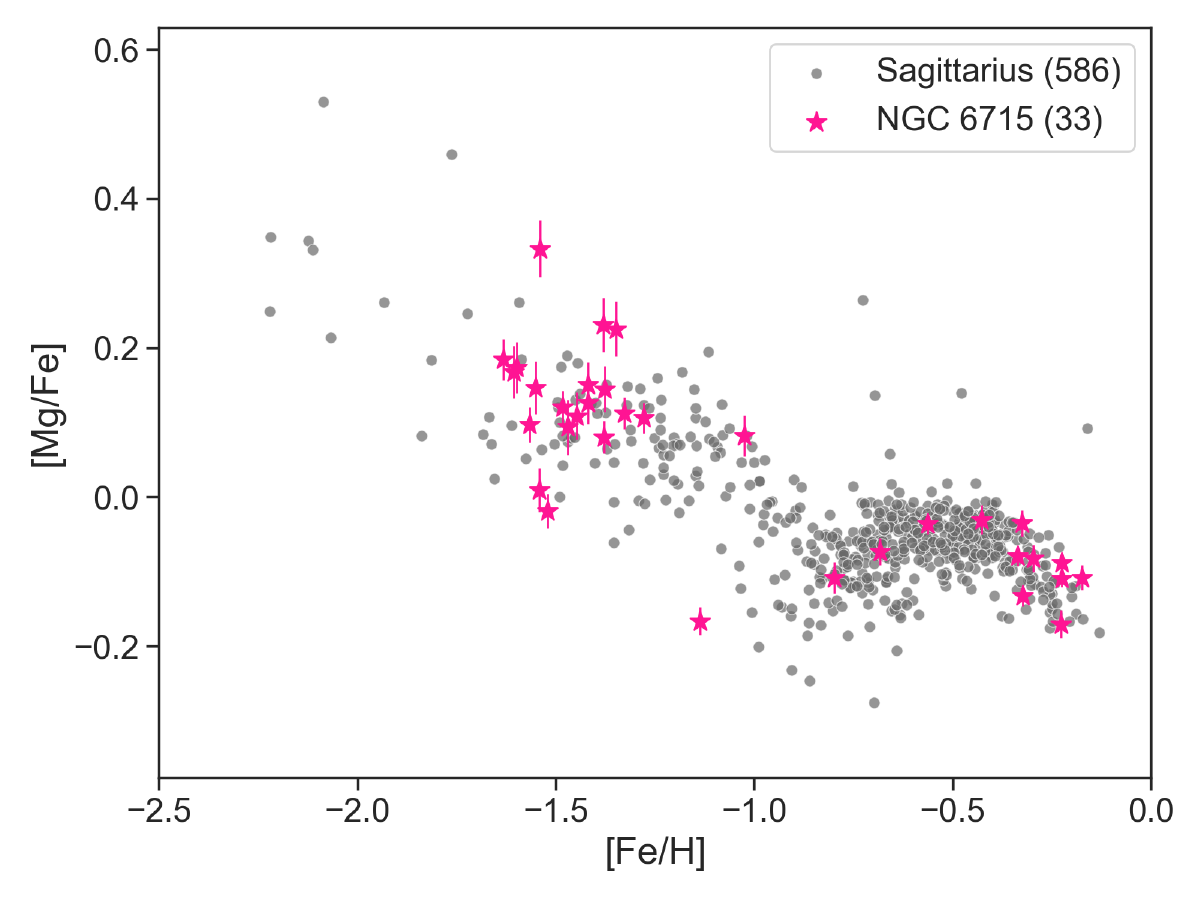}\hspace{-6pt}
\includegraphics[clip=true, trim = 3mm 0mm 0mm 3mm, width=0.69\columnwidth]{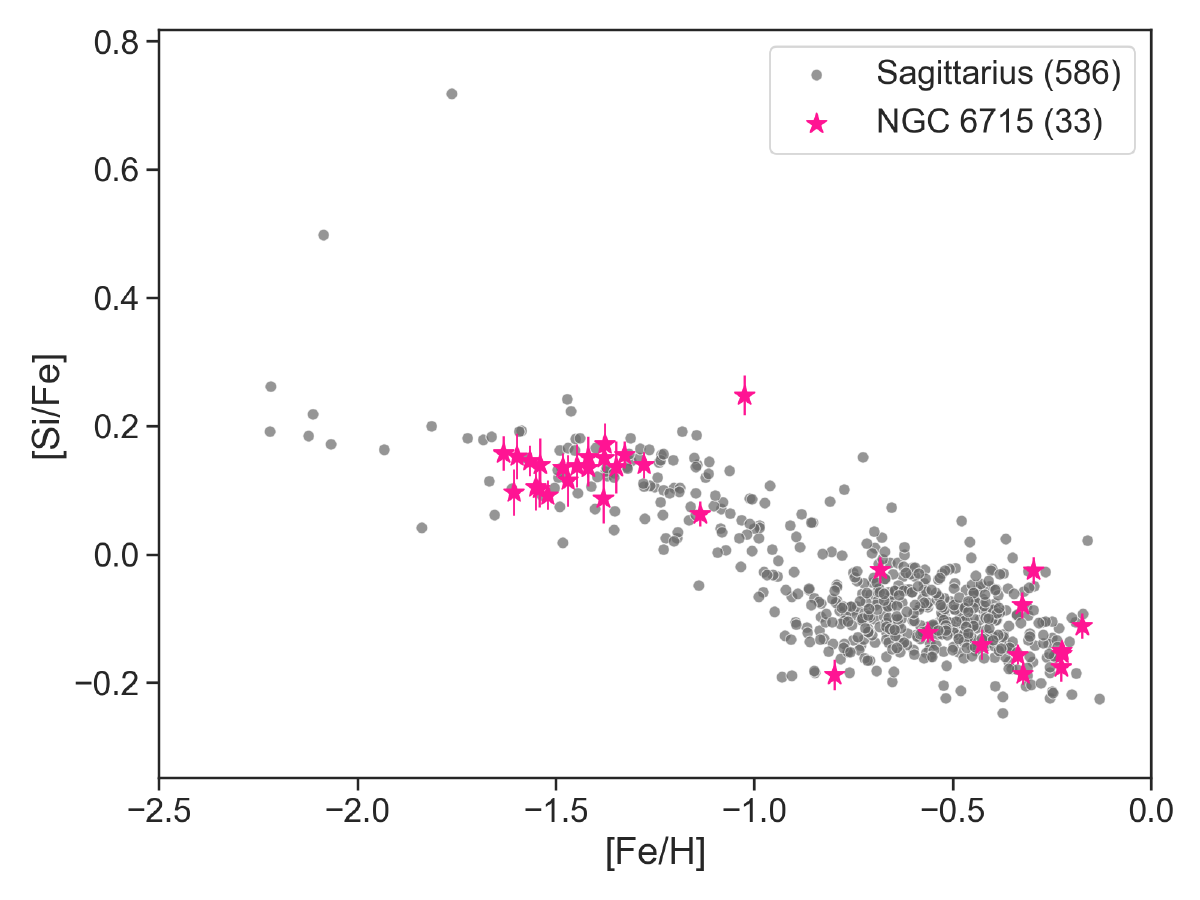}\hspace{-6pt}
\includegraphics[clip=true, trim = 3mm 0mm 0mm 3mm, width=0.69\columnwidth]{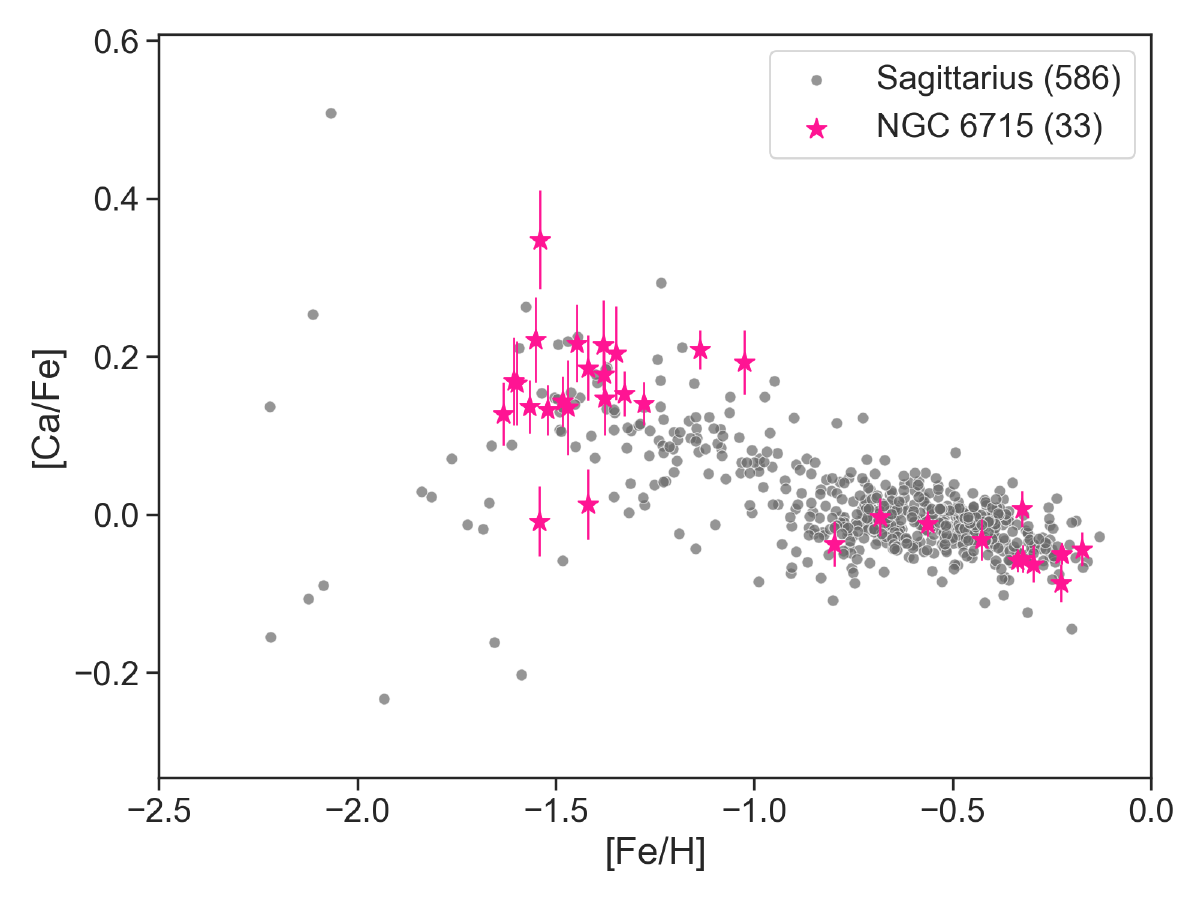}\hspace{-6pt}
\includegraphics[clip=true, trim = 3mm 0mm 0mm 3mm, width=0.69\columnwidth]{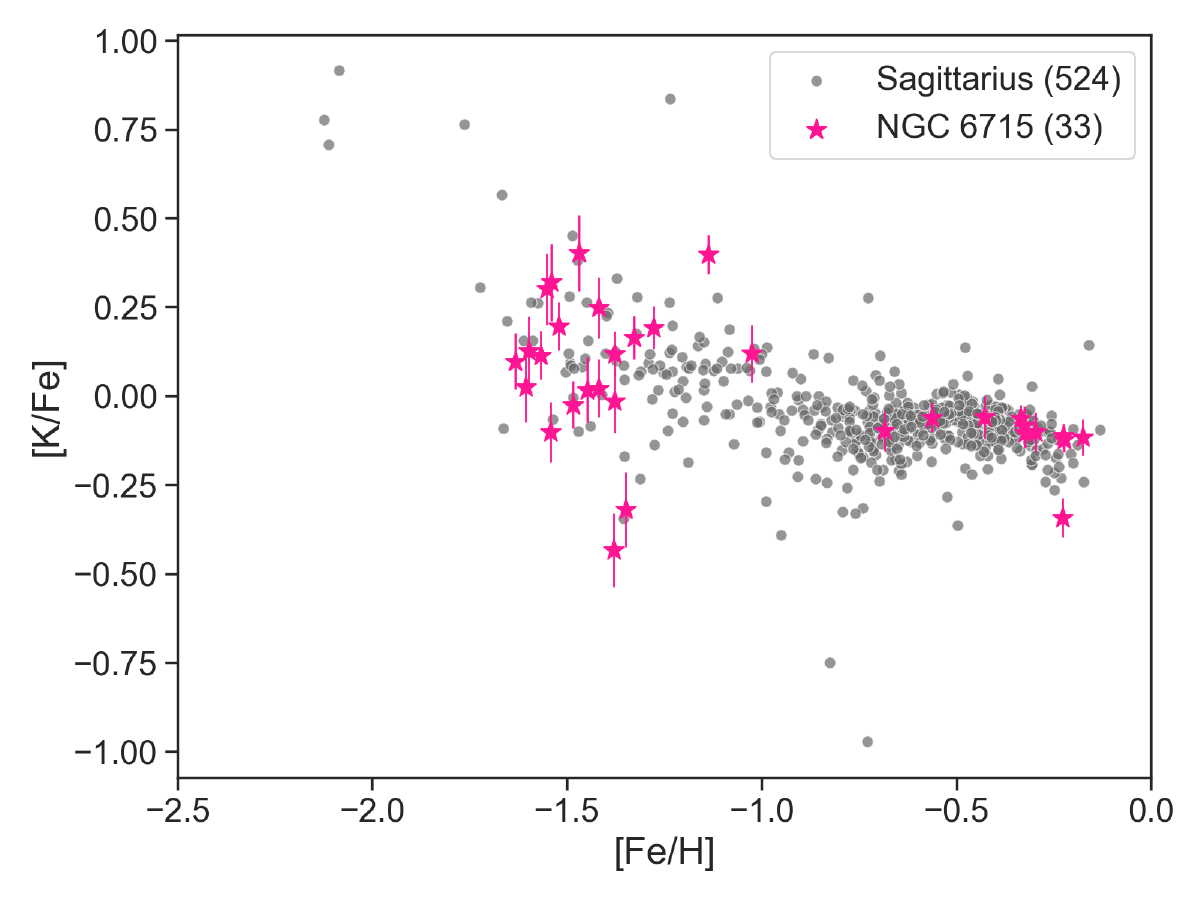}\hspace{-6pt}
\includegraphics[clip=true, trim = 3mm 0mm 0mm 3mm, width=0.69\columnwidth]{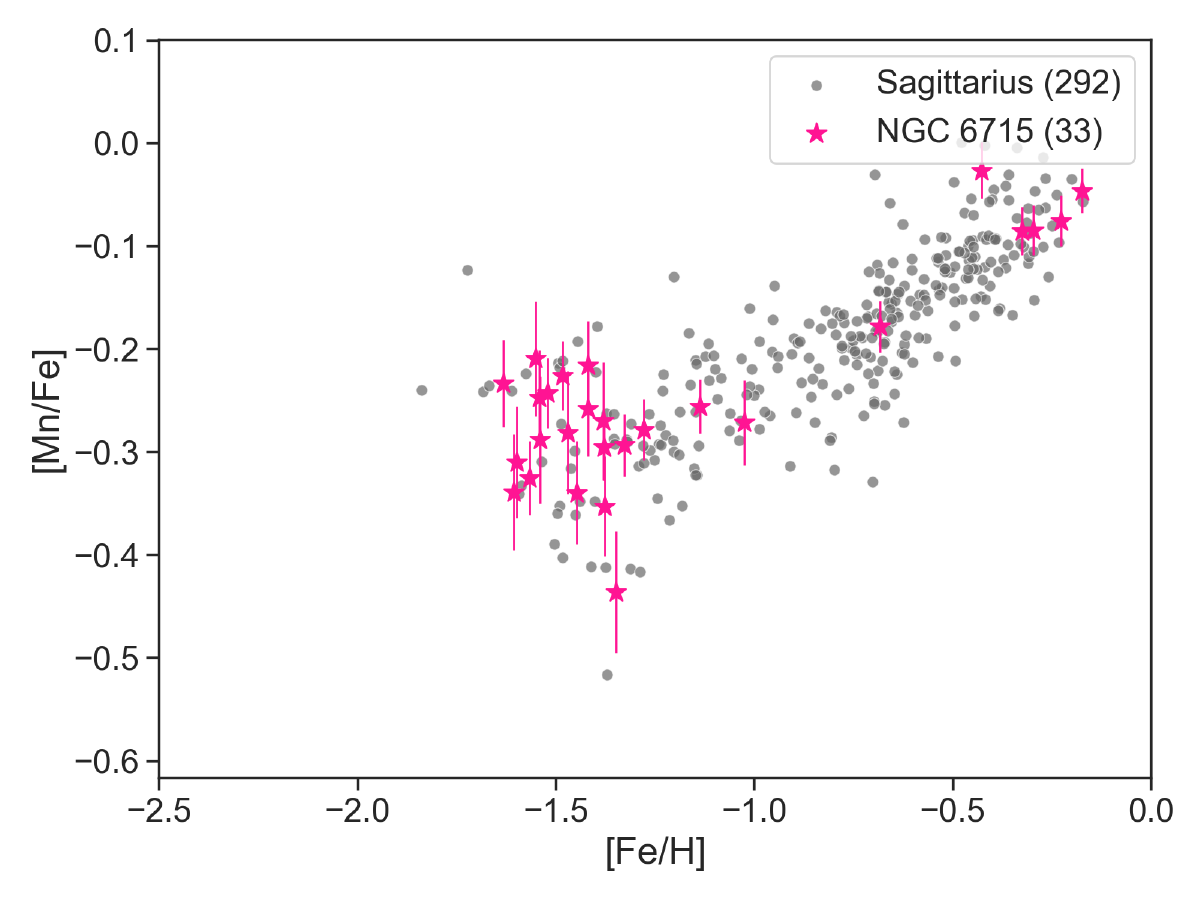}\hspace{-6pt}
\includegraphics[clip=true, trim = 3mm 0mm 0mm 3mm, width=0.69\columnwidth]{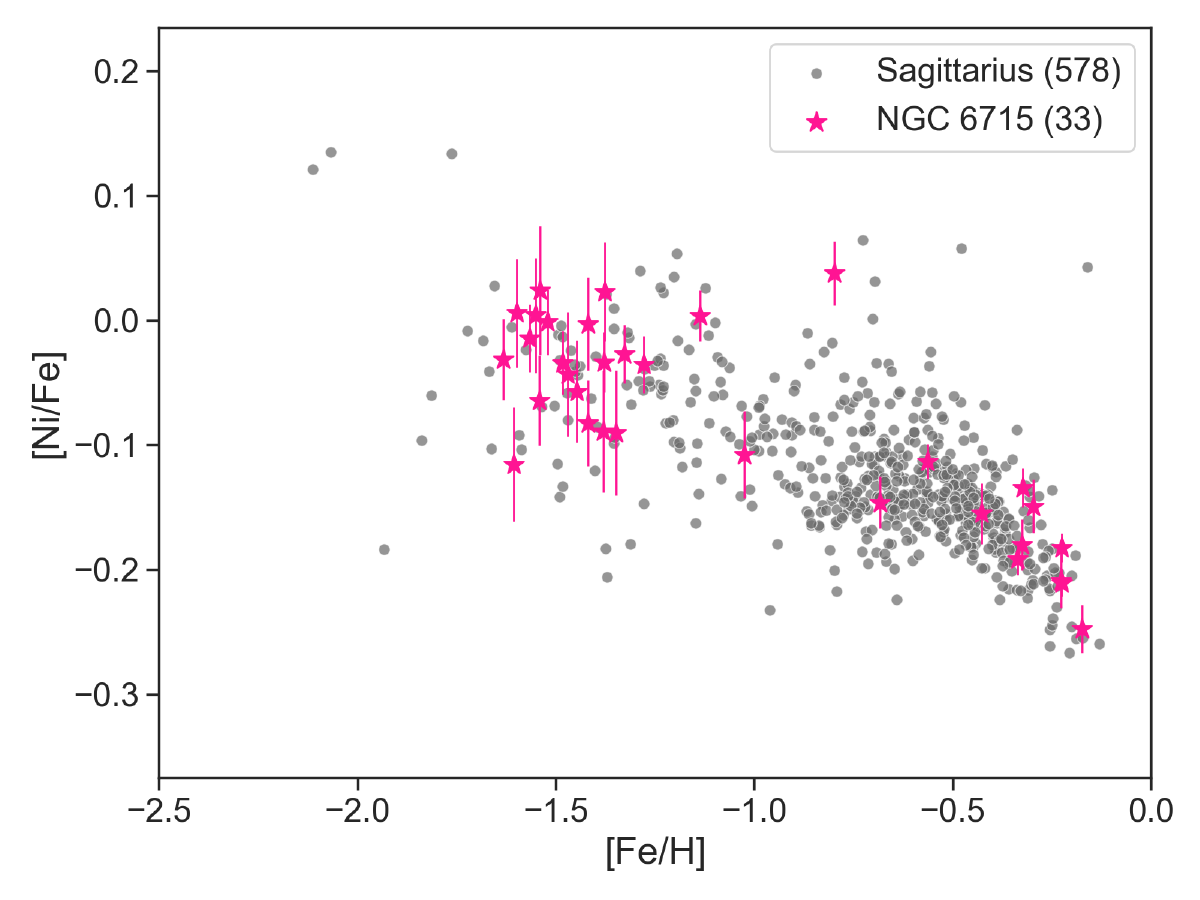}

  \caption{[X/Fe] versus [Fe/H] planes for NGC~6715 (coloured stars), the nuclear star cluster of the Sagittarius dwarf galaxy. For comparison, the distribution of stars belonging to the Sagittarius dwarf is also shown as grey points. In each panel, the number of stars used to make the corresponding plot is given, both for NGC~6715 and for Sagittarius. Note that these numbers can vary from a panel to another, because the APOGEE quality flags on the corresponding abundances may lead to different selections. For each panel, individual error bars on [Fe/H] and [X/Fe] abundances, for NGC~6715, are also reported. }
              \label{M54Sag}%
    \end{figure*}

\begin{figure}[h!]
\centering
\includegraphics[clip=true, trim = 0mm 0mm 0mm 0mm, width=0.85\columnwidth]{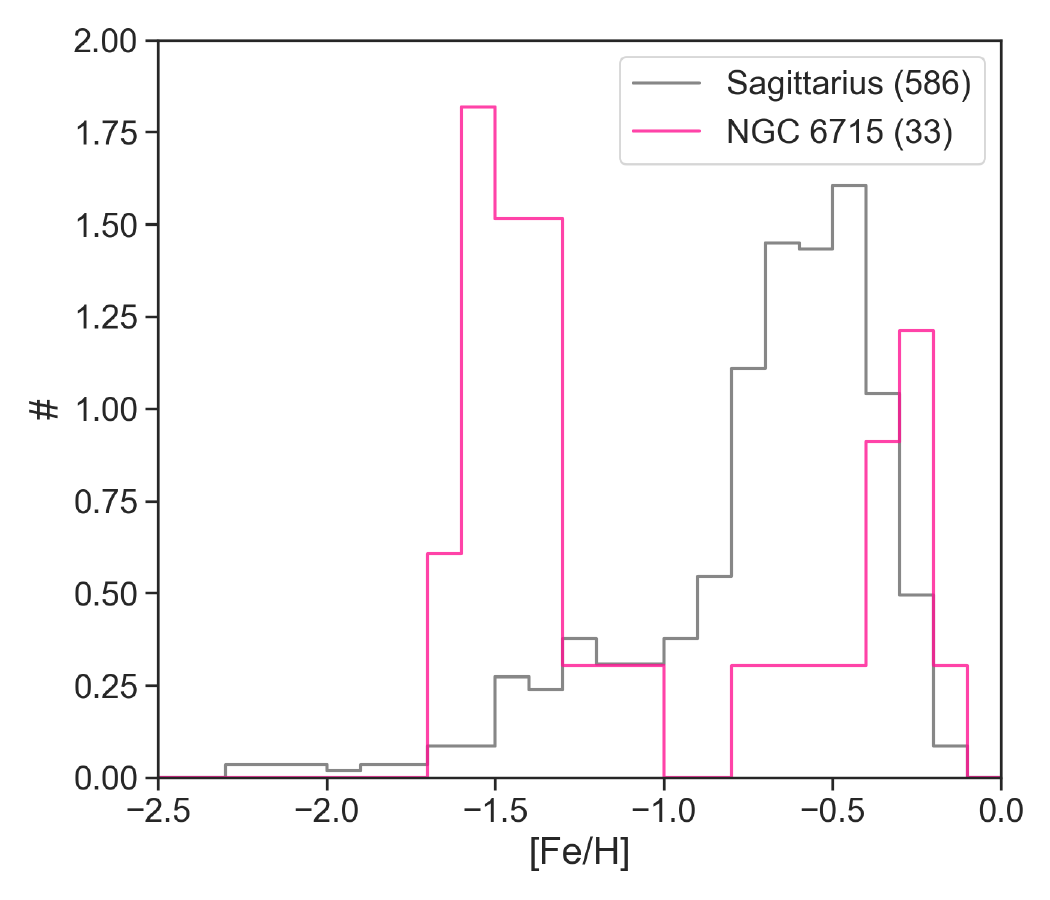}
\caption{[Fe/H] distributions of the analysed APOGEE stars in NGC~6715 (colored histogram) and in the Sagittarius dwarf (grey histogram). Both histograms are normalised so that the integral of the density over the range is equal to one. The number of stars used to trace the two histograms is reported in parenthesis.  }
 \label{M54Sag_MDF}
\end{figure}

Already in the introduction to this paper, we have evoked that - should a nuclear star cluster be representative, at least in part, of the chemical evolution of its host galaxy -  the possibility opens up to use NSC to get a first description of the chemical characteristics of the galaxy to which the NSC belongs (or has belonged). In other words, if we can show that NSC and their host galaxies have similar chemical patterns, we can use the chemical properties of $\omega$~Cen to reconstruct at least part of the chemical patterns of its progenitor galaxy, \textit{Nephele}, now completely dissolved among the Milky Way field stars. \\

In the case of the Milky Way, the comparison between the chemical abundances of the NSC and those of Galactic stellar populations is still ongoing, and is made difficult by the strong extinction present in the central regions of the Galaxy. Fortunately, the APOGEE~DR17 data offer us the opportunity to study another NSC-galaxy host system, the one consisting of the Sagittarius dwarf galaxy and NGC~6715 (M~54), its nuclear star cluster \citep[see also][]{bellazzini2003, bellazzini2008, alfaro-cuello2019, alfaro-cuello2020, bellazzini2020, kacharov2022}{}{}. In this section, we present the results of this comparison. \\
For the Sagittarius galaxy, we used the APOGEE data presented in \citet{hasselquist21}. In particular, their Table~2 provides the identifiers of stars belonging to the Sagittarius dwarf and its streams -- according to their selection -- which we cross-identified with the APOGEE~DR17 catalogue to obtain their chemical abundances. From the retrieved sample, we then selected only stars with:
\begin{enumerate}
\item a signal-to-noise ratio $\tt{SNREV} > 70$;
\item temperatures in the range $\rm 3500\,K < T_{eff} < 5500\,K$ and surface gravities  $\rm logg < 3.6$; 
\item \tt{APOGEE STARFLAG} and \tt{APOGEE STARBAD} $= 0$
\end{enumerate}

As for NGC~6715, we used the \citet{schiavon2023} catalogue already used for all other Galactic globular clusters\footnote{Note that Sagittarius has a few other suspected associated clusters such as Ter 8, Arp 2, NGC 5634, and Ter 7 whose data, however, are not available in the \citet{schiavon2023} catalogue.}, applying the same selections discussed in Sect~\ref{obsdata}. 
Fig.~\ref{M54Sag} shows the result of this comparison. We can see from this figure that in all abundance planes the NGC~6715 chemical patterns as a function of [Fe/H] follow very closely those of the Sagittarius galaxy. In particular:
\begin{itemize}
\item in the Sagittarius dwarf, $\alpha$-elements as [Ca/Fe], [Mg/Fe], [O/Fe] and [Si/Fe]  show a declining trend as a function of [Fe/H], until [Fe/H]$\sim -0.9$~dex; for [Fe/H] values above this limit, these patterns become flatter or with a ``hook"-like shape. As explained by \citet{hasselquist21}, these trends as a function of [Fe/H] can be explained if the Sagittarius dwarf, after an initial burst of star formation occurred early on during its evolution, experienced a second, weaker burst about 10~Gyr after (see Figs.~9, 10 and 11 in their paper). This point is in agreement with what has been found for the age and metallicity of the stellar populations of M~54 by \citet{alfaro-cuello2019}: the metal-rich (young) component of the cluster is compatible with the second starburst of Sagittarius (which likely happened $\sim$3 Gyr ago). Strikingly, our Fig.~\ref{M54Sag} shows that the same trends for $\alpha$-elements as a function of [Fe/H] are present also in NGC~6715, with declining [$\alpha$/Fe] ratios as a function of [Fe/H] up to [Fe/H]$\sim -0.9$~dex, where the ``hooked" profile appears. This indicates that -- whatever the process of formation of NGC~6715 has been -- this cluster has experienced the same star formation evolution, at similar efficiency, as that experienced by its parent galaxy. In particular, at late times, the ISM from which stars of NGC~6715 formed must have been polluted by SNeII, similarly to what happened in the ISM of Sagittarius, that is: the same kind of polluters (SNeII) must have contributed at the same time to enrich the gas reservoir of both systems, in order to produce in both of them the ``hooked" pattern observed in the same metallicity range.
\item More generally, the trends of [X/Fe] versus [Fe/H] observed in the Sagittarius dwarf, compared to those of the Milky Way,  have led to suggest that the chemical enrichment in this galaxy may have been affected by higher Type Ia/Type II SNe ratio than in the MW, or by a  top-light IMF \citep[see, for example][]{mcwilliam13,hasselquist21}. Whatever the sources, or the combination of sources, responsible for the chemical enrichment of this galaxy, the same mechanisms must have been at work in NGC~6715 in order to produce the same trends for all the chemical elements shown in Fig.~\ref{M54Sag}. 
\end{itemize}

From Fig.~\ref{M54Sag},  a few apparent differences are worth being discussed:
\begin{itemize}
\item the presence of a small percentage ($\sim 5\%$) of C-rich ([C/Fe]$\gtrsim -0.2$) stars among the metal-rich ([Fe/H] $\gtrsim$ -0.75) population of Sagittarius. No C-rich counterpart is observed at similar metallicities in NGC~6715, but given the low number of stars with  [Fe/H] $\gtrsim$ -0.75 in this cluster (11 in total), the absence of C-rich stars in this cluster is still statistically compatible with the 5\% fraction observed in Sagittarius.
 \item the presence of a few Al-enhanced ([Al/Fe] $\gtrsim$ 0.25) stars in NGC~6715 (at [Fe/H]$\lesssim -1$, which are not present among the field (i.e. Sagittarius) population. Al-enhanced stars are commonly present in globular clusters. It is possible that a few of these stars may be present also in the Sagittarius galaxy (as the result of the mass loss experienced by some of its globular clusters) but given the overall limited number of Sag stars at [Fe/H]$\lesssim -1$ it is difficult to derive any robust conclusion. 
\end{itemize}

Finally, we would like to draw the reader attention to the fact that, while in all [X/Fe]-[Fe/H] planes reported in Fig.~\ref{M54Sag} the overlap between the chemical patterns of NGC~6715 and those of its host galaxy is impressive, if the comparison is made only on their MDFs, no similarities between the two are found (see Fig.~\ref{M54Sag_MDF}). This should discourage from comparing stellar populations based merely on a comparison of MDFs. Different stellar systems may be affected by different observational biases (which would then lead to changes of the intrinsic MDF of these systems), or different dynamical evolutions may modify the MDFs of the systems over time, modulating them differently. In any case, the comparison between NGC~6715 and Sagittarius offers another example\footnote{To cite another example of this kind, we recall the reader that the MW bulge and thin+thick disc MDFs are different, while their patterns in [X/Fe] versus [Fe/H] planes, for a number of different chemical elements, show a remarkable agreement \citep[see, for example,][]{bensby10, haywood18, bensby21}}  of the limited discriminatory power of MDFs to establish similarities or differences between stellar populations. Populations with the same chemical evolution can have different MDFs (i.e. different fractions of stars in a similar metallicity range) but this is not sufficient to exclude that they had similar chemical evolutions, as evidenced by the NGC~6715-Sagittarius comparison (Fig.~\ref{M54Sag}).

\clearpage
\section{GCs chemically compatible with $\omega$~Cen: other elements}
\label{ocen-like_other_elems}
\begin{figure*}[h!]
\centering
\includegraphics[clip=true, trim = 3mm 0mm 0mm 2mm, width=0.68\columnwidth]{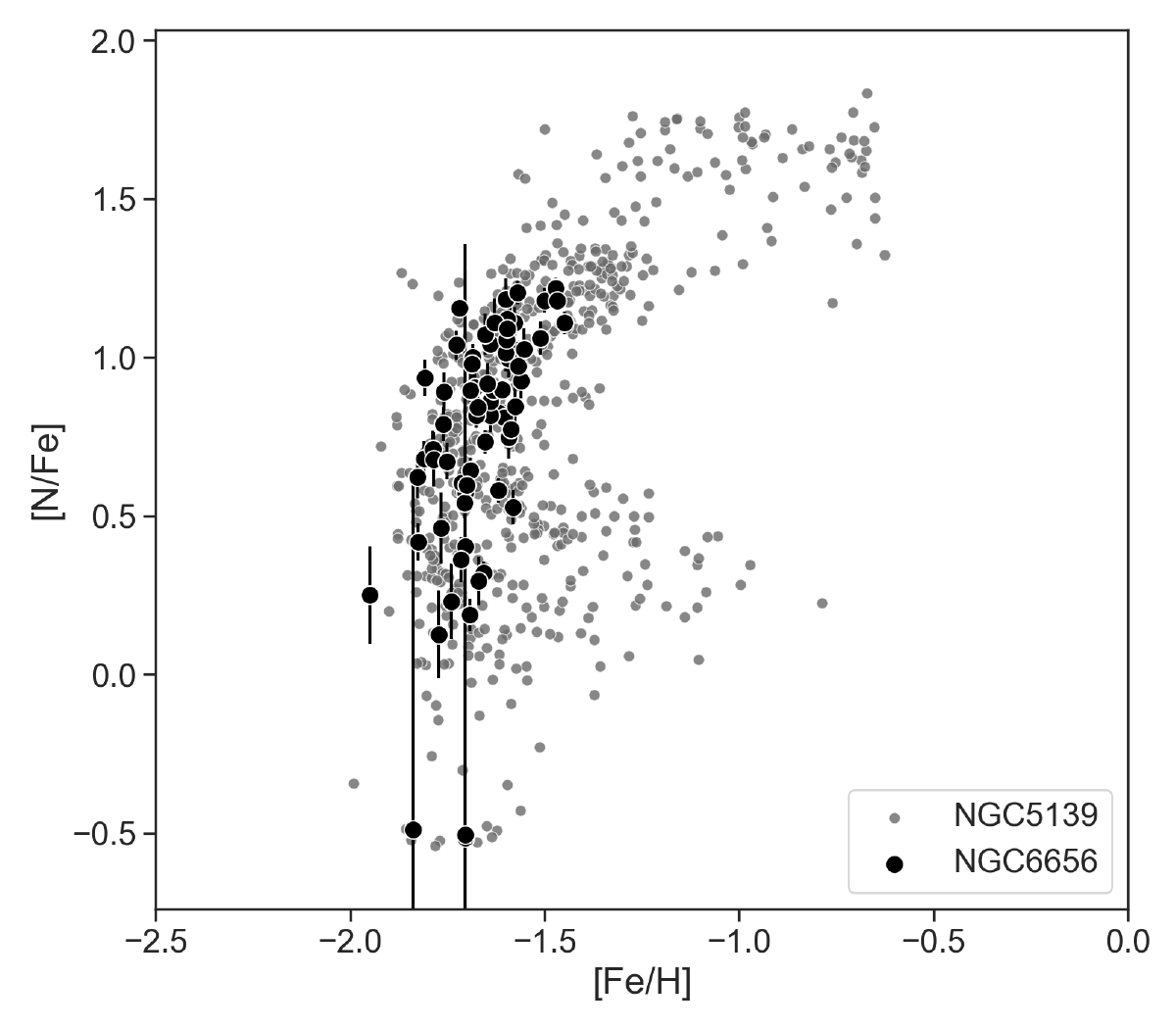}\hspace{-5pt}
\includegraphics[clip=true, trim = 3mm 0mm 0mm 2mm, width=0.68\columnwidth]{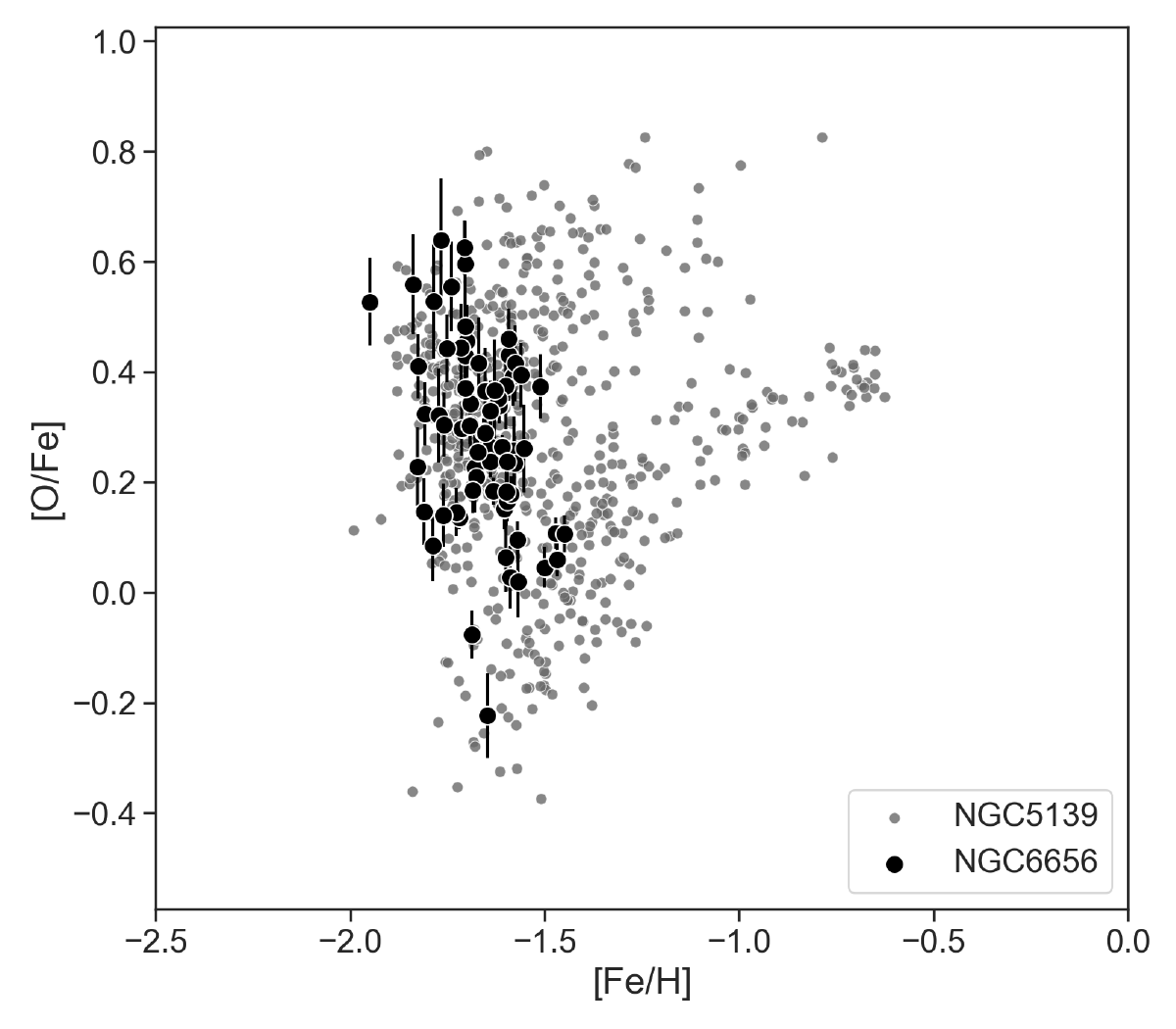}\hspace{-5pt}
\includegraphics[clip=true, trim = 3mm 0mm 0mm 2mm, width=0.68\columnwidth]{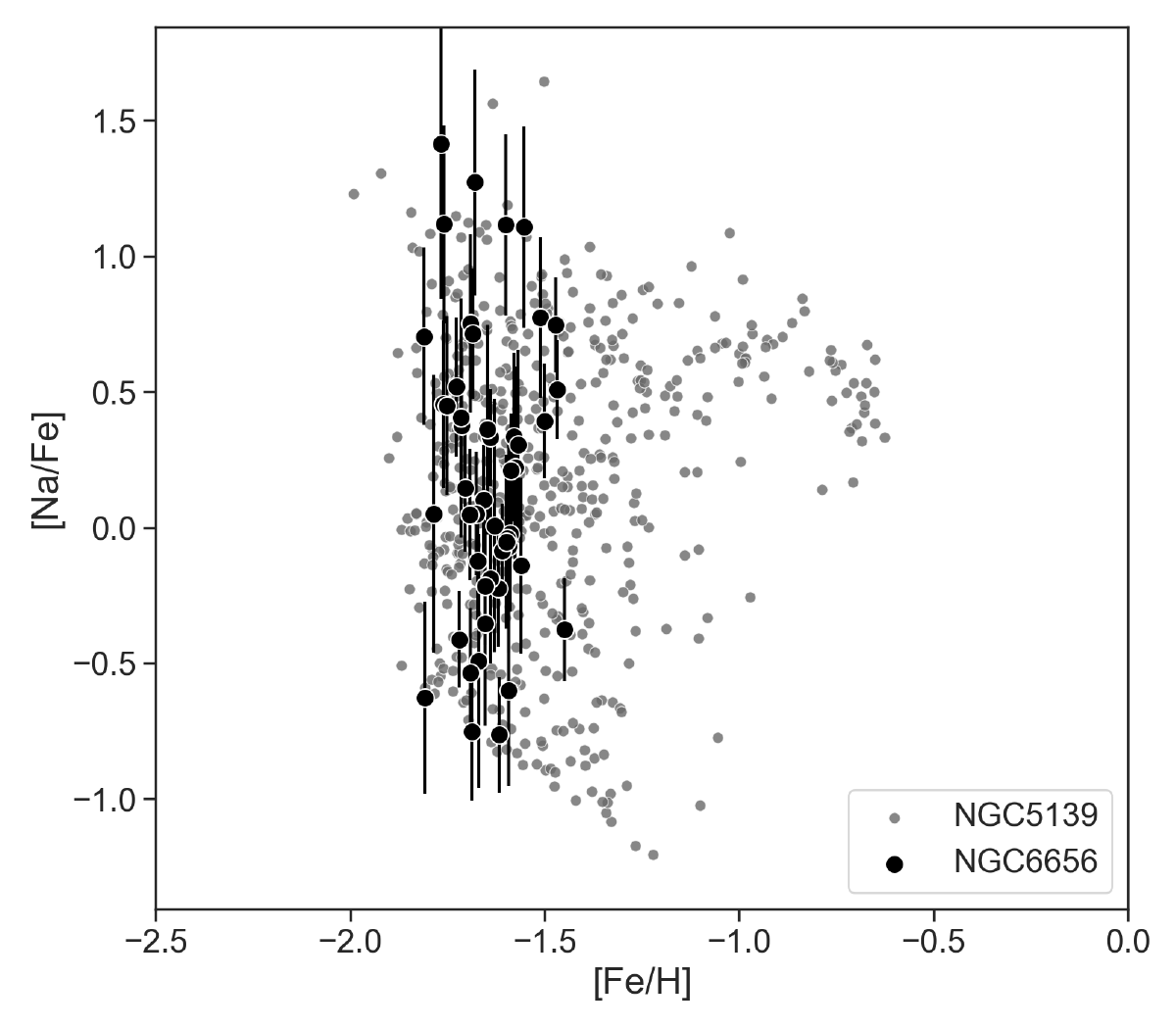}\hspace{-5pt}
\includegraphics[clip=true, trim = 3mm 0mm 0mm 2mm, width=0.68\columnwidth]{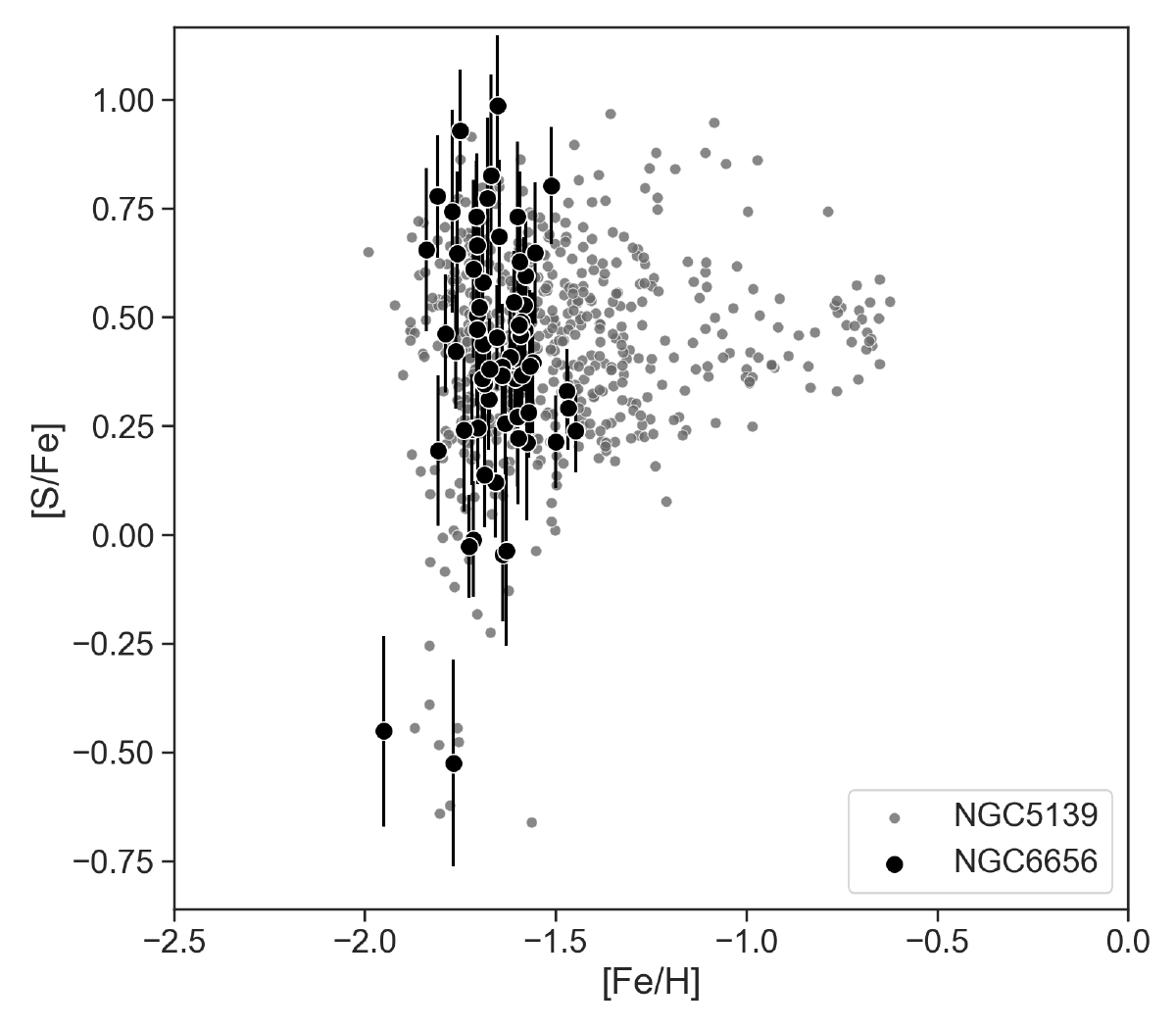}\hspace{-5pt}
\includegraphics[clip=true, trim = 3mm 0mm 0mm 2mm, width=0.68\columnwidth]{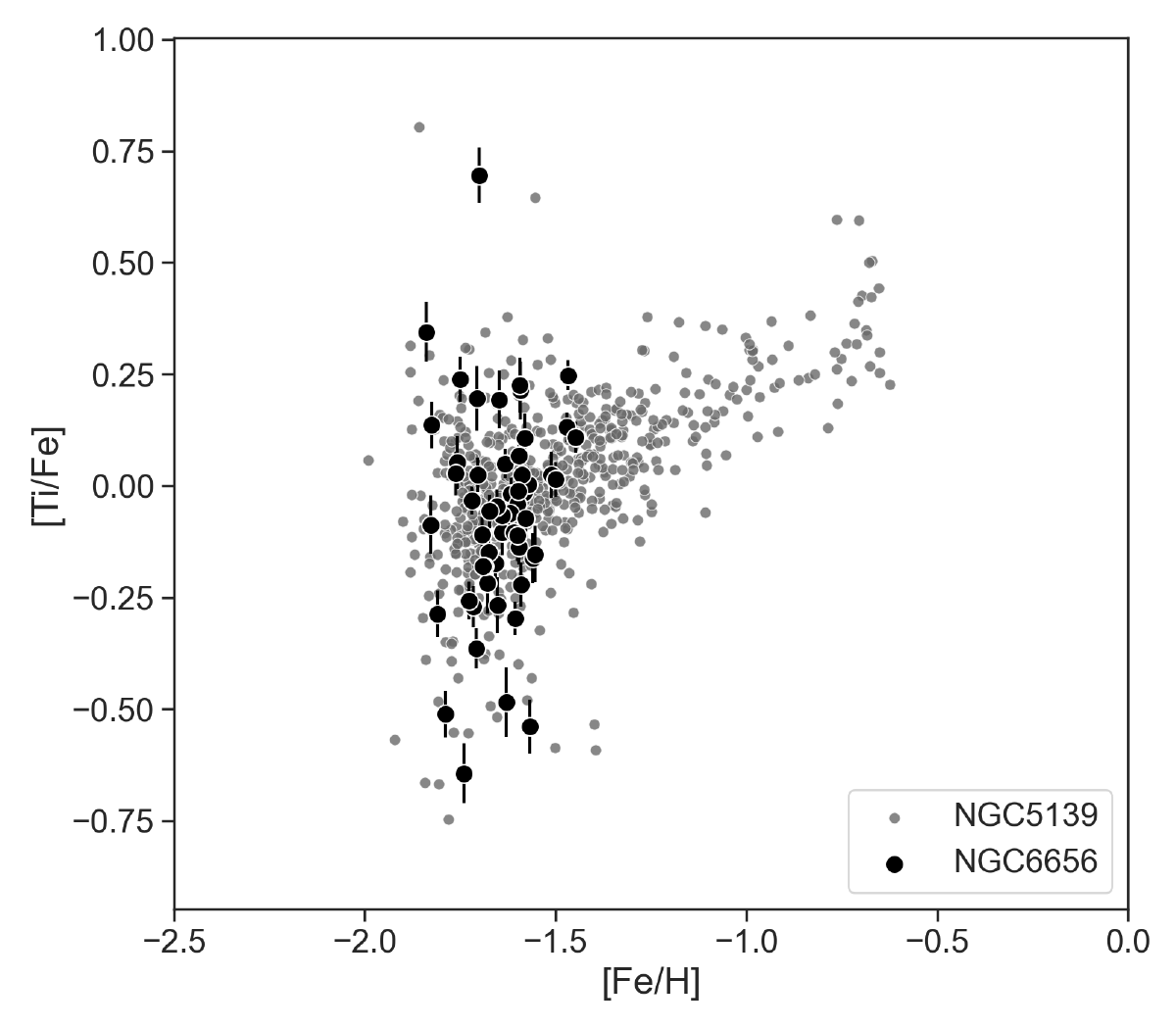}\hspace{-5pt}
\includegraphics[clip=true, trim = 3mm 0mm 0mm 2mm, width=0.68\columnwidth]{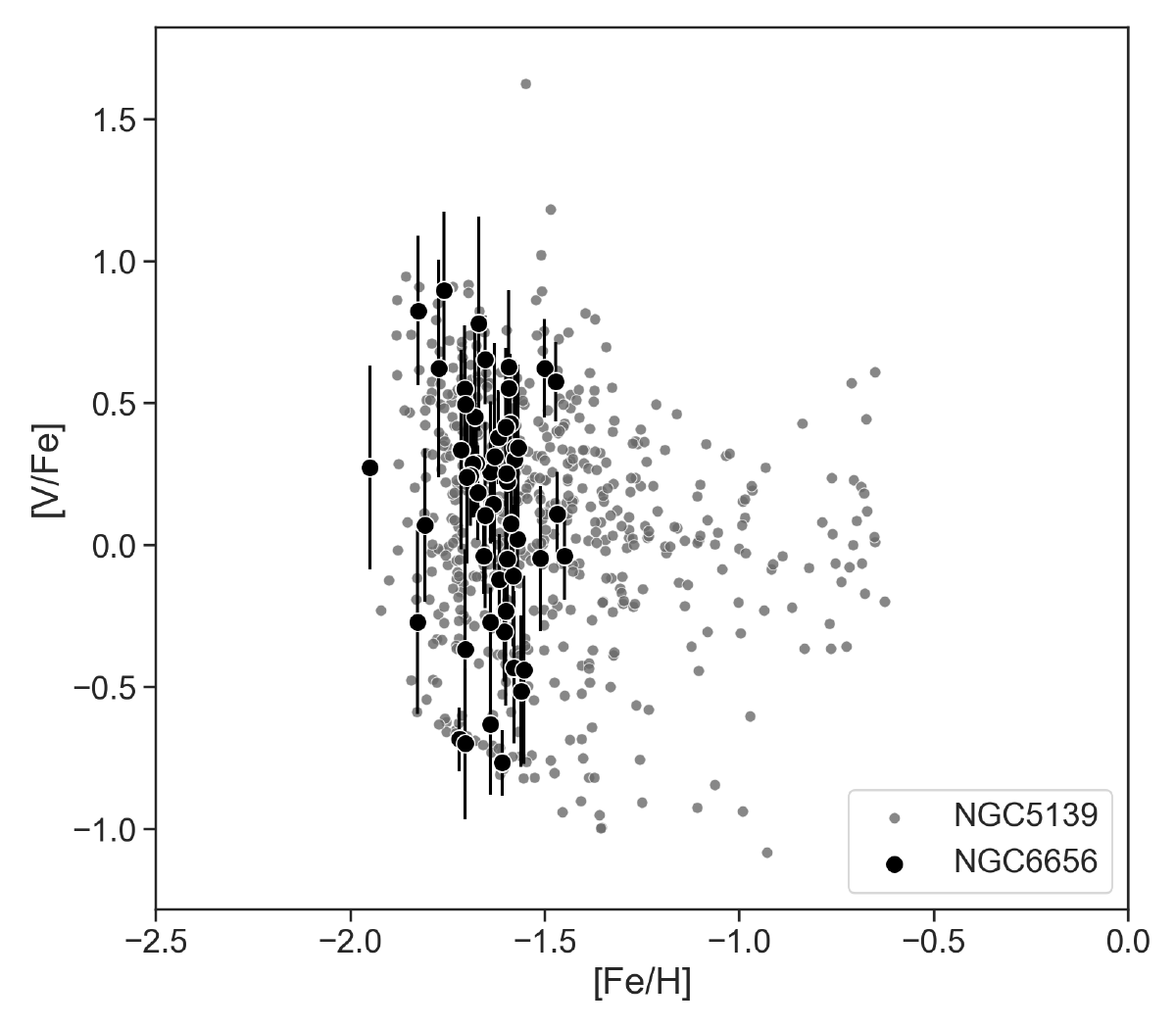}\hspace{-5pt}
\includegraphics[clip=true, trim = 3mm 0mm 0mm 2mm, width=0.68\columnwidth]{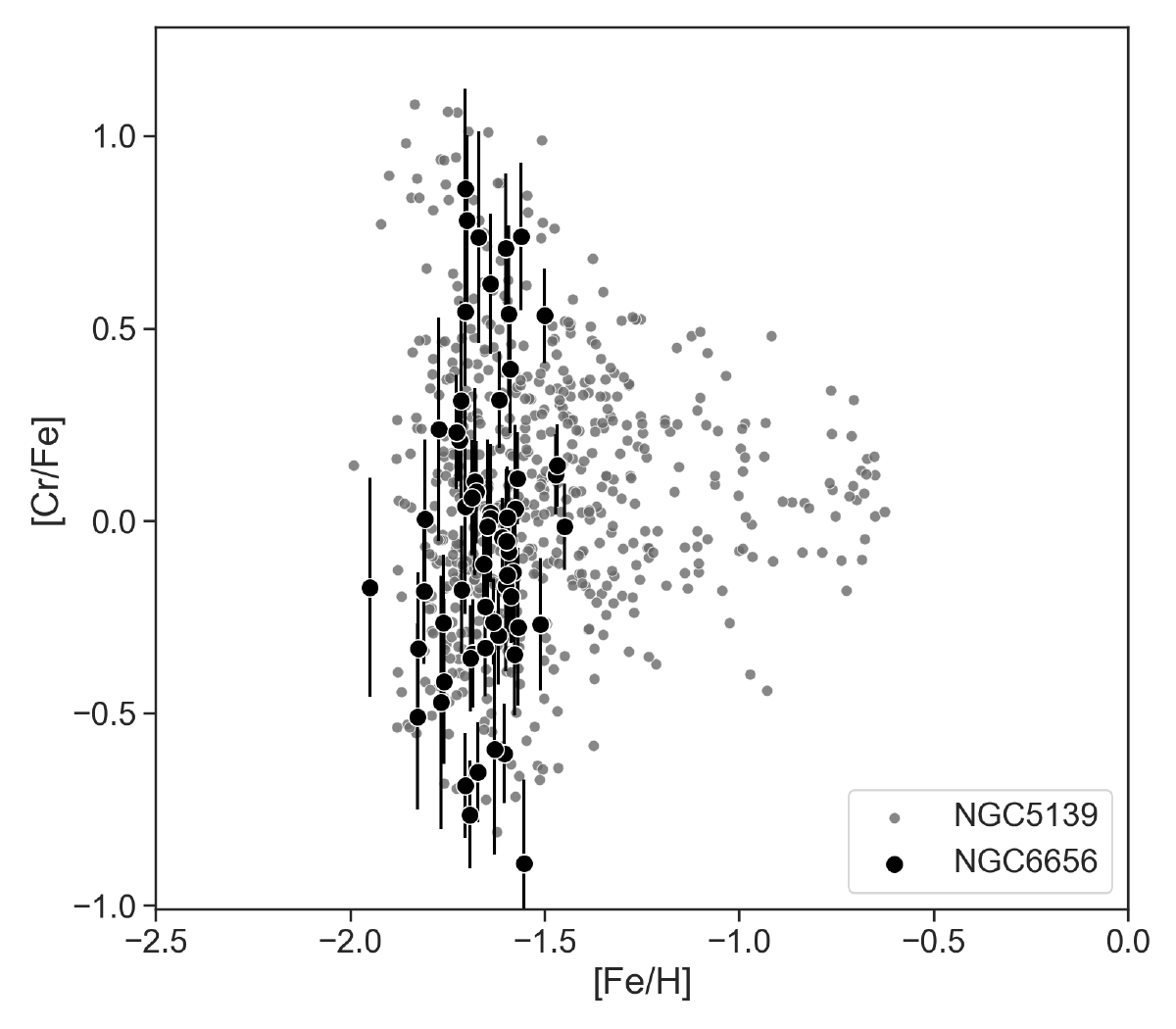}\hspace{-5pt}
\includegraphics[clip=true, trim = 3mm 0mm 0mm 2mm, width=0.68\columnwidth]{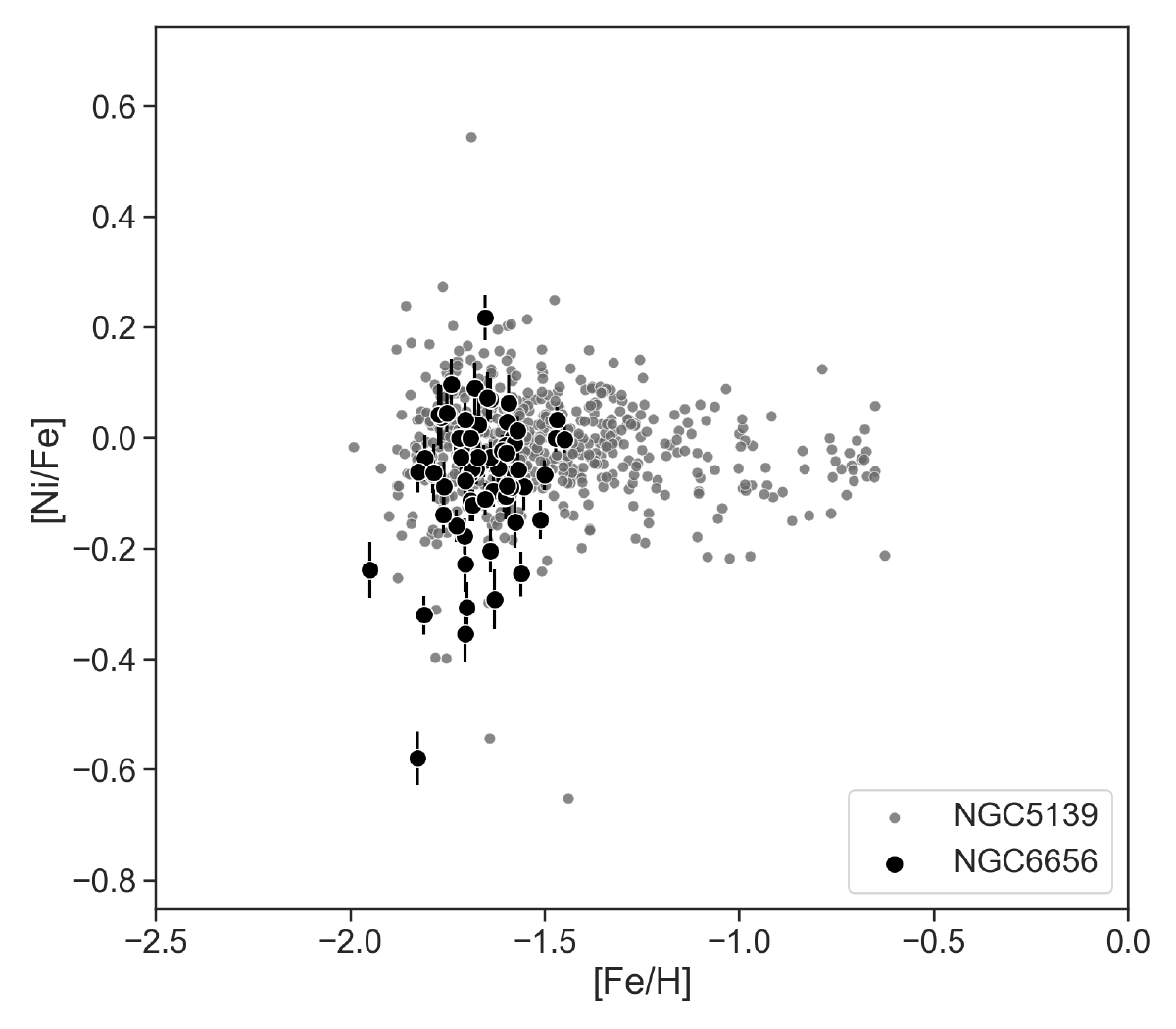}\hspace{-5pt}
\includegraphics[clip=true, trim = 3mm 0mm 0mm 2mm, width=0.68\columnwidth]{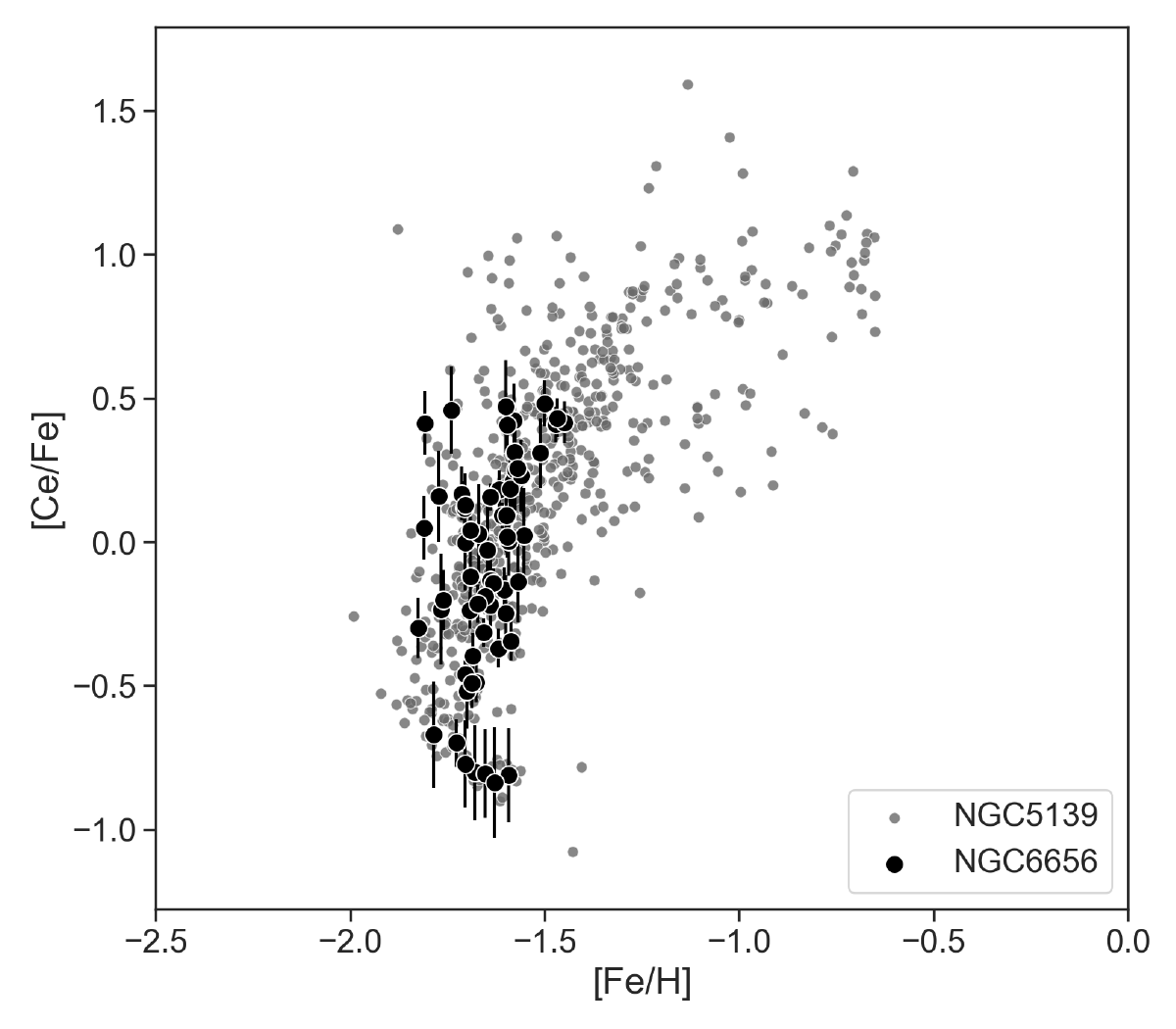}
  \captionof{figure}{[X/Fe] versus [Fe/H] planes for NGC~6656 (black symbols), one of the GCs chemically compatible with $\omega$~Cen. Here, the different X elements are the others provided by APOGEE DR17 but NOT used in the GMM, namely: N, O, Na, S, Ti, V, Cr, Ni, and Ce. For comparison, the distribution of stars belonging to $\omega$~Cen is also shown as grey circles. }
              \label{ngc6656_others}%
    \end{figure*}

\begin{figure*}[h!]
   \centering
\includegraphics[clip=true, trim = 3mm 0mm 0mm 2mm, width=0.68\columnwidth]{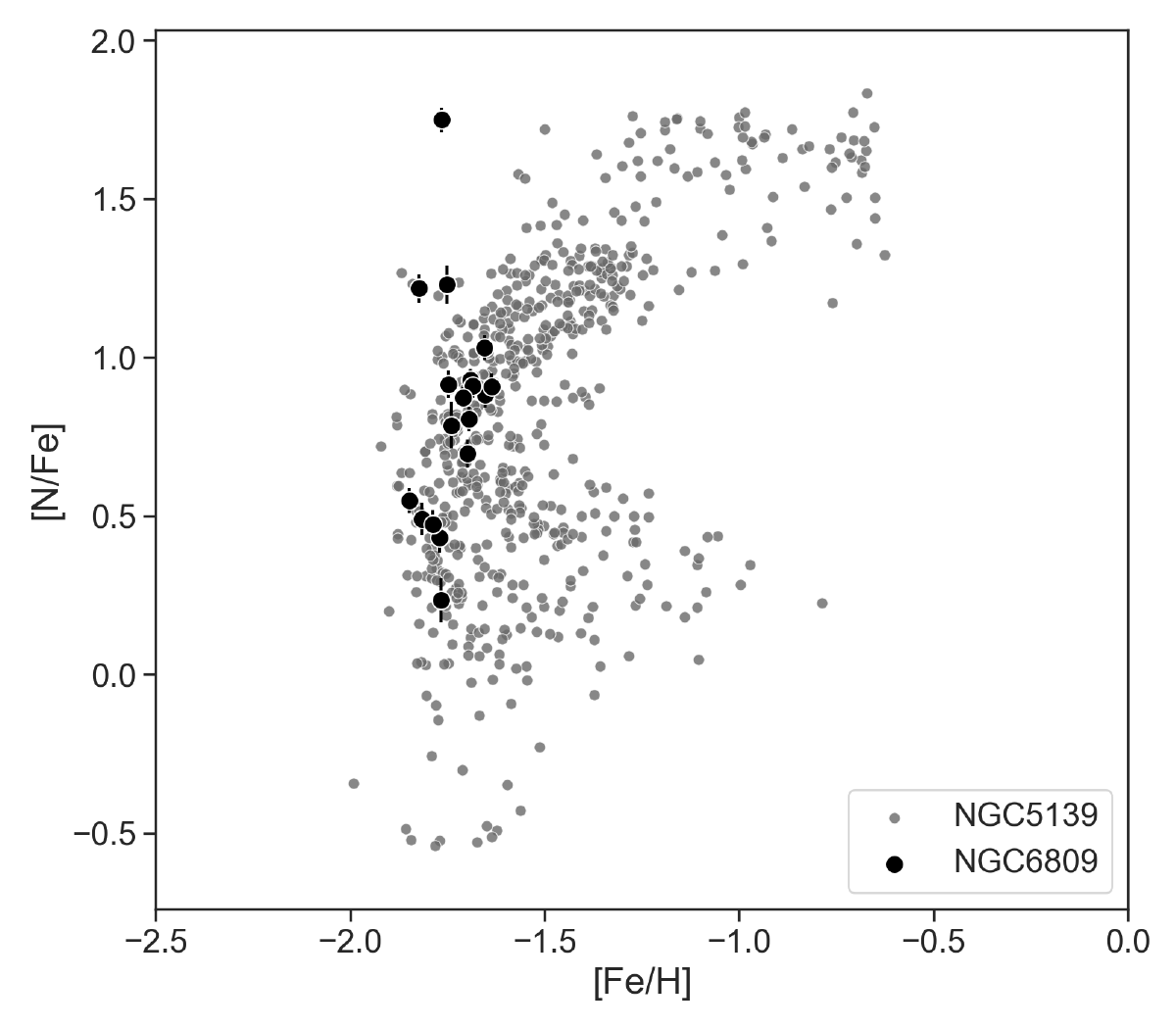}\hspace{-5pt}
\includegraphics[clip=true, trim = 3mm 0mm 0mm 2mm, width=0.68\columnwidth]{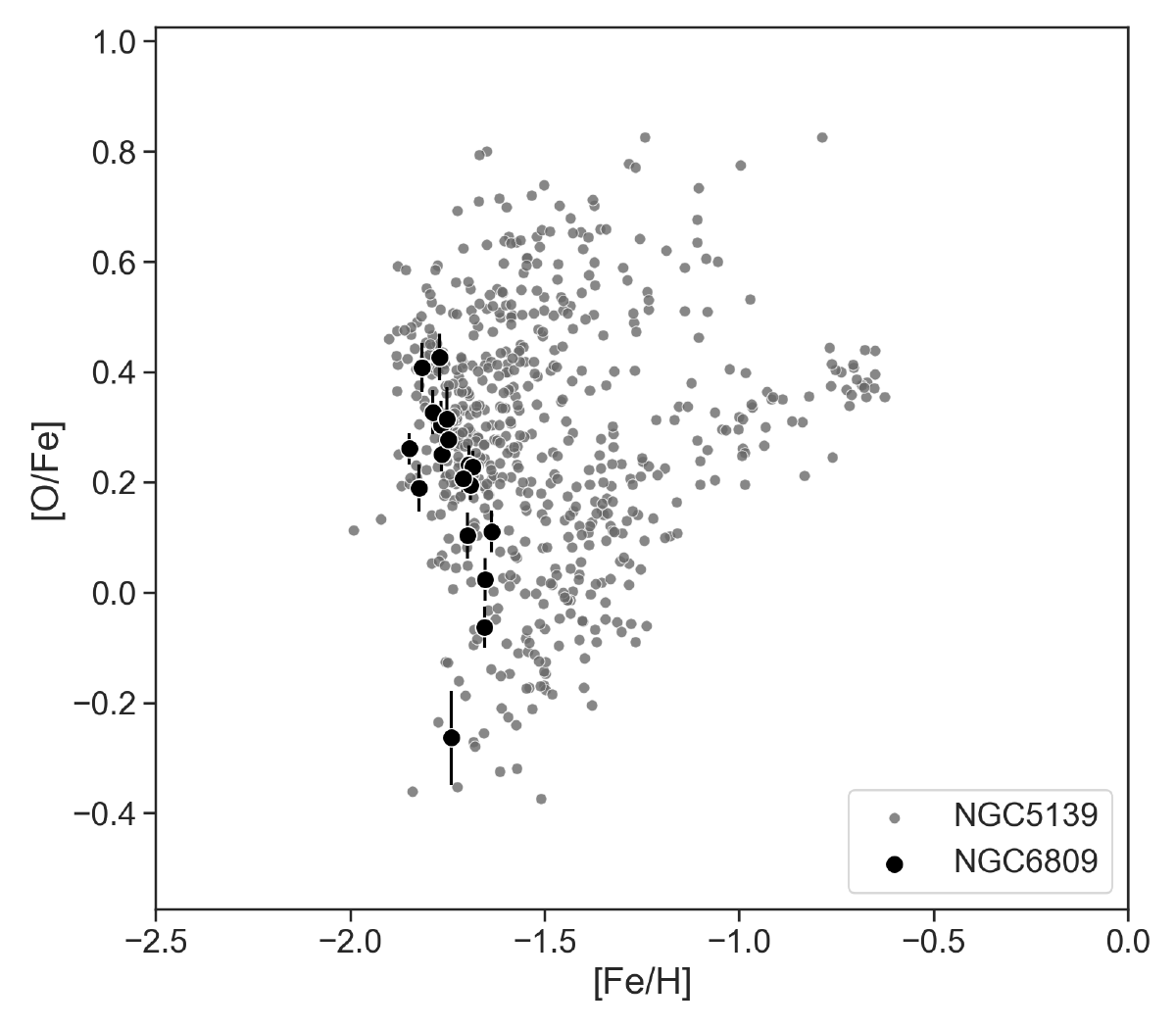}\hspace{-5pt}
\includegraphics[clip=true, trim = 3mm 0mm 0mm 2mm, width=0.68\columnwidth]{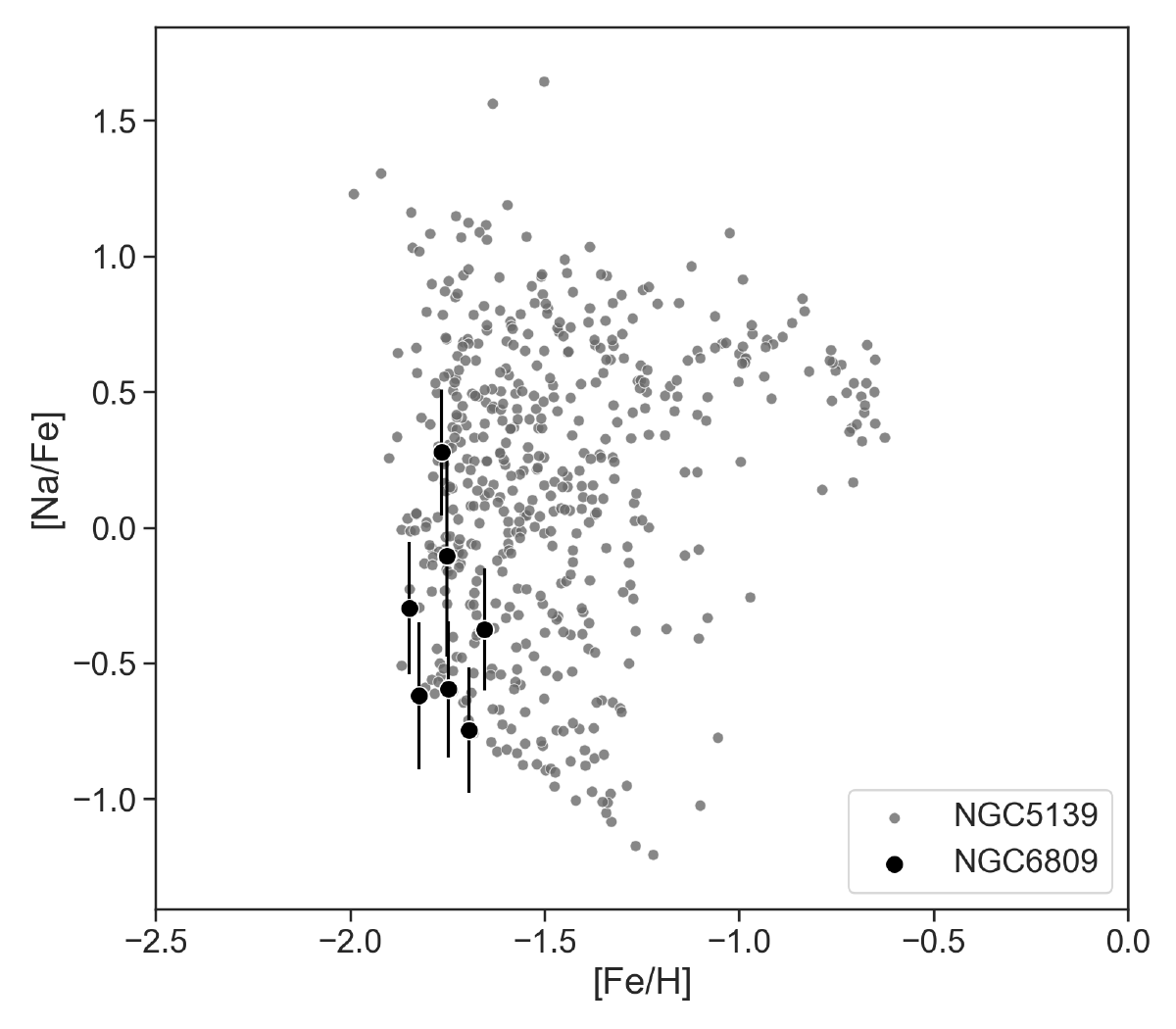}\hspace{-5pt}
\includegraphics[clip=true, trim = 3mm 0mm 0mm 2mm, width=0.68\columnwidth]{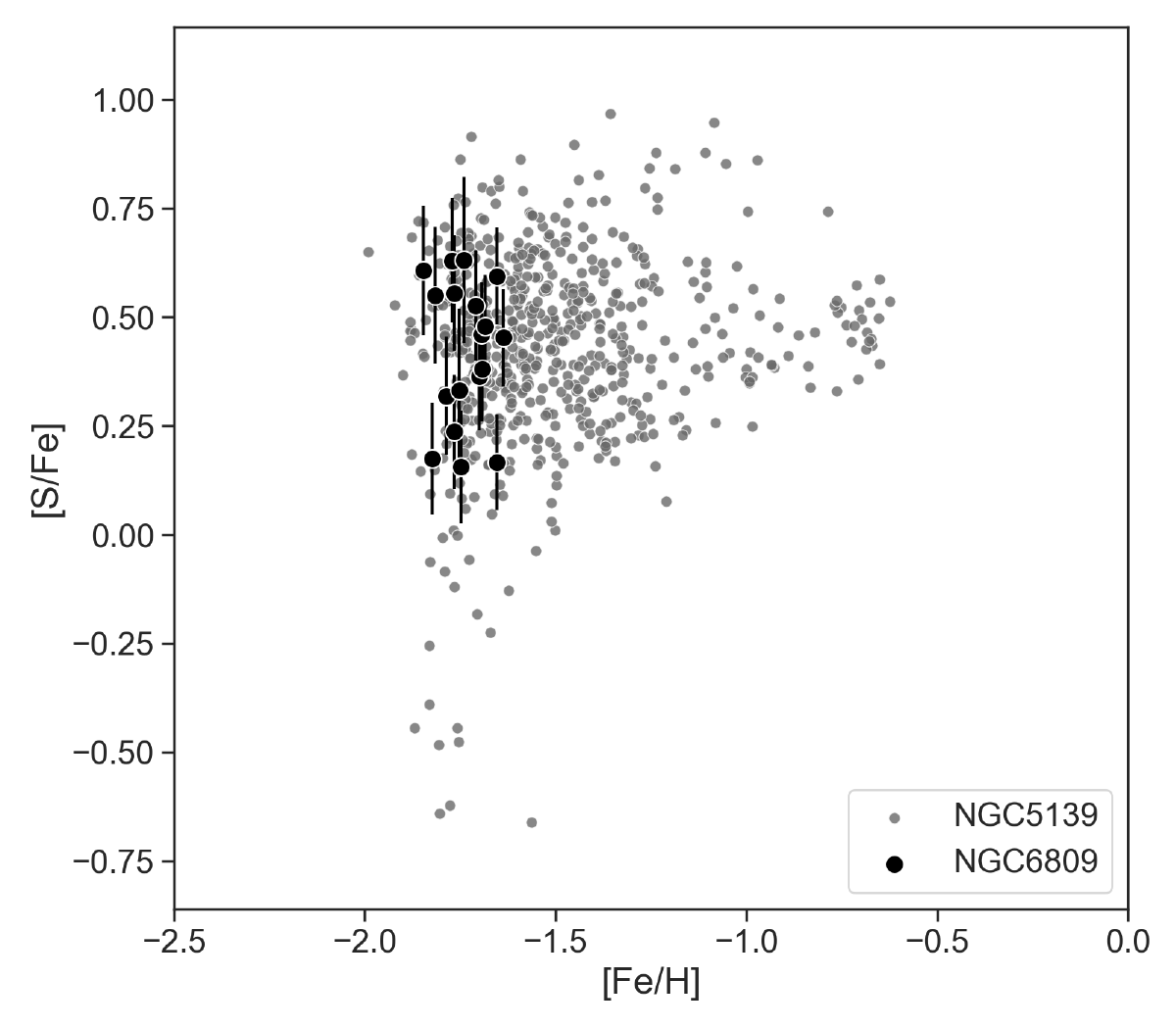}\hspace{-5pt}
\includegraphics[clip=true, trim = 3mm 0mm 0mm 2mm, width=0.68\columnwidth]{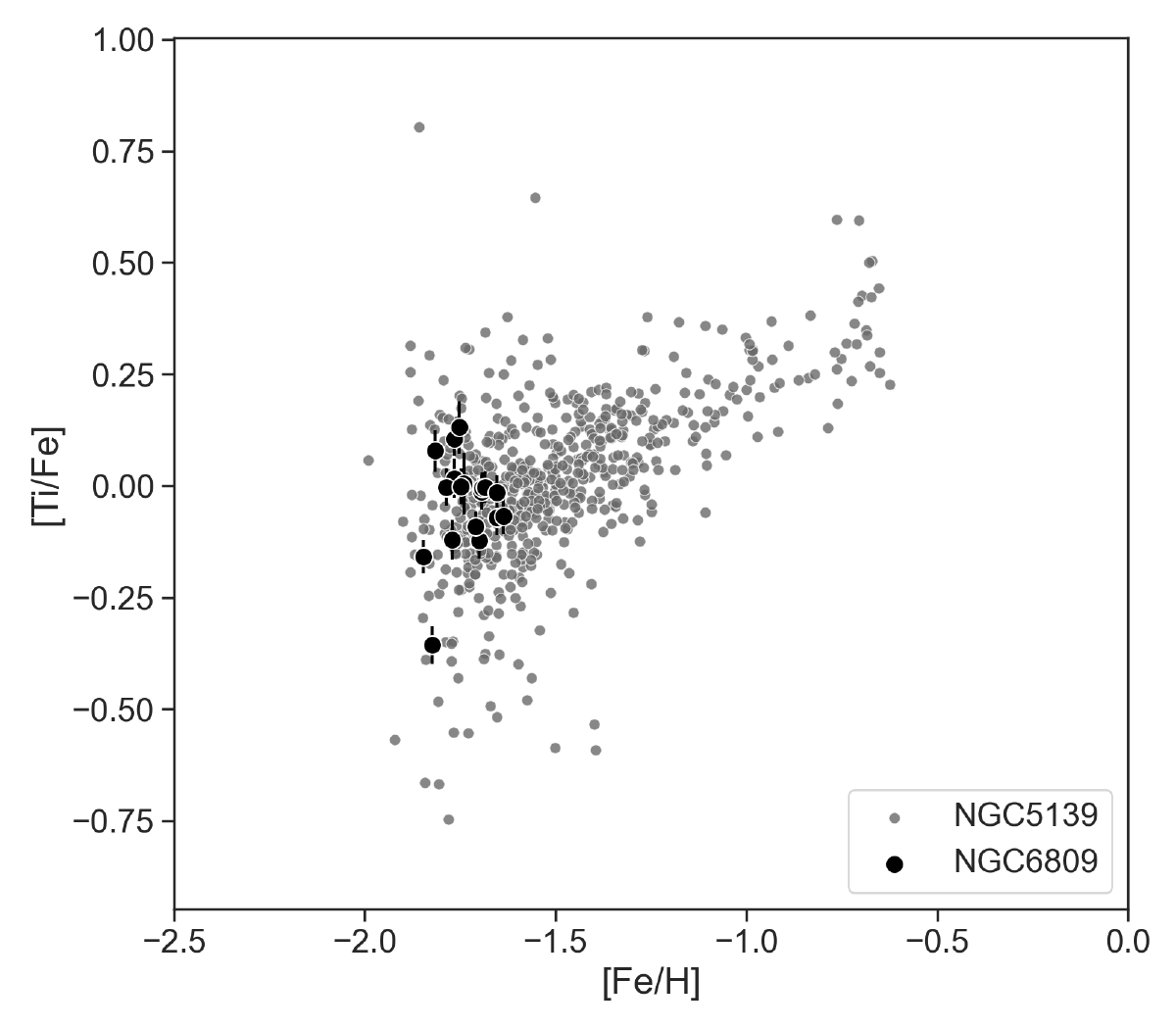}\hspace{-5pt}
\includegraphics[clip=true, trim = 3mm 0mm 0mm 2mm, width=0.68\columnwidth]{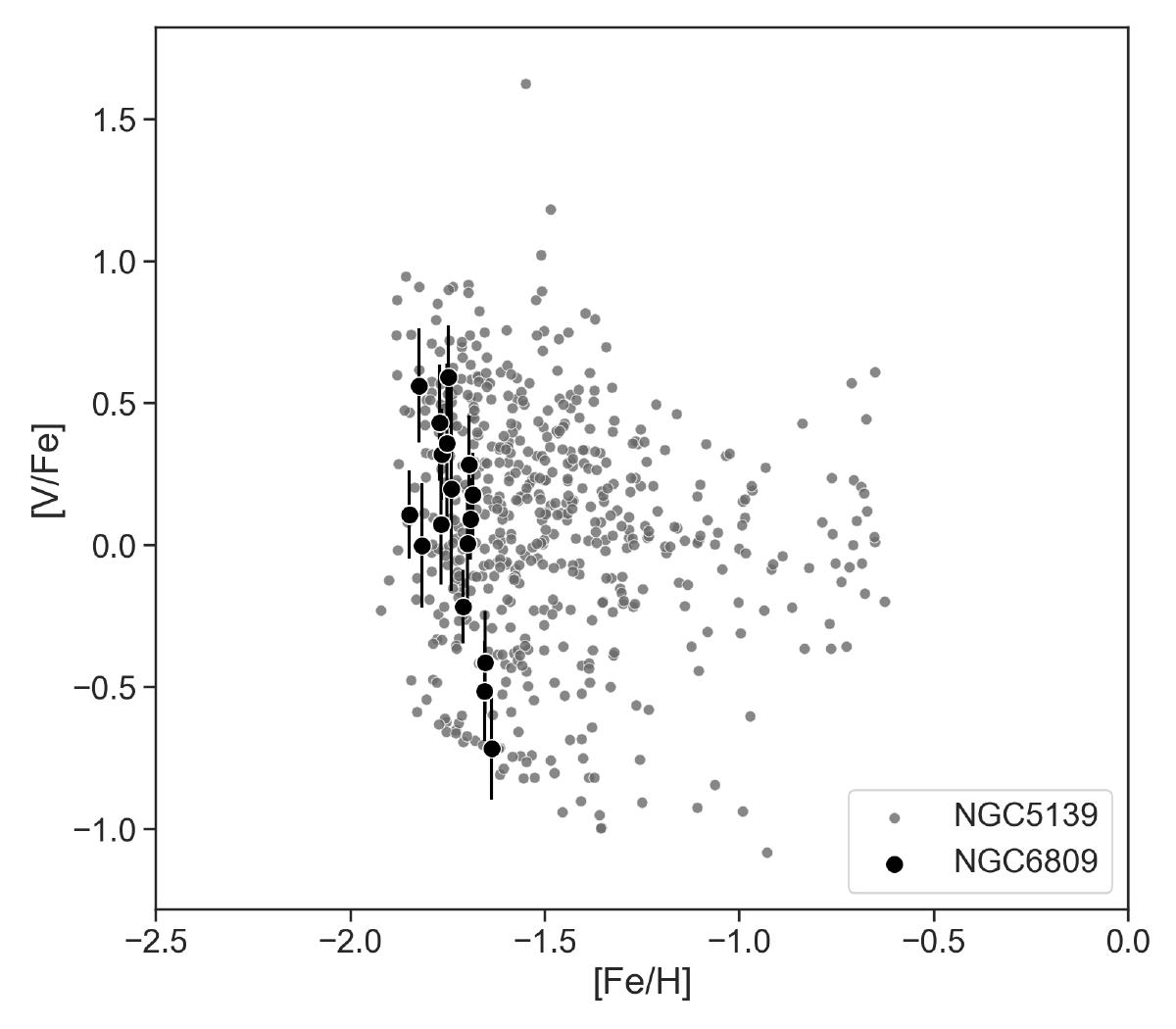}\hspace{-5pt}
\includegraphics[clip=true, trim = 3mm 0mm 0mm 2mm, width=0.68\columnwidth]{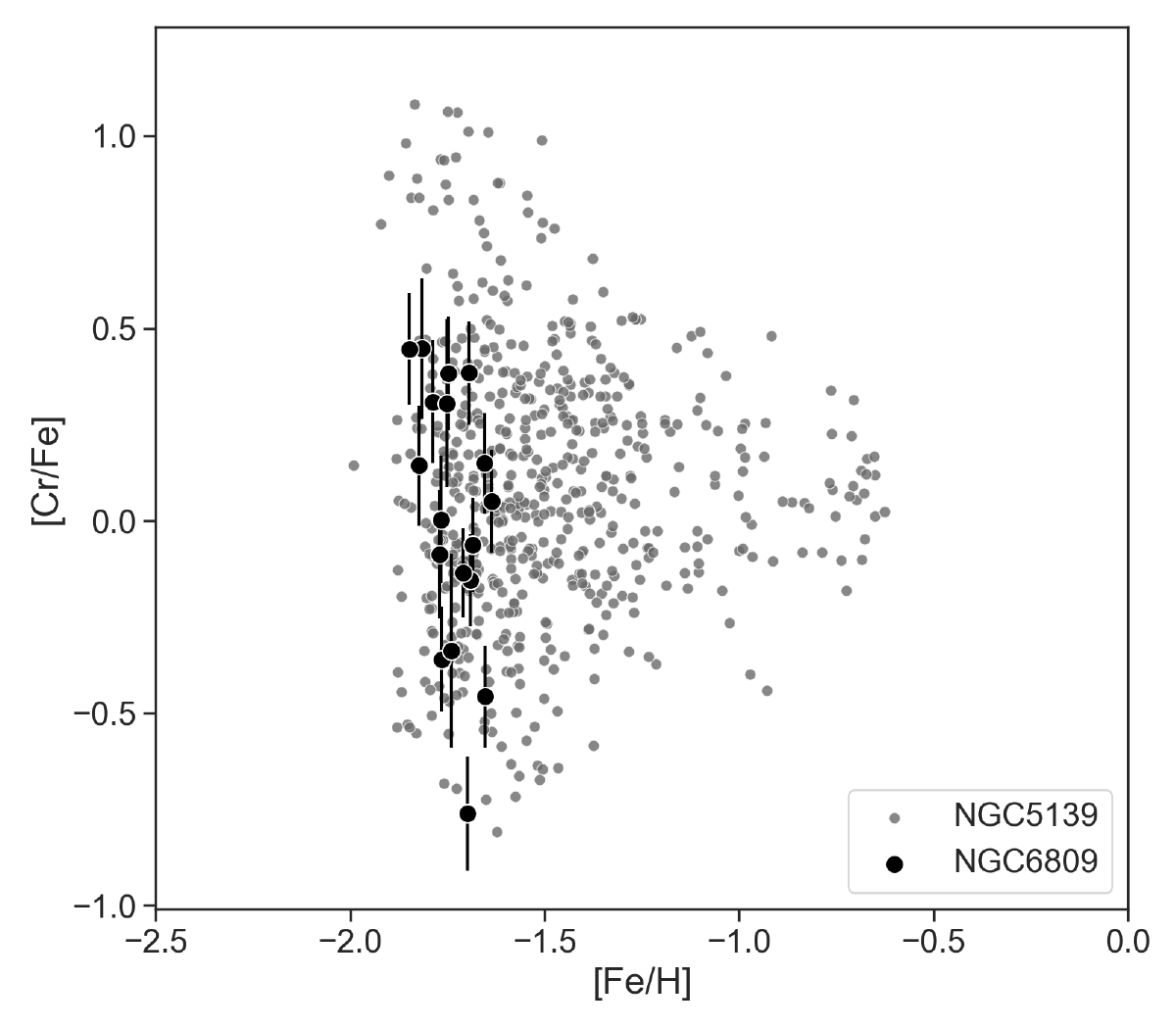}\hspace{-5pt}
\includegraphics[clip=true, trim = 3mm 0mm 0mm 2mm, width=0.68\columnwidth]{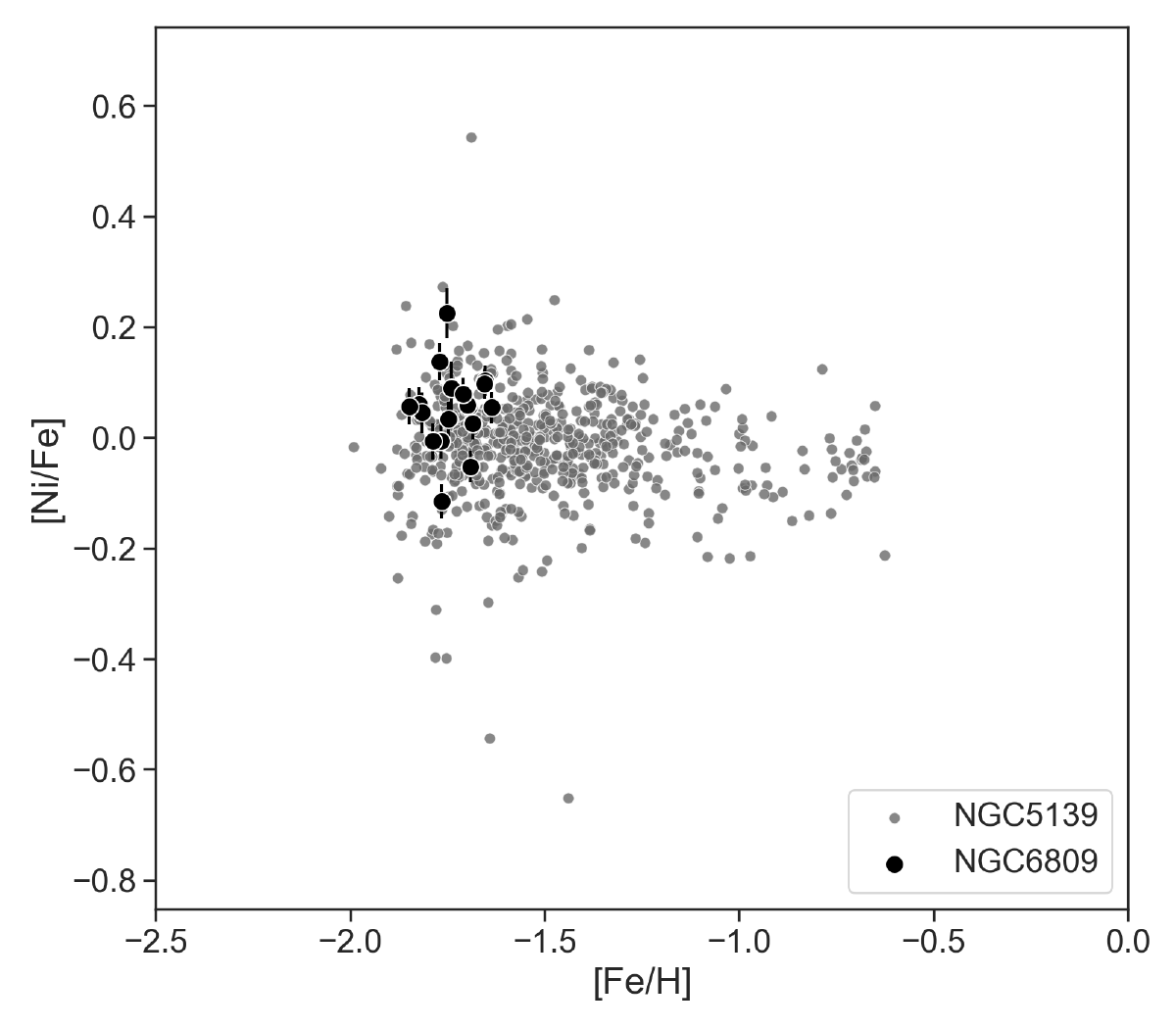}\hspace{-5pt}
\includegraphics[clip=true, trim = 3mm 0mm 0mm 2mm, width=0.68\columnwidth]{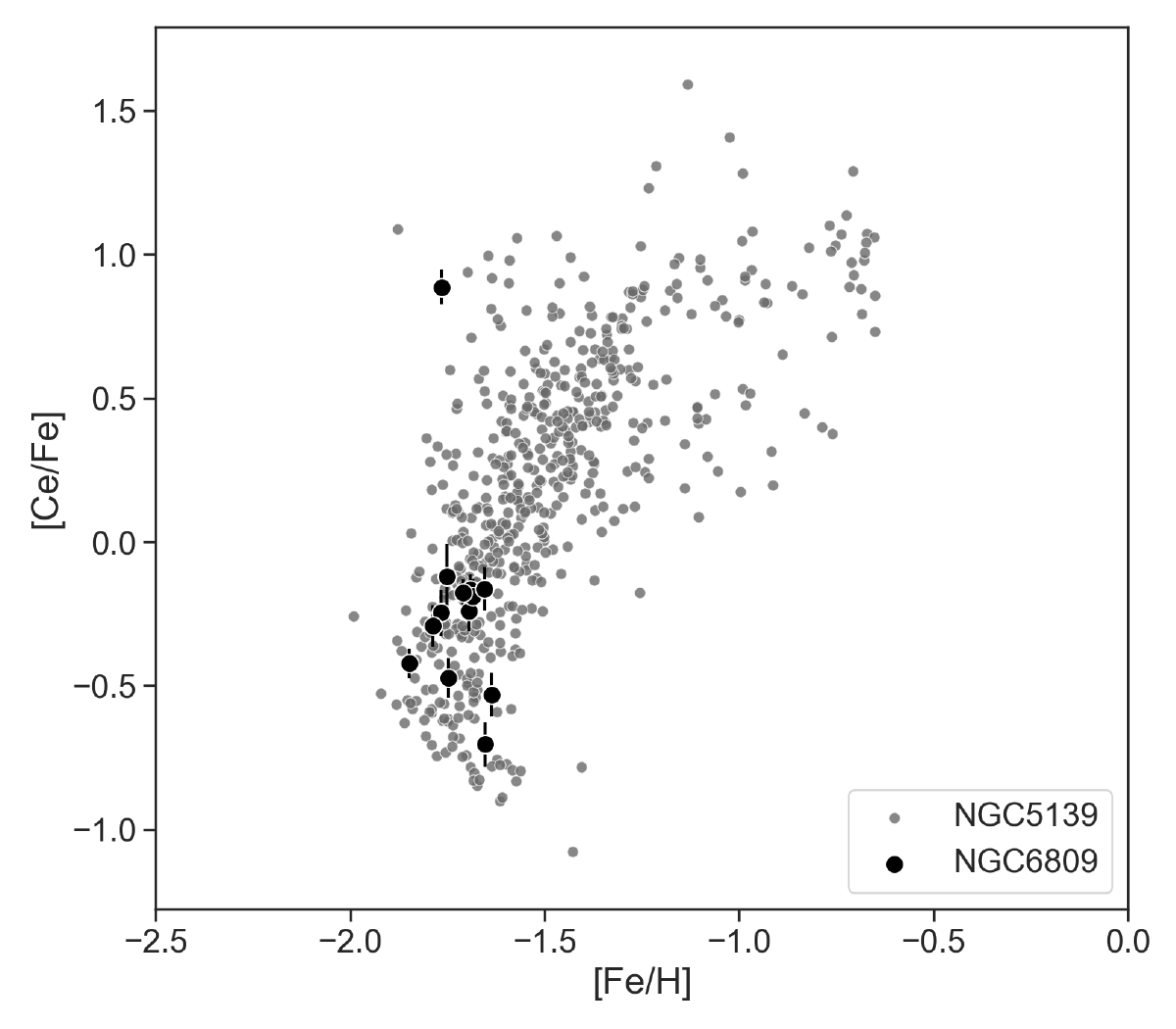}
  \caption{Same as Fig.~\ref{ngc6656_others} for NGC~6809}
              \label{ngc6809_others}%
    \end{figure*}

\begin{figure*}[h!]
   \centering
\includegraphics[clip=true, trim = 3mm 0mm 0mm 2mm, width=0.68\columnwidth]{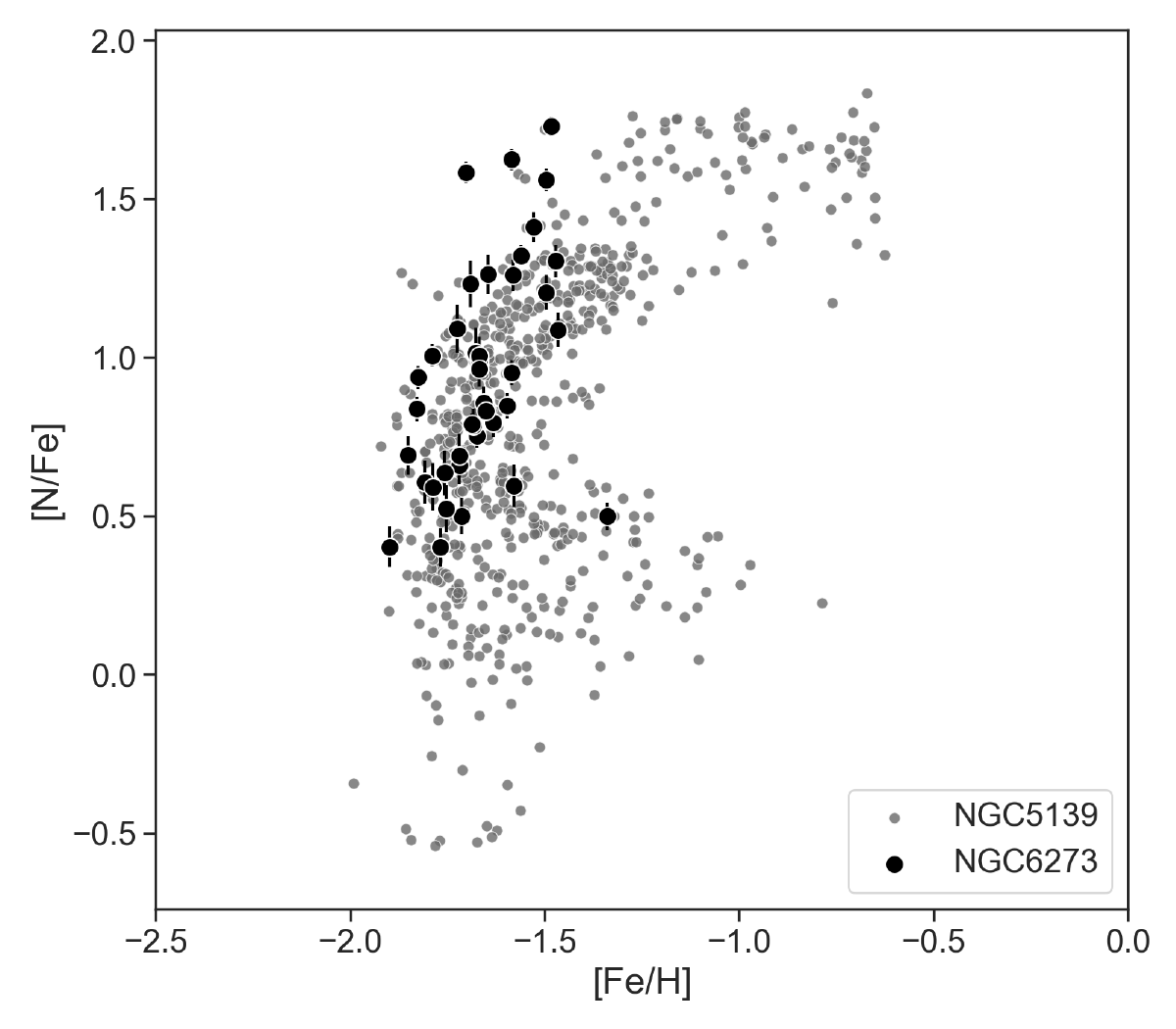}\hspace{-5pt}
\includegraphics[clip=true, trim = 3mm 0mm 0mm 2mm, width=0.68\columnwidth]{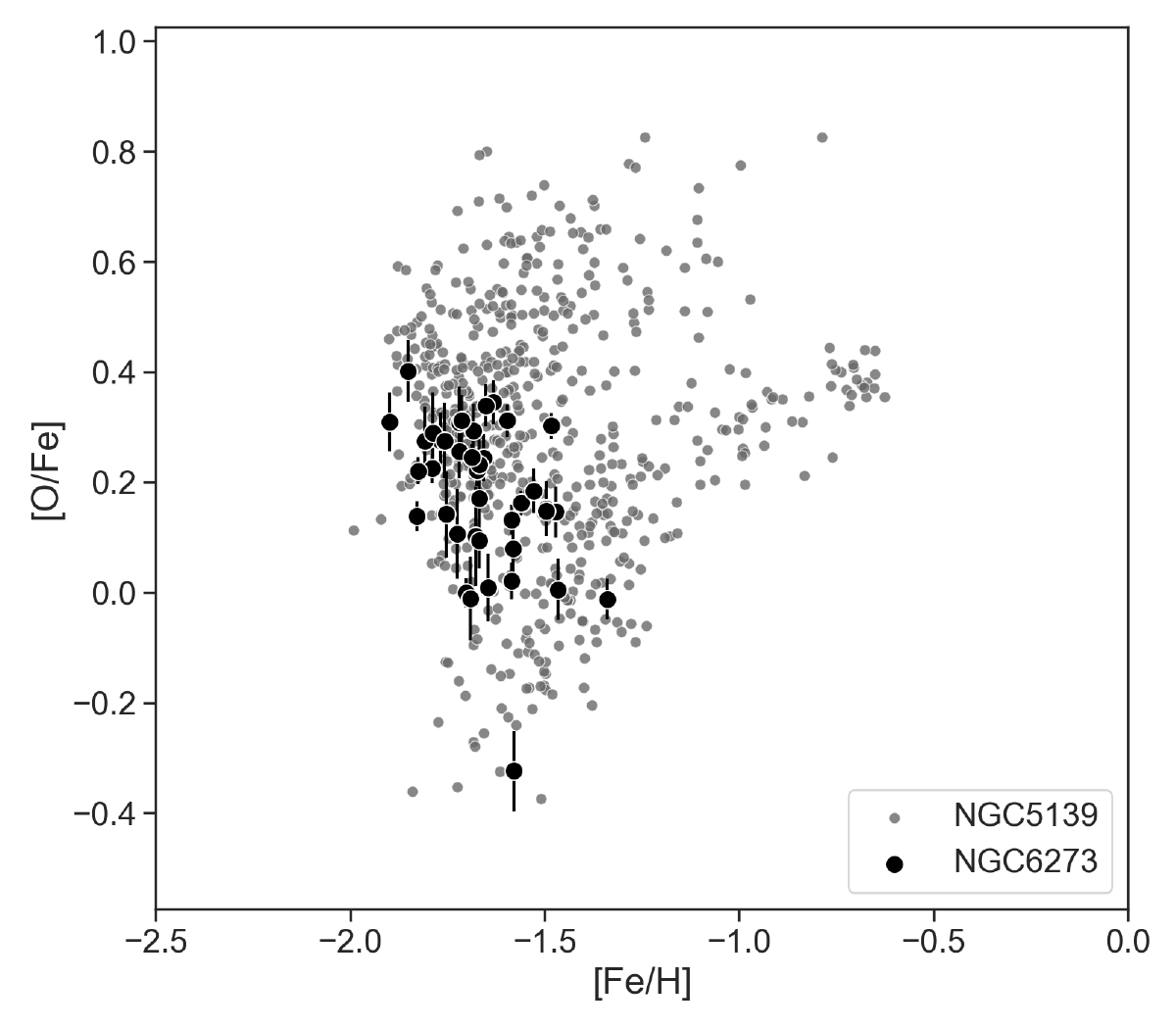}\hspace{-5pt}
\includegraphics[clip=true, trim = 3mm 0mm 0mm 2mm, width=0.68\columnwidth]{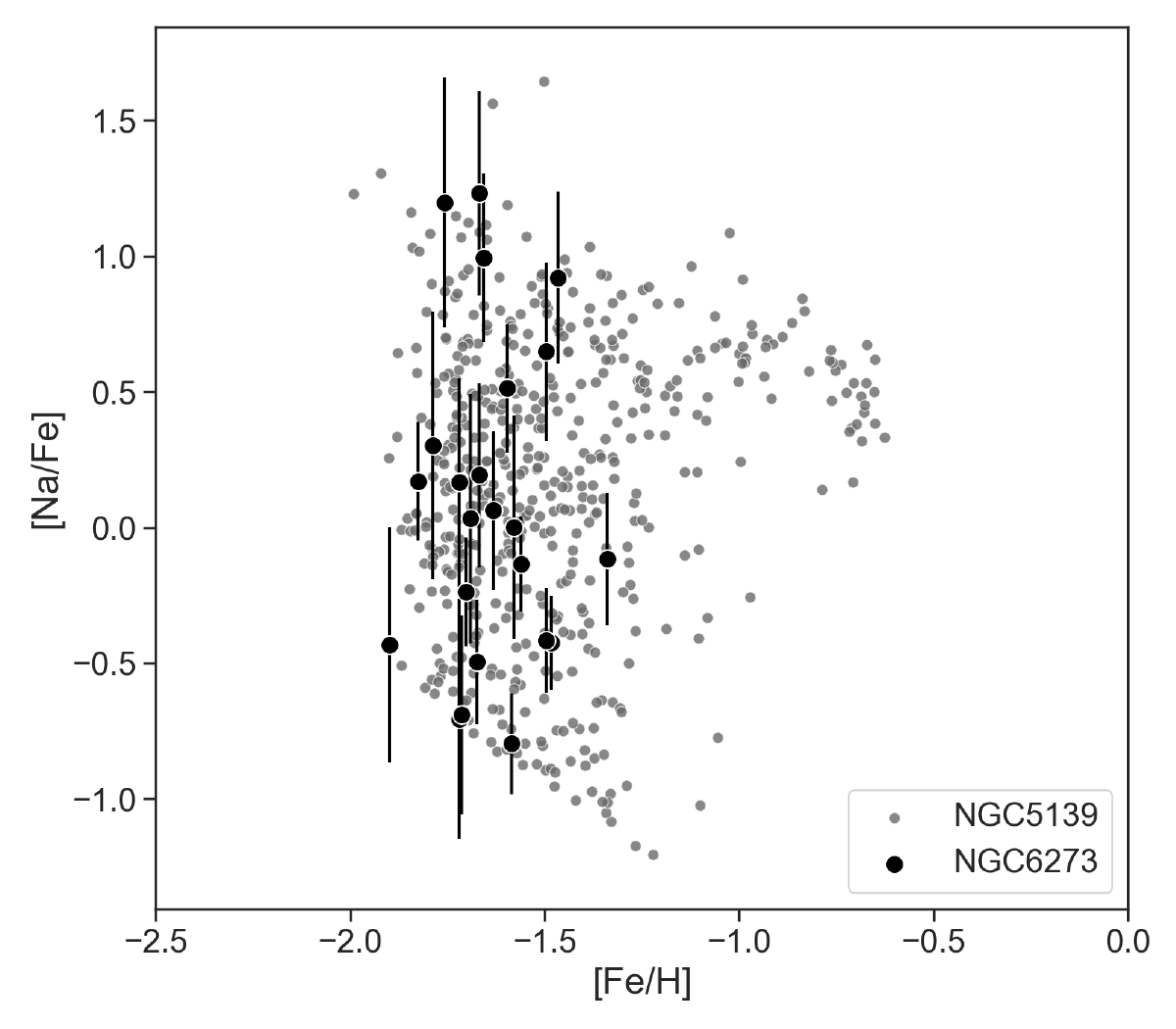}\hspace{-5pt}
\includegraphics[clip=true, trim = 3mm 0mm 0mm 2mm, width=0.68\columnwidth]{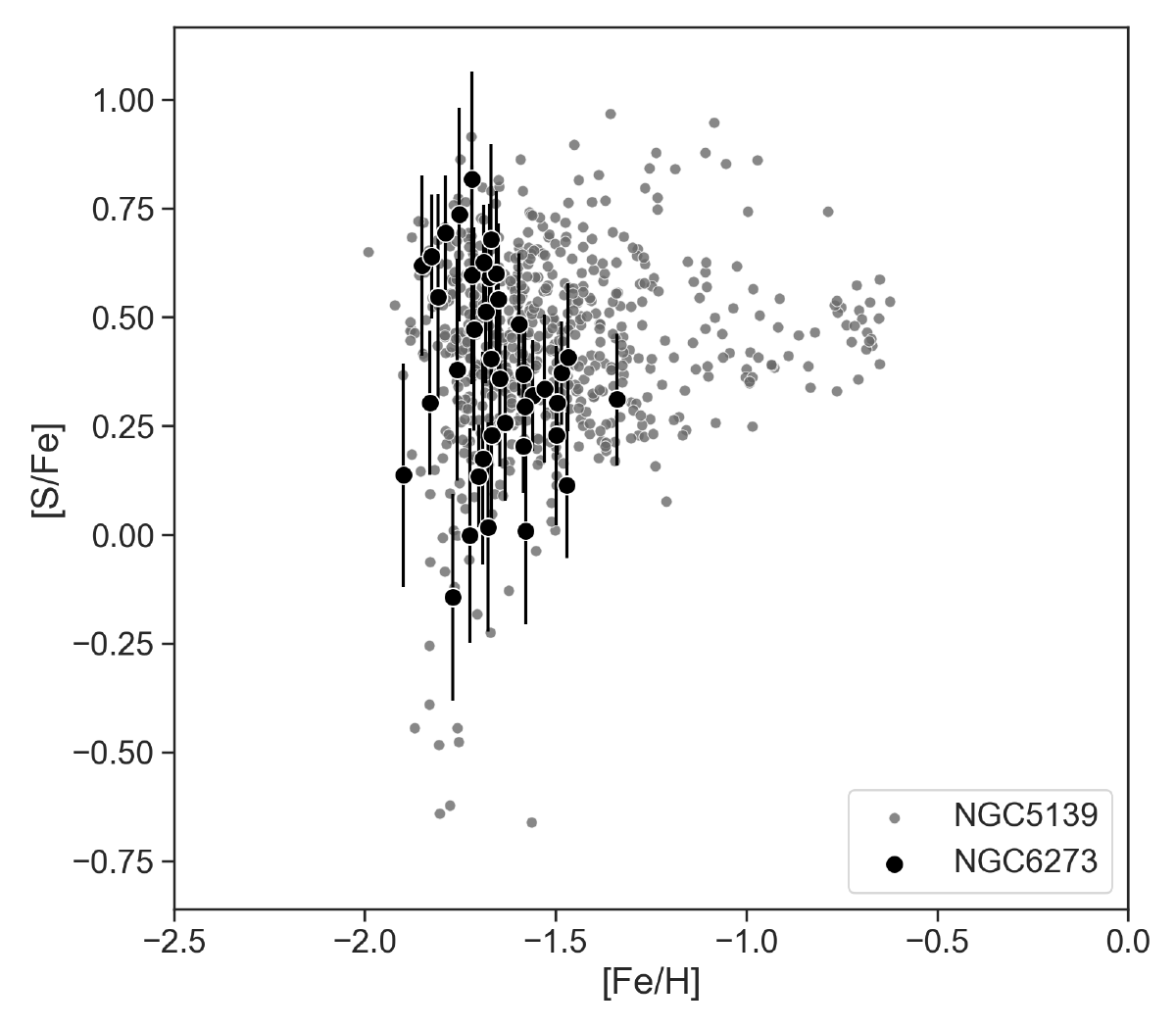}\hspace{-5pt}
\includegraphics[clip=true, trim = 3mm 0mm 0mm 2mm, width=0.68\columnwidth]{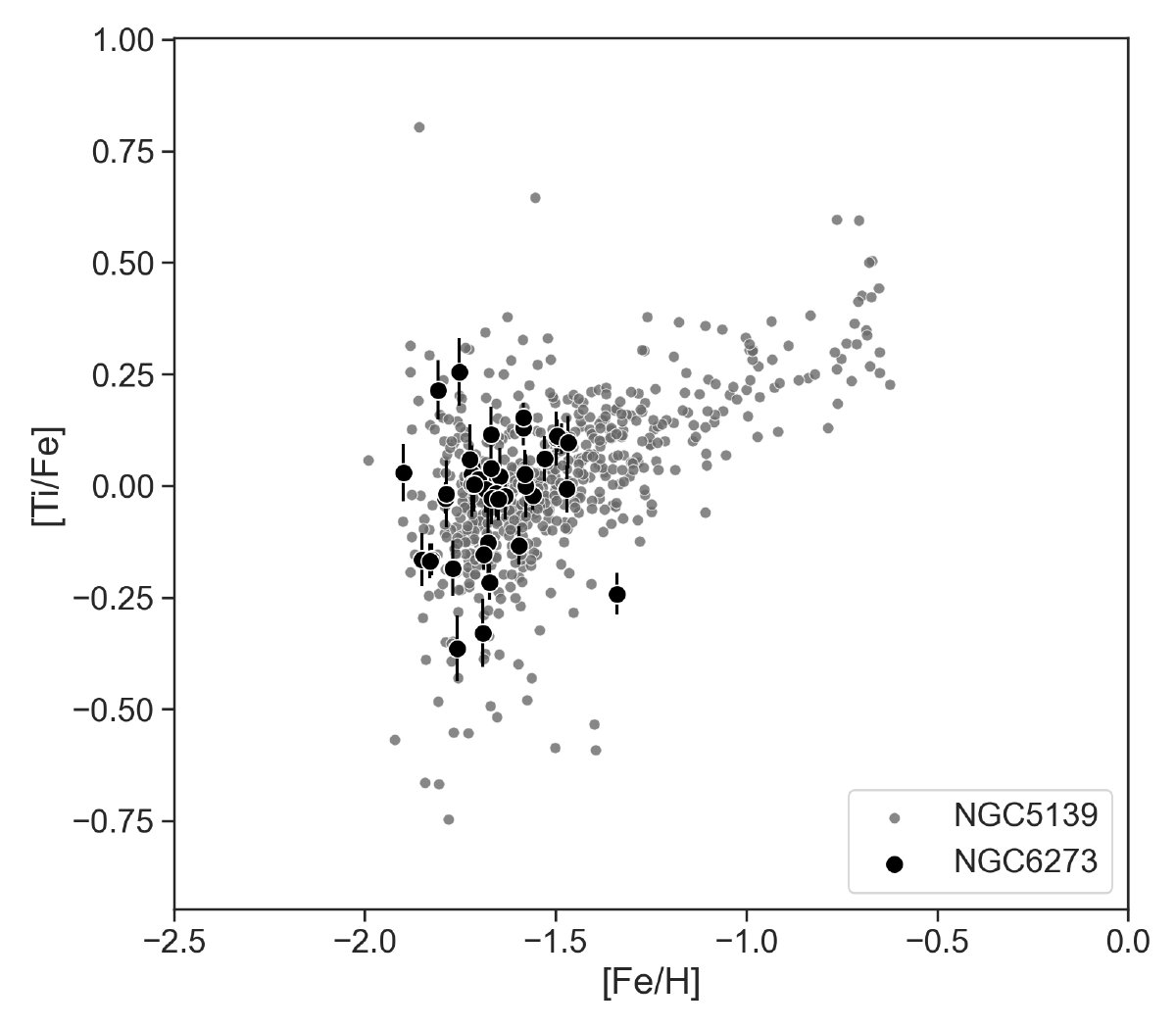}\hspace{-5pt}
\includegraphics[clip=true, trim = 3mm 0mm 0mm 2mm, width=0.68\columnwidth]{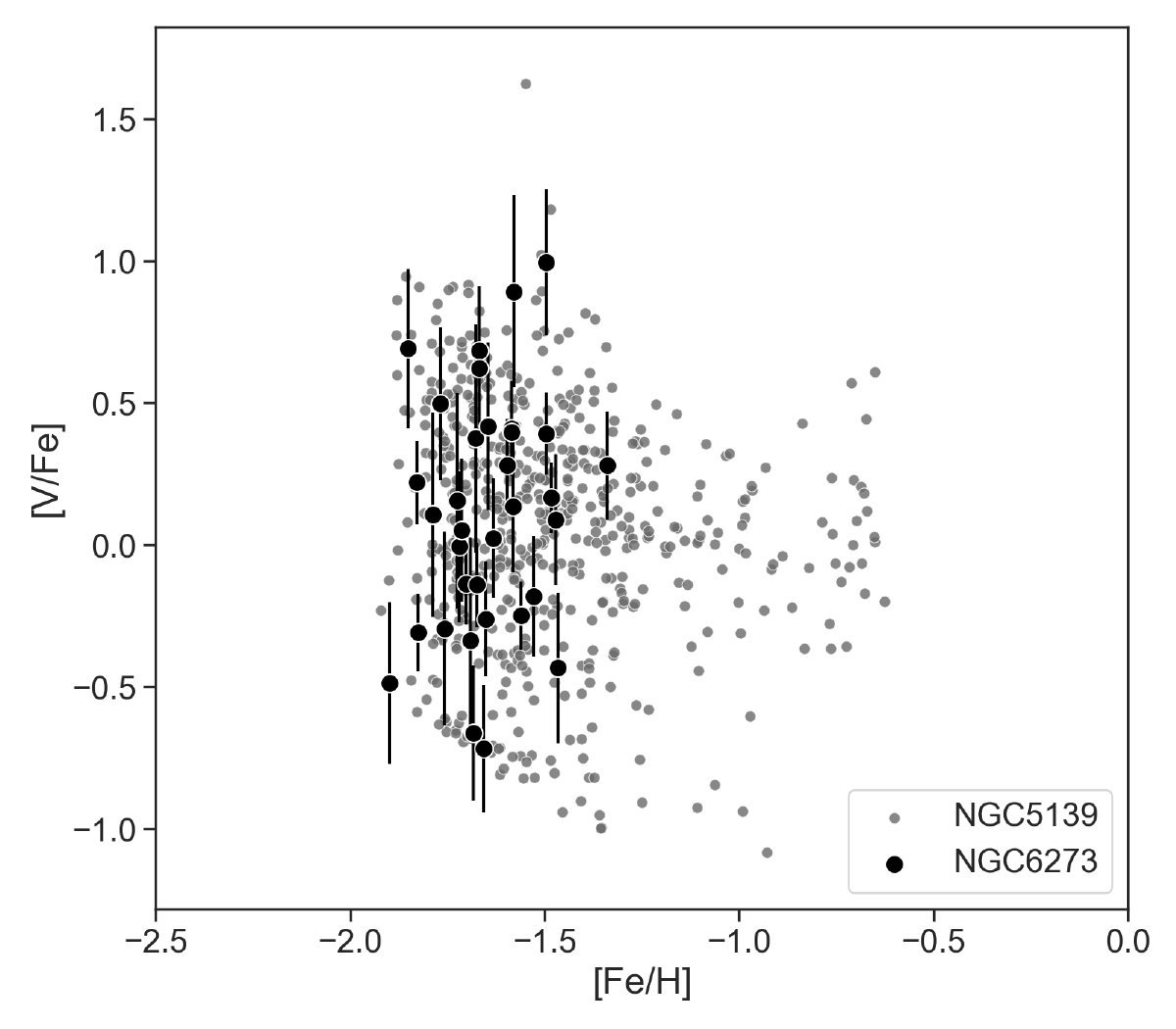}\hspace{-5pt}
\includegraphics[clip=true, trim = 3mm 0mm 0mm 2mm, width=0.68\columnwidth]{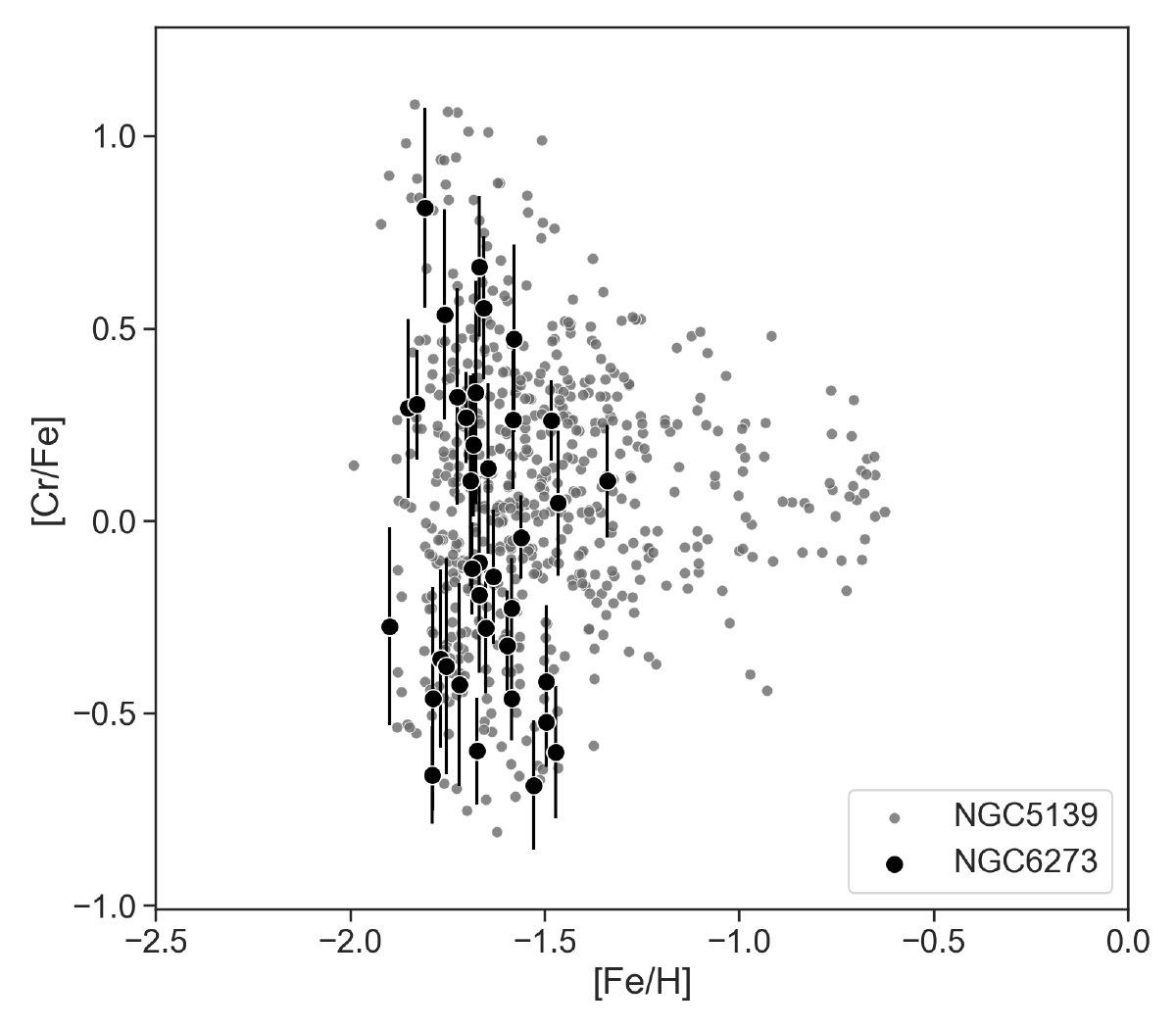}\hspace{-5pt}
\includegraphics[clip=true, trim = 3mm 0mm 0mm 2mm, width=0.68\columnwidth]{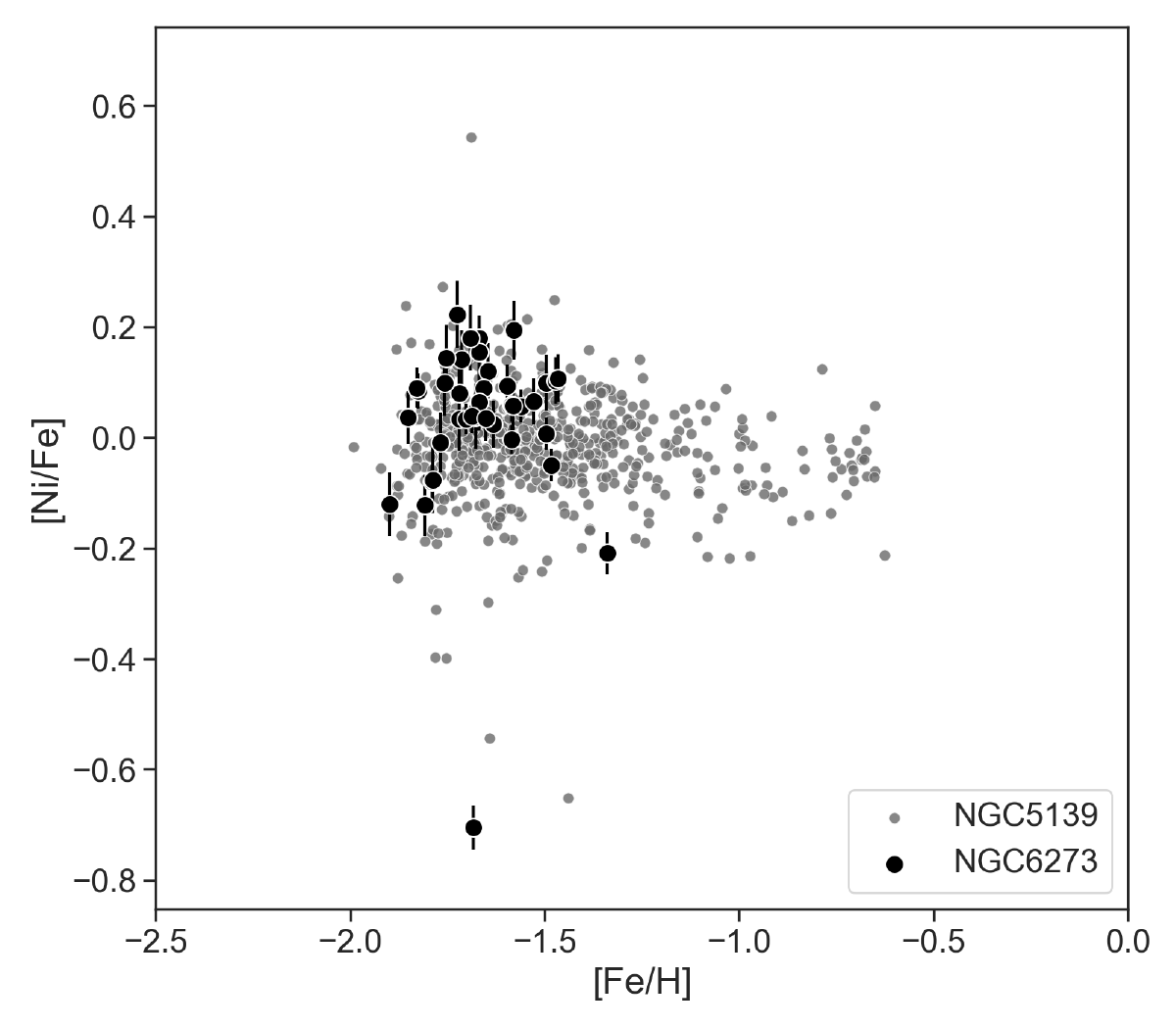}\hspace{-5pt}
\includegraphics[clip=true, trim = 3mm 0mm 0mm 2mm, width=0.68\columnwidth]{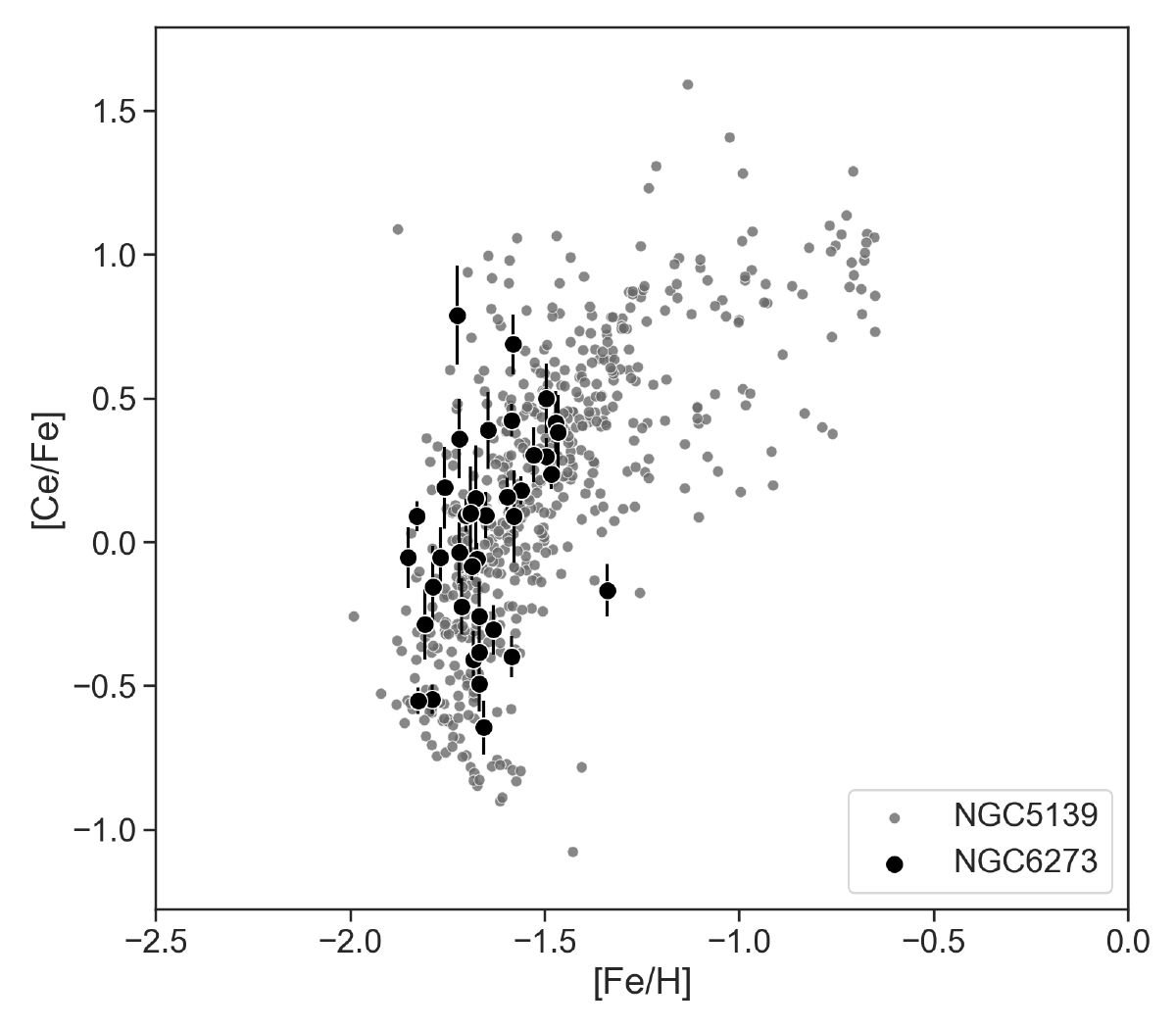}
  \caption{Same as Fig.~\ref{ngc6656_others} for NGC~6273.}
              \label{ngc6273_others}%
    \end{figure*}

\begin{figure*}[h!]
   \centering
\includegraphics[clip=true, trim = 3mm 0mm 0mm 2mm, width=0.68\columnwidth]{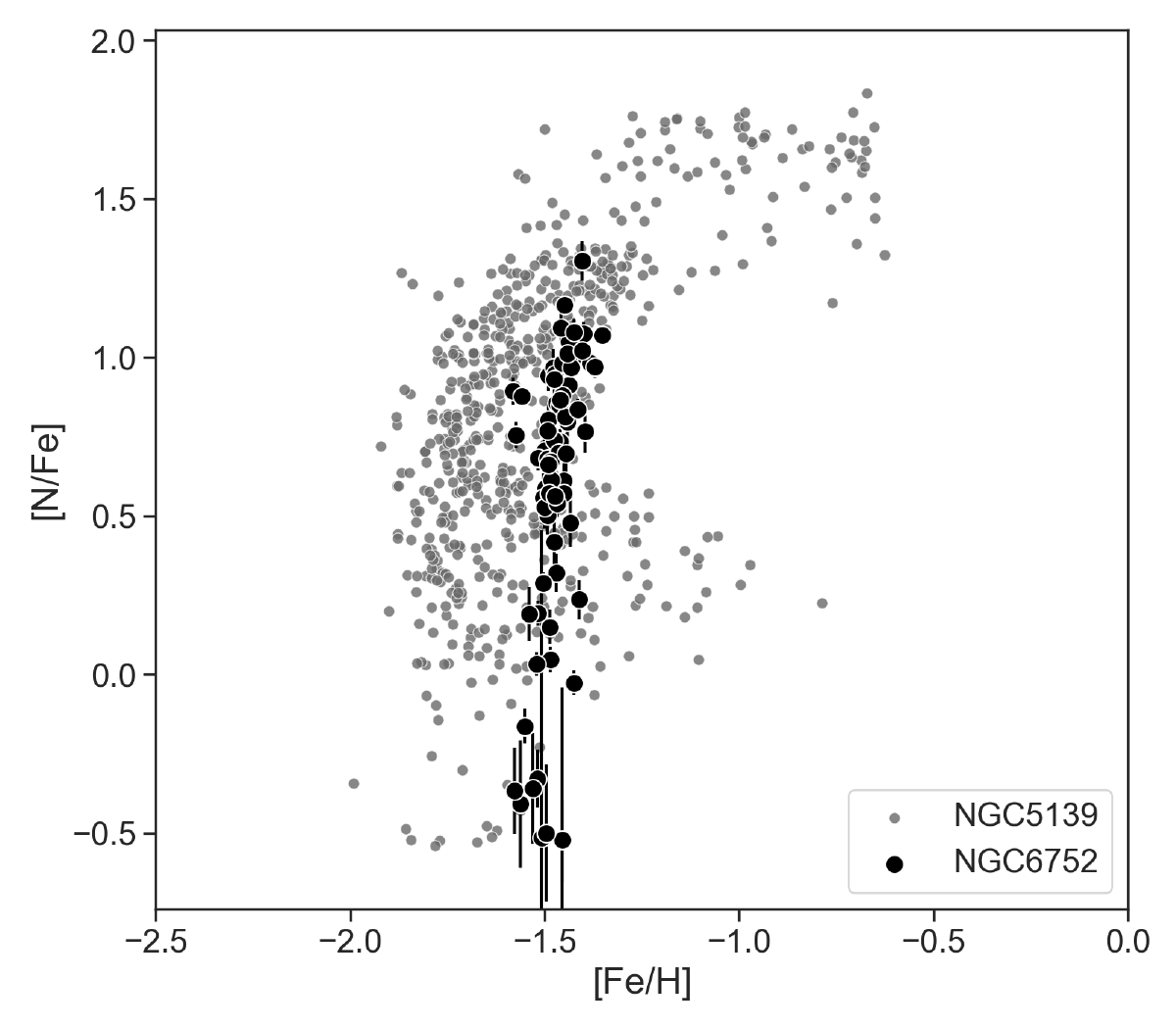}\hspace{-5pt}
\includegraphics[clip=true, trim = 3mm 0mm 0mm 2mm, width=0.68\columnwidth]{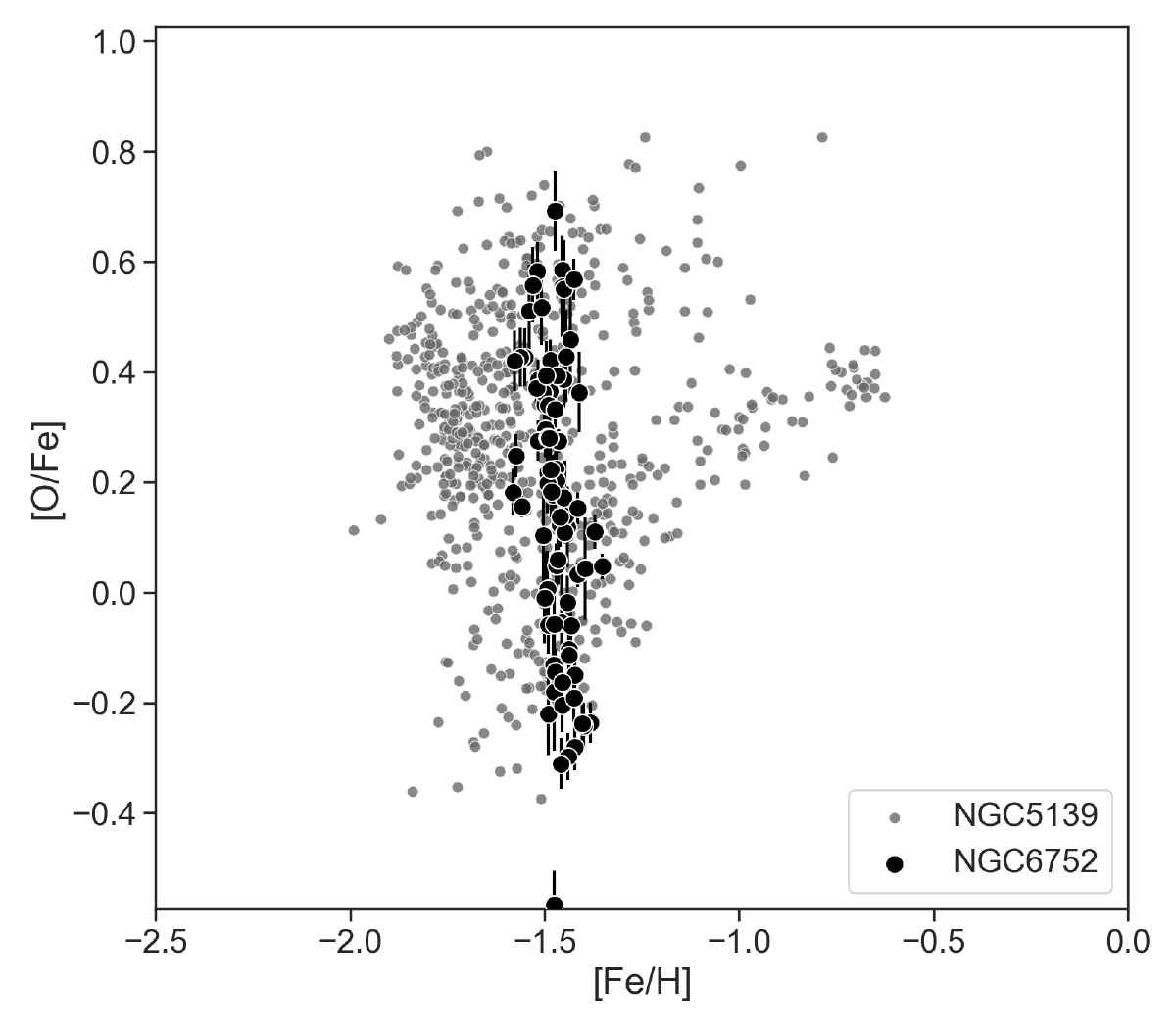}\hspace{-5pt}
\includegraphics[clip=true, trim = 3mm 0mm 0mm 2mm, width=0.68\columnwidth]{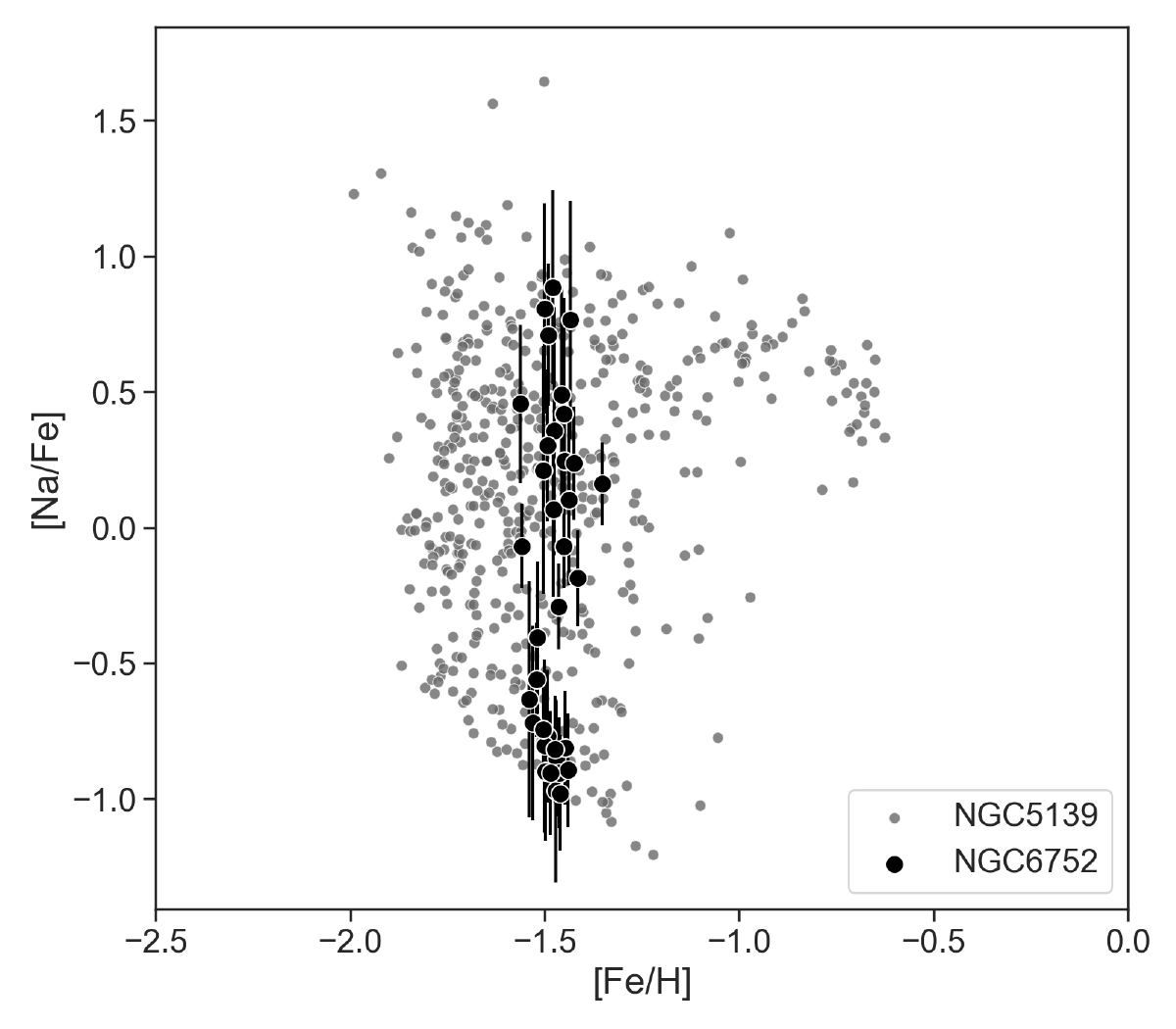}\hspace{-5pt}
\includegraphics[clip=true, trim = 3mm 0mm 0mm 2mm, width=0.68\columnwidth]{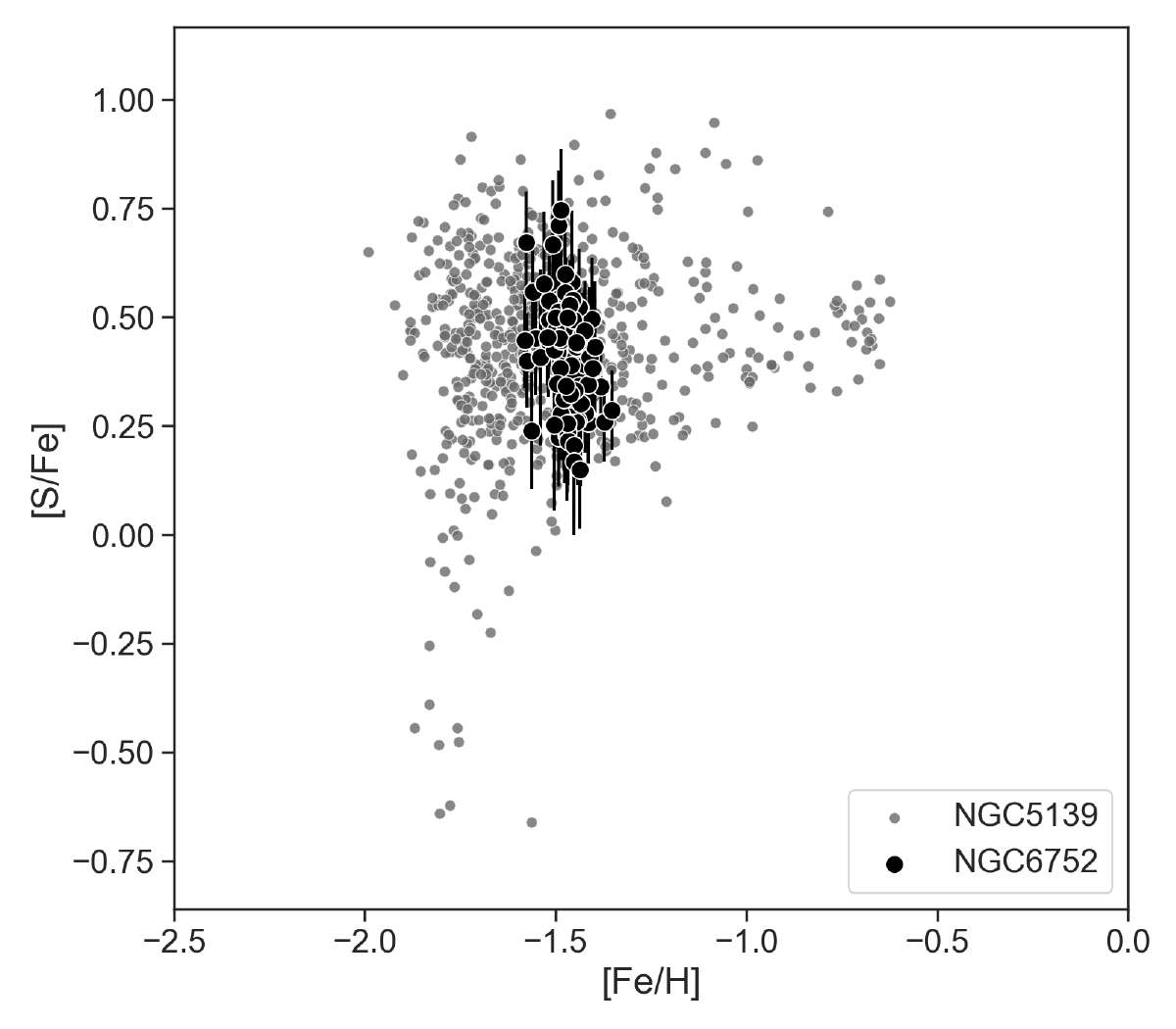}\hspace{-5pt}
\includegraphics[clip=true, trim = 3mm 0mm 0mm 2mm, width=0.68\columnwidth]{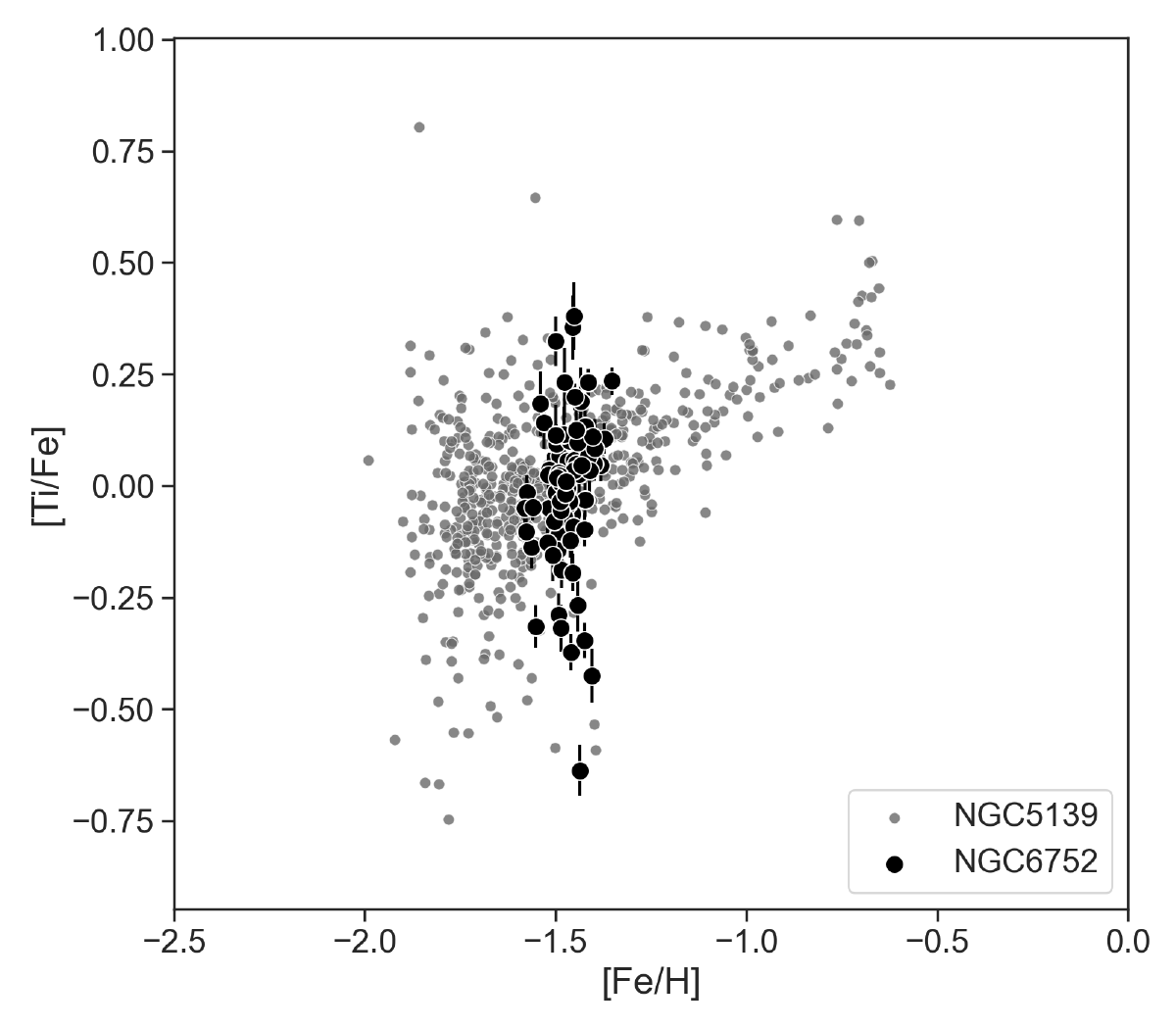}\hspace{-5pt}
\includegraphics[clip=true, trim = 3mm 0mm 0mm 2mm, width=0.68\columnwidth]{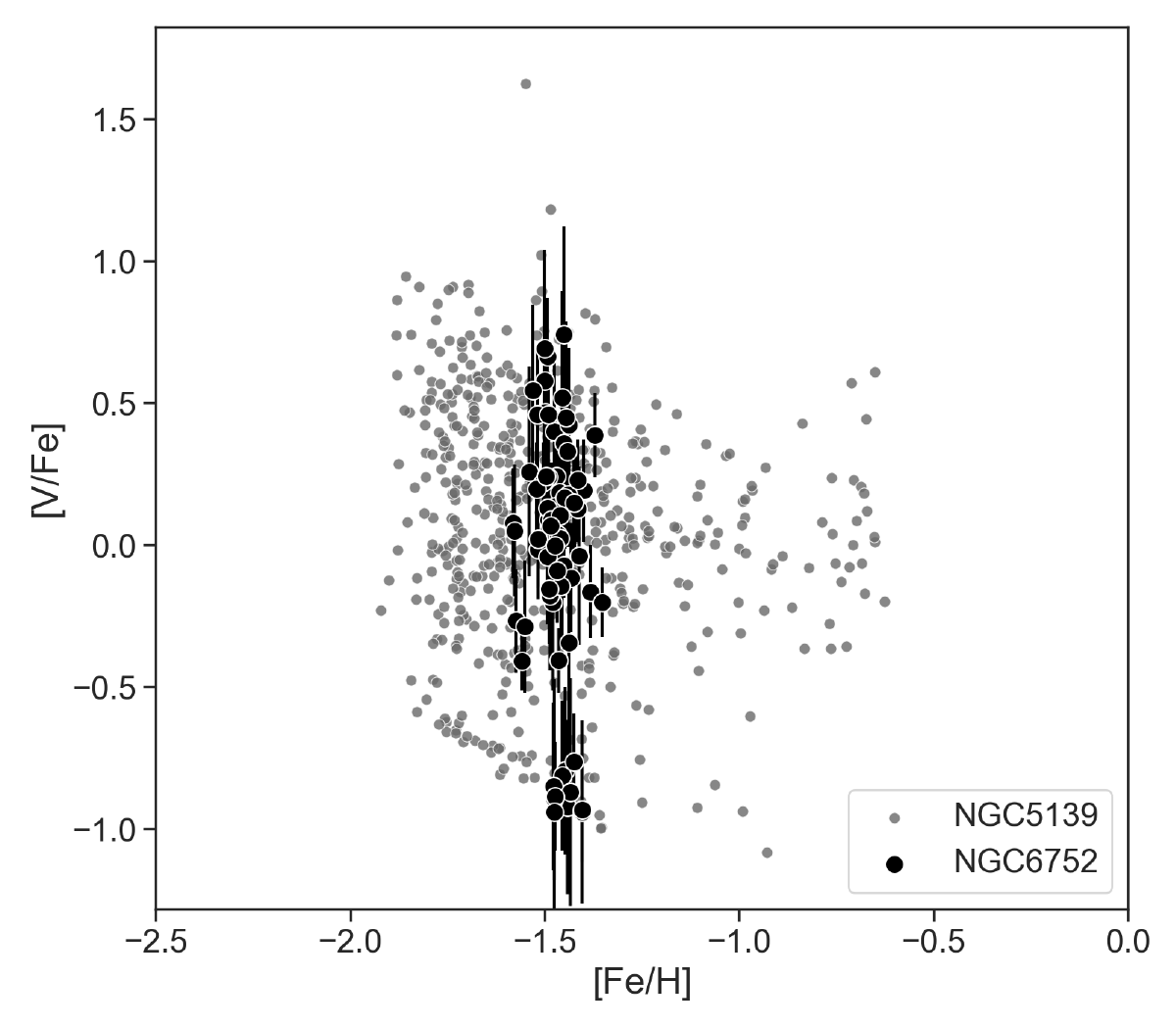}\hspace{-5pt}
\includegraphics[clip=true, trim = 3mm 0mm 0mm 2mm, width=0.68\columnwidth]{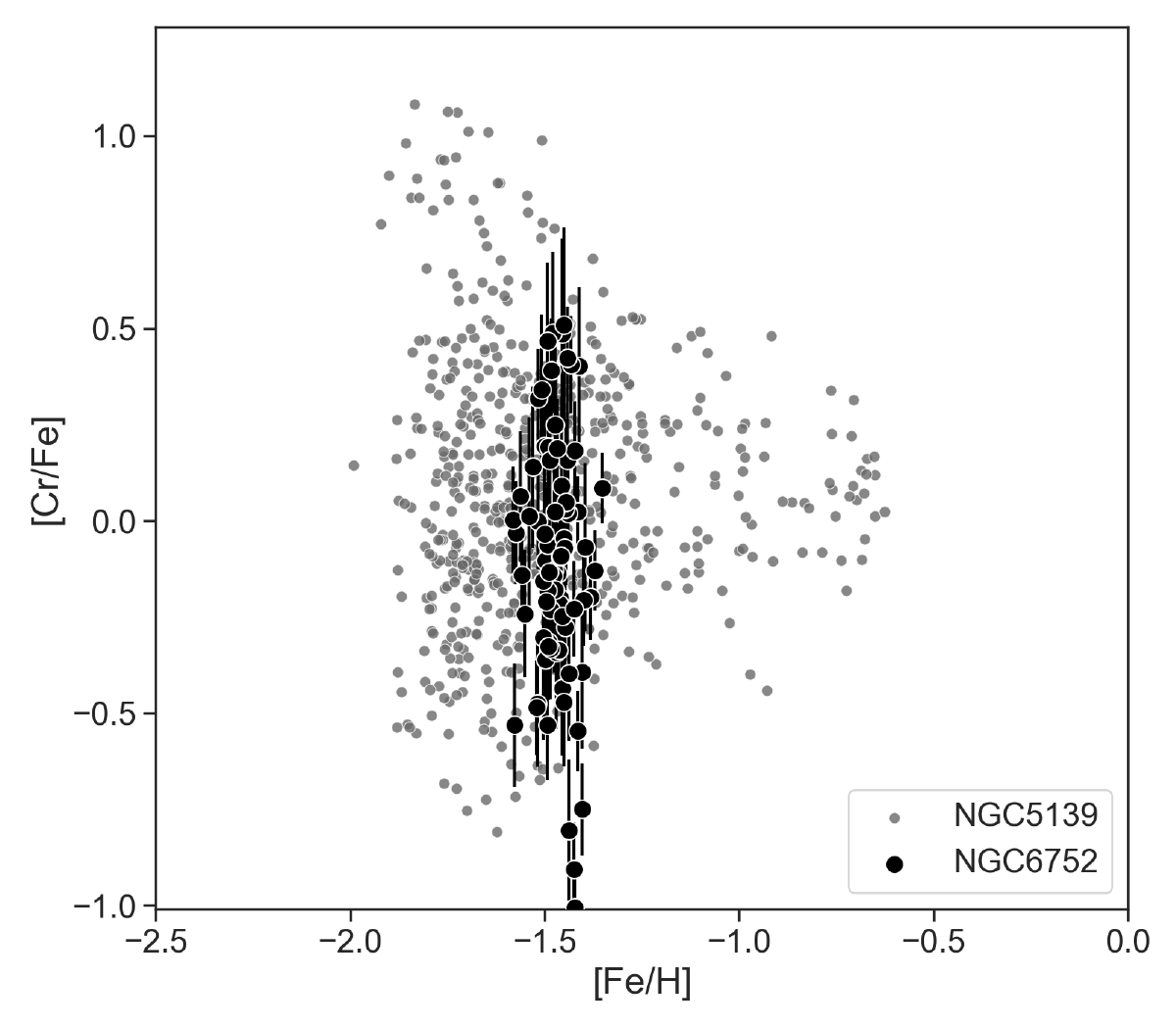}\hspace{-5pt}
\includegraphics[clip=true, trim = 3mm 0mm 0mm 2mm, width=0.68\columnwidth]{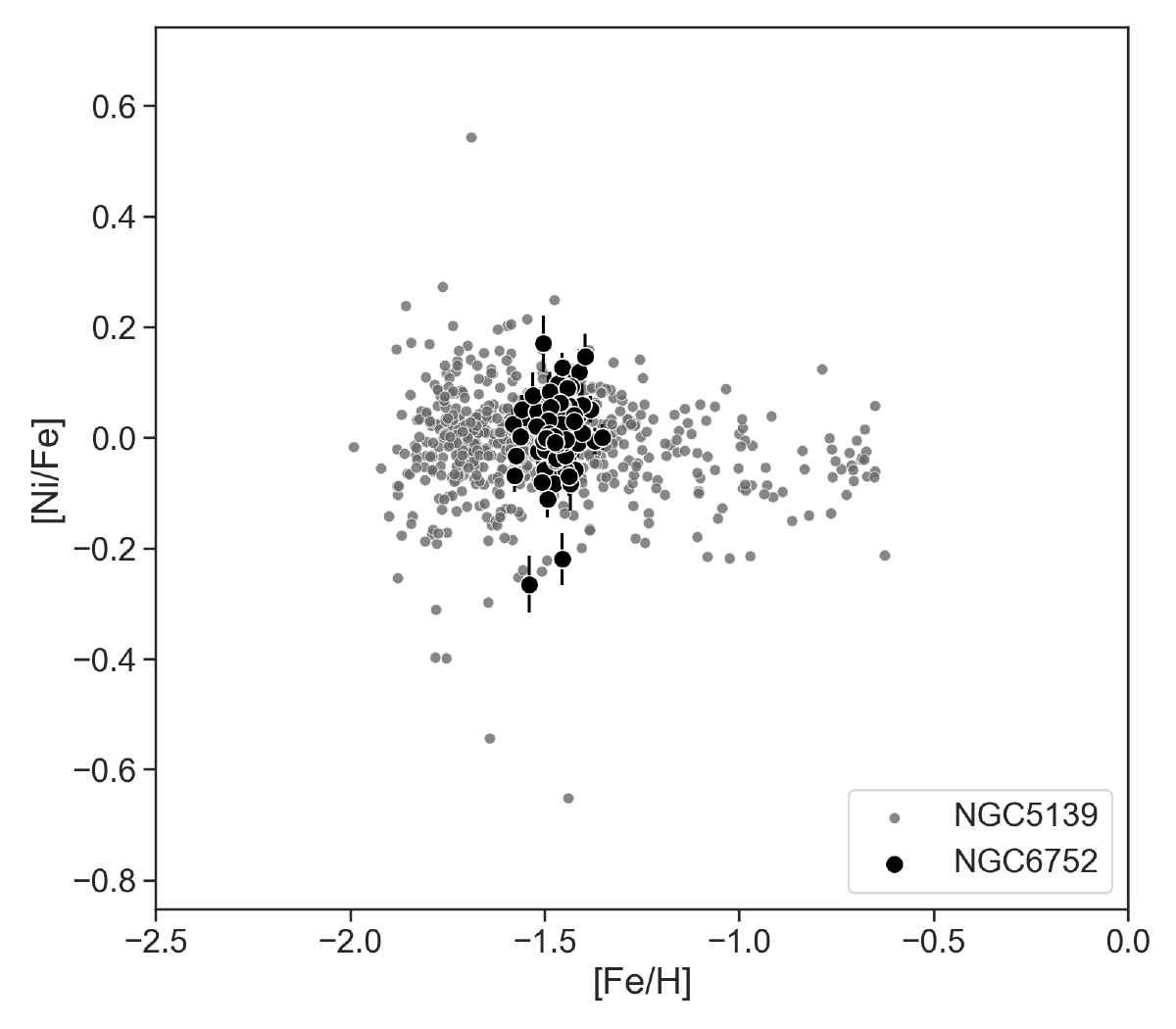}\hspace{-5pt}
\includegraphics[clip=true, trim = 3mm 0mm 0mm 2mm, width=0.68\columnwidth]{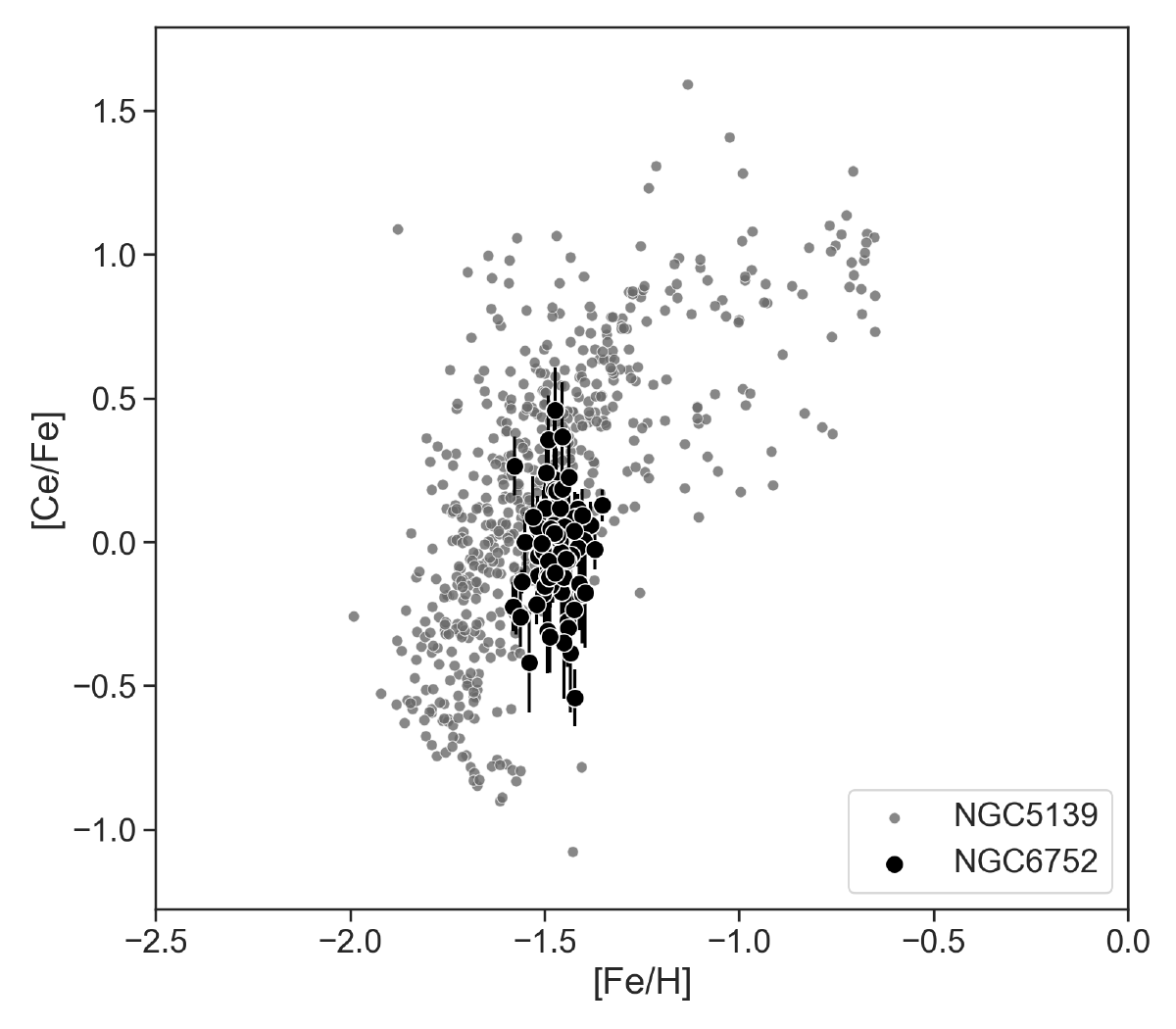}
  \caption{Same as Fig.~\ref{ngc6656_others} for NGC~6752.}
              \label{ngc6752_others}%
    \end{figure*}

\begin{figure*}[h!]
   \centering
\includegraphics[clip=true, trim = 3mm 0mm 0mm 2mm, width=0.68\columnwidth]{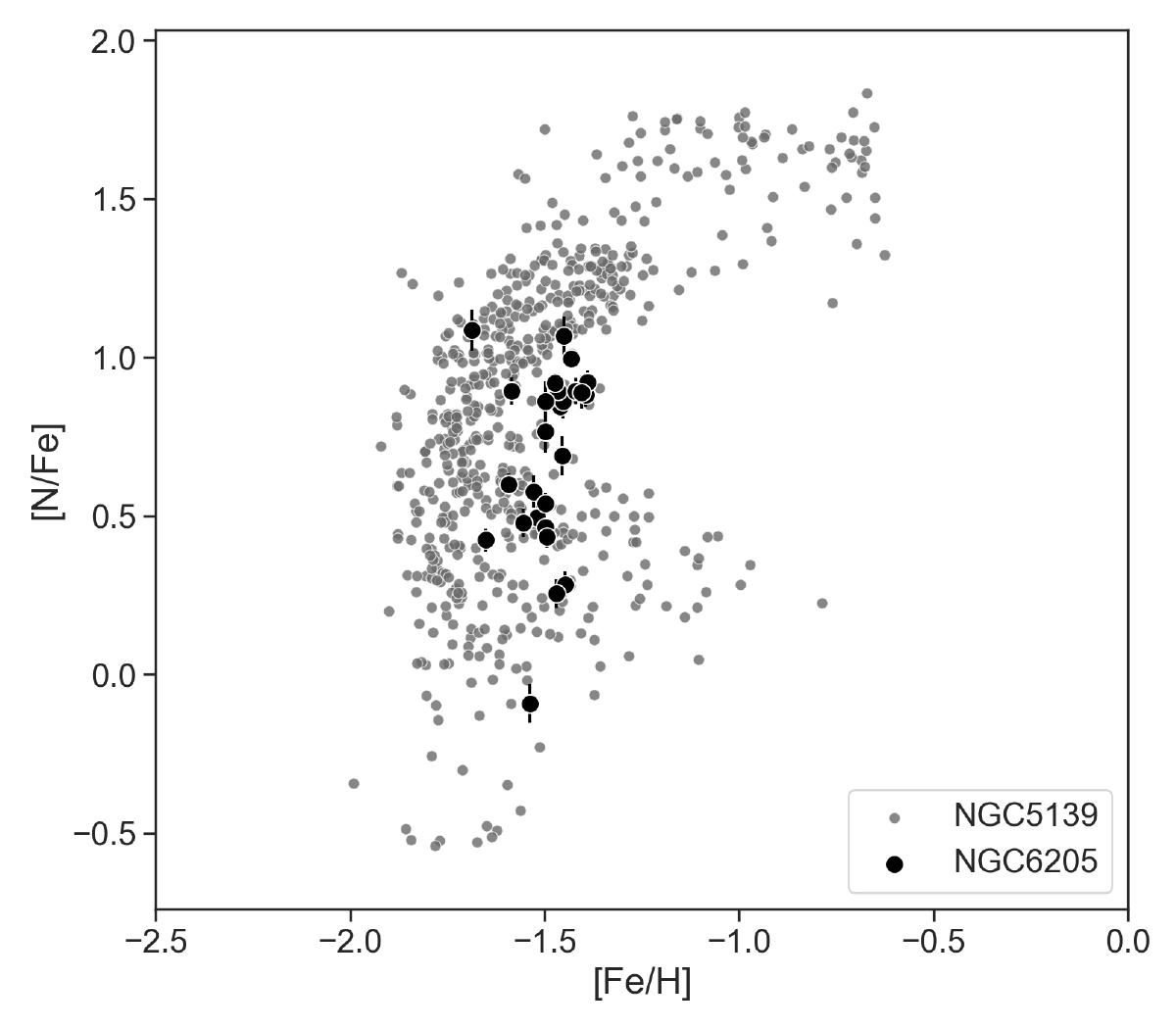}\hspace{-5pt}
\includegraphics[clip=true, trim = 3mm 0mm 0mm 2mm, width=0.68\columnwidth]{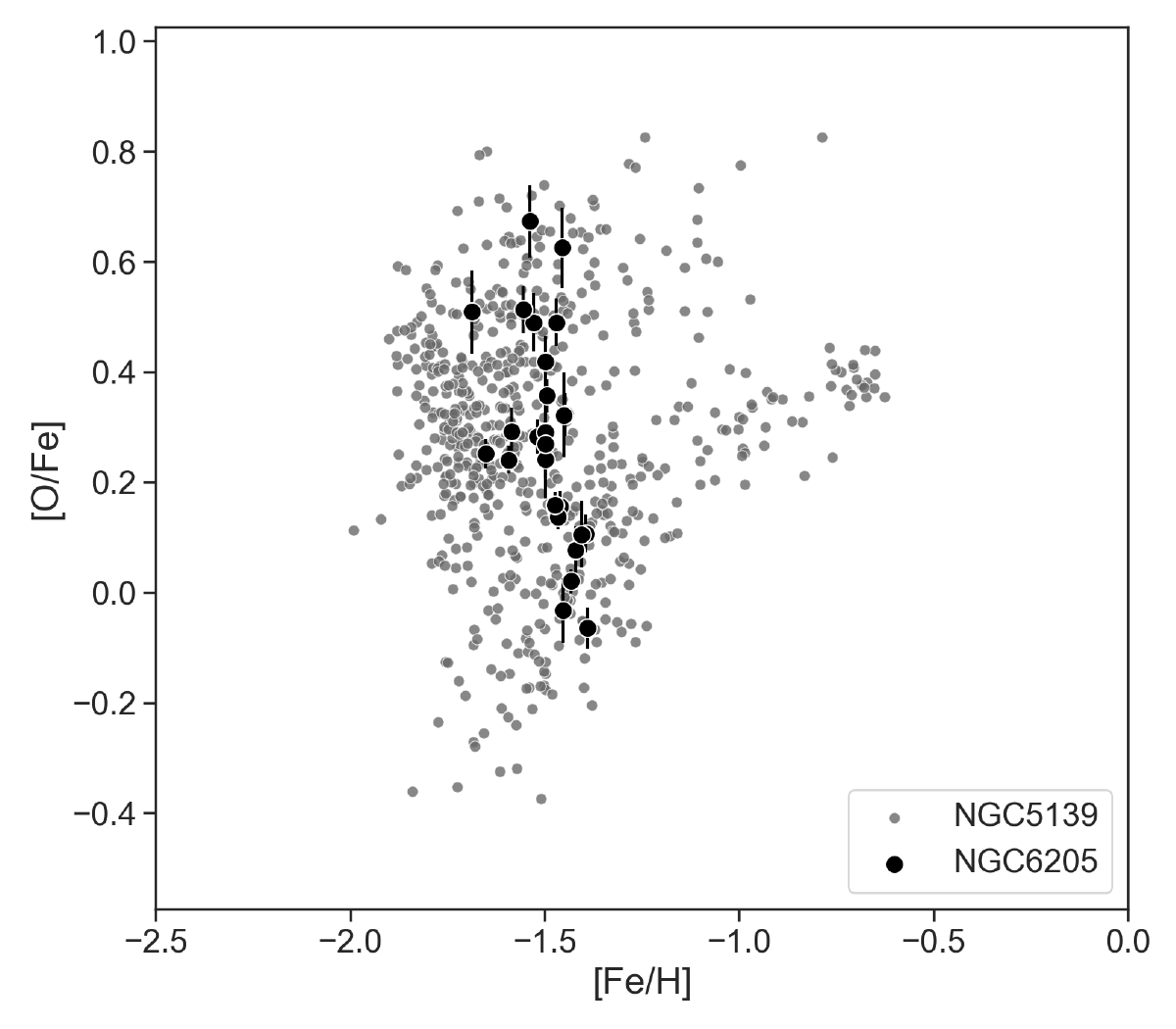}\hspace{-5pt}
\includegraphics[clip=true, trim = 3mm 0mm 0mm 2mm, width=0.68\columnwidth]{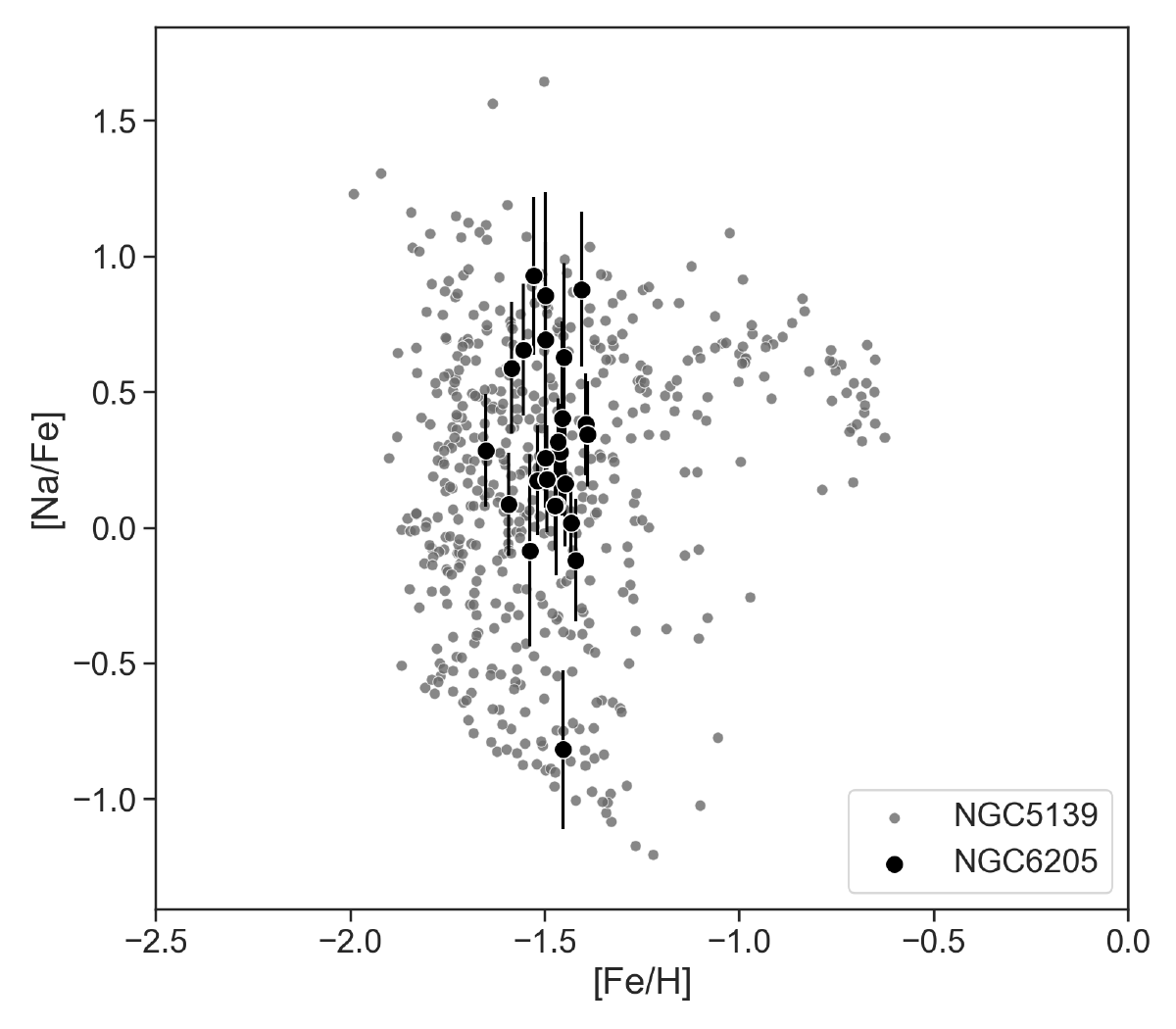}\hspace{-5pt}
\includegraphics[clip=true, trim = 3mm 0mm 0mm 2mm, width=0.68\columnwidth]{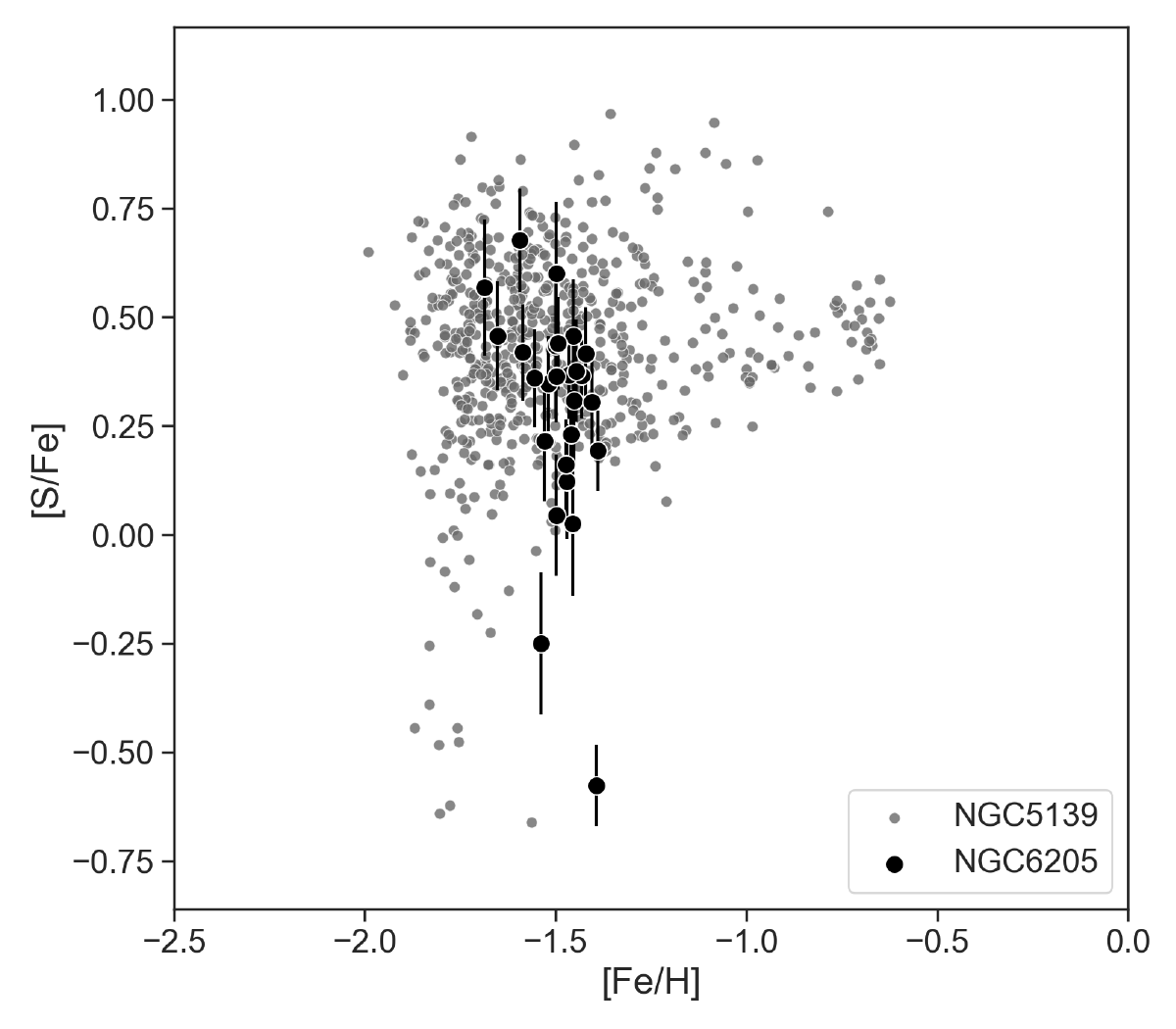}\hspace{-5pt}
\includegraphics[clip=true, trim = 3mm 0mm 0mm 2mm, width=0.68\columnwidth]{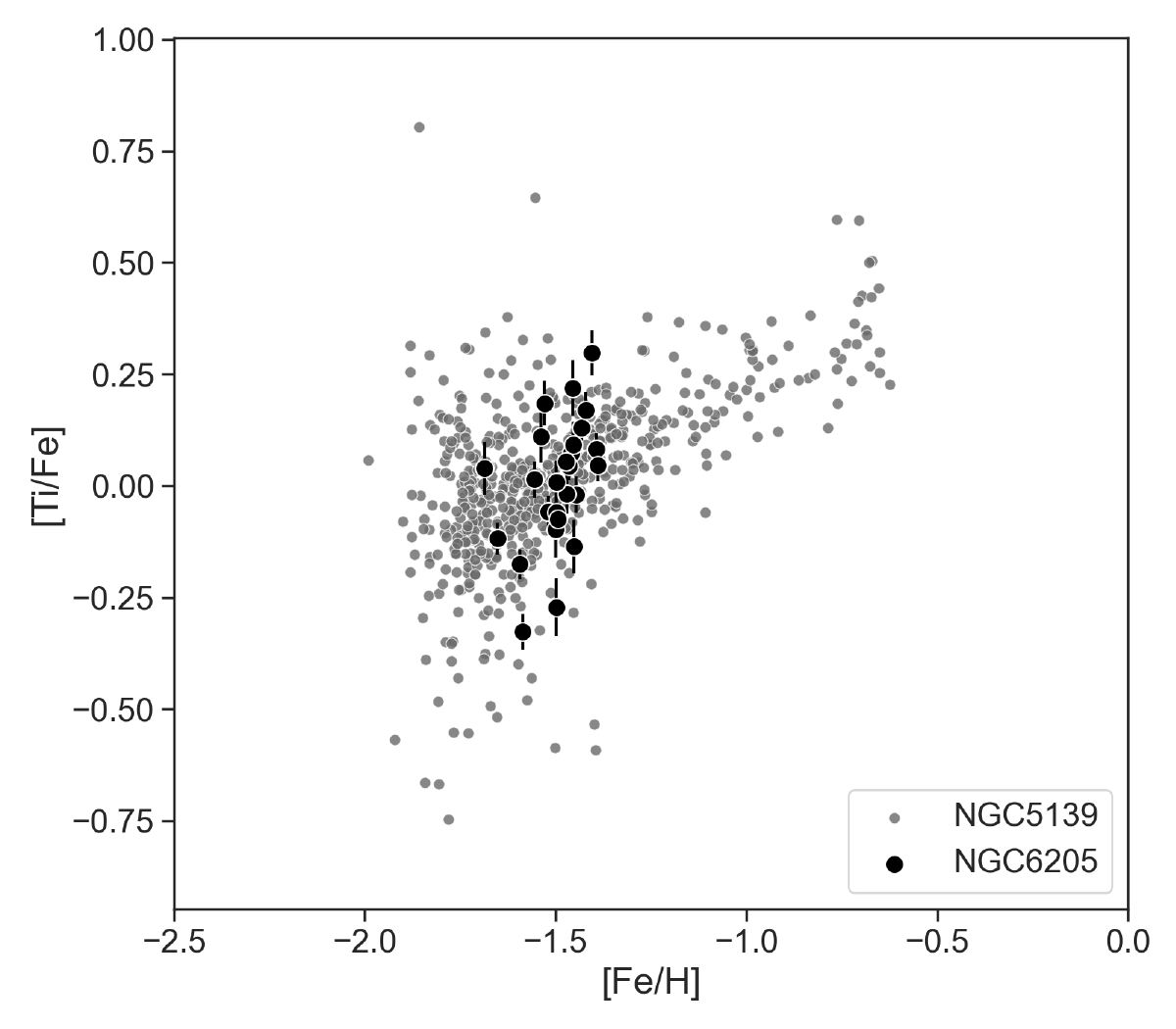}\hspace{-5pt}
\includegraphics[clip=true, trim = 3mm 0mm 0mm 2mm, width=0.68\columnwidth]{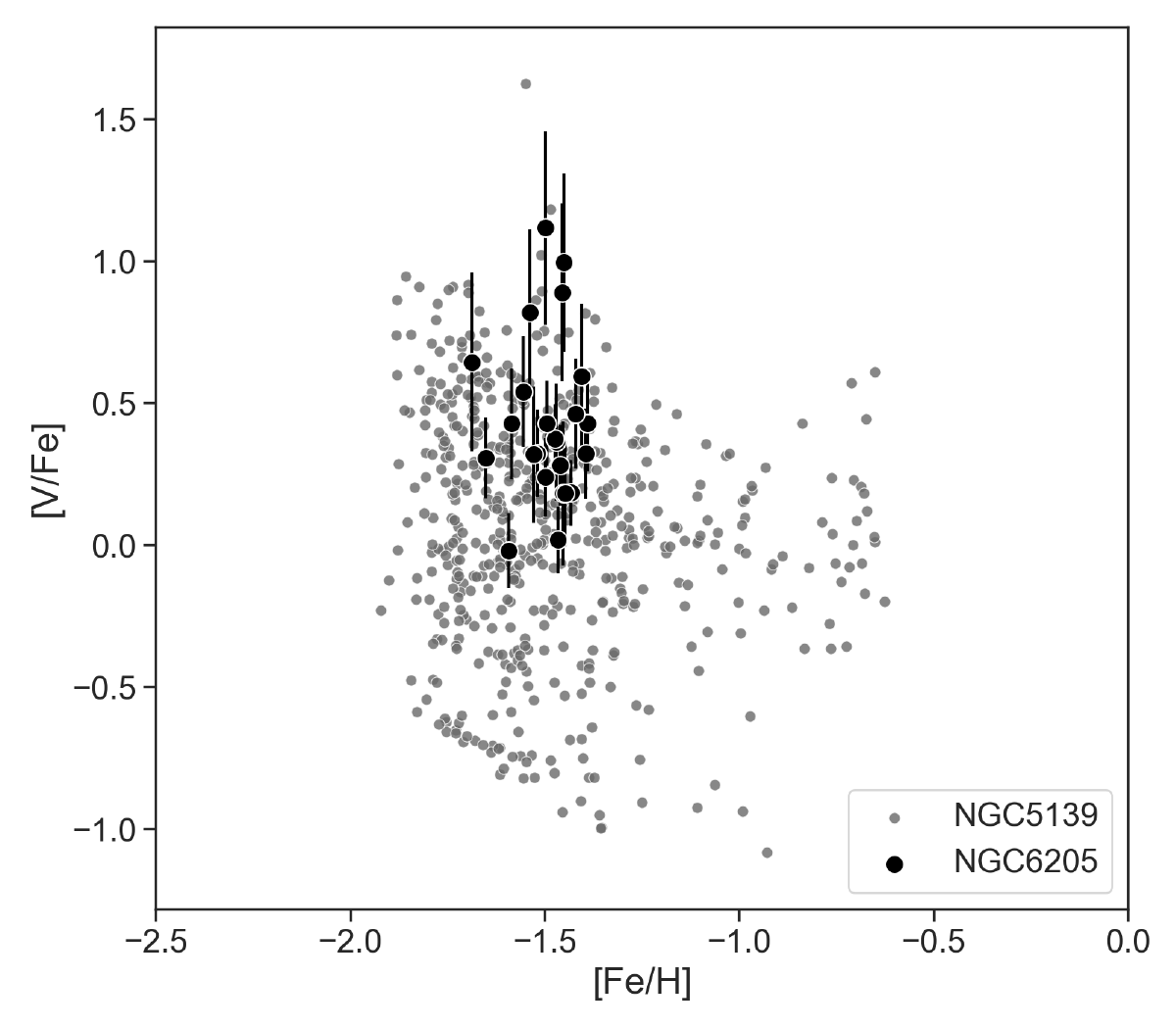}\hspace{-5pt}
\includegraphics[clip=true, trim = 3mm 0mm 0mm 2mm, width=0.68\columnwidth]{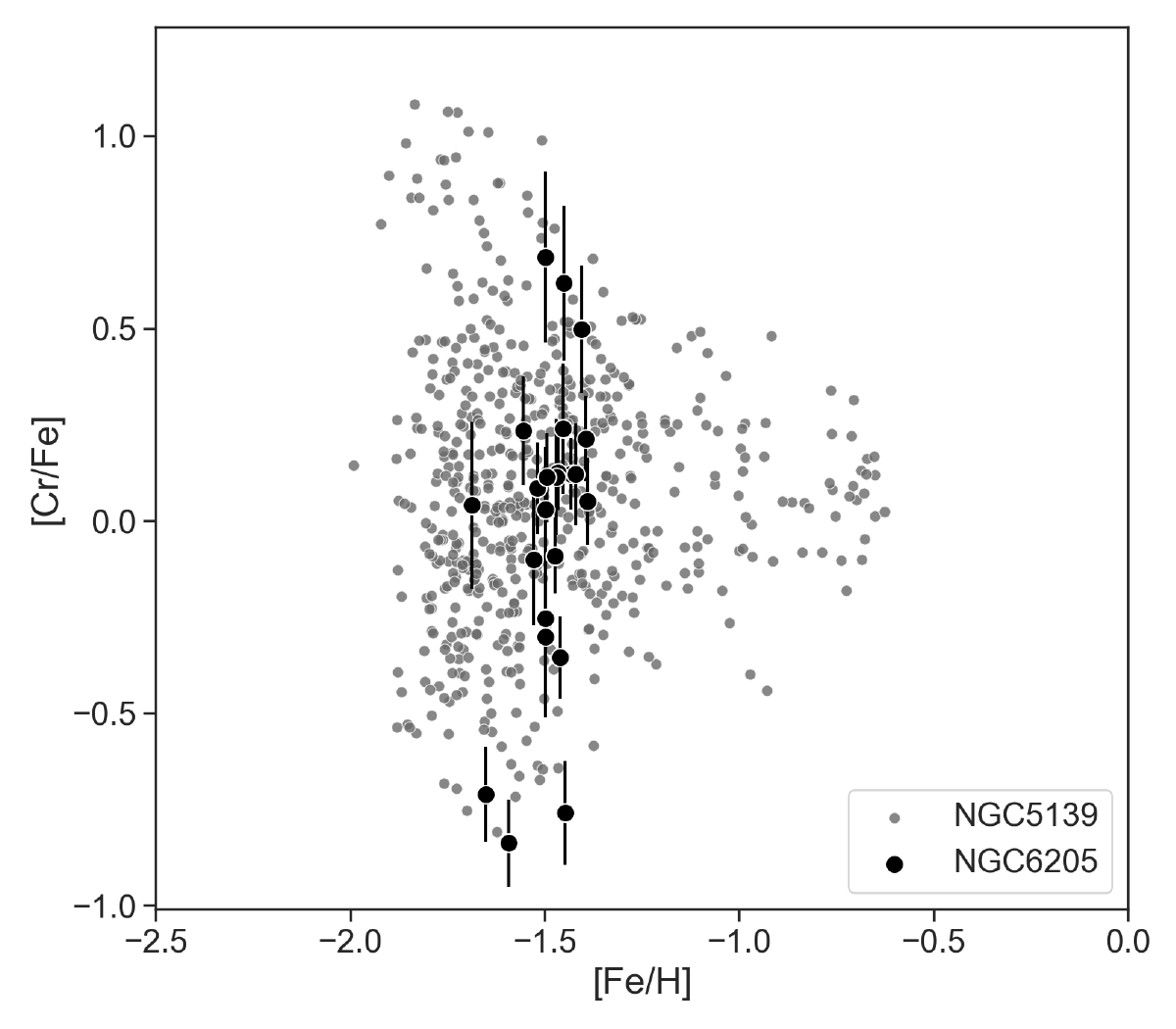}\hspace{-5pt}
\includegraphics[clip=true, trim = 3mm 0mm 0mm 2mm, width=0.68\columnwidth]{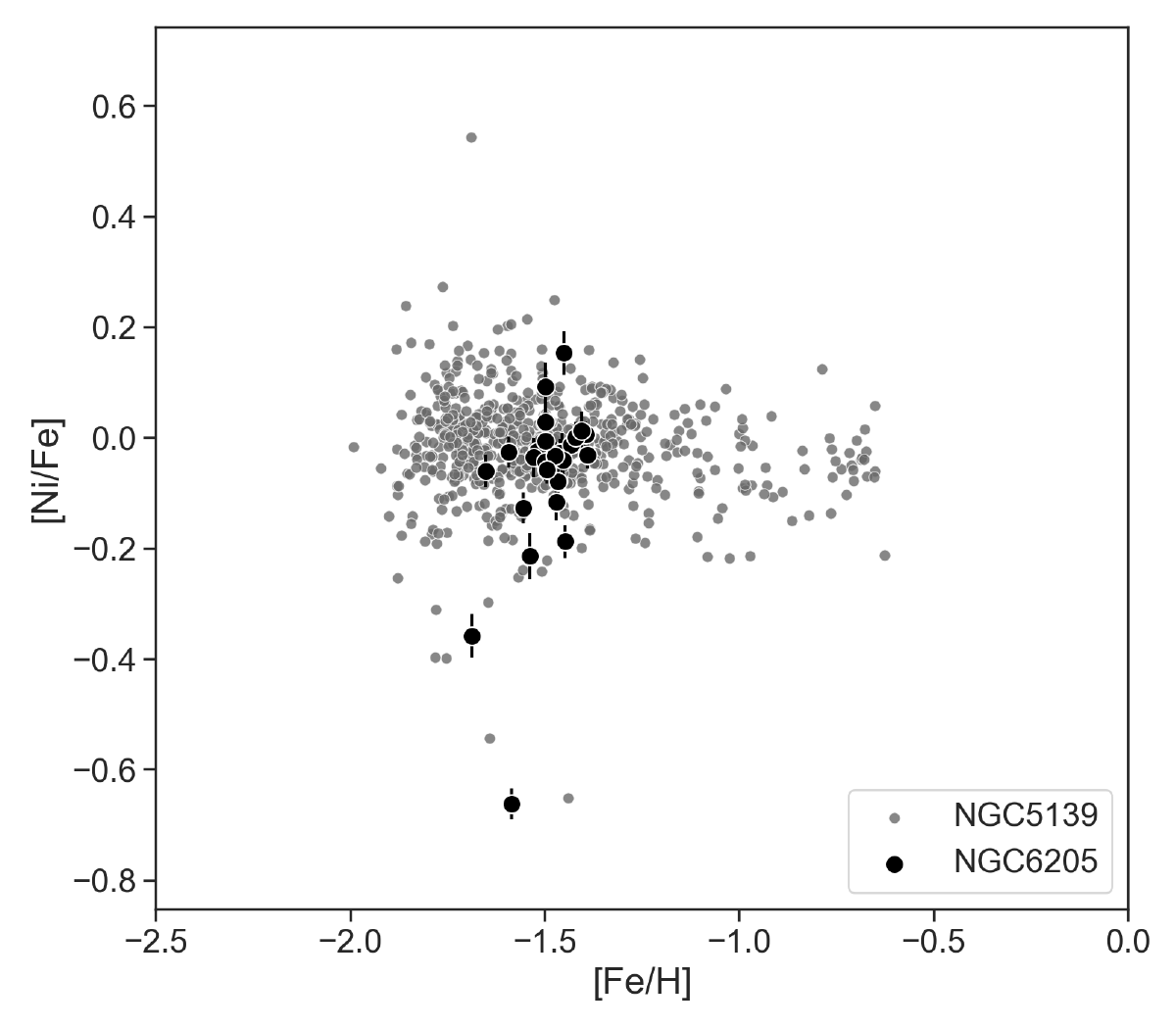}\hspace{-5pt}
\includegraphics[clip=true, trim = 3mm 0mm 0mm 2mm, width=0.68\columnwidth]{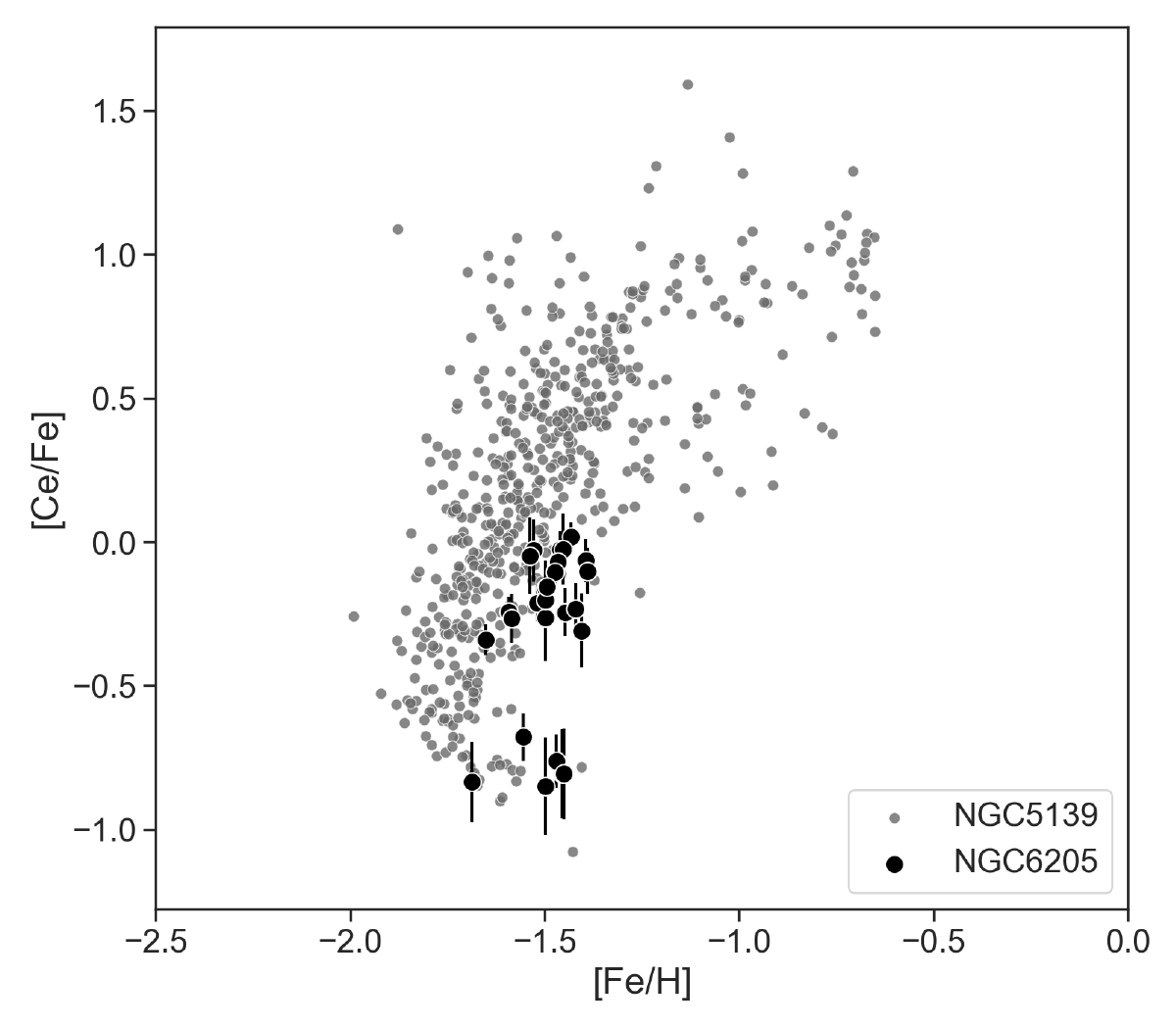}
  \caption{Same as Fig.~\ref{ngc6656_others} for NGC~6205.}
              \label{ngc6205_others}%
    \end{figure*}

\begin{figure*}[h!]
   \centering
\includegraphics[clip=true, trim = 3mm 0mm 0mm 2mm, width=0.68\columnwidth]{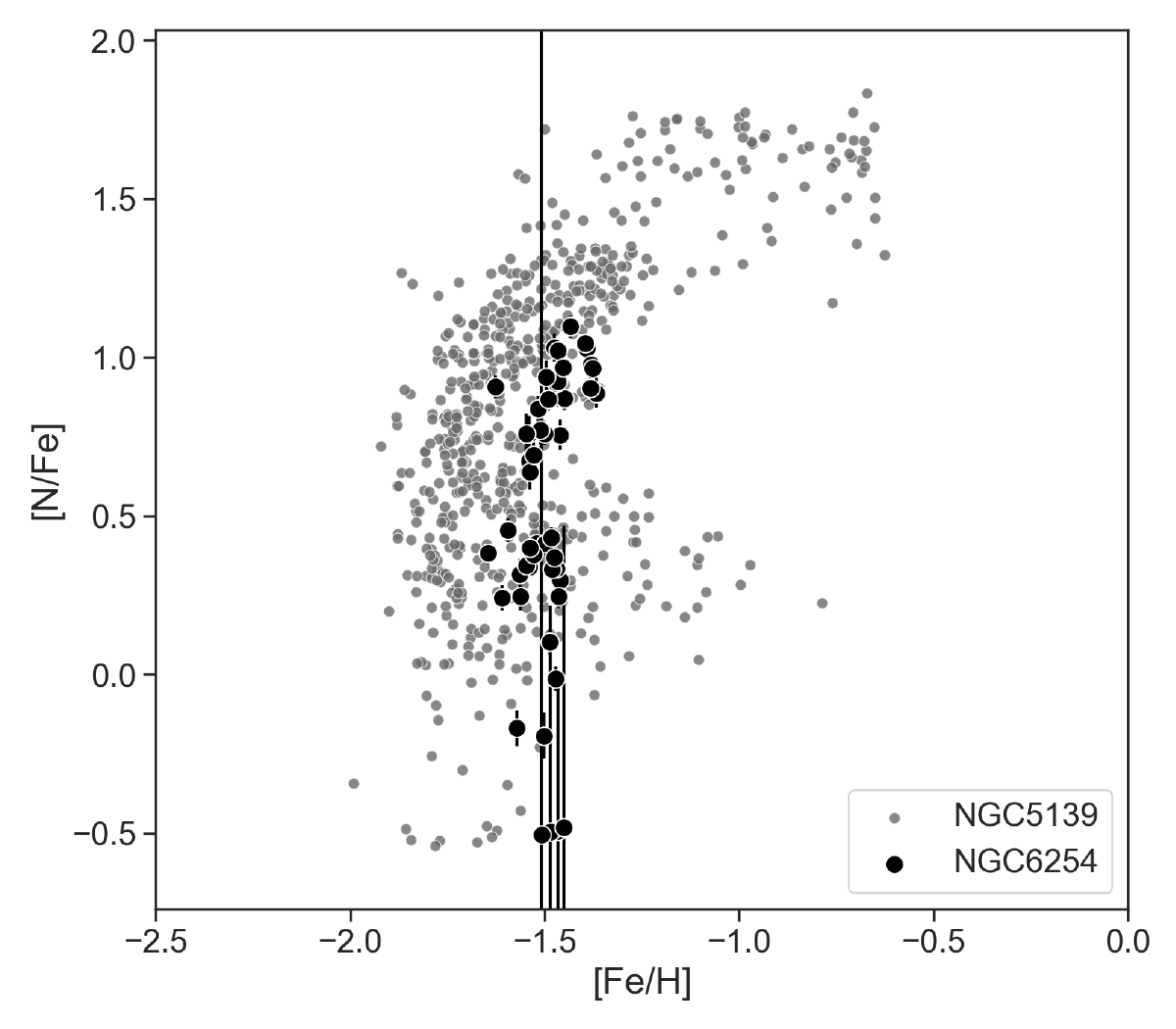}\hspace{-5pt}
\includegraphics[clip=true, trim = 3mm 0mm 0mm 2mm, width=0.68\columnwidth]{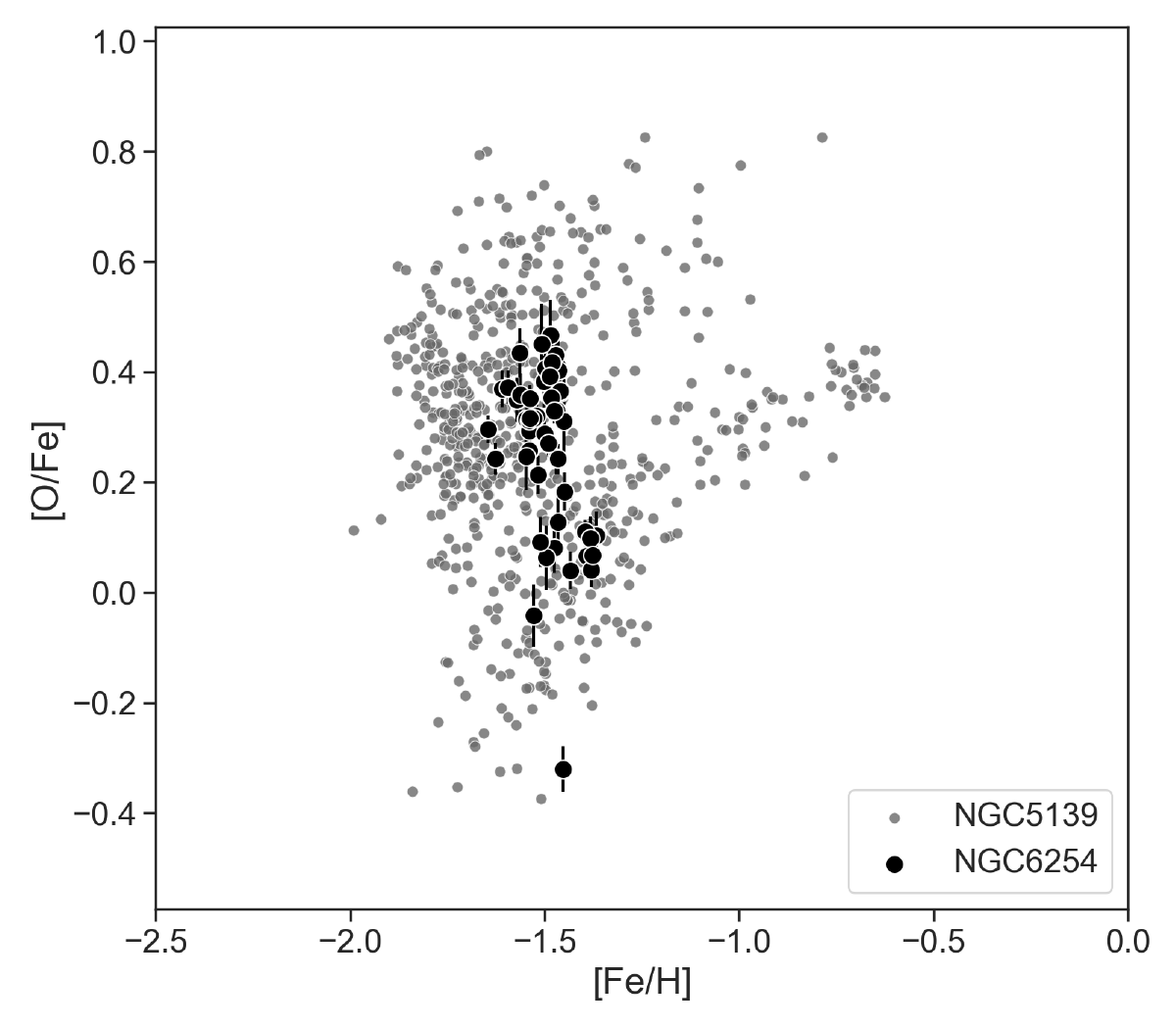}\hspace{-5pt}
\includegraphics[clip=true, trim = 3mm 0mm 0mm 2mm, width=0.68\columnwidth]{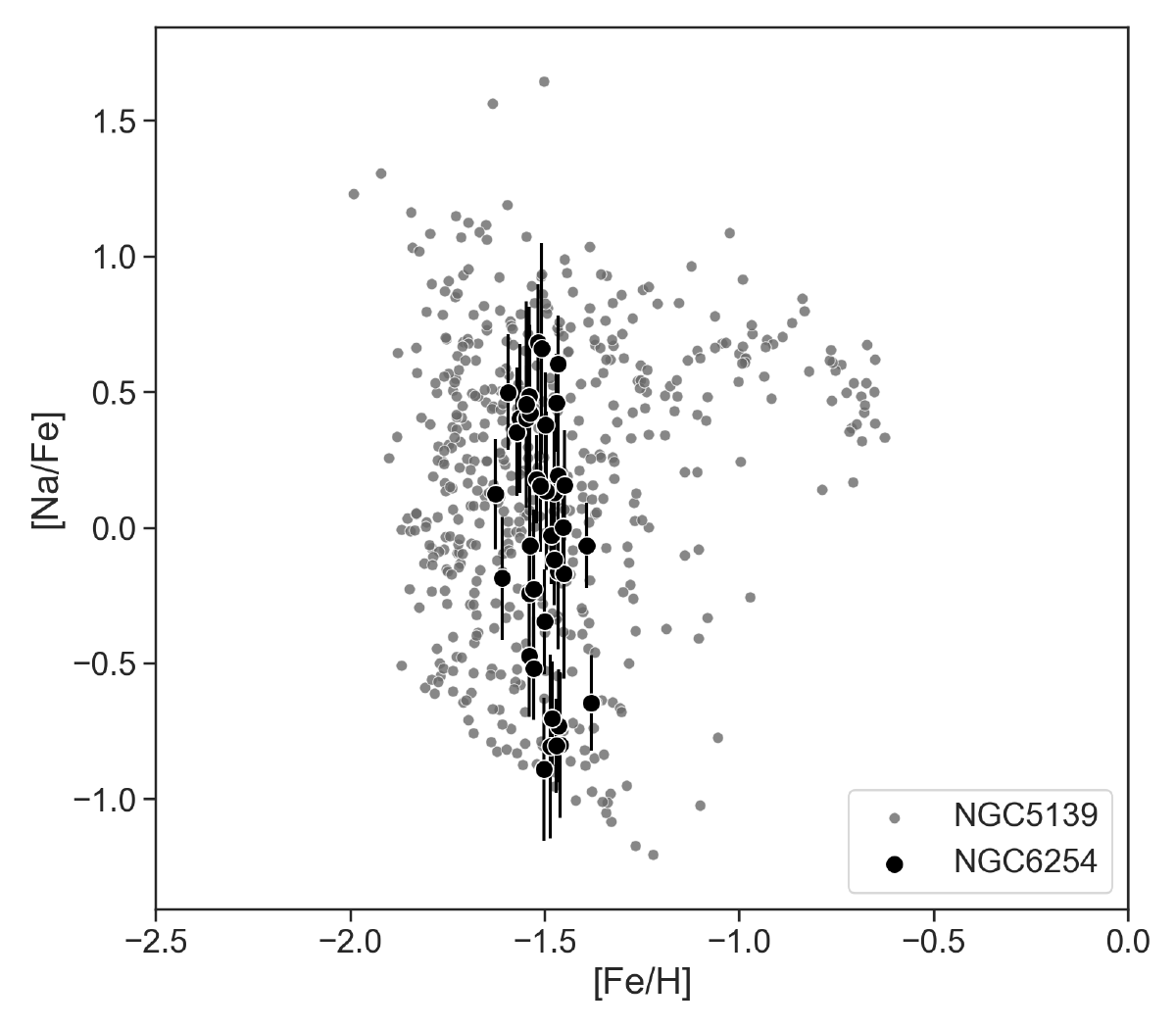}\hspace{-5pt}
\includegraphics[clip=true, trim = 3mm 0mm 0mm 2mm, width=0.68\columnwidth]{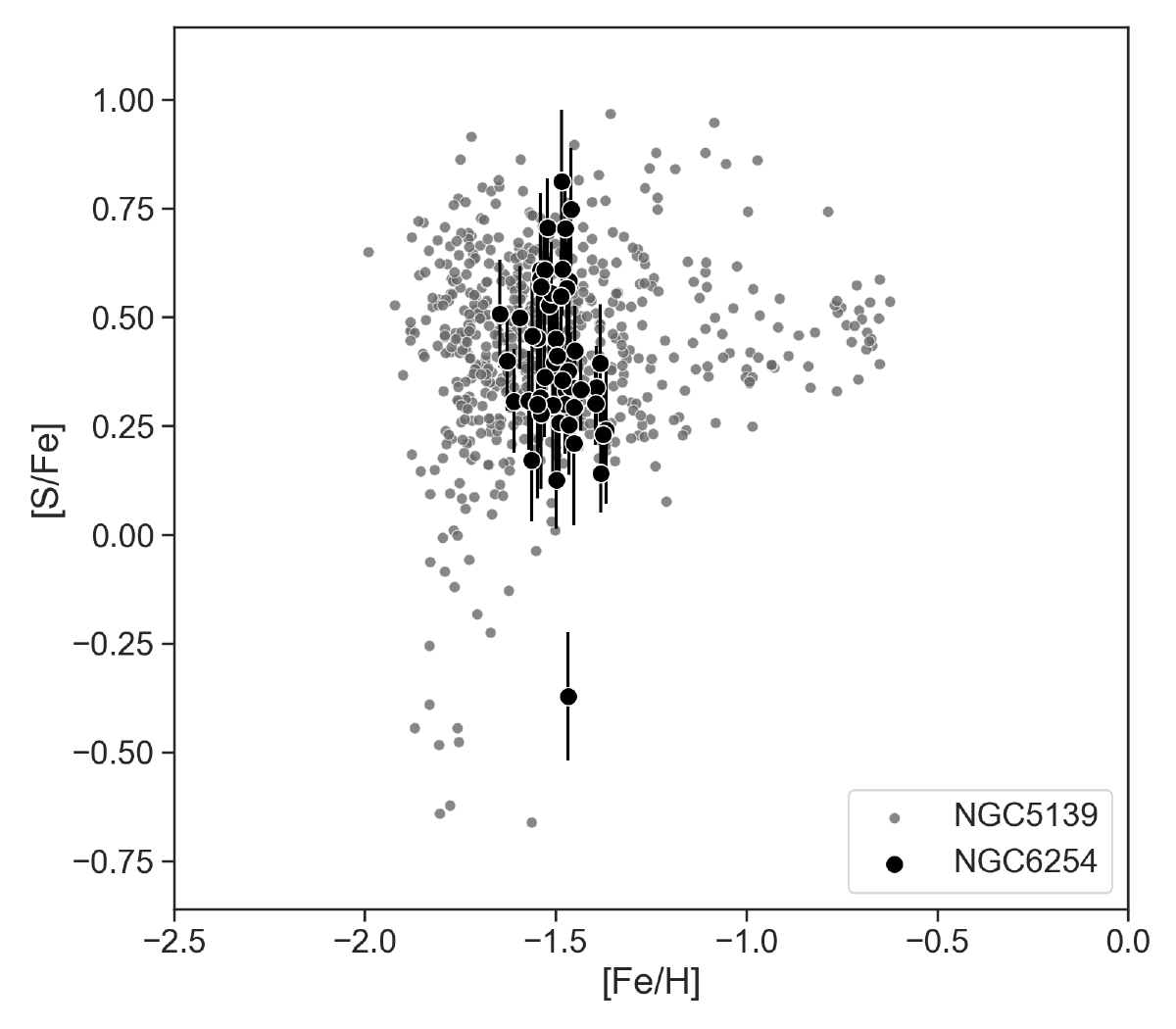}\hspace{-5pt}
\includegraphics[clip=true, trim = 3mm 0mm 0mm 2mm, width=0.68\columnwidth]{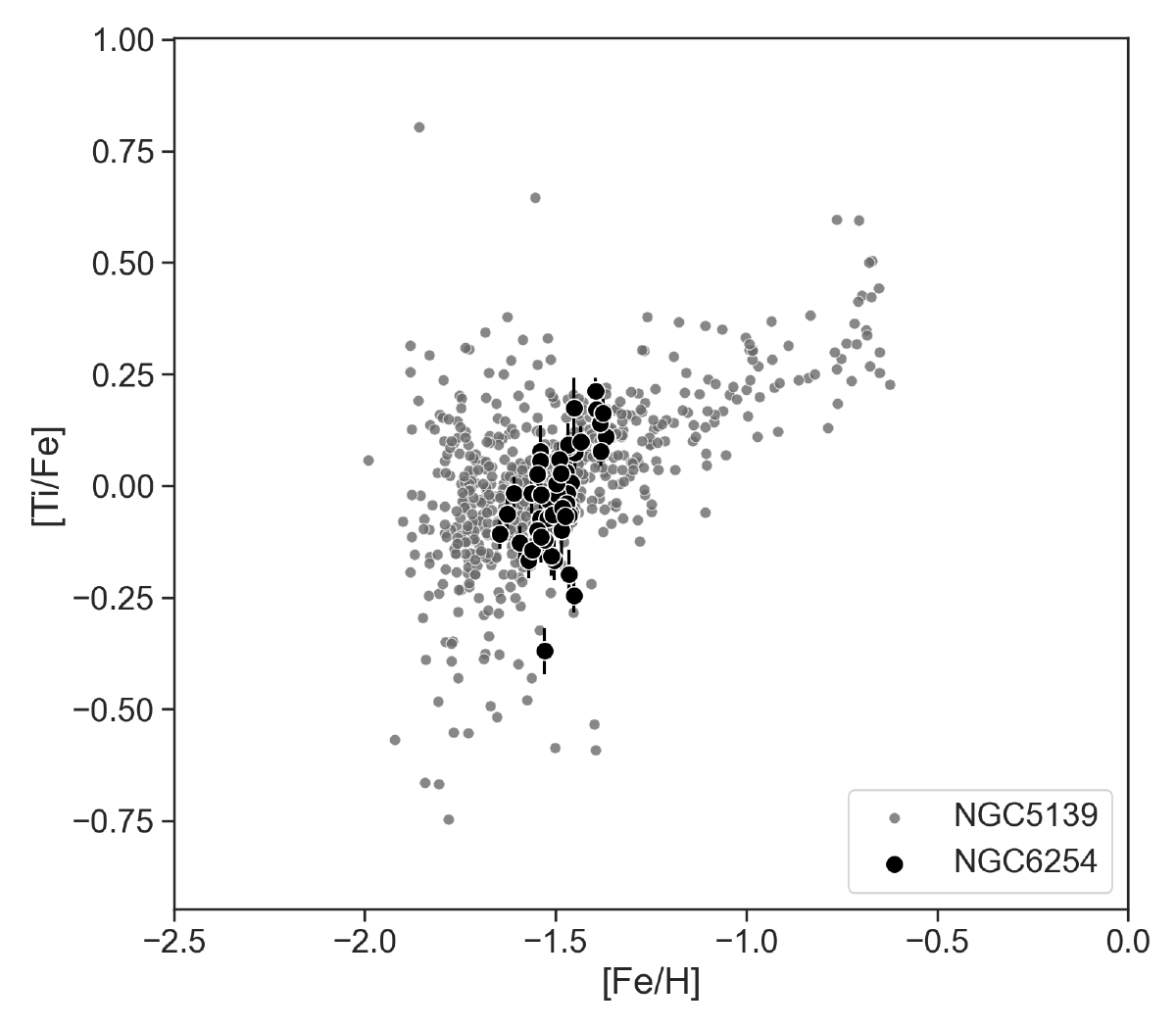}\hspace{-5pt}
\includegraphics[clip=true, trim = 3mm 0mm 0mm 2mm, width=0.68\columnwidth]{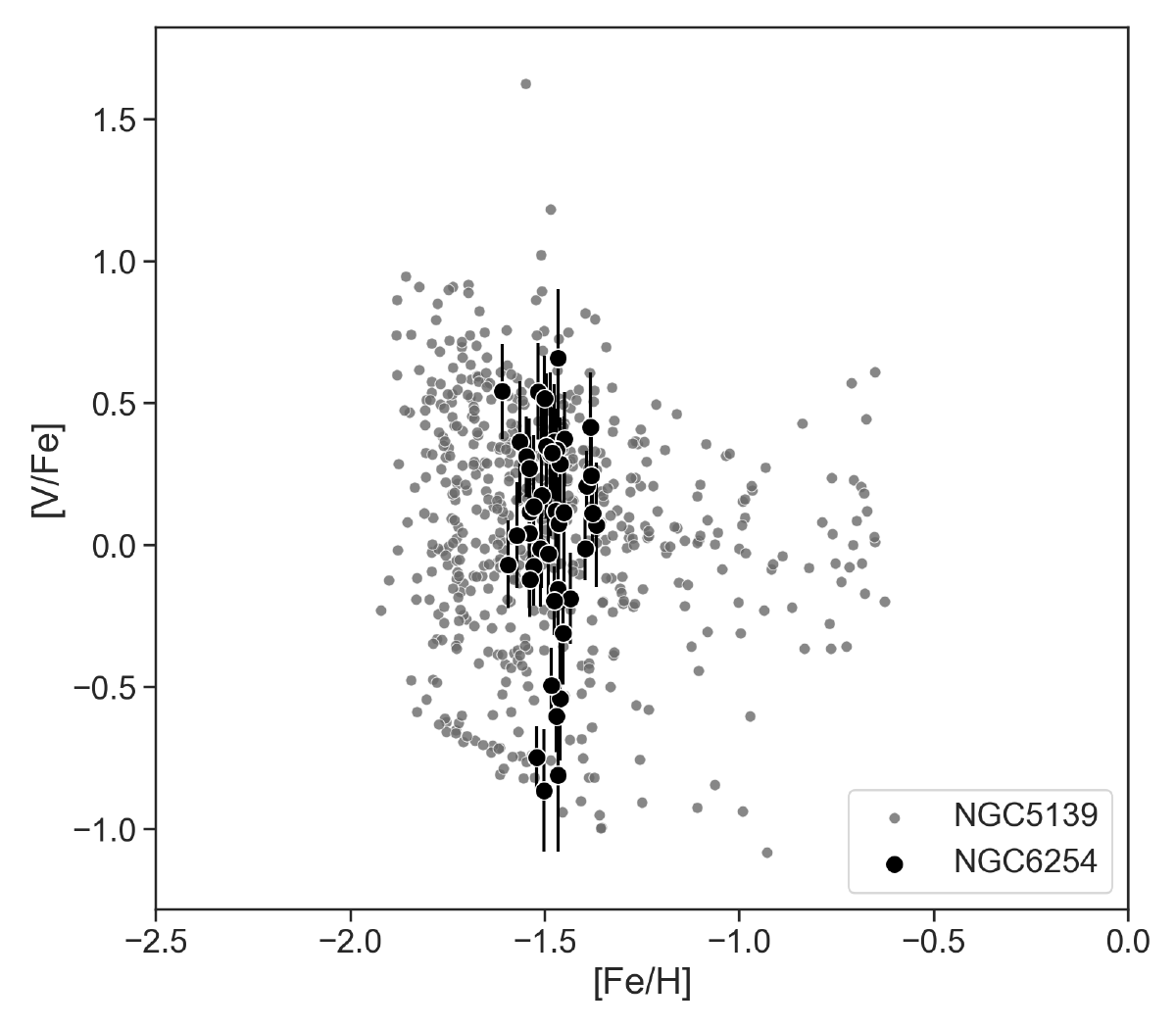}\hspace{-5pt}
\includegraphics[clip=true, trim = 3mm 0mm 0mm 2mm, width=0.68\columnwidth]{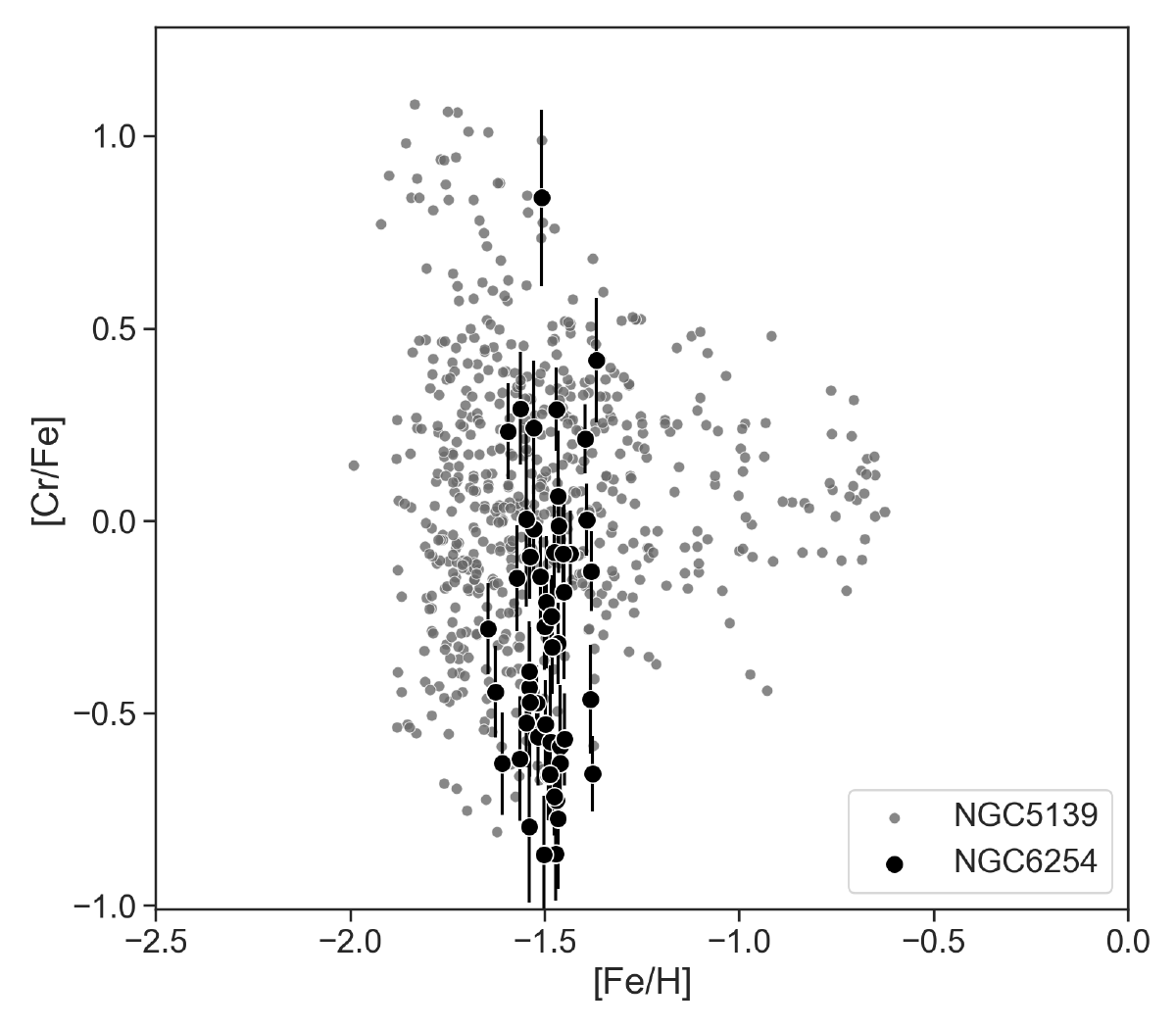}\hspace{-5pt}
\includegraphics[clip=true, trim = 3mm 0mm 0mm 2mm, width=0.68\columnwidth]{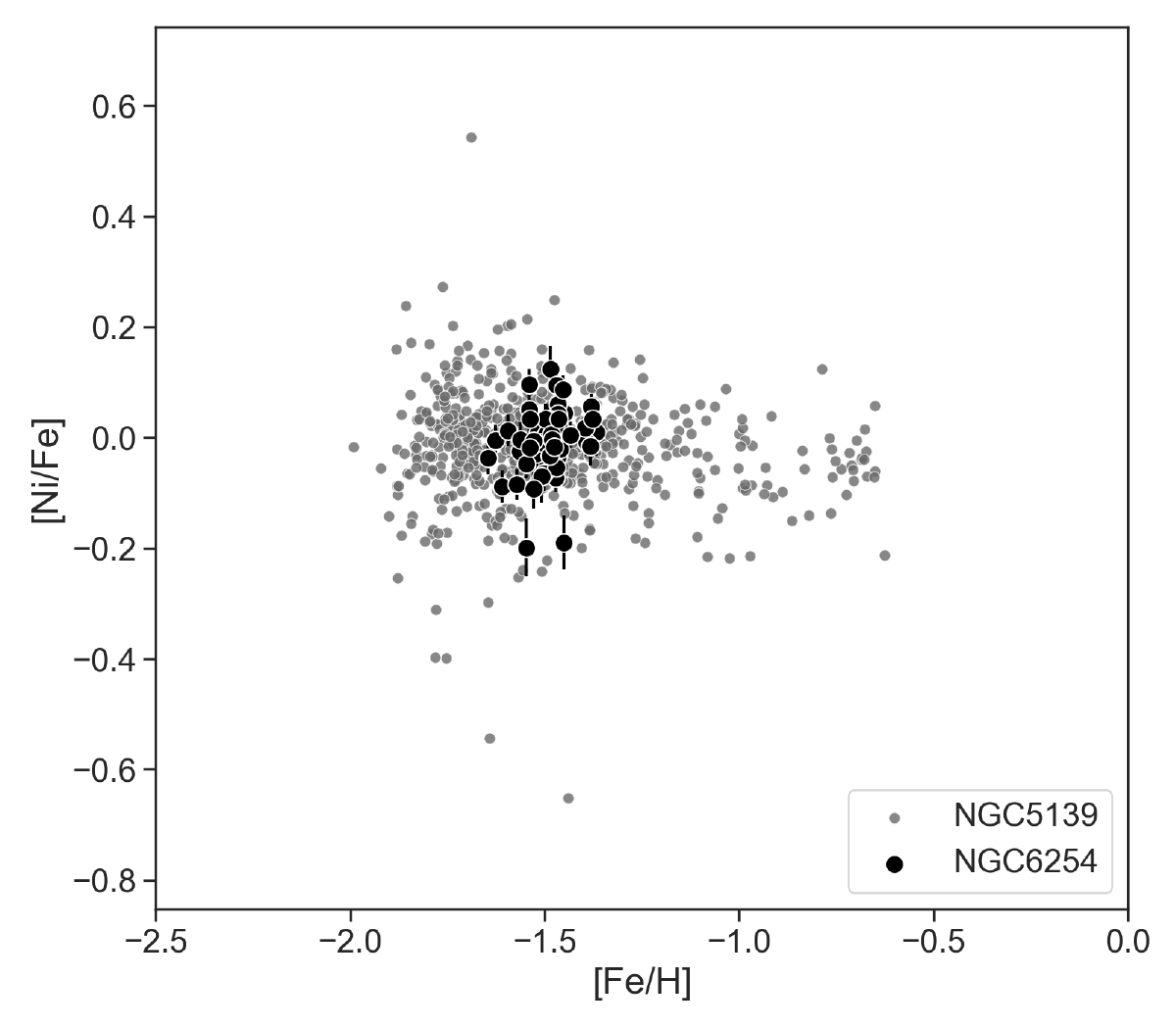}\hspace{-5pt}
\includegraphics[clip=true, trim = 3mm 0mm 0mm 2mm, width=0.68\columnwidth]{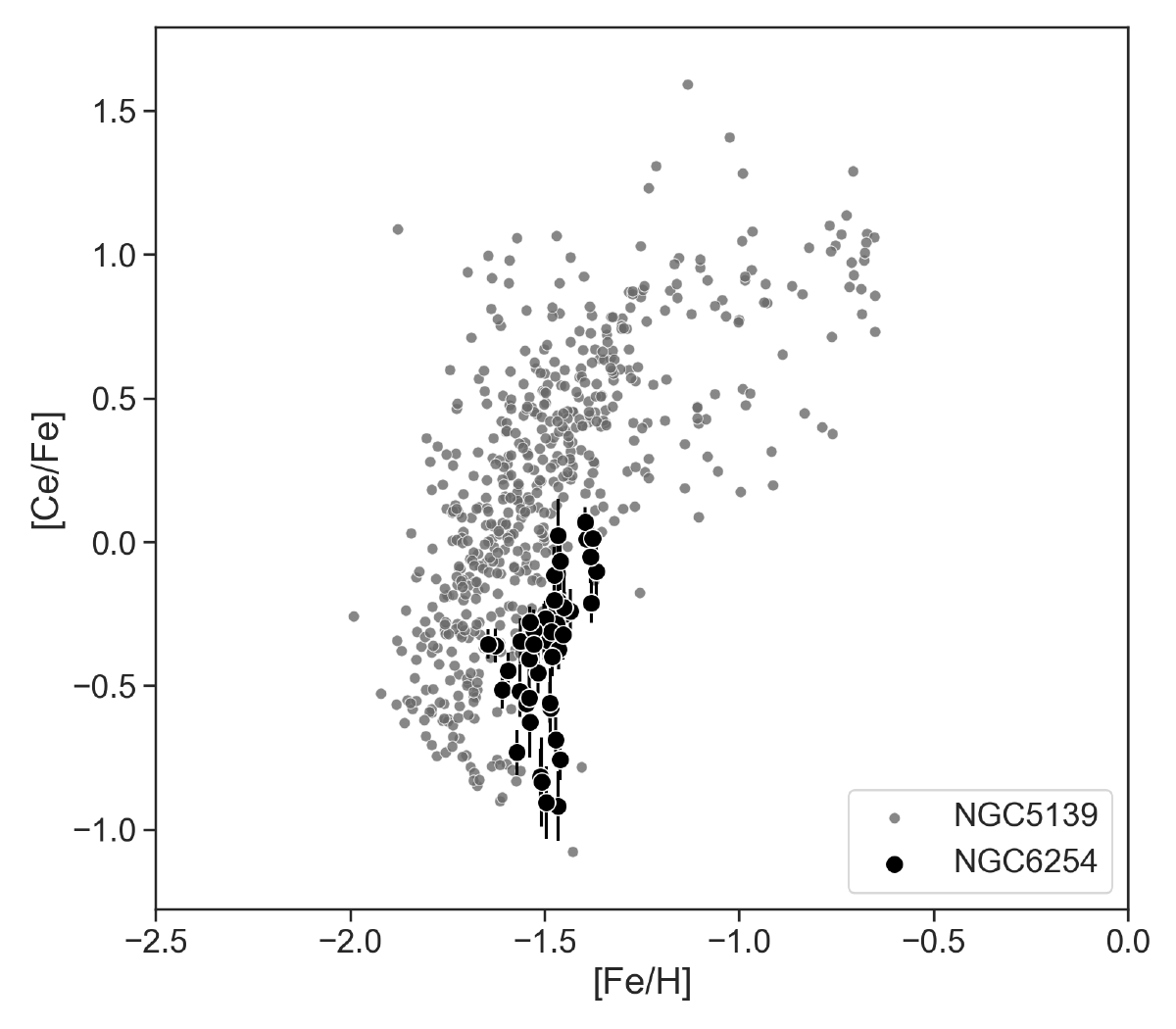}
  \caption{Same as Fig.~\ref{ngc6656_others} for NGC~6254.}
              \label{ngc6254_others}%
    \end{figure*}

Figures \ref{ngc6656_others},\ref{ngc6809_others}, \ref{ngc6273_others}, \ref{ngc6752_others}, \ref{ngc6205_others}, and \ref{ngc6254_others} show for each GC chemically compatible with $\omega$~Cen, the distribution in [X/Fe] versus [Fe/H] planes where the different X elements are the others provided by APOGEE DR17 but not used in the GMM, namely: N, O, Na, S, Ti, V, Cr, Ni, and Ce. Although these abundances have been excluded for the reasons given in Sec \ref{method}, it is nevertheless interesting to see at least qualitatively how they compare with $\omega$~Cen (also shown). In general, for all the GCs, despite the larger uncertainties, the overlap with $\omega$~Cen also clearly occurs in these other chemical spaces. 

The most striking case is that of NGC~6656 since, in addition to the fact that its superposition takes place in regions where $\omega$~Cen has its density peak, it also shows similar features even apart from the bulk of its stars. For instance:
\begin{itemize}
\item in both [N/Fe] and [S/Fe], two stars are detached from the others by being under-abundant ($\sim$ -0.5), thus populating the same region where we find a small group of $\omega$~Cen stars, also detached from the majority of the population;
\item for [Na/Fe] and [V/Fe] values lower than -0.5 both $\omega$~Cen and NGC~6656 seem to trace a kind of demarcation line that decreases as [Fe/H] increases, below which we find no stars (see also NGC~6273);
\item for [Ti/Fe] $\lesssim$ -0.5, we can see a sort of bifurcation into two decreasing patterns (each consisting of two stars) similar to those of $\omega$~Cen but also a star with a Ti-excess comparable with the one observed in two $\omega$~Cen stars;
\item a similar but perhaps more outstanding behaviour is what we observe for [Ce/Fe] $\lesssim$ -0.7, where both NGC~6656 and $\omega$~Cen stars shape the same small declining tail towards higher [Fe/H] (also seen in NGC~6205 and NGC~6254).
\end{itemize}
Analogous peculiarities are also found in the other clusters chemically compatible with $\omega$~Cen.
For instance, by looking at the metal-poor GCs defined in Sect.~\ref{results}, we can see that NGC~6656 and NGC~6273 span the exact same range of [Cr/Fe] and [V/Fe] as $\omega$~Cen at equal [Fe/H], and even NGC~6809, despite its lower statistic compared to the other two, presents V-depleted and Cr-depleted stars like $\omega$~Cen. 
Regarding the metal-rich GCs, each has two Ni-poor stars at [Fe/H] $\sim$ -1.5 and [N/Fe] $\sim$ -0.2 that overlap with the similar Ni-deficient sequence in $\omega$~Cen; NGC~6205 even seems to follow this sequence with 2 stars with [Ni/Fe] < -0.3. Both NGC~6752 and NGC~6254 own a group of stars with [Na/Fe] and [V/Fe] values that fall in correspondence with the lowest boundary sequence drawn by $\omega$~Cen's stars. NGC~6205, on the other hand, features four V-enhanced stars spanning the exact same values as the V-rich stars of $\omega$~Cen in the corresponding [Fe/H] range ([Fe/H]$\sim$ -1.5).
In contrast, a clear under-abundance seen in this GC compatible with that observed in $\omega$~Cen is represented by three stars at [Cr/Fe]< -0.5. Finally, both NGC~6205 and NGC~6254 show a decreasing trend of [O/Fe] with increasing [Fe/H] that resembles that observed for the more metal-poor stars in $\omega$~Cen.\\

\clearpage
\section{On the intrinsic dispersions of abundance ratios for the Galactic GCs chemically compatible with $\omega$~Cen}
\label{spreads}    
In Sect.~\ref{mdf}, we have seen that all the Galactic GCs for which we find chemical compatibility with $\omega$~Cen show significant intrinsic dispersions, unless the ASPCAP uncertainties on [Fe/H] had been underestimated of a factor between 4 and 5, at the metallicities of these clusters. In Table~\ref{XFE_spreads} we report a similar analysis for all the other abundance ratios that we have used for building the GMM of $\omega$~Cen. As in Sect.~\ref{mdf}, for each cluster, we estimate the total dispersion of the generic [X/Fe] abundance,  $\sigma_{\rm{[X/Fe], tot}}$, the median of [X/Fe] ASPCAP uncertainties for stars in the cluster, $\epsilon_{\rm{[X/Fe]}}$, and the corresponding  $\sigma_{\rm{[X/Fe], int}}$ defined as $\sqrt{\sigma_{\rm{[X/Fe], tot}}^2 - \epsilon_{\rm{[X/Fe]}}^2}$.\\ In general we notice that for all these GCs, the internal dispersions range between 0.04~dex and 0.46~dex, the smallest dispersions being generally found for [Si/Fe], and the largest ones for [Al/Fe]. While the [Si/Fe] dispersions are probably not significant (it is sufficient for ASPCAP uncertainties on this abundance ratio to have been underestimated by 30\% to 50\%, see parameter $A$ in Table~\ref{XFE_spreads}, besides [Al/Fe], which is known to show an extended range of values in Galactic GCs, other abundance ratios seem to be characterised by not null internal dispersions. This is the case of [C/Fe], which shows intrinsic dispersions between 0.2 and 0.3~dex for all these GCs, except for NGC~6254. Note that these dispersions would be not statistically significant if the ASPCAP uncertainties on [C/Fe] had been underestimated of a factor of 3 at least. [Ca/Fe] and [K/Fe] are also elements for which intrinsic dispersions are found for these GCs, unless the corresponding ASPCAP uncertainties had been underestimated by a factor between 2 and 3. Note that such underestimations would lead to a not statistically significant dispersion in [Ca/Fe] and [K/Fe]  also for NGC~5139.
\begin{table}
\centering
\caption{Total and intrinsic [X/Fe] dispersions for the metal-poor, the metal-rich globular clusters chemically compatible with $\omega$~Centauri, and $\omega$~Cen.}\label{XFE_spreads}
\resizebox{1.\columnwidth}{!}{    
 \begin{tabular}{llrrrrr}
\toprule   
     GCname & [X/Fe] &  nstars &   $\sigma_{\rm{[X/Fe], tot}}$ &   $\epsilon_{\rm{[X/Fe]}}$  &  $\sigma_{\rm{[X/Fe], int}}$ &  A \\
\midrule

NGC6656 &   [Mg/Fe] &     230 &      0.09 & 0.03 &      0.08 &   2.04 \\
NGC6656 &   [Al/Fe] &     230 &      0.40 & 0.03 &      0.39 &   8.55 \\
NGC6656 &   [C/Fe] &     211 &      0.32 & 0.06 &      0.32 &   3.75 \\
NGC6656 &   [Si/Fe] &     230 &      0.06 & 0.03 &      0.05 &   1.36 \\
NGC6656 &   [Ca/Fe] &     219 &      0.17 & 0.05 &      0.17 &   2.54 \\
NGC6656 &   [Mn/Fe] &      84 &      0.22 & 0.05 &      0.21 &   3.25 \\
NGC6656 &    [K/Fe] &     201 &      0.25 & 0.09 &      0.24 &   2.06 \\
\vspace{1\baselineskip}\\

NGC6809 &   [Mg/Fe] &      58 &      0.10 & 0.03 &      0.09 &   2.34 \\
NGC6809 &   [Al/Fe] &      58 &      0.40 & 0.03 &      0.40 &  10.37 \\
NGC6809 &   [C/Fe] &      51 &      0.27 & 0.05 &      0.26 &   3.79 \\
NGC6809 &   [Si/Fe] &      58 &      0.05 & 0.03 &      0.04 &   1.27 \\
NGC6809 &   [Ca/Fe] &      52 &      0.17 & 0.04 &      0.17 &   2.92 \\
NGC6809 &   [Mn/Fe] &      24 &      0.13 & 0.04 &      0.12 &   2.15 \\
NGC6809 &    [K/Fe] &      57 &      0.24 & 0.08 &      0.23 &   2.06 \\
\vspace{1\baselineskip}\\

NGC6273 &   [Mg/Fe] &      56 &      0.12 & 0.04 &      0.12 &   2.38 \\
NGC6273 &   [Al/Fe] &      56 &      0.42 & 0.04 &      0.41 &   6.90 \\
NGC6273 &   [C/Fe] &      56 &      0.29 & 0.06 &      0.28 &   3.50 \\
NGC6273 &   [Si/Fe] &      56 &      0.08 & 0.04 &      0.07 &   1.47 \\
NGC6273 &   [Ca/Fe] &      54 &      0.21 & 0.06 &      0.20 &   2.57 \\
NGC6273 &   [Mn/Fe] &      46 &      0.15 & 0.06 &      0.14 &   1.97 \\
NGC6273 &    [K/Fe] &      51 &      0.29 & 0.10 &      0.27 &   2.08 \\
\vspace{1\baselineskip}\\

NGC6752 &   [Mg/Fe] &     117 &      0.11 & 0.03 &      0.11 &   3.10 \\
NGC6752 &   [Al/Fe] &     117 &      0.44 & 0.03 &      0.44 &  11.31 \\
NGC6752 &   [C/Fe] &     117 &      0.22 & 0.05 &      0.22 &   3.08 \\
NGC6752 &   [Si/Fe] &     117 &      0.05 & 0.03 &      0.04 &   1.38 \\
NGC6752 &   [Ca/Fe] &     115 &      0.13 & 0.04 &      0.12 &   2.29 \\
NGC6752 &   [Mn/Fe] &      89 &      0.16 & 0.04 &      0.16 &   2.95 \\
NGC6752 &    [K/Fe] &     109 &      0.27 & 0.08 &      0.26 &   2.58 \\
\vspace{1\baselineskip}\\

NGC6205 &   [Mg/Fe] &      34 &      0.11 & 0.03 &      0.11 &   3.04 \\
NGC6205 &   [Al/Fe] &      34 &      0.44 & 0.03 &      0.44 &  10.70 \\
NGC6205 &   [C/Fe] &      34 &      0.26 & 0.05 &      0.26 &   3.91 \\
NGC6205 &   [Si/Fe] &      34 &      0.06 & 0.03 &      0.06 &   1.62 \\
NGC6205 &   [Ca/Fe] &      34 &      0.11 & 0.04 &      0.11 &   2.02 \\
NGC6205 &   [Mn/Fe] &      26 &      0.13 & 0.04 &      0.12 &   2.33 \\
NGC6205 &    [K/Fe] &      34 &      0.16 & 0.08 &      0.14 &   1.49 \\
\vspace{1\baselineskip}\\

NGC6254 &   [Mg/Fe] &      58 &      0.12 & 0.02 &      0.11 &   3.29 \\
NGC6254 &   [Al/Fe] &      58 &      0.46 & 0.03 &      0.46 &  12.49 \\
NGC6254 &    [C/Fe] &      58 &      0.17 & 0.04 &      0.16 &   3.05 \\
NGC6254 &   [Si/Fe] &      58 &      0.05 & 0.02 &      0.05 &   1.50 \\
NGC6254 &   [Ca/Fe] &      57 &      0.10 & 0.04 &      0.09 &   1.96 \\
NGC6254 &   [Mn/Fe] &      56 &      0.13 & 0.04 &      0.13 &   2.47 \\
NGC6254 &    [K/Fe] &      53 &      0.23 & 0.07 &      0.22 &   2.28 \\
\vspace{1\baselineskip}\\

NGC5139 &   [Mg/Fe] &    1175 &      0.19 & 0.03 &      0.19 &   4.59 \\
NGC5139 &   [Al/Fe] &    1195 &      0.51 & 0.03 &      0.51 &  11.47 \\
NGC5139 &   [C/Fe] &    1129 &      0.32 & 0.05 &      0.31 &   4.32 \\
NGC5139 &   [Si/Fe] &    1201 &      0.08 & 0.03 &      0.08 &   1.94 \\
NGC5139 &   [Ca/Fe] &    1158 &      0.17 & 0.04 &      0.17 &   2.79 \\
NGC5139 &   [Mn/Fe] &     721 &      0.22 & 0.04 &      0.21 &   3.56 \\
NGC5139 &    [K/Fe] &    1071 &      0.28 & 0.08 &      0.27 &   2.42 \\
\bottomrule
\end{tabular}}\tablefoot{For each cluster, we also report the median of the [X/Fe] uncertainties of its stars, $\epsilon_{\rm{[X/Fe]}}$, and the factor $A$ by which these uncertainties would have to be underestimated for the intrinsic dispersions not to be statistically significant.}
\end{table}

\clearpage
\section{On the chemical compatibility of $\omega$~Cen with the most massive satellites of the Milky Way}\label{SatvsoCen}
\begin{figure*}\centering
\hspace{-20pt}\includegraphics[clip=true, trim = 3mm 0mm 0mm 3mm, width=0.7\columnwidth]{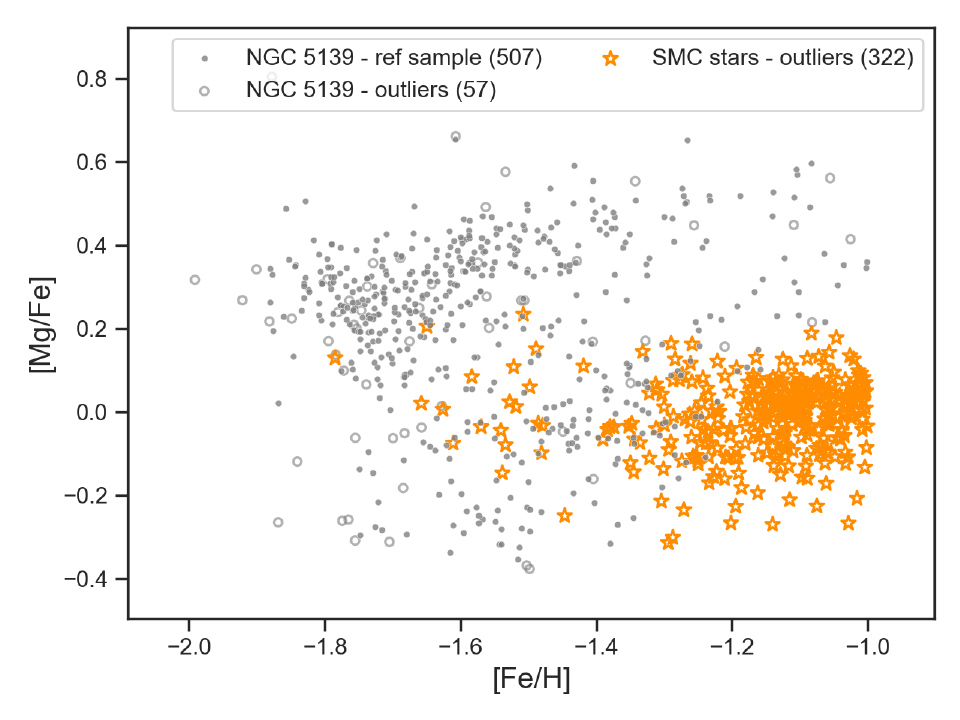}
\includegraphics[clip=true, trim = 3mm 0mm 0mm 3mm, width=0.7\columnwidth]{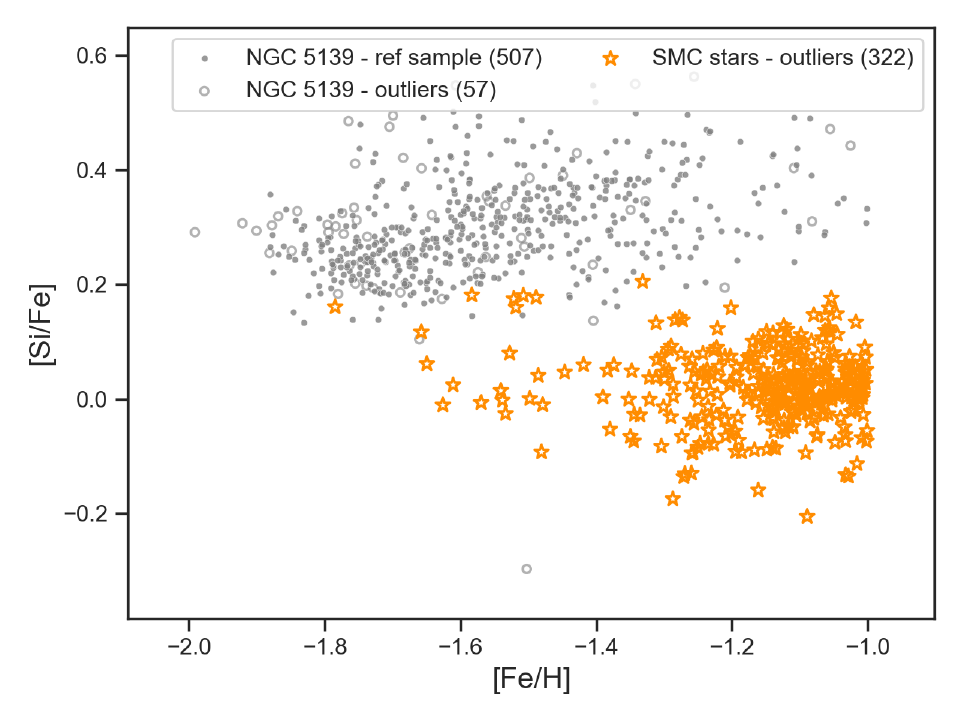}
\includegraphics[clip=true, trim = 3mm 0mm 0mm 3mm, width=0.7\columnwidth]{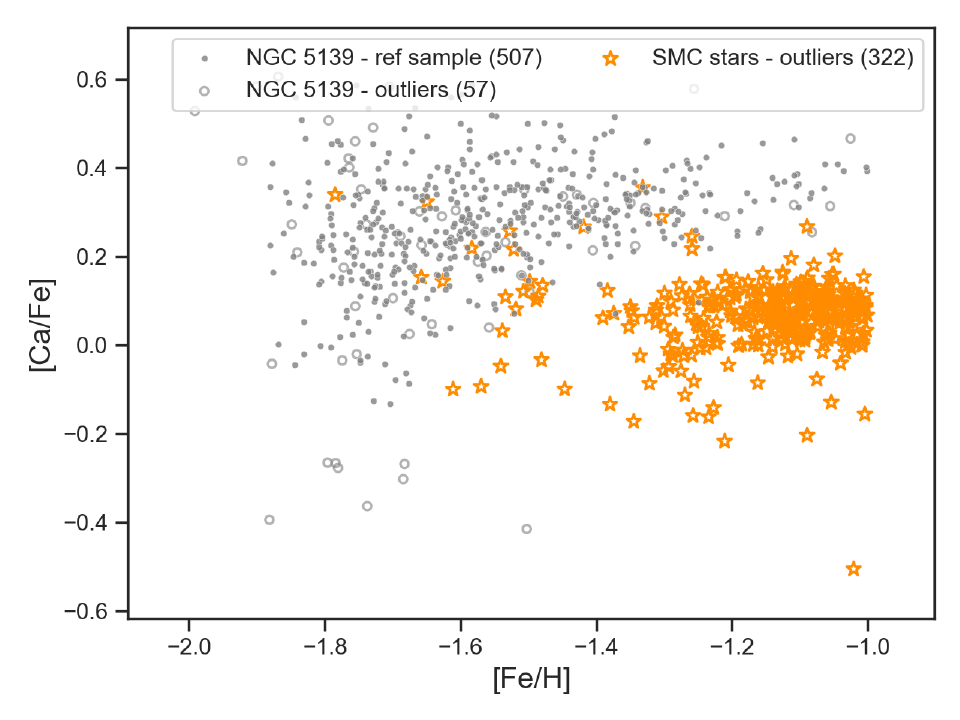}\par
\hspace{-25pt}\includegraphics[clip=true, trim = 1mm 0mm 0mm 2mm, width=0.7\columnwidth]{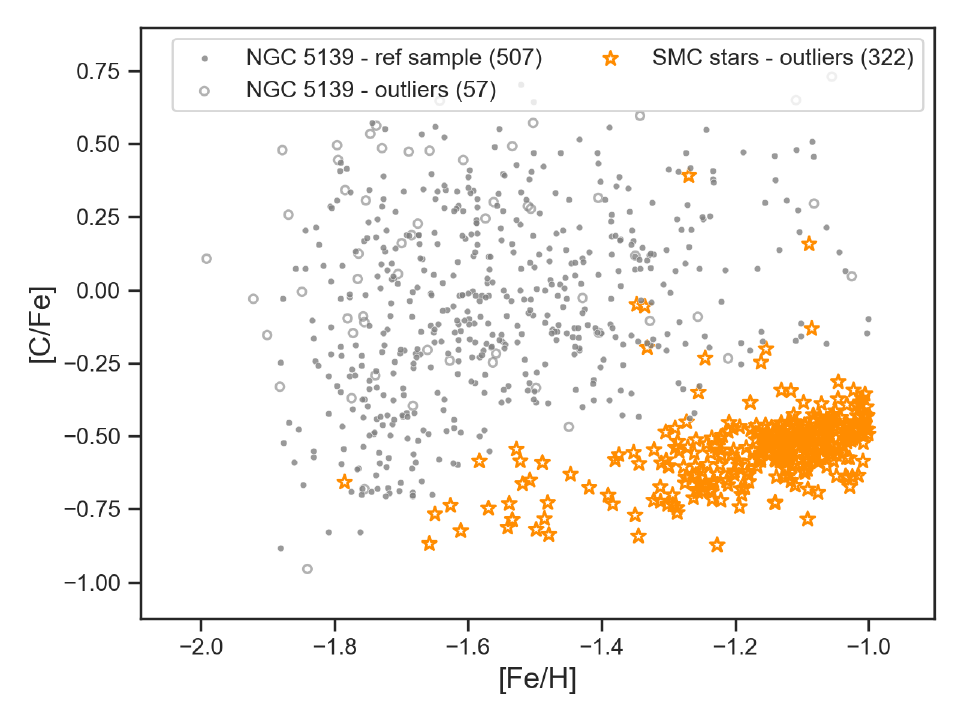}
\includegraphics[clip=true, trim = 1mm 0mm 0mm 1mm, width=0.7\columnwidth]{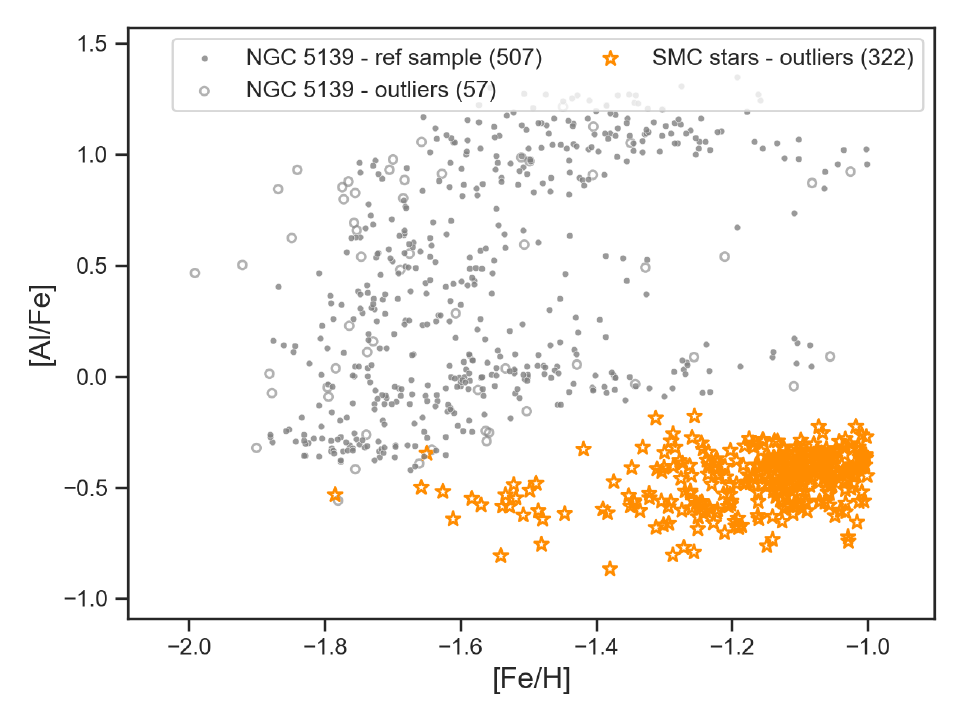}
\includegraphics[clip=true, trim = 1mm 0mm 0mm 1mm, width=0.7\columnwidth]{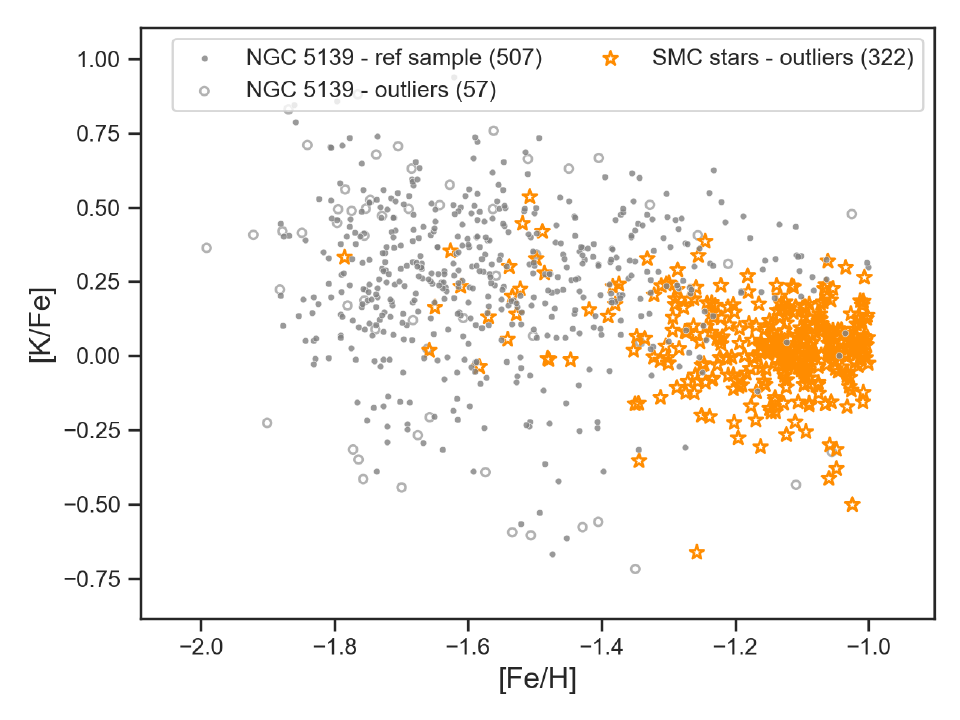}\par
\hspace{-27pt}
\includegraphics[clip=true, trim = 1mm 0mm 0mm 1mm, width=0.7\columnwidth]{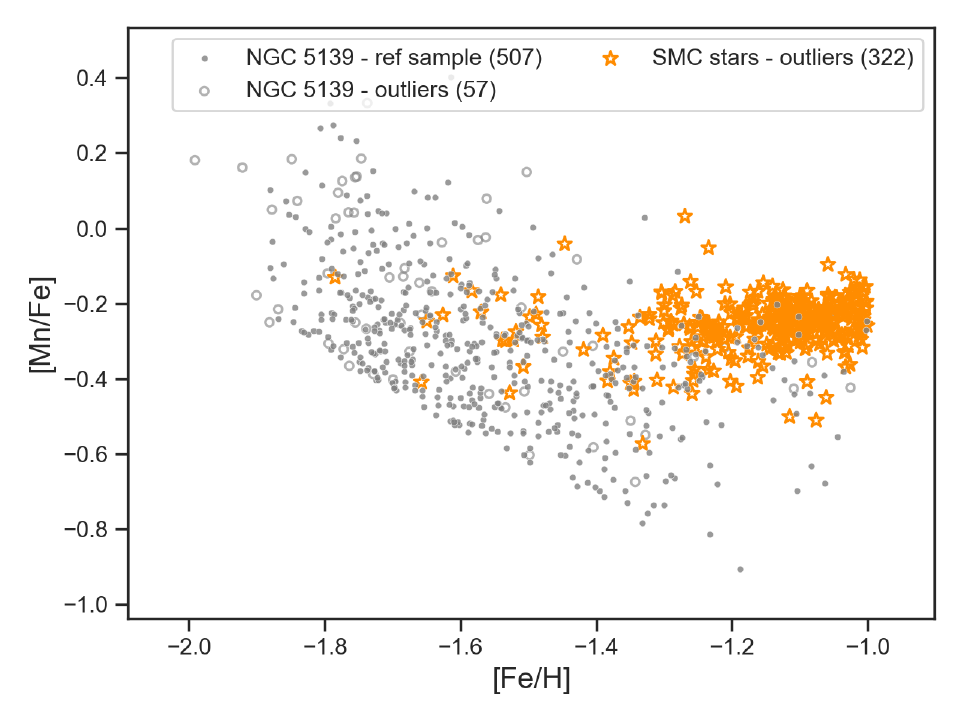}
\caption{Same as Fig.~\ref{LMCoCen}, but for the SMC. }\label{SMCoCen}
\end{figure*}

\begin{figure*}[h!]\centering
\hspace{-20pt}\includegraphics[clip=true, trim = 3mm 0mm 0mm 3mm, width=0.7\columnwidth]{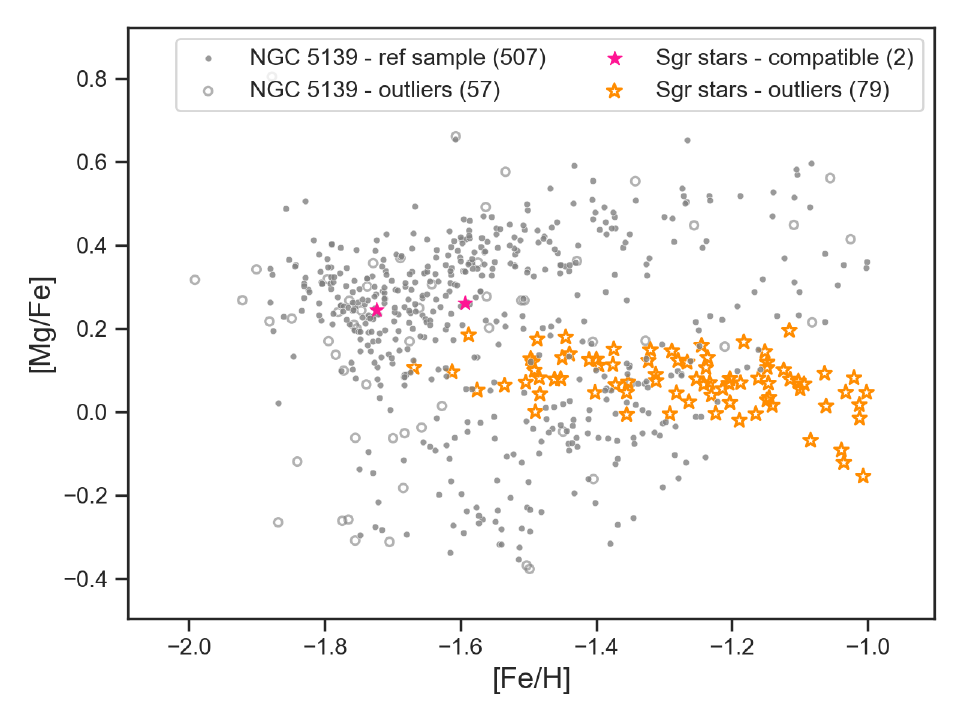}
\includegraphics[clip=true, trim = 3mm 0mm 0mm 3mm, width=0.7\columnwidth]{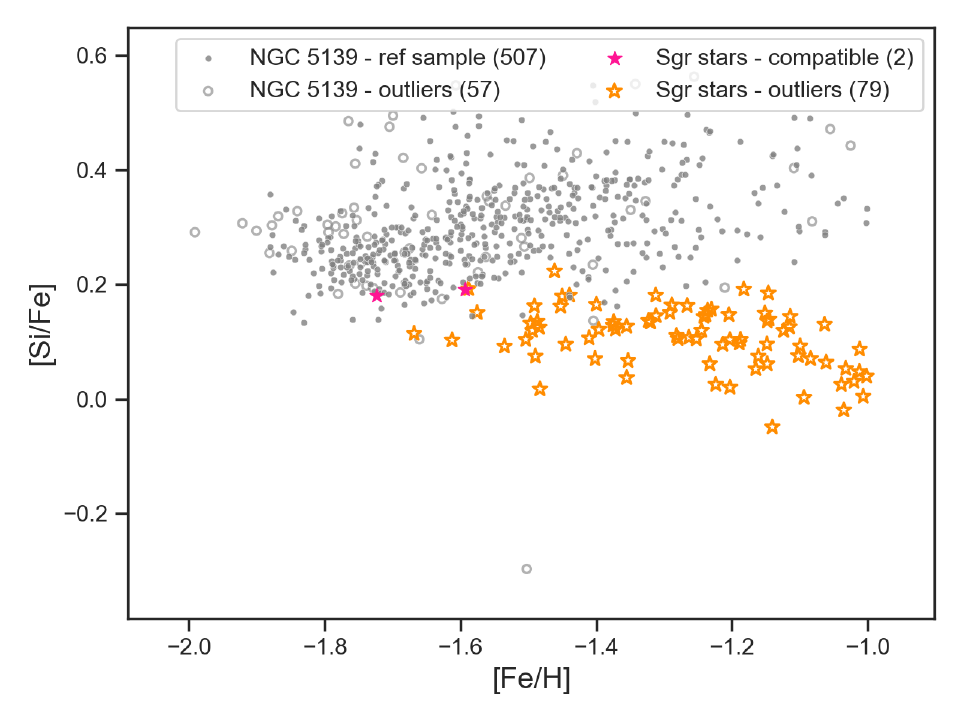}
\includegraphics[clip=true, trim = 3mm 0mm 0mm 3mm, width=0.7\columnwidth]{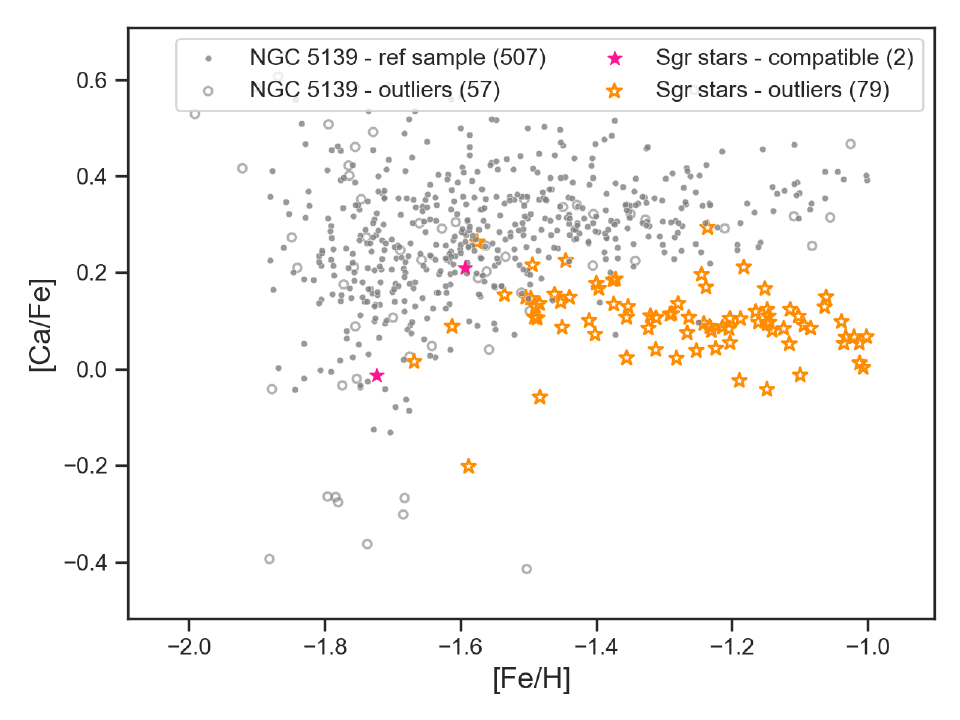}\par
\hspace{-25pt}\includegraphics[clip=true, trim = 1mm 0mm 0mm 2mm, width=0.7\columnwidth]{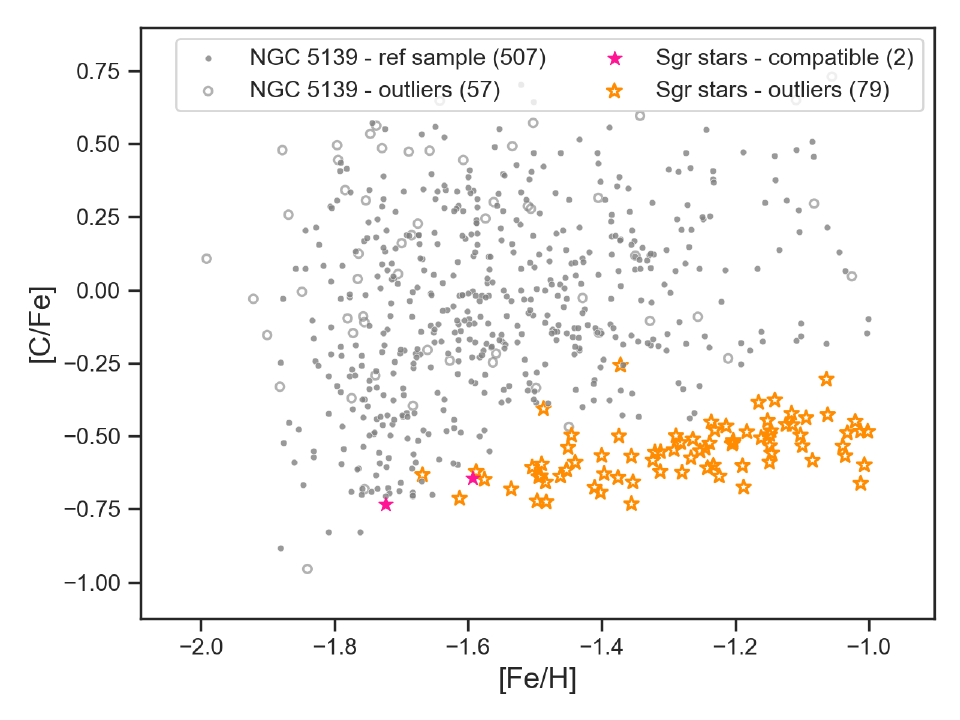}
\includegraphics[clip=true, trim = 1mm 0mm 0mm 1mm, width=0.7\columnwidth]{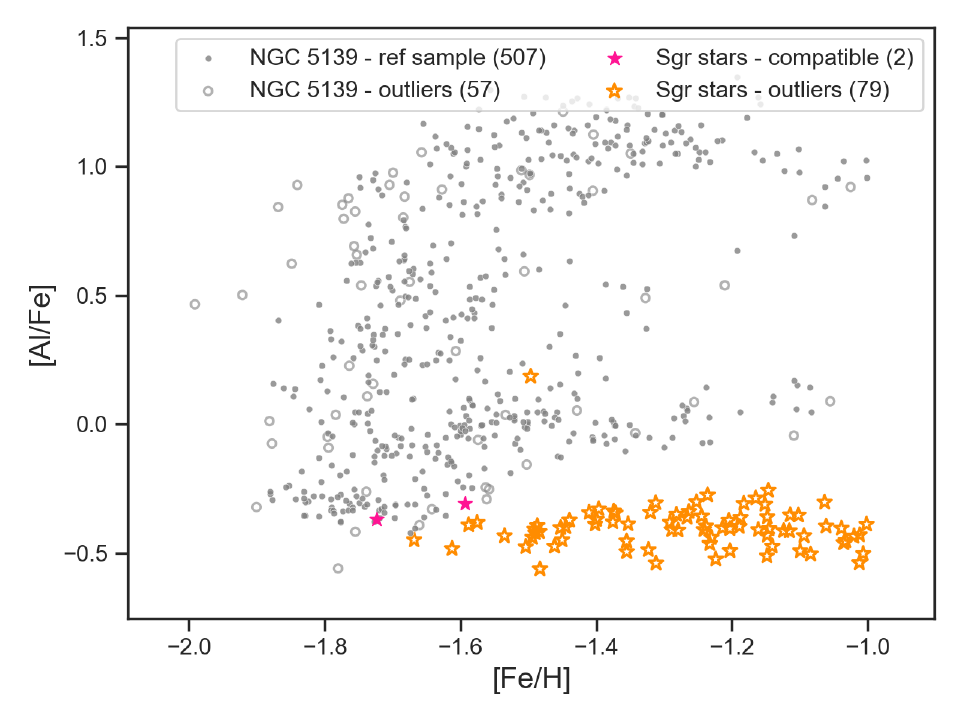}
\includegraphics[clip=true, trim = 1mm 0mm 0mm 1mm, width=0.7\columnwidth]{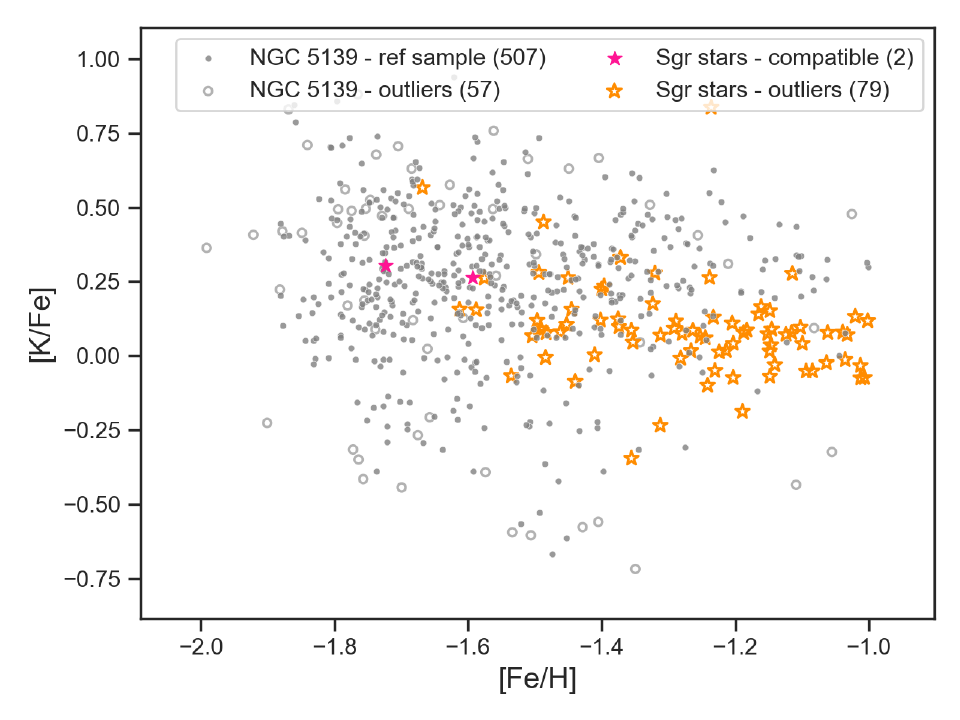}\par
\hspace{-27pt}
\includegraphics[clip=true, trim = 1mm 0mm 0mm 1mm, width=0.7\columnwidth]{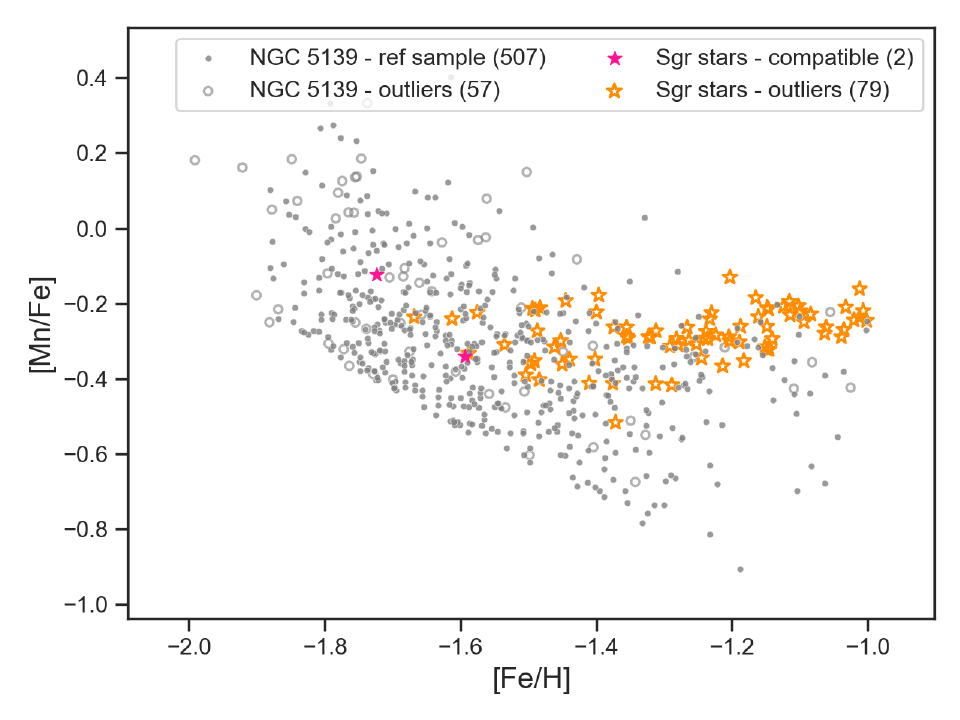}
\caption{Same as Fig.~\ref{LMCoCen} and \ref{SMCoCen} but for Sagittarius. }\label{SagoCen}
\end{figure*}

\begin{figure*}\centering
\hspace{-20pt}\includegraphics[clip=true, trim = 3mm 0mm 0mm 3mm, width=0.7\columnwidth]{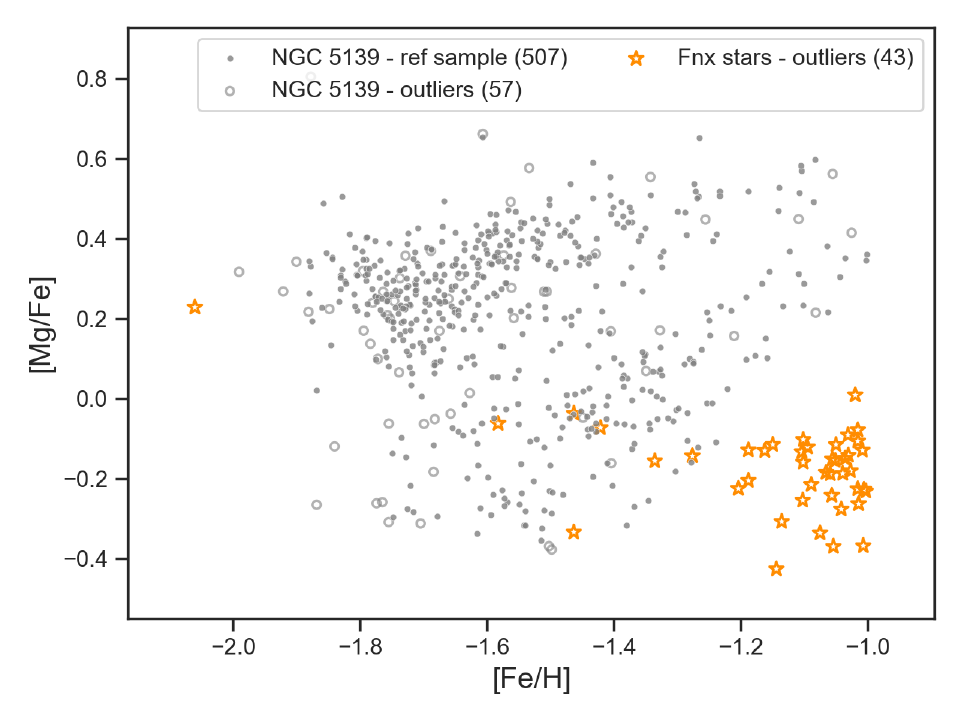}
\includegraphics[clip=true, trim = 3mm 0mm 0mm 3mm, width=0.7\columnwidth]{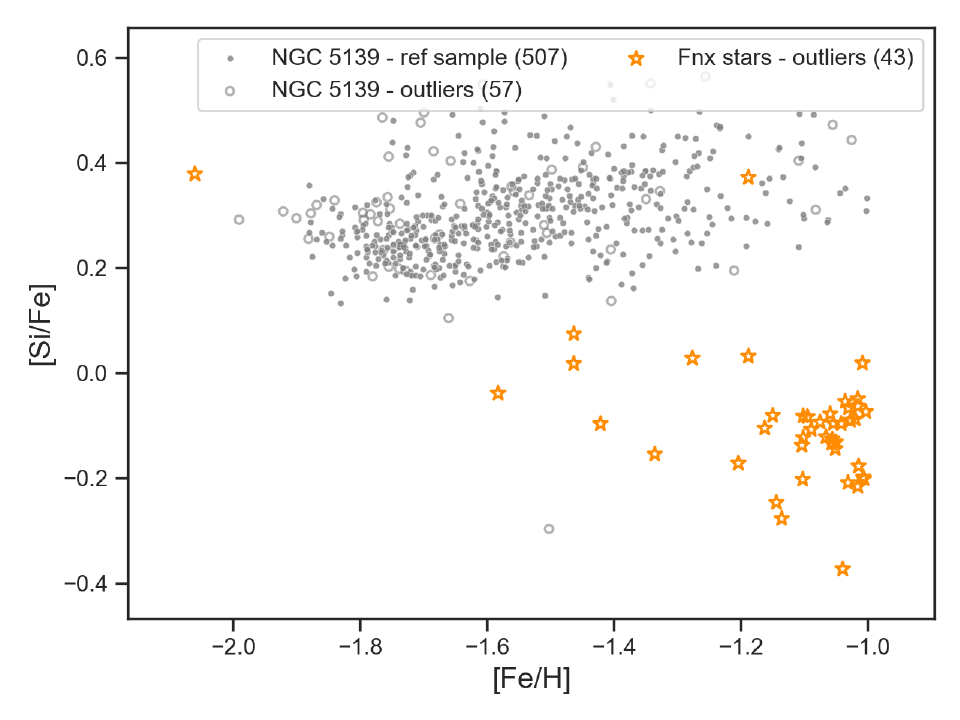}
\includegraphics[clip=true, trim = 3mm 0mm 0mm 3mm, width=0.7\columnwidth]{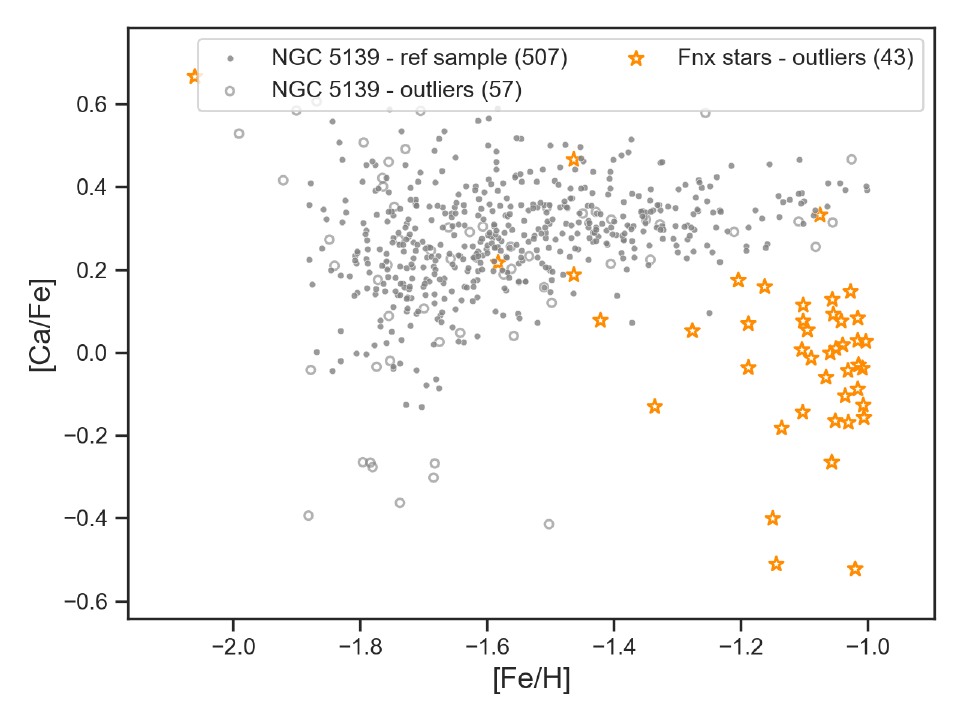}\par
\hspace{-25pt}\includegraphics[clip=true, trim = 1mm 0mm 0mm 2mm, width=0.7\columnwidth]{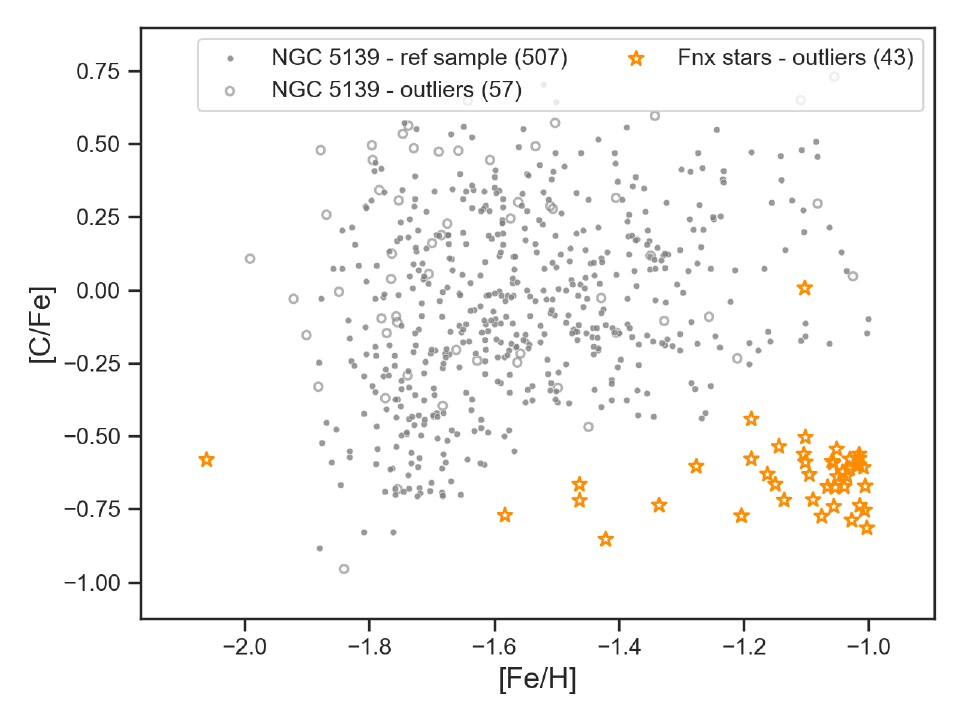}
\includegraphics[clip=true, trim = 1mm 0mm 0mm 1mm, width=0.7\columnwidth]{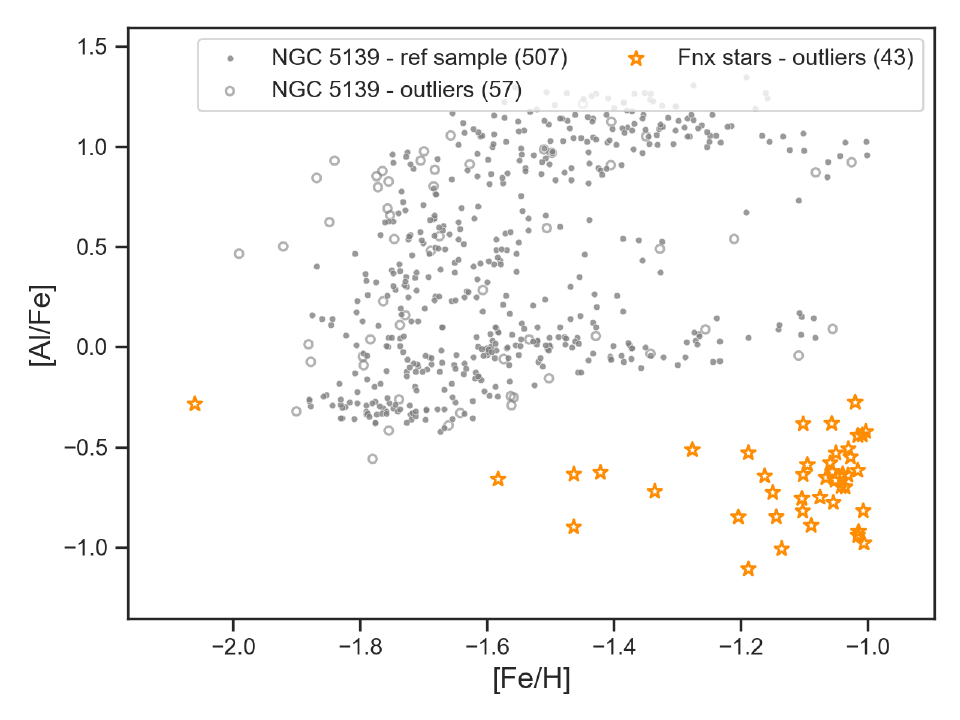}
\includegraphics[clip=true, trim = 1mm 0mm 0mm 1mm, width=0.7\columnwidth]{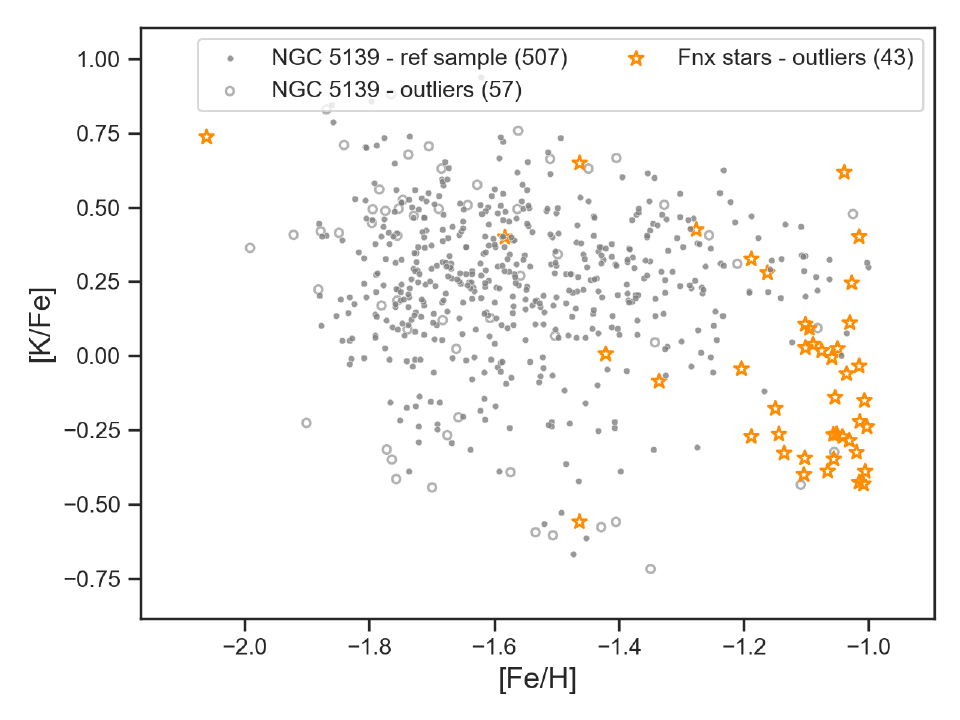}\par
\hspace{-27pt}
\includegraphics[clip=true, trim = 1mm 0mm 0mm 1mm, width=0.7\columnwidth]{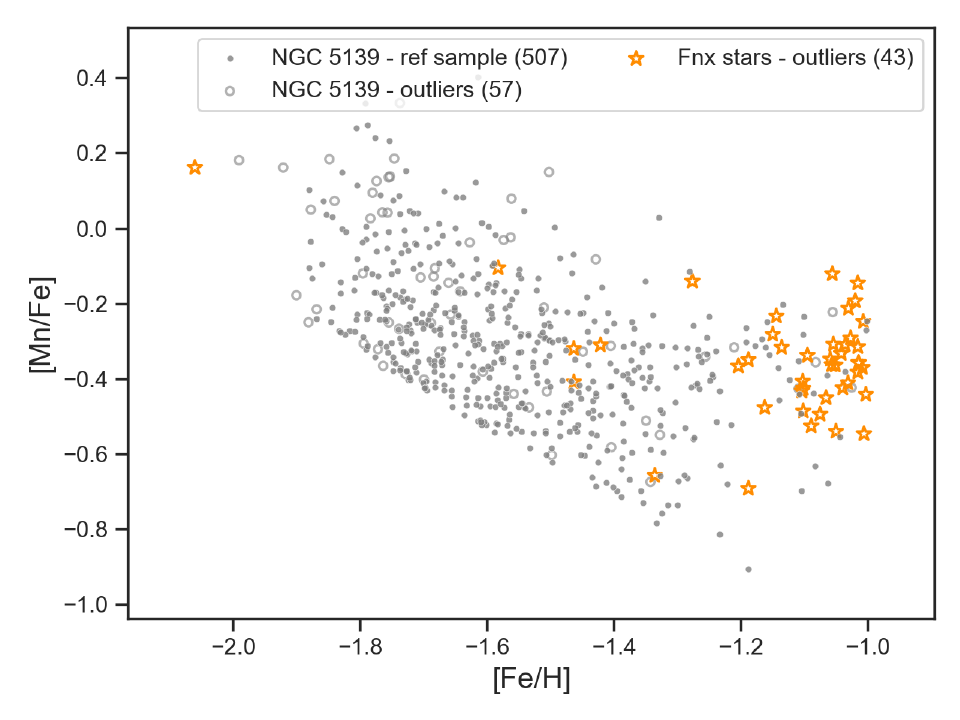}
\caption{Same as Figs.~\ref{LMCoCen}, \ref{SMCoCen} and \ref{SagoCen} but for Fornax. }\label{FnxoCen}
\end{figure*}

In Figs.~\ref{LMCoCen}, \ref{SMCoCen}, \ref{SagoCen} and \ref{FnxoCen}, we show the comparison of the chemical patterns of $\omega$~Cen with those of the Large and Small Magellanic Clouds, Sagittarius and Fornax. In all plots, magenta symbols indicate stars of the aforementioned dwarfs which are chemically compatible with $\omega$~Centauri, while orange symbols indicate stars which are outliers of the $\omega$~Cen chemical domain, according to the GMM model. \\

\begin{minipage}{\textwidth}
\centering
\hspace{-35pt}\includegraphics[clip=true, trim = 3mm 0mm 0mm 3mm, width=0.35\textwidth]{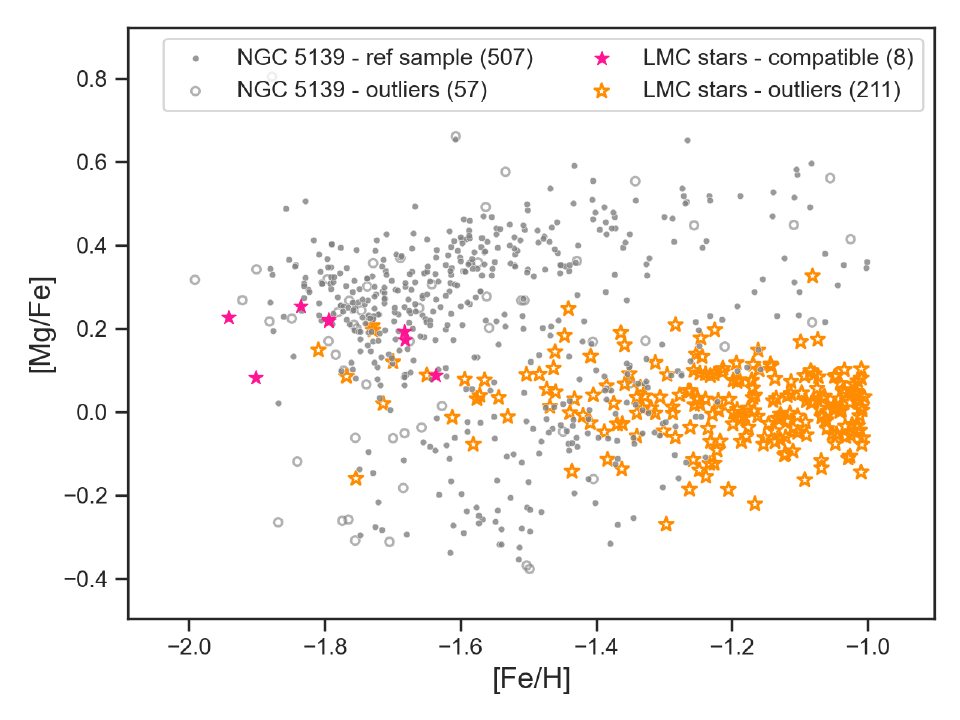}
\includegraphics[clip=true, trim = 3mm 0mm 0mm 3mm, width=0.35\textwidth]{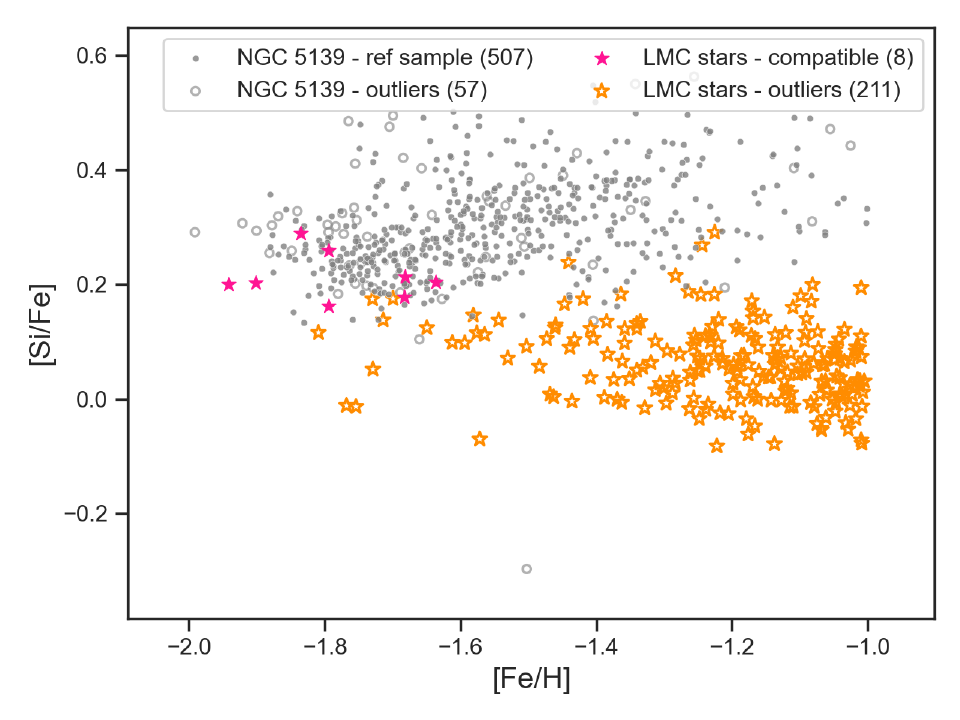}
\includegraphics[clip=true, trim = 3mm 0mm 0mm 3mm, width=0.35\textwidth]{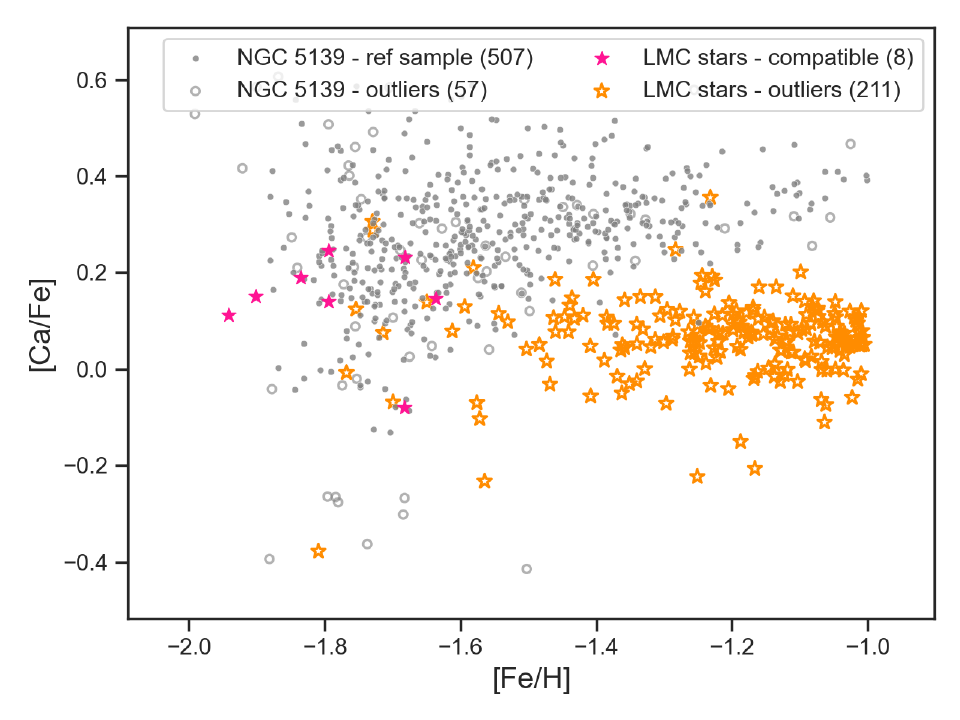}\par
\hspace{-35pt}\includegraphics[clip=true, trim = 1mm 0mm 0mm 2mm, width=0.35\textwidth]{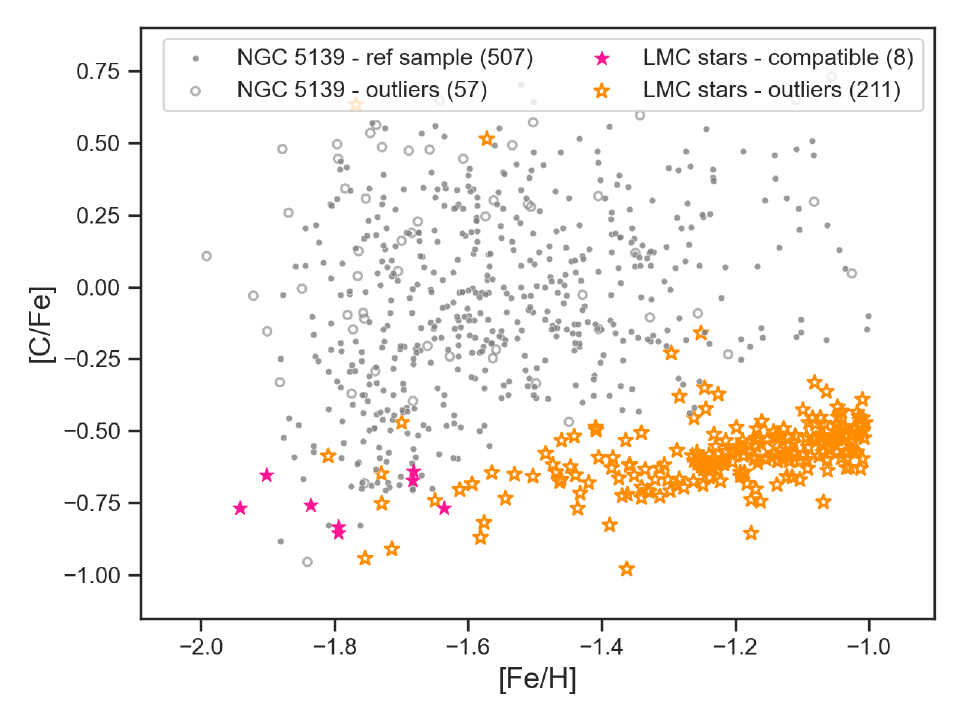}
\includegraphics[clip=true, trim = 1mm 0mm 0mm 1mm, width=0.35\textwidth]{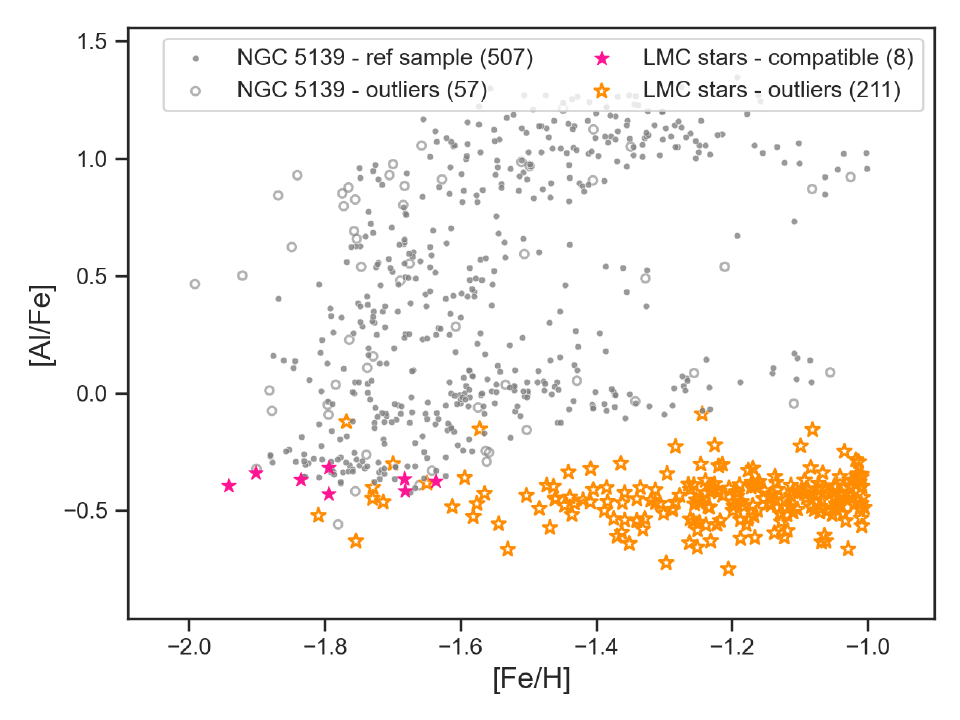}
\includegraphics[clip=true, trim = 1mm 0mm 0mm 1mm, width=0.35\textwidth]{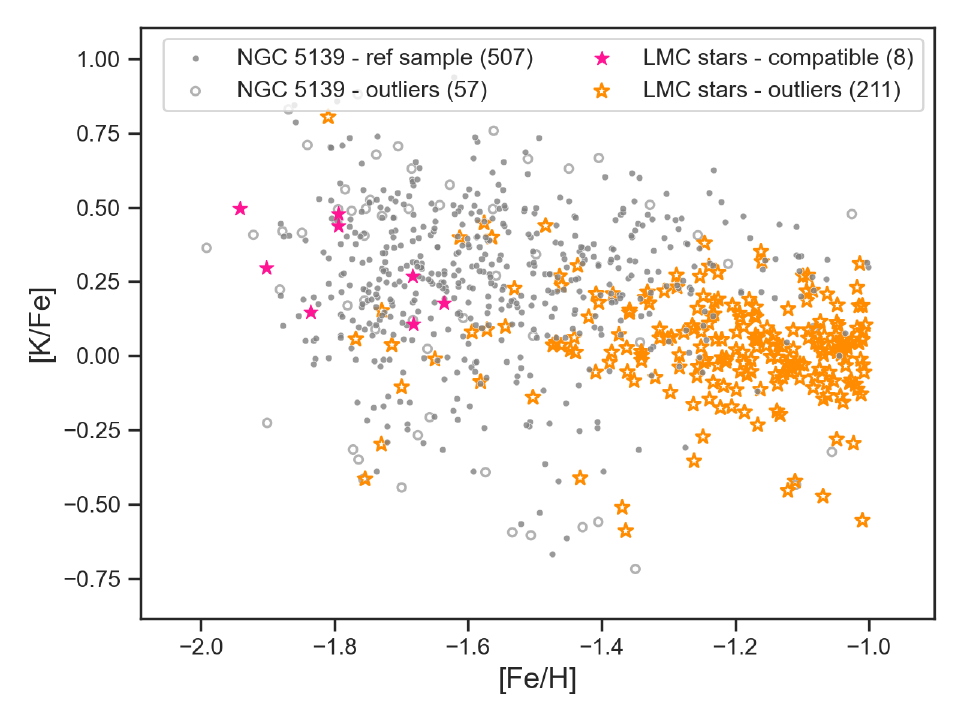}\par
\hspace{-38pt}
\includegraphics[clip=true, trim = 1mm 0mm 0mm 1mm, width=0.35\textwidth]{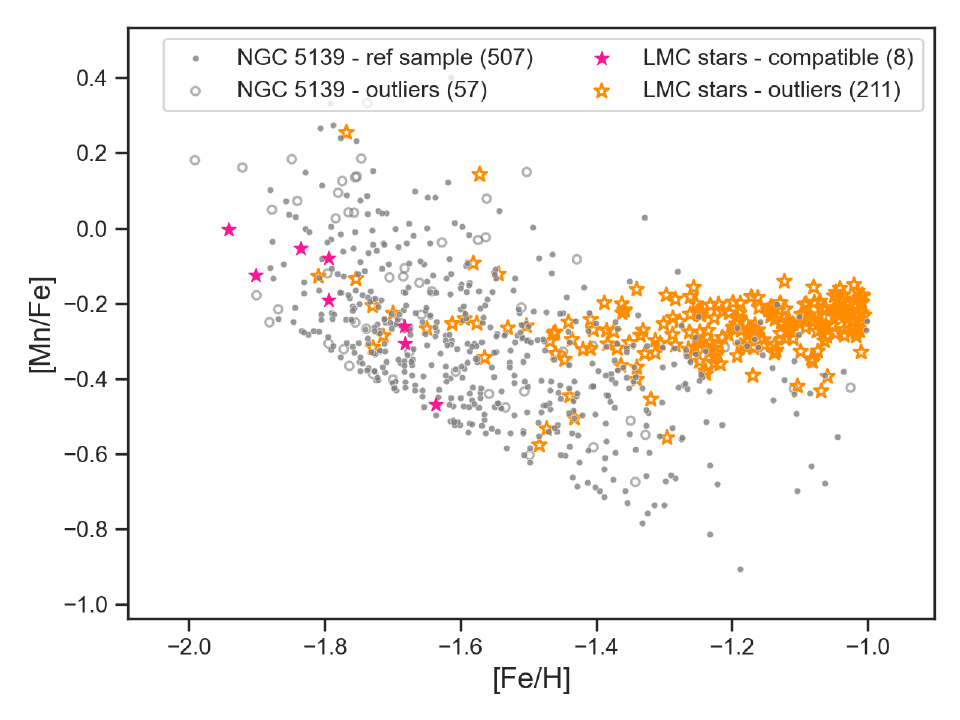}
\captionof{figure}{Chemical abundance relations for members of LMC (colour) and $\omega$~Cen (grey). The filled symbols show the reference sample of $\omega$~Cen (grey) and the stars of LMC (magenta) chemically compatible with it according to the GMM (see Sect. \ref{results}), while the empty ones (grey and orange colours) correspond to their outliers. The number of stars in each category is reported in parentheses. For clarity, only stars with [Fe/H]$<-1$ are shown. }\label{LMCoCen}
\end{minipage}


\clearpage    
\section{Median abundances of GCs chemically compatible with $\omega$~Cen}\label{median_abundances}
\begin{figure*}
\includegraphics[clip=true, trim = 3mm 0mm 0mm 3mm, width=0.69\columnwidth]{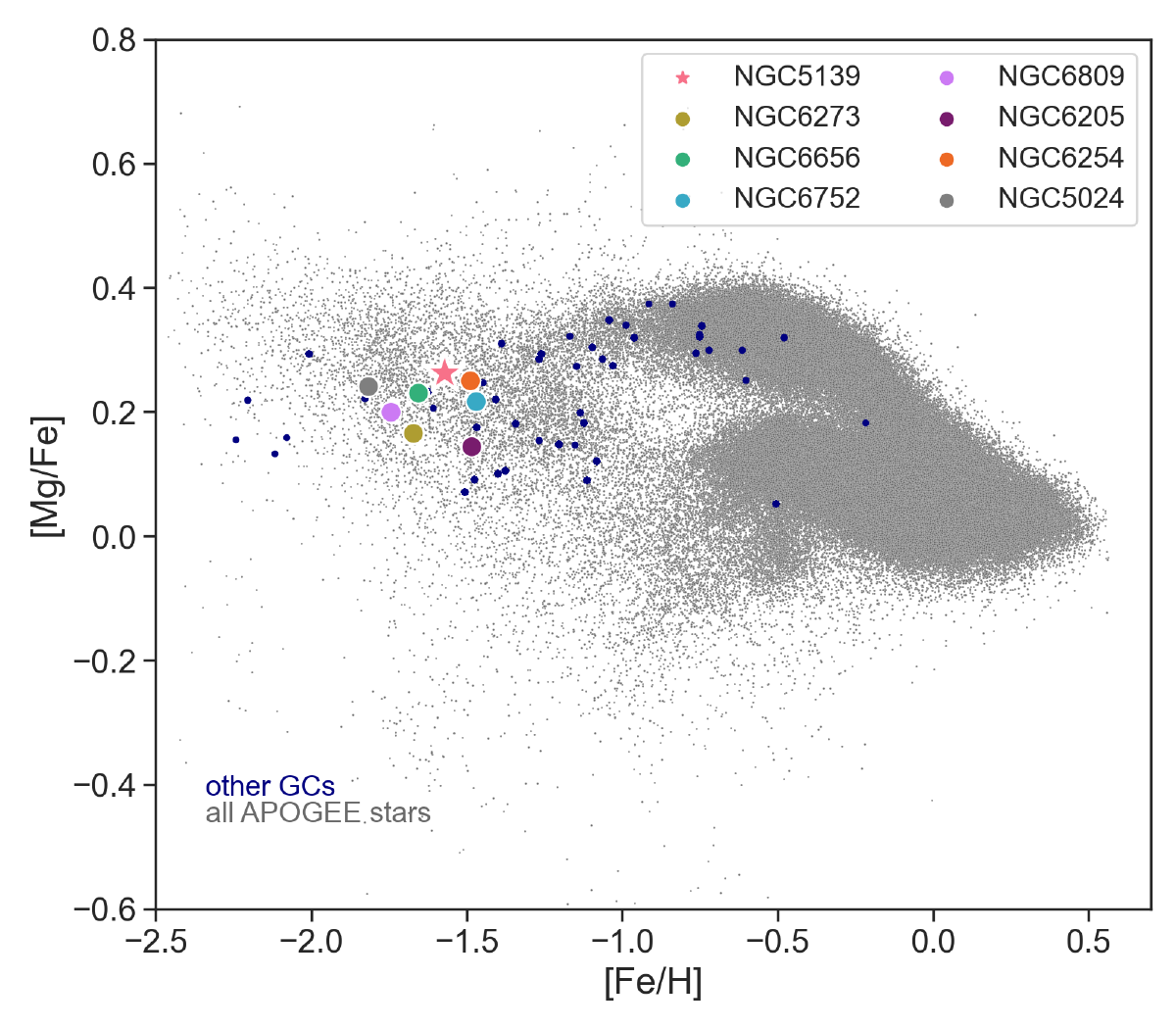}\hspace{-5pt}
\includegraphics[clip=true, trim = 3mm 0mm 0mm 3mm, width=0.69\columnwidth]{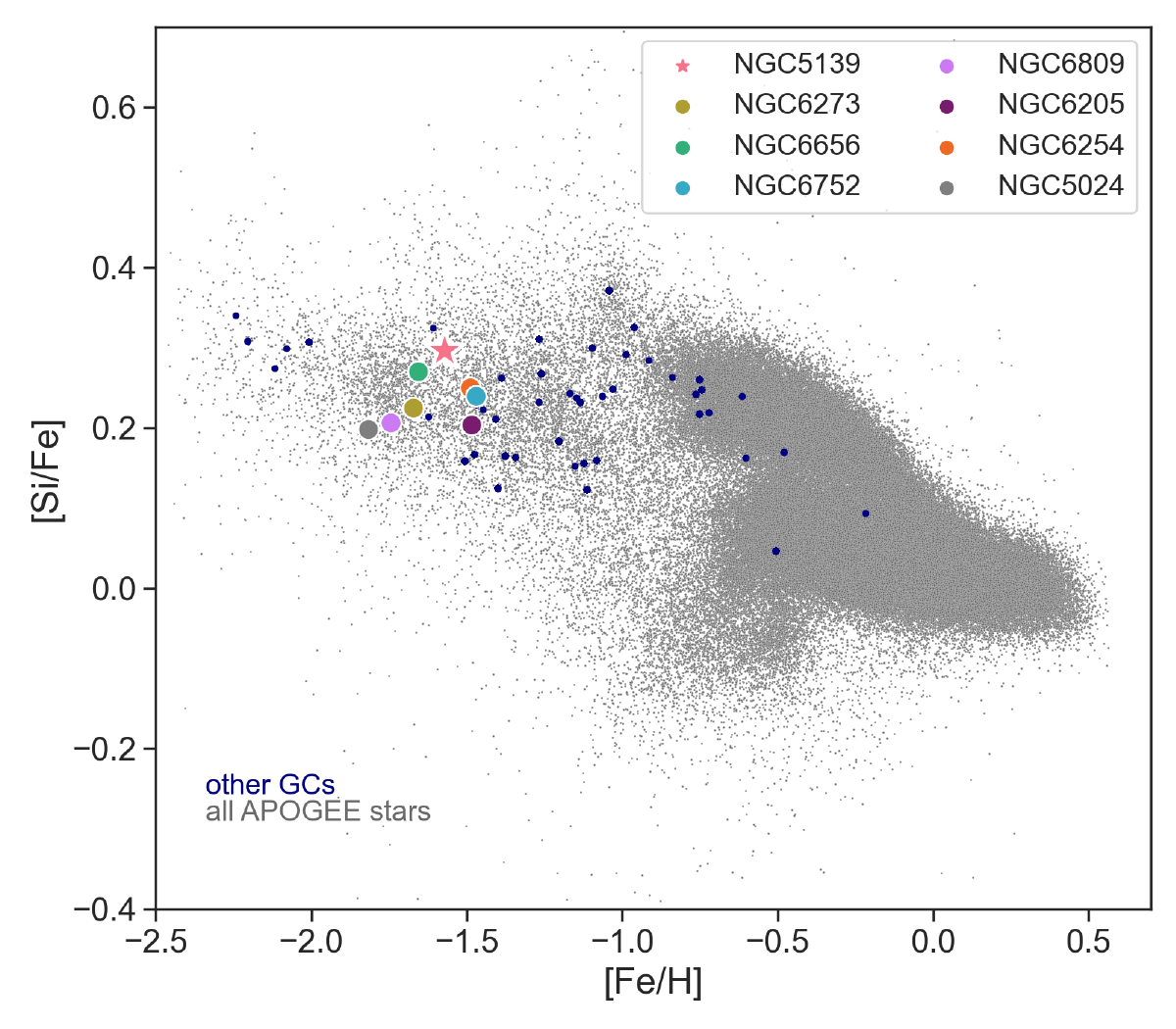}\hspace{-5pt}
\includegraphics[clip=true, trim = 3mm 0mm 0mm 3mm, width=0.69\columnwidth]{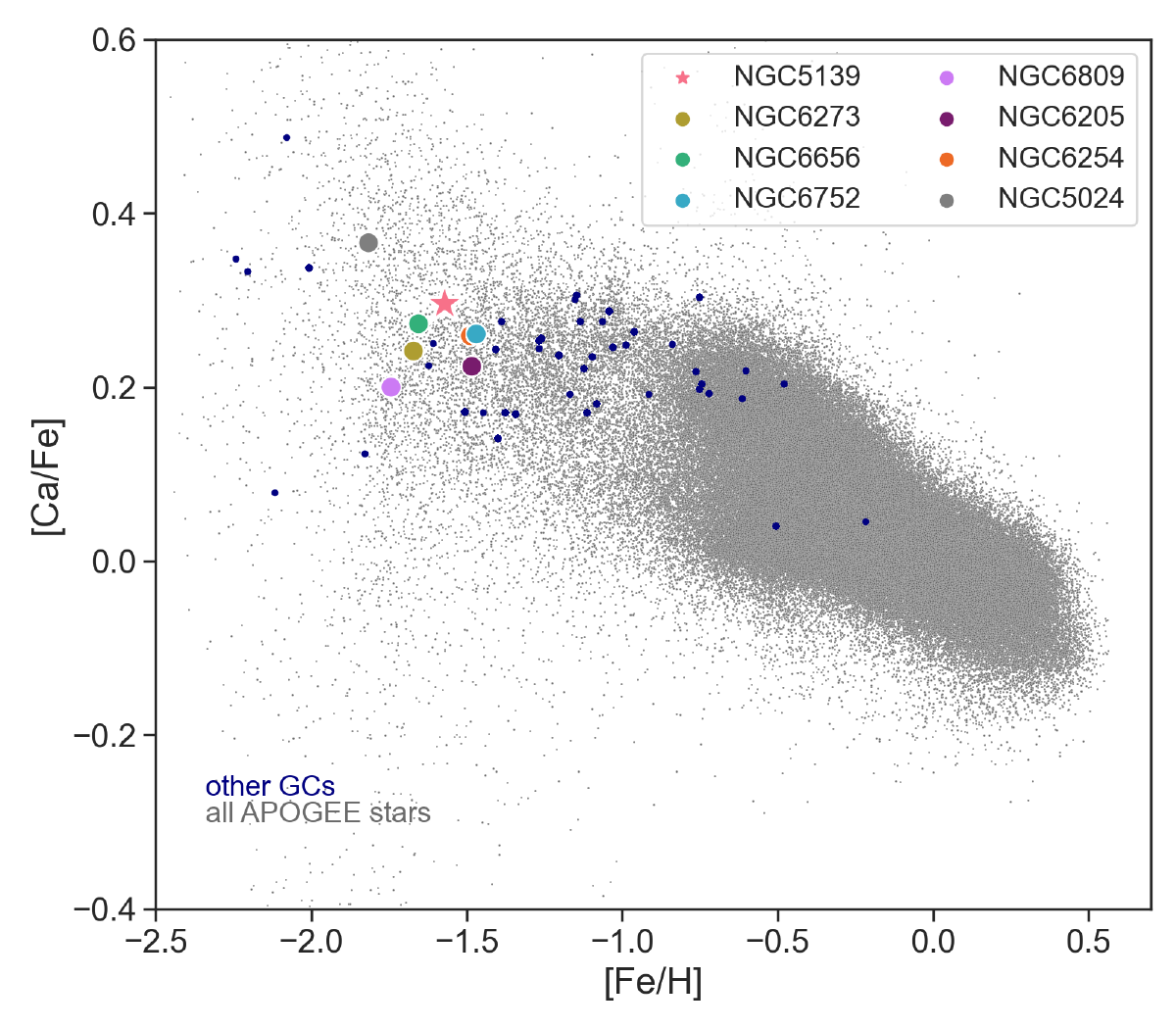}\hspace{-5pt}
\includegraphics[clip=true, trim = 3mm 0mm 0mm 3mm, width=0.69\columnwidth]{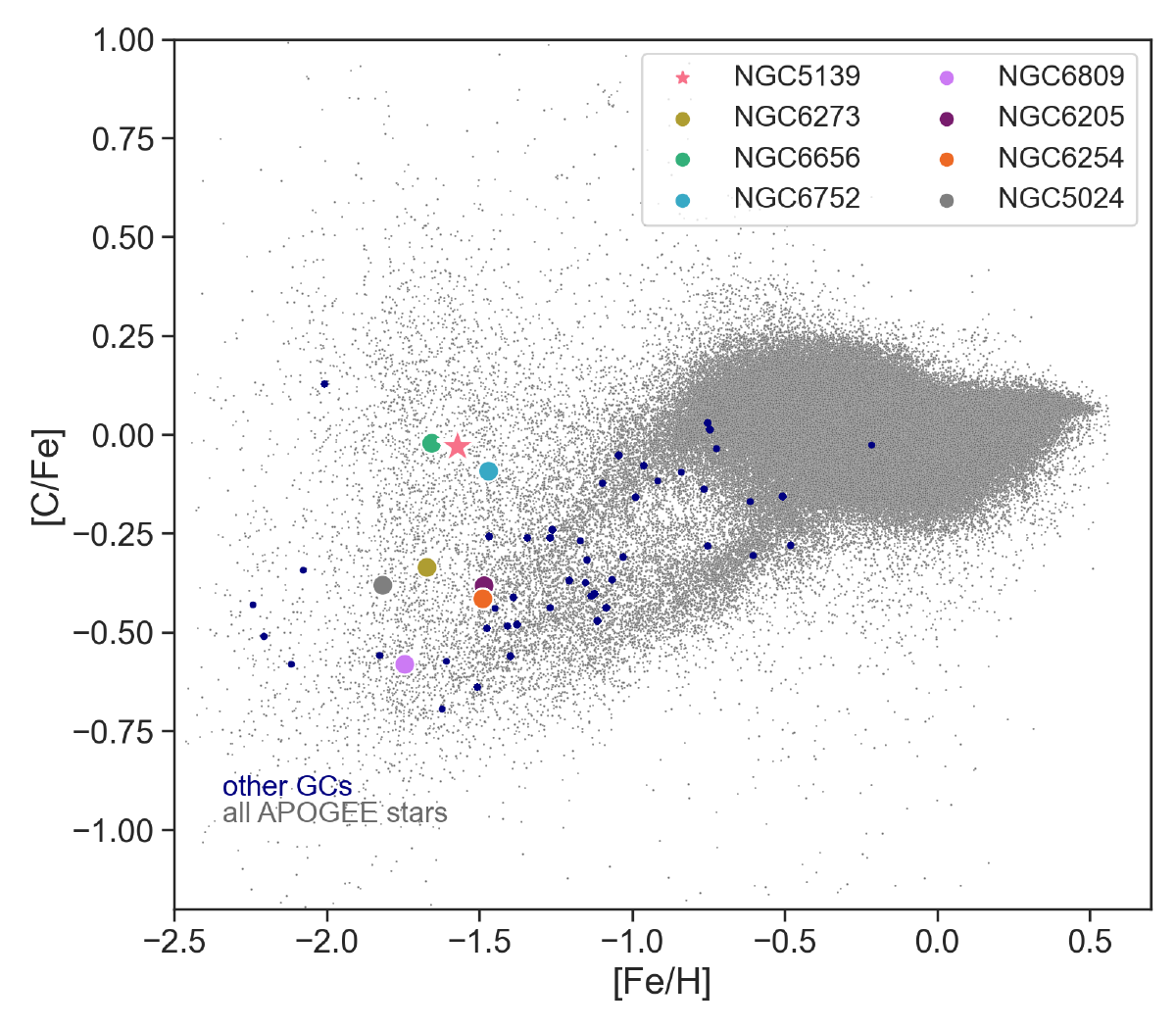}\hspace{-5pt}
\includegraphics[clip=true, trim = 3mm 0mm 0mm 3mm, width=0.69\columnwidth]{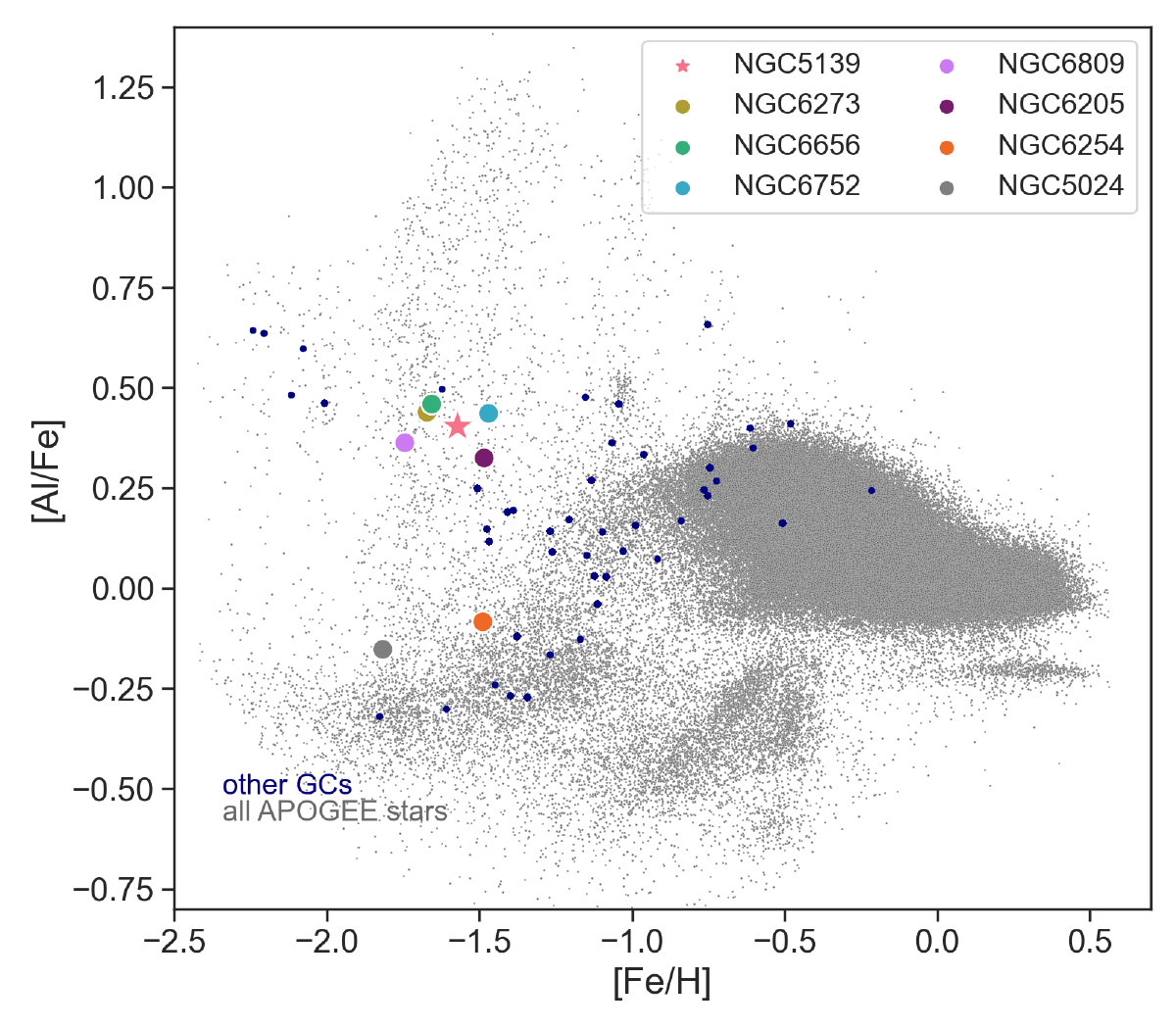}\hspace{-5pt}
\includegraphics[clip=true, trim = 3mm 0mm 0mm 3mm, width=0.69\columnwidth]{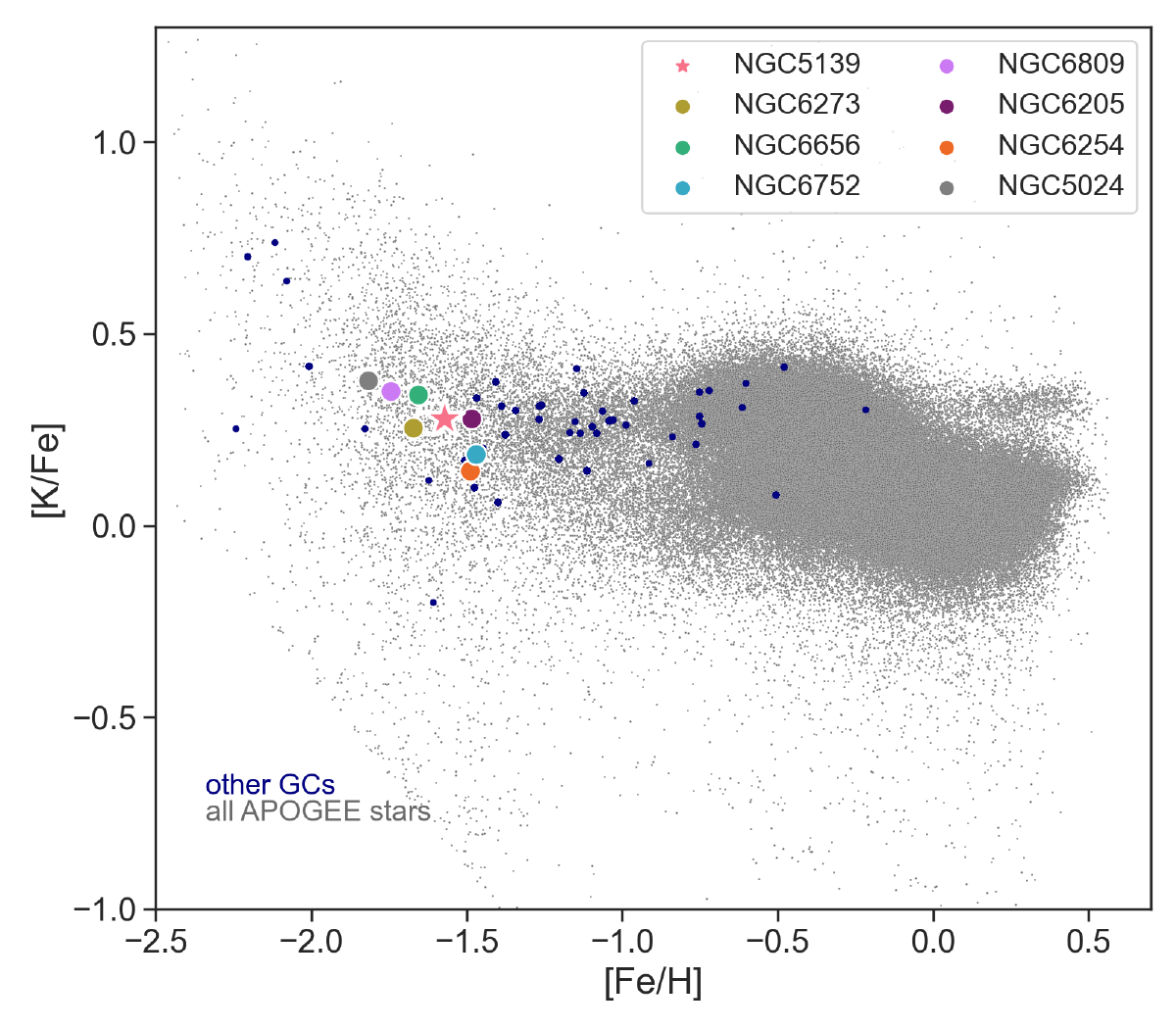}\par
\centering\includegraphics[clip=true, trim = 3mm 0mm 0mm 3mm, width=0.69\columnwidth]{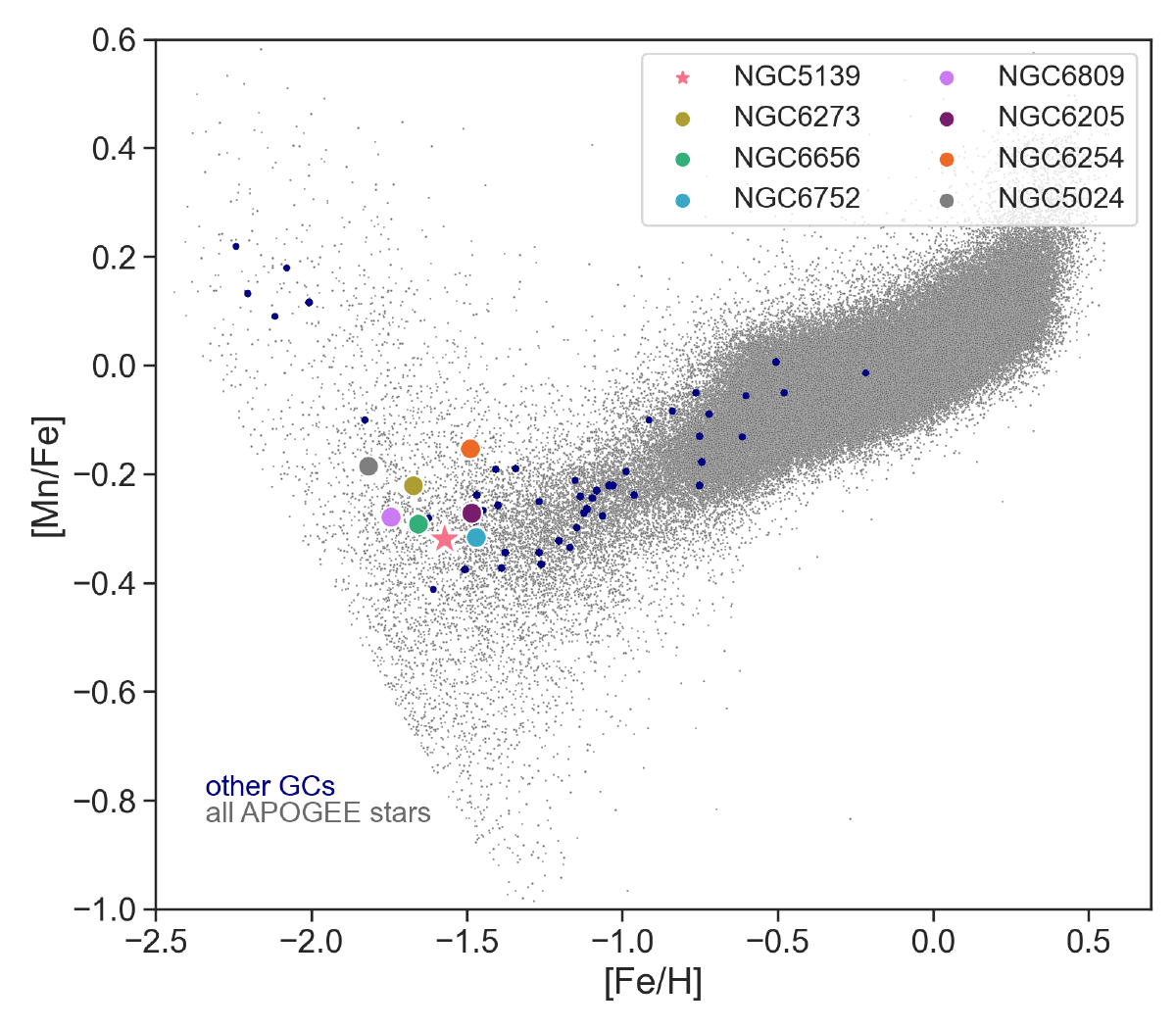}
  \caption{Median [X/Fe] versus [Fe/H] for GCs chemically compatible - with at least 60\% of their stars - with $\omega$~Cen; the different X elements are the ones used in the GMM, namely: Mg, Si, Ca, C, Al, K, and Mn. For comparison, other GCs in our sample (blue points) and all APOGEE stars (GCs' stars included,  grey points) are also shown.}
              \label{median_abund_ocenlike}%
    \end{figure*}

\begin{figure*}
\includegraphics[clip=true, trim = 3mm 0mm 0mm 3mm, width=0.69\columnwidth]{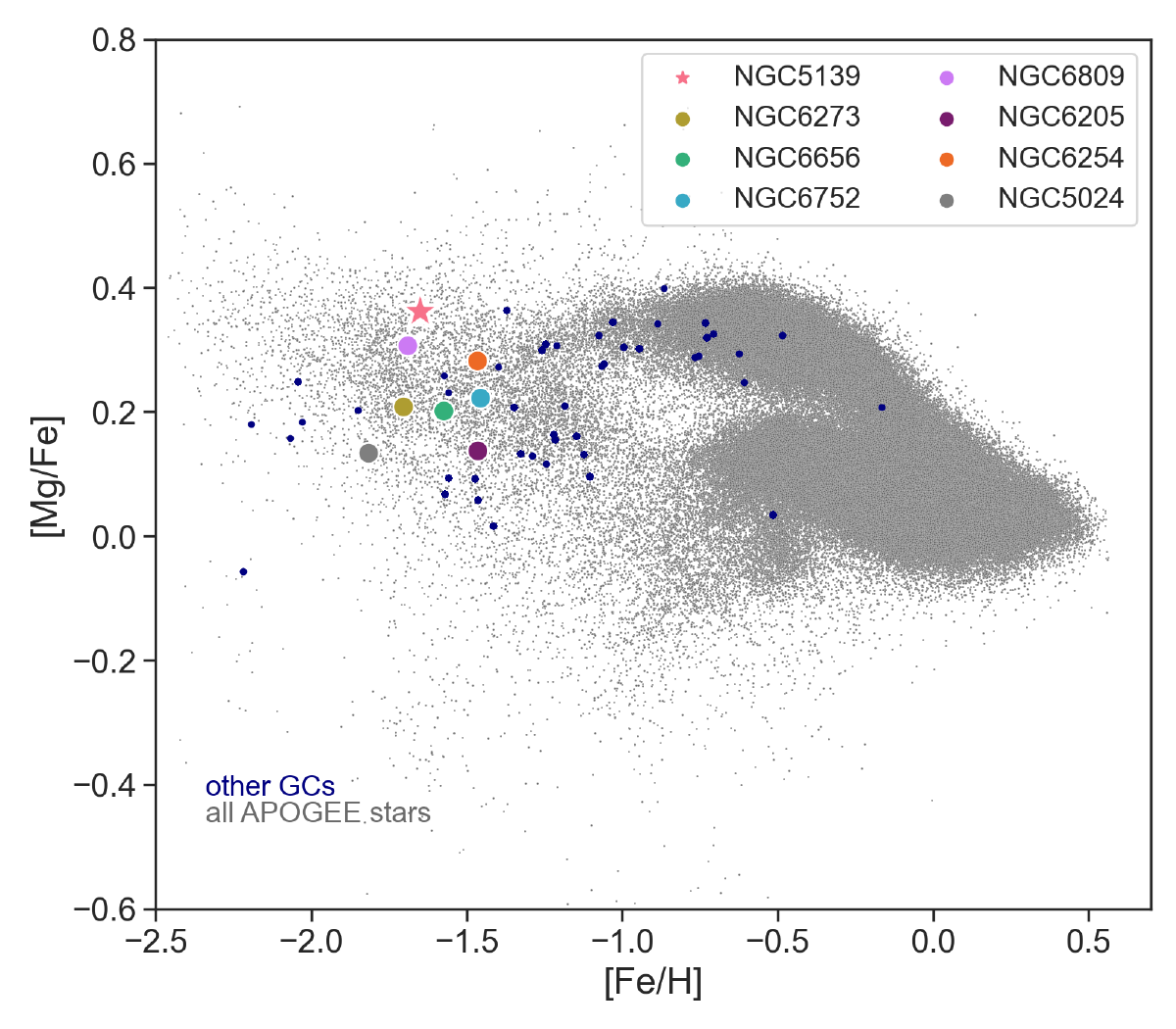}\hspace{-5pt}
\includegraphics[clip=true, trim = 3mm 0mm 0mm 3mm, width=0.69\columnwidth]{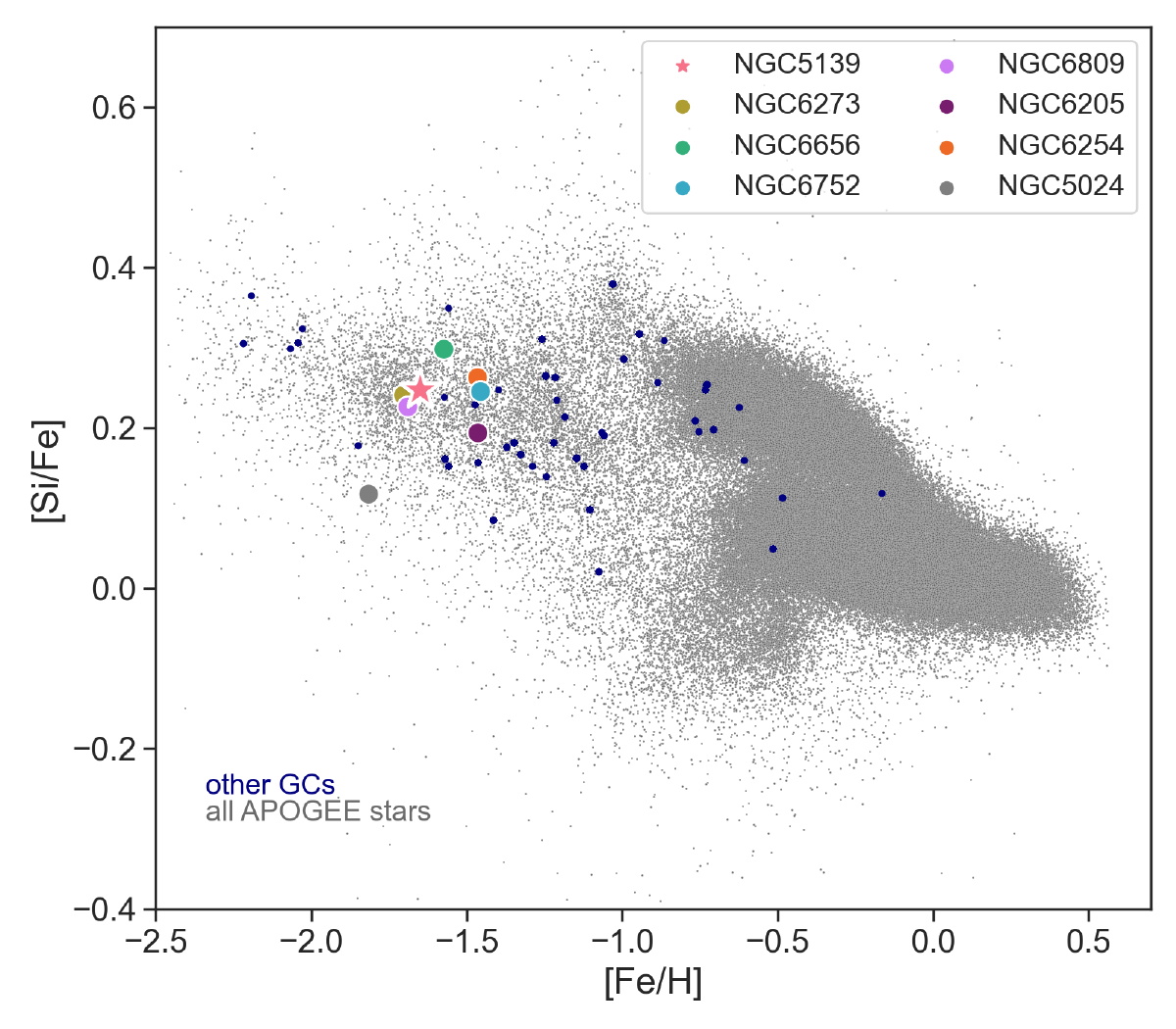}\hspace{-5pt}
\includegraphics[clip=true, trim = 3mm 0mm 0mm 3mm, width=0.69\columnwidth]{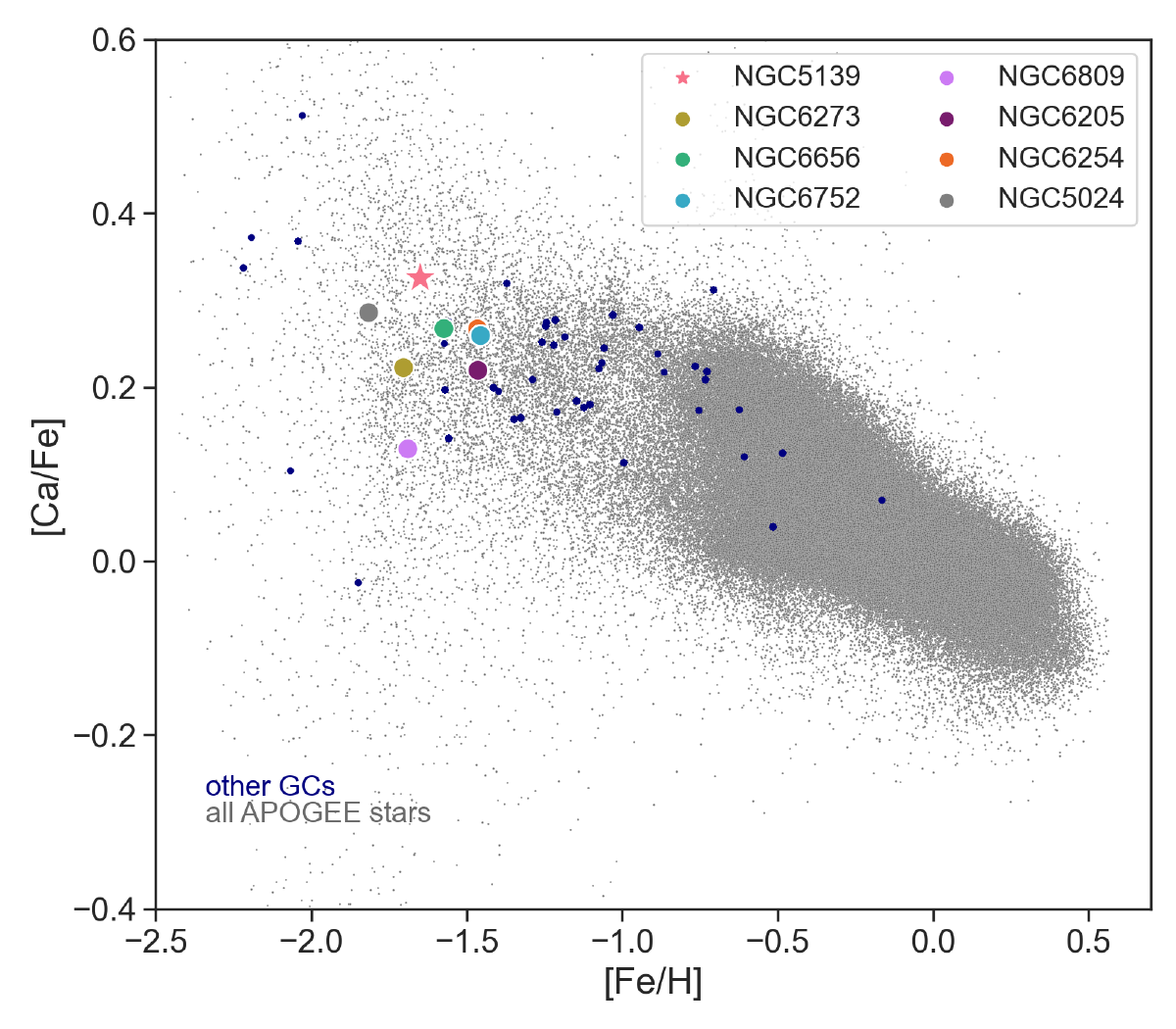}\hspace{-5pt}
\includegraphics[clip=true, trim = 3mm 0mm 0mm 3mm, width=0.69\columnwidth]{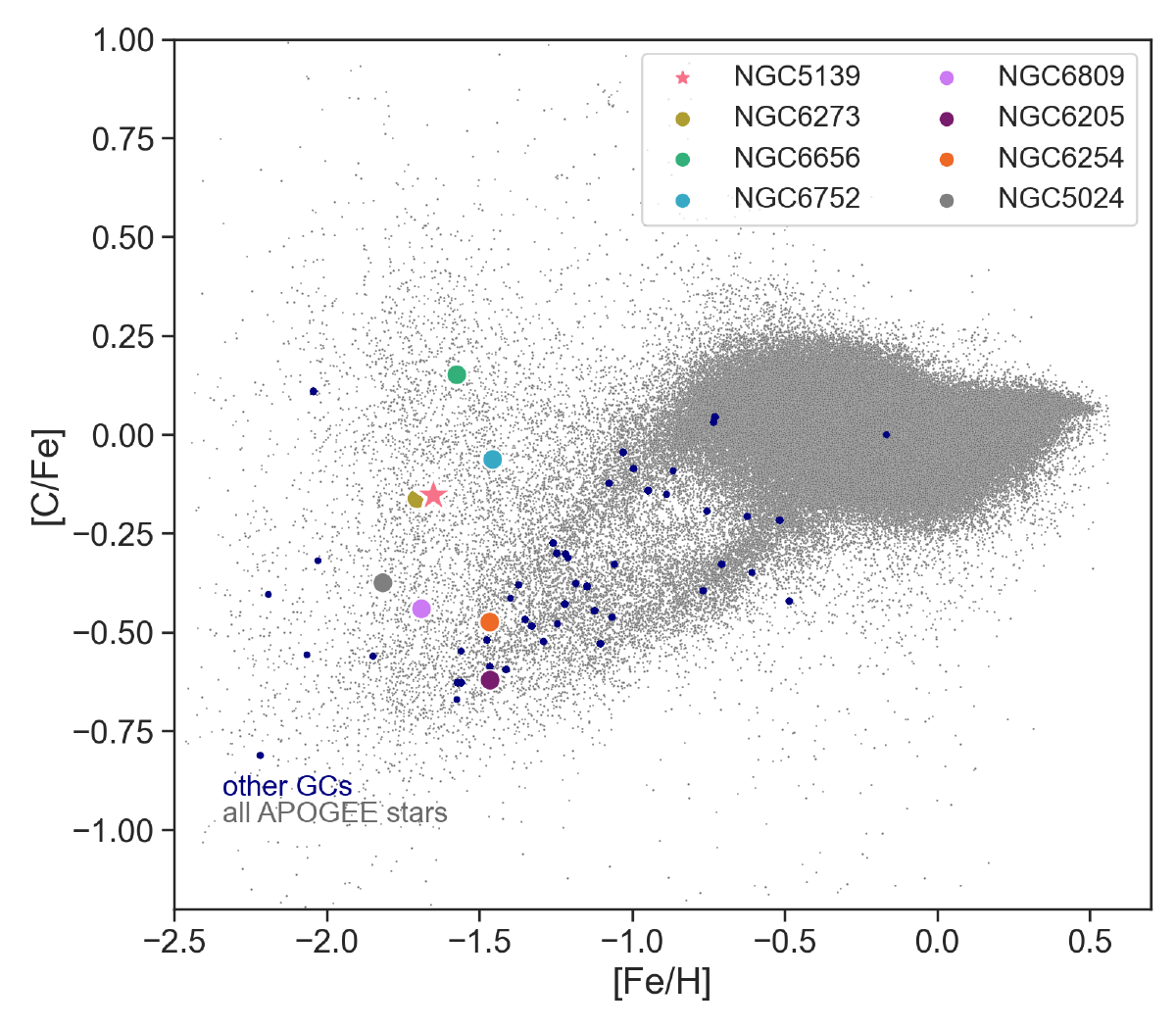}\hspace{-5pt}
\includegraphics[clip=true, trim = 3mm 0mm 0mm 3mm, width=0.69\columnwidth]{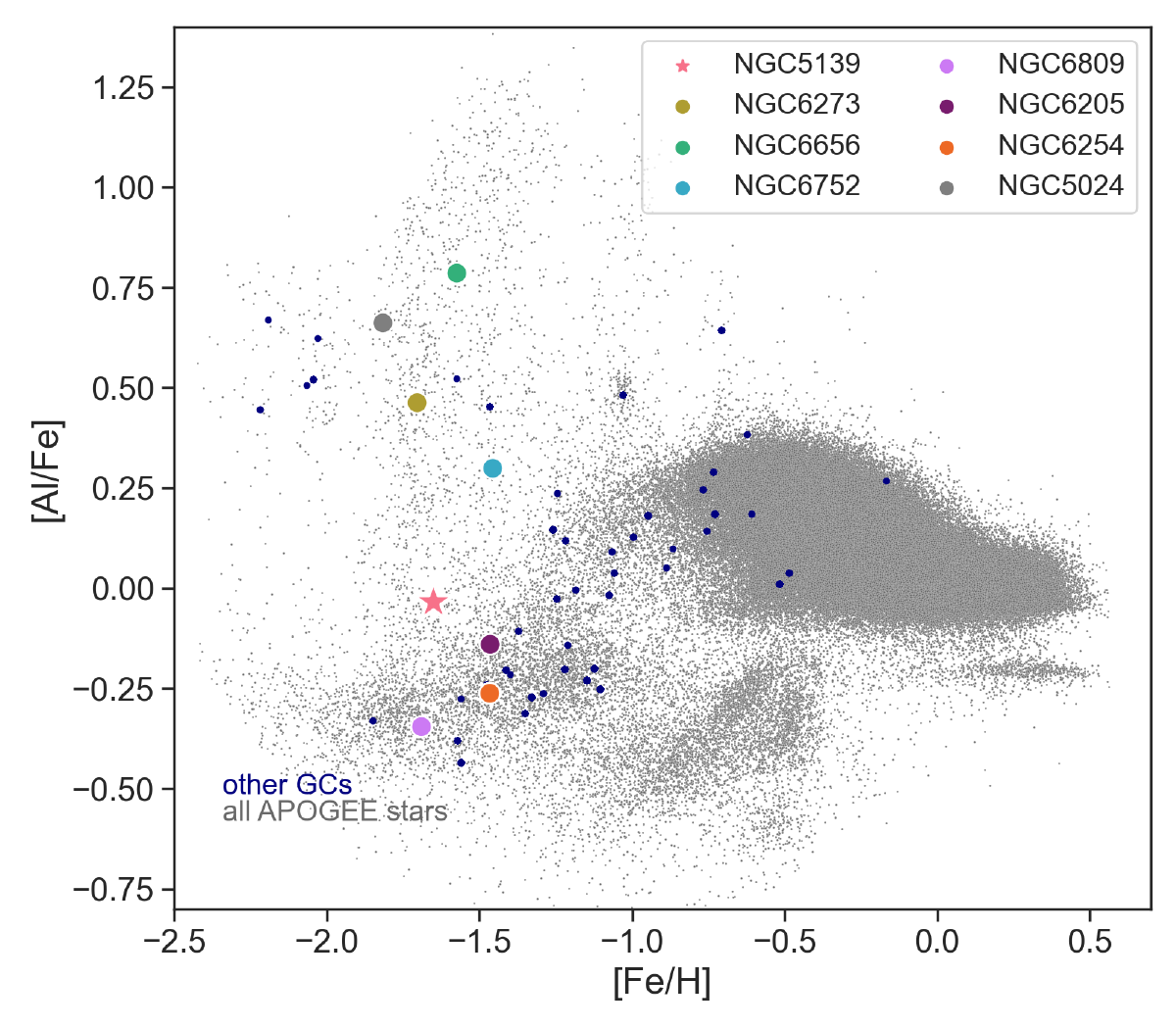}\hspace{-5pt}
\includegraphics[clip=true, trim = 3mm 0mm 0mm 3mm, width=0.69\columnwidth]{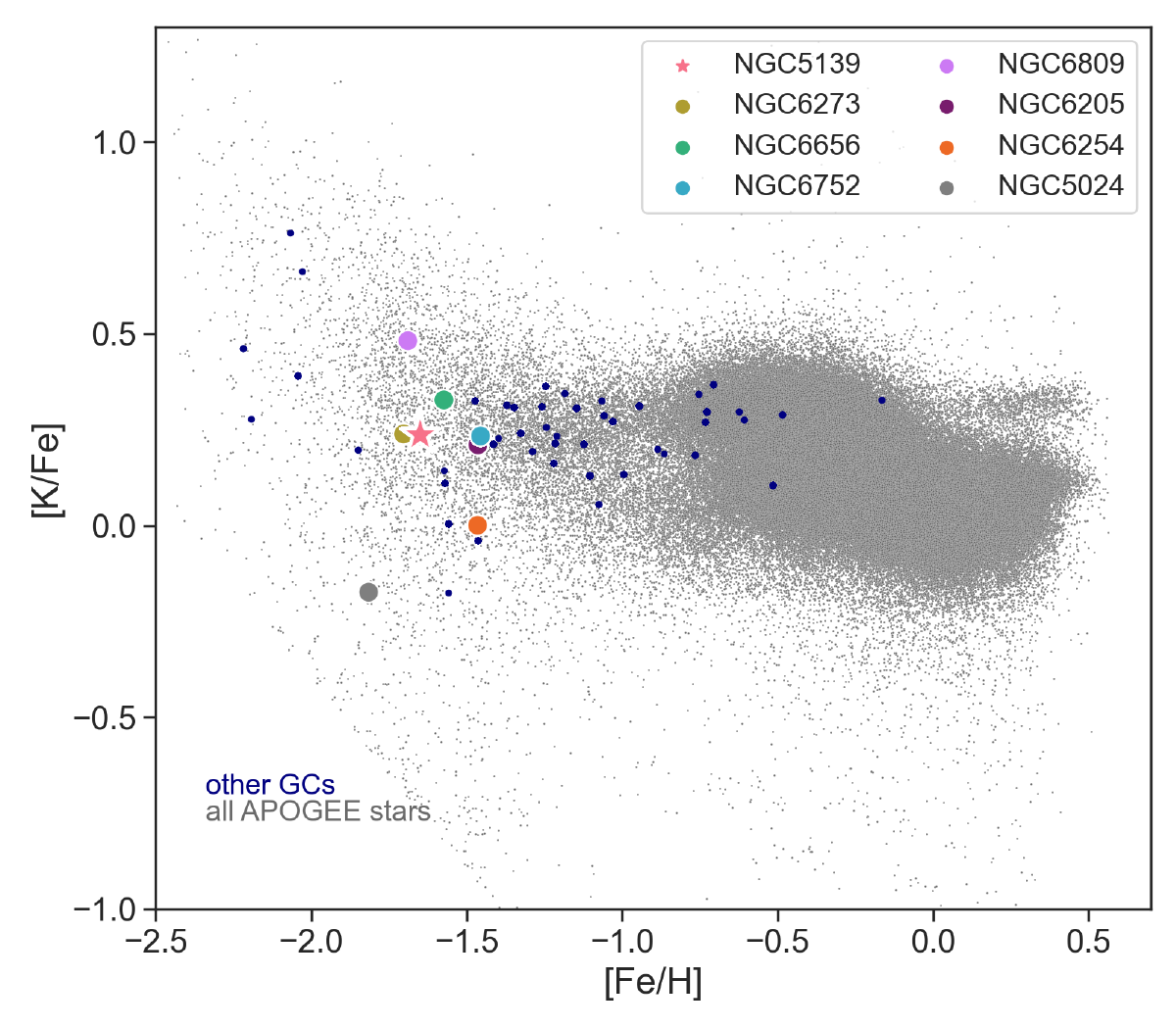}\par
\centering\includegraphics[clip=true, trim = 3mm 0mm 0mm 3mm, width=0.69\columnwidth]{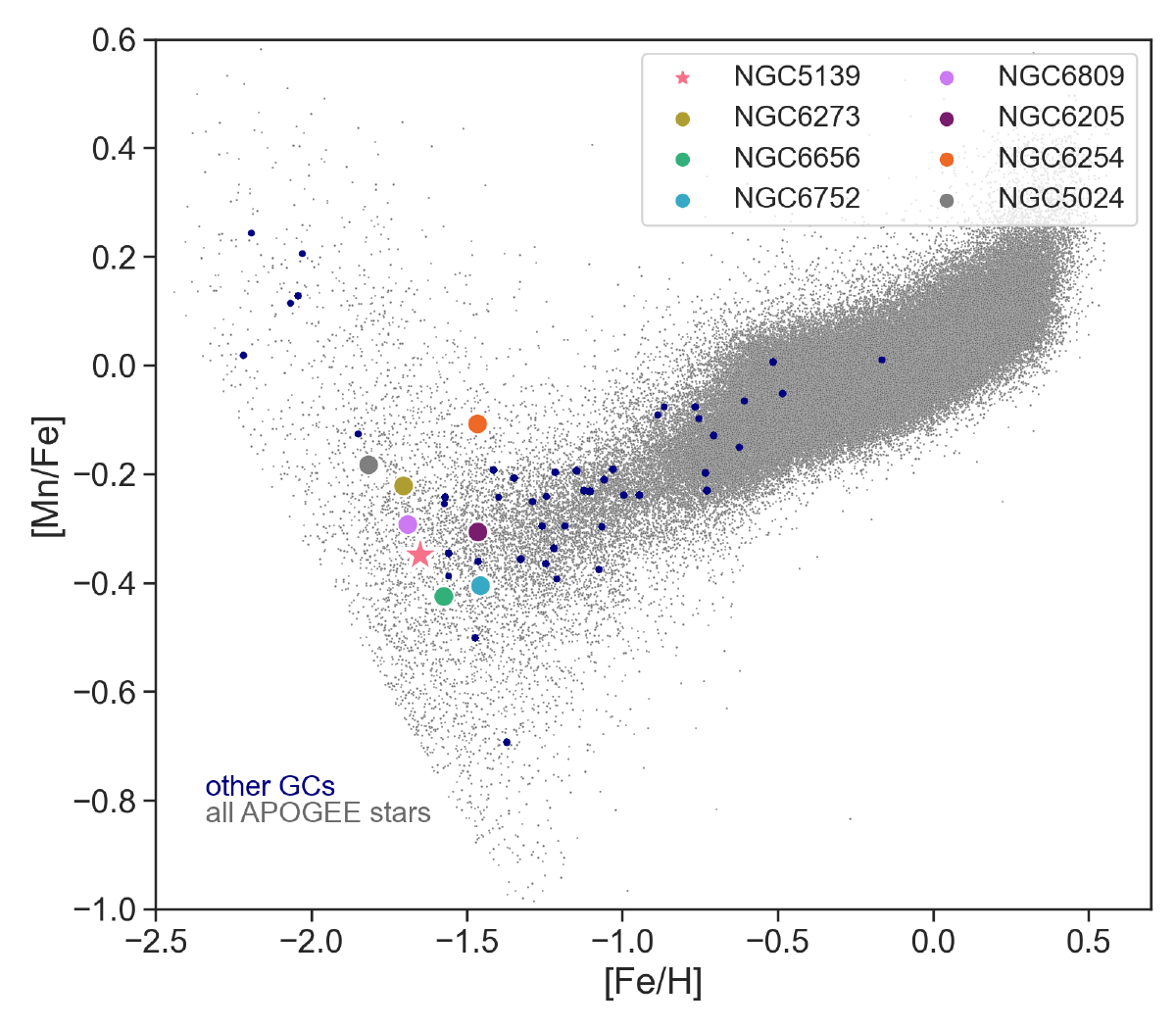}
  \caption{Mode [X/Fe] versus [Fe/H] for GCs chemically compatible - with at least 60\% of their stars - with $\omega$~Cen; the different X elements are the ones used in the GMM, namely: Mg, Si, Ca, C, Al, K, and Mn. For comparison, other GCs in our sample (blue points) and all APOGEE stars (GCs' stars included, grey points) are also shown.}
              \label{mode_abund_ocenlike}%
    \end{figure*}

Regarding GCs chemically compatible - with at least 60\% of their stars - with $\omega$~Cen, Figures \ref{median_abund_ocenlike} and \ref{mode_abund_ocenlike} show respectively their median and mode values of [X/Fe] versus [Fe/H] for the elements used in the GMM. For comparison, other GCs in our sample and all APOGEE stars are also shown. We show these quantities to stress the need to consider for each GC the complete distribution of stars in these chemical spaces when the aim is to find similarities between clusters linked to a common origin. In fact, as we can see considering medians or modes, even though $\omega$~Cen and clusters compatible with it share the same range of [Fe/H] (-2 < [Fe/H] $\lesssim$ -1.5), they do not show any distinct patterns from the rest of the GCs in the other elements. Furthermore, for some elements for which the spread is larger and the distribution of individual stars is more complex, for instance, Al, C, and K, depending on whether one considers means, medians, or modes, substantially different results are obtained. Thus, in addition to the overlap between $\omega$~Cen-compatible and non-$\omega$~Cen-compatible clusters, the choice of whether to use the median or the mode also impacts the results: if the distribution in a certain element has a density peak, the mode may be more indicative, but if the distribution is continuous or multimodal, the median should be preferred. Therefore, to obtain more robust results and to capture all the different peculiarities in the chemical evolution of a globular cluster, we believe that the most robust approach is to analyse star individual abundances.

\clearpage
\section{Testing the reliability of our classification}
In this section we test whether the classification shown in Table~\ref{OCenGCs_table_VAC} is robust if [Al/Fe] or [C/Fe] is excluded in the GMM. In fact [Al/Fe] is the element most affected by a possible chemical evolution within the cluster while the value of [C/Fe] depends on the evolutionary phase of the star.

Table~\ref{tab:class_noal} shows, for each cluster, the fraction of stars chemically compatible with $\omega$~Cen as in Tab.~\ref{OCenGCs_table_VAC} when considering a 7-dimensional abundance space in the GMM defined by [Fe/H], [Mg/Fe], [Si/Fe],[Ca/Fe], [C/Fe], [K/Fe], and [Mn/Fe]. As we can see, the fractions in this case are in general all much higher than the classification made when also considering [Al/Fe] having 12 clusters (with a total number of stars greater than 15) with compatibility higher than 60\% instead of 6. As explained in Sec~\ref{method}, this should be expected since, by decreasing the dimension of the chemical space, fewer similarity constraints are imposed between clusters so the number of spurious associations increases. Despite this, also in this case all the \textit{Nephele}'s clusters result strongly chemically compatible with $\omega$~Cen, having a fraction higher than 70\%. Other clusters that share such a high fraction are NGC~6544 (with 15 stars in total), NGC~6218, NGC~288, NGC~6121, and NGC~7089 (with 15 stars in total). We refrain from associating them with \textit{Nephele} since their similarity is less obvious when considering all the 8 elements in this study (see Tab.~\ref{OCenGCs_table_VAC}) but also when retaining [Al/Fe] and removing [C/Fe] (see Tab.~\ref{tab:class_noc}).

Table~\ref{tab:class_noc} indeed shows, for each cluster, the same fraction of stars chemically compatible with $\omega$~Cen when considering a 7-dimensional abundance space defined by [Fe/H], [Mg/Fe], [Si/Fe],[Ca/Fe], [Al/Fe], [K/Fe], and [Mn/Fe]. In this case, we retrieve that the clusters (with a total number of stars greater than 15) with the highest fraction of stars compatible with $\omega$~Cen are the \textit{Nephele}'s GCs, namely: NGC~6656, NGC~6809, NGC~6752, NGC~6254, NGC~6273, NGC~6205.
This is the reason why we believe that the association of clusters with \textit{Nephele} proposed in this study, despite being the result of a more conservative classification, is the most robust. 
Indeed, finding a similarity between clusters also in [Al/Fe] and [C/Fe] suggests that this group of clusters did not only have to form in an ISM with similar initial composition but also that their IMFs, as well as the dilution they experienced, must have been similar to give rise to such similar final abundance distributions.

\begin{table}
\centering
\caption{Fraction of stars chemically compatible with $\omega$~Cen when removing [Al/Fe] in the GMM.}\label{tab:class_noal}\resizebox{0.6\columnwidth}{!}{
\begin{tabular}{lrr}
\toprule
GC name &  Fraction (\%) &  \# of stars \\
\midrule
  Ter10 &   96  $\pm$  20 &   1 \\
\textbf{NGC5139} &   \textbf{90  $\pm$   3} & \textbf{607} \\
\textbf{NGC6656} &   \textbf{90  $\pm$   5} &  \textbf{68} \\
NGC2298 &   90  $\pm$ 25 &   2 \\
\textbf{NGC6752} &   \textbf{88  $\pm$   6} &  \textbf{83} \\
\textbf{NGC6809} &   \textbf{84  $\pm$  11} &  \textbf{18} \\
\textbf{NGC6205} &   \textbf{84  $\pm$  11} &  \textbf{26} \\
NGC6522 &   83 $\pm$  34 &   2 \\
NGC6558 &   81  $\pm$  32 &   3 \\
Djorg\_2 &   80 $\pm$   31 &   4 \\
NGC6544 &   79  $\pm$  14 &  15 \\
\textbf{NGC6273} &   \textbf{79  $\pm$  8} &  \textbf{40} \\
NGC6218 &   76 $\pm$   14 &  40 \\
NGC0288 &   71  $\pm$  13 &  37 \\
\textbf{NGC6254} &   \textbf{70  $\pm$  14} &  \textbf{50} \\
NGC6121 &   70 $\pm$   14 & 169 \\
NGC7089 &   70 $\pm$   17 &  15 \\
NGC5904 &   67  $\pm$  15 &  79 \\
   Ter4 &   65 $\pm$   48 &   1 \\
NGC6229 &   64 $\pm$   37 &   3 \\
NGC5024 &   63 $\pm$   26 &   5 \\
NGC6723 &   62 $\pm$   30 &   7 \\
NGC6642 &   62  $\pm$  26 &   6 \\
NGC6171 &   61 $\pm$   20 &  23 \\
NGC1904 &   60  $\pm$  15 &  26 \\
NGC6093 &   59  $\pm$  49 &   1 \\
NGC5272 &   54 $\pm$   18 &  71 \\
    HP1 &   52 $\pm$   21 &  10 \\
NGC6569 &   52 $\pm$   28 &   6 \\
FSR1758 &   50 $\pm$   26 &   7 \\
NGC6380 &   46 $\pm$   38 &   9 \\
NGC6717 &   45 $\pm$   45 &   2 \\
NGC1851 &   41 $\pm$   20 &  31 \\
   Ter2 &   40 $\pm$   46 &   2 \\
   Ter9 &   36 $\pm$   24 &   9 \\
NGC3201 &   36 $\pm$   14 &  98 \\
NGC6838 &   35 $\pm$   21 &  45 \\
NGC0104 &   31 $\pm$   24 & 224 \\
NGC0362 &   26 $\pm$   19 &  48 \\
NGC6715 &   25 $\pm$   12 &  26 \\
NGC2808 &   24 $\pm$   16 &  98 \\
NGC6397 &   20 $\pm$   20 &  11 \\
   Ter5 &   15 $\pm$   31 &   2 \\
NGC6316 &   14 $\pm$   25 &   6 \\
NGC6760 &   12 $\pm$   27 &   3 \\
   Pal6 &   10 $\pm$   30 &   1 \\
   Ton2 &    4 $\pm$   20 &   2 \\
NGC6304 &    1 $\pm$    4 &   5 \\
NGC7078 &    0  $\pm$   0 &   1 \\
NGC6553 &    0 $\pm$    0 &   1 \\
NGC6388 &    0  $\pm$   0 &  24 \\
NGC6293 &    0  $\pm$   0 &   1 \\
NGC4590 &    0 $\pm$    0 &   1 \\
NGC6341 &    0  $\pm$   0 &   3 \\
\bottomrule
\end{tabular}}\tablefoot{\textit{Nephele}'s clusters are marked in bold. For each cluster, the number of stars used for the analysis is also shown.}
\end{table}

\begin{table}
\centering
\caption{Fraction of stars chemically compatible with $\omega$~Cen when removing [C/Fe] in the GMM.}\label{tab:class_noc}\resizebox{0.6\columnwidth}{!}{
\begin{tabular}{lrrr}
\toprule
  GC name &  Fraction (\%) &  \# of stars \\
\midrule
  Ter10 &   99  $\pm$  10 &   1 \\
\textbf{NGC5139} &  \textbf{ 90 $\pm$    3} & \textbf{629 }\\
\textbf{NGC6656} &   \textbf{89 $\pm$    6} &  \textbf{71} \\
\textbf{NGC6809} &   \textbf{88  $\pm$   9} &  \textbf{22} \\
FSR1758 &   85 $\pm$   18 &   7 \\
\textbf{NGC6752} &   \textbf{84  $\pm$   6} &  \textbf{83} \\
\textbf{NGC6254} &   \textbf{82 $\pm$   10} &  \textbf{50} \\
\textbf{NGC6273} &   \textbf{81 $\pm$   10} &  \textbf{40} \\
   Ter4 &   79 $\pm$   41 &   1 \\
NGC5024 &   74 $\pm$   25 &   5 \\
NGC6093 &   73  $\pm$  45 &   1 \\
\textbf{NGC6205} &   \textbf{72 $\pm$   12} &  \textbf{26} \\
NGC2298 &   62 $\pm$   31 &   3 \\
NGC6544 &   58  $\pm$  18 &  15 \\
NGC1904 &   57 $\pm$   18 &  28 \\
NGC0288 &   53 $\pm$   20 &  37 \\
NGC6218 &   48 $\pm$   17 &  40 \\
NGC7089 &   43 $\pm$   22 &  15 \\
Djorg\_2 &   43 $\pm$   39 &   4 \\
   Ter9 &   37 $\pm$   24 &   9 \\
NGC6121 &   36 $\pm$   26 & 169 \\
NGC6171 &   31 $\pm$   23 &  23 \\
NGC6380 &   28 $\pm$   28 &   9 \\
    HP1 &   23 $\pm$   26 &  10 \\
NGC6715 &   22 $\pm$   12 &  27 \\
NGC6558 &   22 $\pm$   31 &   3 \\
NGC6397 &   22 $\pm$   14 &  15 \\
NGC0104 &   20 $\pm$   18 & 224 \\
NGC6569 &   20 $\pm$   23 &   6 \\
NGC6522 &   18 $\pm$   33 &   2 \\
NGC5272 &   17 $\pm$   12 &  72 \\
NGC6838 &   16  $\pm$  16 &  45 \\
NGC6723 &   14 $\pm$   20 &   7 \\
NGC6642 &   13 $\pm$   24 &   6 \\
NGC6717 &   13 $\pm$   31 &   2 \\
NGC3201 &    9 $\pm$   10 & 101 \\
NGC6229 &    7 $\pm$   20 &   3 \\
NGC5904 &    6 $\pm$    8 &  80 \\
   Ter2 &    5 $\pm$   21 &   2 \\
NGC6293 &    5 $\pm$   22 &   1 \\
NGC4590 &    4 $\pm$   20 &   1 \\
   Pal6 &    2 $\pm$   14 &   1 \\
   Ton2 &    2 $\pm$   14 &   2 \\
NGC6760 &    1 $\pm$   10 &   3 \\
NGC6316 &    1 $\pm$    8 &   6 \\
NGC1851 &    1  $\pm$   2 &  31 \\
NGC6553 &    0 $\pm$    0 &   1 \\
NGC7078 &    0 $\pm$    0 &   2 \\
NGC6388 &    0 $\pm$    0 &  24 \\
NGC6304 &    0 $\pm$    0 &   5 \\
NGC2808 &    0 $\pm$    0 &  98 \\
   Ter5 &    0 $\pm$    0 &   2 \\
NGC0362 &    0 $\pm$    1 &  48 \\
NGC6341 &    0 $\pm$    0 &   4 \\
\bottomrule
\end{tabular}}\tablefoot{\textit{Nephele}'s clusters are marked in bold. For each cluster, the number of stars used for the analysis is also shown.}
\end{table}

\newpage
\end{appendix}

\end{document}